\documentclass[a4paper,11pt]{article}
\usepackage{jheppub} 
\usepackage{lineno}
\usepackage{dsfont}
\usepackage{calc}

\allowdisplaybreaks

\newcommand{\normal}[1]{{\hspace{-0.1ex}:\hspace{-0.6ex} #1 \hspace{-0.6ex}:\hspace{0.2ex}}}
\newcommand{\barred}[1]{\hspace{0.2ex}\overline{\hspace{-0.2ex} #1}}

\newlength{\scriptlength}
\newcommand{\scriptshift}[2]{\settowidth{\scriptlength}{$\scriptstyle #1$}\hspace{-\scriptlength}#2}
\newlength{\firstlength}
\newlength{\secondlength}
\newlength{\spacelength}
\newcommand{\up}[4]{\settowidth{\firstlength}{$\scriptstyle #2$}\raisebox{#1}{$\scriptstyle #2$}\settowidth{\secondlength}{$\scriptstyle #4$}\raisebox{#3}{$\scriptstyle\scriptshift{#2}{#4}$}\hspace{-\secondlength}\setlength{\spacelength}{\maxof{\firstlength}{\secondlength}}\hspace{\spacelength}}

\title{\boldmath Massive Type IIB Superstrings \\ Part I: 3- and 4-Point Amplitudes}

\author{Nicholas Agia}
\affiliation{Center for the Fundamental Laws of Nature, Harvard University,
Cambridge, MA, USA}

\emailAdd{nagia@g.harvard.edu}

\abstract{We present the 3- and 4-point tree-level scattering amplitudes involving first-massive states in closed Type IIB superstring theory, computed in the Minkowski vacuum. In particular, the 3-point amplitudes are computed for all possible combinations of first-massive and massless string states, and the 4-point amplitudes are computed for all such combinations involving at most two massive string states. We verify that unitarity is satisfied by checking the massless and first-massive poles of a 4-point amplitude. This paper is intended to serve as a reference providing all these amplitudes explicitly in consistent conventions as well as worldsheet correlators necessary to perform further computations beyond those listed here.}

\begin{document}

\maketitle
\flushbottom

\section{Introduction and Conventions}\label{intro}

Superstring perturbation theory is a well-understood topic with a long history \cite{Friedan85,Friedan86,DHoker88,Polchinski982,Witten12,Witten13}. Concerning explicit expressions for scattering amplitudes, much of what is known pertains to massless string states. In this note, we present flat space tree-level scattering amplitudes involving closed Type IIB superstring states at the first mass level, with $m^2 = \frac{4}{\alpha'}$. The vertex operators associated to these first-massive states have been known for a long time \cite{Friedan86,Koh87}, so it is a straightforward exercise to compute such scattering amplitudes in the standard NSR formalism. The aim of this paper is simply to present these amplitudes in one self-contained reference, paying special attention to relative coefficients and signs. The appendices contain the main identities, OPEs and correlators needed to obtain these results.

For completeness, we also include the relevant amplitudes involving only massless strings, which were first computed long ago. Some of the other results below have also appeared before, but usually limited to three-point amplitudes or to leading order in $\alpha'$; many of the results are new, particularly those involving the massive RR states. For concrete appearances of massive string scattering in the literature, see \cite{Liu87,Feng10,Park11,Guillen21}.

The main application we have in mind is to construct the target spacetime effective action beyond the massless supergravity limit. In the context of AdS/CFT, understanding massive strings in the bulk provides information about unprotected observables in the dual theory. Importantly, even flat space scattering in the NSR formalism has access to such quantities, despite the difficulties in adapting the NSR framework to work directly in RR backgrounds.

In the remainder of this section, we shall establish our conventions used below. In Section \ref{vertex operators}, the relevant worldsheet vertex operators are briefly reviewed together with the propagators of the corresponding target spacetime fields. In Section \ref{3-point}, all 3-point amplitudes are given between any combination of massless and first-massive states. In Section \ref{4-point}, all 4-point amplitudes are given involving at most two first-massive states; these amplitudes suffice to determine the coupling of the massive states to any supergravity background. Since Sections \ref{3-point} and \ref{4-point} simply list results, they will be largely free of text. In Section \ref{unitarity}, we explicitly verify the factorization of one of the 4-point amplitudes onto the relevant 3-point amplitudes on its massless and first-massive poles, as required by unitarity. In addition to providing consistency checks on the amplitudes themselves, this section verifies the normalization of the vertex operators given in Section \ref{vertex operators}. The results needed to compute any desired worldsheet correlator are relegated to the appendices.

We use the canonical normalization of the matter worldsheet fields, namely so that
\begin{align}
X^{\mu}(z_1,\barred{z}_1)X^{\nu}(z_2,\barred{z}_2) & = -\frac{\alpha'}{2}\eta^{\mu\nu}\ln|z_{12}|^2 + \normal{X^{\mu}(z_1,\barred{z}_1)X^{\nu}(z_2,\barred{z}_2)}
\\ \psi^{\mu}(z_1)\psi^{\nu}(z_2) & = \frac{\eta^{\mu\nu}}{z_{12}} + \normal{\psi^{\mu}(z_1)\psi^{\nu}(z_2)},
\end{align}
with $\mu,\nu,\dotsc\in\{0,1,\dotsc,9\}$ spacetime vector indices. For spinors, we work in the Weyl basis with indices $\alpha,\beta,\dotsc\in\{1,2,\dotsc,16\}$; we use only undotted spinor indices, with any lower index being chiral and any upper index being antichiral. The 10d gamma matrices with two chiral indices or two antichiral indices are $\gamma^{\mu}_{\alpha\beta}$ and $\gamma^{\mu\alpha\beta}$, respectively, which obey the defining anticommutator
\begin{equation}
\{\gamma^{\mu},\gamma^{\nu}\}^{\alpha}_{\phantom{\alpha}\beta} = 2\eta^{\mu\nu}\delta^{\alpha}_{\phantom{\alpha}\beta}.
\end{equation}
In 10d, the gamma matrices are symmetric, $\gamma^{\mu}_{\alpha\beta} = \gamma^{\mu}_{\beta\alpha}$ and $\gamma^{\mu\alpha\beta} = \gamma^{\mu\beta\alpha}$. We follow the conventions of \cite{Sen21}, in which $\gamma^0_{\alpha\beta} = -\delta_{\alpha\beta}$ and $\gamma^{0\alpha\beta} = \delta^{\alpha\beta}$. The complete bispinor basis is formed by the totally antisymmetrized gamma matrices, $\gamma^{\mu_1\mu_2\cdots\mu_n} \equiv \gamma^{[\mu_1}\gamma^{\mu_2}\cdots\gamma^{\mu_n]}$ for $n = 0,1,\dotsc,5$; the odd-rank gamma matrices have like-chirality indices whereas the even-rank gamma matrices have opposite-chirality indices. In general, $[\cdots]$ denotes antisymmetrization of the enclosed indices, $(\cdots)$ likewise denotes symmetrization and $\{\cdots\}$ denotes traceless-symmetrization. All (anti)symmetrization is performed with unit weight. 

Indices related to the $i^{\text{th}}$ particle in a scattering amplitude will themselves be labeled by a subscript `$i$'. The vertical positions of vector indices will be flexible to accommodate antisymmetrizing multiple sets of indices, and for a symmetric tensor like $\eta^{\mu\nu}$ may even be stacked on top of each other. For instance, the totally antisymmetrized product of three Minkowski metrics may be written as $\eta\up{8pt}{[\mu_1}{2pt}{\mu_2}\eta\up{8pt}{\nu_1}{2pt}{\nu_2}\eta\up{8pt}{\rho_1]}{2pt}{\rho_2}$. For momenta, we use the notation $k_{ij}^{\mu} \equiv k_i^{\mu} - k_j^{\mu}$ where convenient.

The worldsheet chiral (resp.~antichiral) spin field is denoted $S_{\alpha}$ (resp.~$S^{\alpha}$), which is not a local operator and hence always appears with an appropriate exponential of the picture field $\phi$. The combinations
\begin{align}
\mathcal{S}_{\alpha} & \equiv \normal{e^{-\frac{\phi}{2}}}S_{\alpha}
\\ \mathcal{S}^{\alpha} & \equiv \normal{e^{-\frac{3\phi}{2}}}S^{\alpha}
\end{align}
are both local operators of weight one, which are nevertheless Grassmann odd. Similarly, $\normal{e^{-\phi}}\psi^{\mu}$ is also a local operator of weight one but is Grassmann even. In these conventions, the fundamental OPE of the fermion with the $(-\frac{1}{2})$-picture chiral spin field is
\begin{equation}
\normal{e^{-\phi(z)}}\psi^{\mu}(z)\mathcal{S}_{\alpha}(0) \sim -\frac{\gamma^{\mu}_{\alpha\beta}}{\sqrt{2}z}\mathcal{S}^{\beta}(0).
\end{equation}
It is the operator $\mathcal{S}_{\alpha}$ which is kept under the Type IIB GSO projection \cite{Gliozzi76,Polchinski982}.

The spacetime Lorentz algebra is
\begin{equation}
[M^{\mu\nu},M^{\rho\lambda}] = 4i\eta^{\mu][\rho}M^{\lambda][\nu}.
\end{equation}
By definition, an operator $\mathcal{O}^a$ transforms in a representation $R$ of the spacetime Lorentz algebra if
\begin{equation}
[M^{\mu\nu},\mathcal{O}^a] = -\left(\mathcal{M}^{\mu\nu}_{R}\right)^a_{\phantom{a}b}\mathcal{O}^b,
\end{equation}
where the $R$-representation matrices obey the matrix commutator
\begin{equation}
[\mathcal{M}_{R}^{\mu\nu},\mathcal{M}_{R}^{\rho\lambda}]^a_{\phantom{a}b} = 4i\eta^{\mu][\rho}\big(\mathcal{M}_{R}^{\lambda][\nu}\big)^a_{\phantom{a}b}.
\end{equation}
The representation matrices in the vector representation, the (chiral) spinor representation and the antichiral spinor representation are therefore
\begin{align}
\left(\mathcal{M}_{\text{V}}^{\mu\nu}\right)^{\rho}_{\phantom{\rho}\lambda} & = 2i\eta^{\rho[\mu}\delta^{\nu]}_{\phantom{\nu]}\lambda}
\\* \left(\mathcal{M}_{\text{S}}^{\mu\nu}\right)_{\alpha}^{\phantom{\alpha}\beta} & = \frac{i}{2}\left(\gamma^{\mu\nu}\right)_{\alpha}^{\phantom{\alpha}\beta}
\\* \left(\mathcal{M}_{\text{A}}^{\mu\nu}\right)^{\alpha}_{\phantom{\alpha}\beta} & = \frac{i}{2}\left(\gamma^{\mu\nu}\right)^{\alpha}_{\phantom{\alpha}\beta}.
\end{align}
From the fundamental fermion OPE $\psi^{\mu}(z)\psi^{\nu}(0) \sim \frac{\eta^{\mu\nu}}{z}$ and the fact that the vector representation matrices are $\left(\mathcal{M}_{\text{V}}^{\mu\nu}\right)^{\rho}_{\phantom{\rho}\lambda} = 2i\eta^{\rho[\mu}\delta^{\nu]}_{\phantom{\nu]}\lambda}$, we conclude that the holomorphic worldsheet Lorentz current is
\begin{equation}
j^{\mu\nu} = i\hspace{2pt}\normal{\psi^{\mu}\psi^{\nu}}.
\end{equation}
Therefore, the OPE of the Lorentz current with the spin field is
\begin{equation}
j^{\mu\nu}(z)\mathcal{S}_{\alpha}(0) \sim -\frac{i(\gamma^{\mu\nu})_{\alpha}^{\phantom{\alpha}\beta}}{2z}\mathcal{S}_{\beta}(0).
\end{equation}
Analogous statements apply for the antiholomorphic fermion $\widetilde{\psi}^{\mu}$ and spin field $\widetilde{\mathcal{S}}_{\alpha}$. The normal-ordered product of two holomorphic operators is defined as usual by \cite{DiFrancesco97}
\begin{equation}
\normal{\mathcal{O}_1\mathcal{O}_2}(z_2) \equiv \oint_{z_2}\frac{dz_1}{2\pi iz_{12}}\mathcal{O}_1(z_1)\mathcal{O}_2(z_2),
\end{equation}
which is necessary to use for products involving $\mathcal{S}_{\alpha}$, which is not a free field.

Finally, the anomalous picture number current dictates that the only nonvanishing sphere correlators have picture number $-2$ on both the holomorphic and the antiholomorphic sides. Then, the basic nonvanishing worldsheet sphere correlator in the full matter-plus-ghosts-plus-superghosts theory is
\begin{equation}\label{string normalization}
\left\langle \normal{e^{-2\phi(z_0)-2\widetilde{\phi}(\barred{z}_0)}}c\widetilde{c}(z_1,\barred{z}_1)c\widetilde{c}(z_2,\barred{z}_2)c\widetilde{c}(z_3,\barred{z}_3)\right\rangle_{S^2} = \frac{8\pi}{\alpha'g_{\text{s}}^2}|z_{12}z_{13}z_{23}|^2,
\end{equation}
where $g_{\text{s}}$ is the string coupling. The precise numerical coefficient is either computed directly from, say, zeta function regularization or indirectly via unitarity of the string S-matrix \cite{Polchinski981}.

\section{Vertex Operators and Propagators}\label{vertex operators}

Each string state vertex operator is constructed from a holomorphic NS- or R-sector operator and an antiholomorphic NS- or R-sector operator at the same level. At mass level $m^2 = \frac{4n}{\alpha'}$, we let $\text{NS}_n^j$ and $\text{R}_n^j$ denote the sets of operators on a single side, the value of $j$ enumerating the distinct operator structures at that level; at the massless level each sector has a unique operator (so $j$ is omitted in that case), and at the first-massive level the NS sector has two operator structures while the R again has one (for which the $j$ is again omitted). All polarization tensors will be denoted by the symbol $\varepsilon$, but they are distinguished by their index structure and (in amplitudes) particle number. We shall keep the polarizations completely generic for each type of state instead of decomposing them into irreducible components. For example, the label $\mathrm{NS_0 NS_0}$ encompasses the graviton, Kalb-Ramond field and dilaton. Obtaining the amplitudes associated to the irreducible states is simply a matter of projecting the polarizations onto the relevant subspaces. Moreover, here we explicitly record the expressions for the states where the holomorphic and antiholomorphic sides have the same operator type, since the amplitudes for all other states are obtained trivially via holomorphic factorization.

We take the BRST cohomology representatives for the massless bosonic Type IIB vertex operators in their natural pictures to be
\begin{align}\label{NS0 operator}
\mathcal{V}^{(-1,-1)}_{\mathrm{NS_0 NS_0}}(\varepsilon,k) & = g_{\text{s}}\varepsilon_{\mu\barred{\mu}}\hspace{2pt}\normal{e^{-\phi}}\psi^{\mu}\hspace{2pt}\normal{e^{-\widetilde{\phi}}}\widetilde{\psi}^{\barred{\mu}}\hspace{2pt}\normal{e^{ik\cdot X}}
\\ \label{R0 operator} \mathcal{V}^{(-\frac{1}{2},-\frac{1}{2})}_{\mathrm{R_0 R_0}}(\varepsilon,k) & = \frac{i\sqrt{\alpha'}}{2}g_{\text{s}}\varepsilon^{\alpha\barred{\alpha}}\hspace{2pt}\mathcal{S}_{\alpha}\hspace{2pt}\widetilde{\mathcal{S}}_{\barred{\alpha}}\hspace{2pt}\normal{e^{ik\cdot X}},
\end{align}
where $k^2 = 0$ and the polarizations satisfy the transversality conditions
\begin{align}
k^{\mu}\varepsilon_{\mu\barred{\mu}} & = 0
\\ k_{\mu}\gamma^{\mu}_{\alpha\beta}\varepsilon^{\beta\barred{\alpha}} & = 0,
\end{align}
together with the analogous expressions for the antiholomorphic indices. In expressions such as these, it is always understood that the momentum vector which appears is that associated to the corresponding polarization. The normalization of a vertex operator can always be changed by rescaling the polarization tensor, so the overall coefficient is physical only once we have specified the normalization of the polarization tensor via the propagator in the spacetime effective field theory. Writing a general propagator as $-\frac{i\Pi}{k^2+m^2-i\epsilon}$, the normalizations in \eqref{NS0 operator} and \eqref{R0 operator} correspond to the respective numerators
\begin{align}\label{NS0 propagator}
\text{$\mathrm{NS_0 NS_0}$:} \qquad \Pi_{\mu\barred{\mu},\nu\barred{\nu}}(k) & = \eta_{\mu\nu}\eta_{\barred{\mu}\barred{\nu}}
\\ \label{R0 propagator} \text{$\mathrm{R_0 R_0}$:} \qquad \Pi^{\alpha\barred{\alpha},\beta\barred{\beta}}(k) & = k^{\mu}\gamma_{\mu}^{\alpha\beta}k^{\barred{\mu}}\gamma_{\barred{\mu}}^{\barred{\alpha}\barred{\beta}}.
\end{align}
Of course, what we really mean by \eqref{NS0 propagator} and \eqref{R0 propagator} are the only physically relevant parts of the massless field propagators; computing an explicit propagator for them requires fixing a gauge, but ultimately the gauge-noninvariant contributions to any physical amplitude due to terms beyond \eqref{NS0 propagator} and \eqref{R0 propagator} are guaranteed to vanish by gauge invariance.

The picture-changing operator (PCO) is $\mathcal{X}\widetilde{\mathcal{X}}$, where \cite{Sen21,Polchinski982}
\begin{equation}
\mathcal{X} = i\sqrt{\frac{2}{\alpha'}}\hspace{2pt}\normal{e^{\phi}}\psi_{\mu}\partial X^{\mu} + c\partial\xi + b\partial\eta\hspace{2pt}\normal{e^{2\phi}} + \partial\left(b\eta\hspace{2pt}\normal{e^{2\phi}}\right).
\end{equation}
With the BRST cohomology representatives chosen here, the pure-ghost contributions to any one PCO are irrelevant. For example, the massless bosonic NSNS vertex operator in the $(0,0)$-picture is
\begin{equation}
\mathcal{V}^{(0,0)}_{\mathrm{NS_0 NS_0}}(\varepsilon,k) = -\frac{2g_{\text{s}}}{\alpha'}\varepsilon_{\mu\barred{\mu}}\hspace{2pt}\normal{\left(\partial X^{\mu} + \frac{\alpha'}{2}k_{\nu}j^{\mu\nu}\right)\left(\barred{\partial}X^{\barred{\mu}} + \frac{\alpha'}{2}k_{\barred{\nu}}\widetilde{\jmath}^{\hspace{2pt}\barred{\mu}\barred{\nu}}\right)\hspace{-2pt}e^{ik\cdot X}},
\end{equation}
where $j^{\mu\nu} = i\hspace{2pt}\normal{\psi^{\mu}\psi^{\nu}}$ is the Lorentz current.

At the first-massive level $m^2 = \frac{4}{\alpha'}$, there are two types of NS-sector operators and one type of R-sector operator \cite{Koh87}. We take the BRST cohomology representatives for the first-massive NSNS states in their natural pictures to be
\begin{align}\label{NS11 operator}
\mathcal{V}^{(-1,-1)}_{\mathrm{NS_1^1 NS_1^1}}(\varepsilon,k) & = -\frac{2g_{\text{s}}}{\alpha'}\varepsilon_{\mu\nu\barred{\mu}\barred{\nu}}\hspace{2pt}\normal{e^{-\phi}}\psi^{\mu}\hspace{2pt}\normal{e^{-\widetilde{\phi}}}\widetilde{\psi}^{\barred{\mu}}\hspace{2pt}\normal{\partial X^{\nu}\barred{\partial}X^{\barred{\nu}}e^{ik\cdot X}}
\\ \label{NS12 operator} \mathcal{V}^{(-1,-1)}_{\mathrm{NS_1^2 NS_1^2}}(\varepsilon,k) & = -\frac{g_{\text{s}}}{3!}\varepsilon_{\mu\nu\rho\barred{\mu}\barred{\nu}\barred{\rho}}\hspace{2pt}\normal{e^{-\phi}}\normal{\psi^{\mu}\psi^{\nu}\psi^{\rho}}\hspace{2pt}\normal{e^{-\widetilde{\phi}}}\normal{\widetilde{\psi}^{\barred{\mu}}\widetilde{\psi}^{\barred{\nu}}\widetilde{\psi}^{\barred{\rho}}}\hspace{2pt}\normal{e^{ik\cdot X}},
\end{align}
where $k^2 = -\frac{4}{\alpha'}$, $\varepsilon_{\mu\nu\barred{\mu}\barred{\nu}}$ is traceless-symmetric in $\mu\nu$ (and in $\barred{\mu}\barred{\nu}$), $\varepsilon_{\mu\nu\rho\barred{\mu}\barred{\nu}\barred{\rho}}$ is totally antisymmetric in $\mu\nu\rho$ (and in $\barred{\mu}\barred{\nu}\barred{\rho}$) and all polarization indices are transverse to their respective momentum. By definition, the normal-ordered product of three fermions is totally antisymmetric,
\begin{equation}
\normal{\psi^{\mu}\psi^{\nu}\psi^{\rho}} \equiv \normal{\psi^{[\mu}\normal{\psi^{\nu}\psi^{\rho]}}}.
\end{equation}
The results for the other first-massive NSNS states $\mathcal{V}^{(-1,-1)}_{\mathrm{NS_1^1 NS_1^2}}$ and $\mathcal{V}^{(-1,-1)}_{\mathrm{NS_1^2 NS_1^1}}$ are obtained as usual from holomorphic factorization.

We take the BRST cohomology representative of the first-massive RR vertex operator in its natural picture to be
\begin{equation}\label{R1 operator}
\mathcal{V}_{\mathrm{R_1 R_1}}^{(-\frac{1}{2},-\frac{1}{2})}(\varepsilon,k) = \frac{8ig_{\text{s}}}{9\sqrt{\alpha'}}\varepsilon_{\mu\barred{\mu}}^{\alpha\barred{\alpha}}\hspace{2pt}\normal{\mathcal{J}^{\mu}_{\alpha}\widetilde{\mathcal{J}}^{\barred{\mu}}_{\barred{\alpha}}e^{ik\cdot X}},
\end{equation}
where
\begin{equation}
\mathcal{J}^{\mu}_{\alpha} \equiv \partial X^{\mu}\mathcal{S}_{\alpha} + \frac{\alpha'}{16}k^{\rho}(\gamma_{\rho}\gamma_{\nu})_{\alpha}^{\phantom{\alpha}\beta}\normal{j^{\mu\nu}\mathcal{S}_{\beta}},
\end{equation}
and where the polarization bi-vector-spinor satisfies
\begin{align}
k^{\mu}\varepsilon_{\mu\barred{\mu}}^{\alpha\barred{\alpha}} & = 0
\\ \gamma^{\mu}_{\alpha\beta}\varepsilon_{\mu\barred{\mu}}^{\beta\barred{\alpha}} & = 0,
\end{align}
and likewise for the antiholomorphic indices. It is important to note that the operator $k^{\rho}(\gamma_{\rho}\gamma_{\nu})_{\alpha}^{\phantom{\alpha}\beta}\normal{j^{\mu\nu}\mathcal{S}_{\beta}}$ is a primary only after being contracted with a polarization vector-spinor satisfying these physicality conditions.

For the massive propagators, define the transverse projection of the metric as
\begin{equation}
\hat{\eta}_{\mu\nu}(k) \equiv \eta_{\mu\nu} - \frac{k_{\mu}k_{\nu}}{k^2},
\end{equation}
which obeys $\hat{\eta}_{\mu\nu}(k)\hat{\eta}^{\nu}_{\phantom{\nu}\rho}(k) = \hat{\eta}_{\mu\rho}(k)$. Then, the first-massive NSNS transverse propagator numerators corresponding to the normalizations \eqref{NS11 operator} and \eqref{NS12 operator} are
\begin{align}\label{N11 propagator}
\text{$\mathrm{NS_1^1 NS_1^1}$:} \hspace{17pt}\quad \Pi^{\mu_1\nu_1\barred{\mu}_1\barred{\nu}_1}_{\mu_2\nu_2\barred{\mu}_2\barred{\nu}_2}(k) & = \left(\hat{\eta}^{(\mu_1}_{\mu_2}\hat{\eta}^{\nu_1)}_{\nu_2} - \frac{1}{9}\hat{\eta}^{\mu_1\nu_1}\hat{\eta}_{\mu_2\nu_2}\right)\left(\hat{\eta}^{(\barred{\mu}_1}_{\barred{\mu}_2}\hat{\eta}^{\barred{\nu}_1)}_{\barred{\nu}_2} - \frac{1}{9}\hat{\eta}^{\barred{\mu}_1\barred{\nu}_1}\hat{\eta}_{\barred{\mu}_2\barred{\nu}_2}\right)
\\ \label{N12 propagator} \text{$\mathrm{NS_1^2 NS_1^2}$:} \quad \Pi^{\mu_1\nu_1\rho_1\barred{\mu}_1\barred{\nu}_1\barred{\rho}_1}_{\mu_2\nu_2\rho_2\barred{\mu}_2\barred{\nu}_2\barred{\rho}_2}(k) & = \hat{\eta}^{[\mu_1}_{\mu_2}\hat{\eta}^{\nu_1}_{\nu_2}\hat{\eta}^{\rho_1]}_{\rho_2}\hat{\eta}^{[\barred{\mu}_1}_{\barred{\mu}_2}\hat{\eta}^{\barred{\nu}_1}_{\barred{\nu}_2}\hat{\eta}^{\barred{\rho}_1]}_{\barred{\rho}_2},
\end{align}
where we momentarily leave the dependence on $k$ implicit for readability. The transverse propagator numerator associated to the first-massive RR vertex operator \eqref{R1 operator} is
\begin{equation}\label{R1 propagator}
\text{$\mathrm{R_1 R_1}$:} \qquad \Pi^{\mu\barred{\mu},\nu\barred{\nu}}_{\alpha\barred{\alpha},\beta\barred{\beta}}(k) = \left[\hat{\eta}^{\mu\nu}(k)(k\cdot\gamma)_{\alpha\beta} - \frac{1}{8}k^{\rho}(\gamma_{\rho}^{\phantom{\rho}\mu\nu})_{\alpha\beta}\right]\left[\hat{\eta}^{\barred{\mu}\barred{\nu}}(k)(k\cdot\gamma)_{\barred{\alpha}\barred{\beta}} - \frac{1}{8}k^{\barred{\rho}}(\gamma_{\barred{\rho}}^{\phantom{\barred{\rho}}\barred{\mu}\barred{\nu}})_{\barred{\alpha}\barred{\beta}}\right].
\end{equation}
That the normalization of \eqref{R1 operator} is then correct will be verified in Section \ref{unitarity}.

\section{3-Point Amplitudes}\label{3-point}

\subsection{All Massless String States}

Known since antiquity.
\begin{multline}
\mathcal{A}_3[\mathrm{NS_0 NS_0};\mathrm{NS_0 NS_0};\mathrm{NS_0 NS_0}] = \pi ig_{\text{s}}(2\pi)^{10}\delta^{10}(k_1+k_2+k_3)\varepsilon_{1,\mu_1\barred{\mu}_1}\varepsilon_{2,\mu_2\barred{\mu}_2}\varepsilon_{3,\mu_3\barred{\mu}_3}
\\ \times \Big(\eta^{\mu_1\mu_2}k_{12}^{\mu_3} + \eta^{\mu_2\mu_3}k_{23}^{\mu_1} + \eta^{\mu_3\mu_1}k_{31}^{\mu_2}\Big)\Big(\eta^{\barred{\mu}_1\barred{\mu}_2}k_{12}^{\barred{\mu}_3} + \eta^{\barred{\mu}_2\barred{\mu}_3}k_{23}^{\barred{\mu}_1} + \eta^{\barred{\mu}_3\barred{\mu}_1}k_{31}^{\barred{\mu}_2}\Big)
\end{multline}

\begin{equation}
\mathcal{A}_3[\mathrm{NS_0 NS_0};\mathrm{R_0 R_0};\mathrm{R_0 R_0}] = \pi ig_{\text{s}}(2\pi)^{10}\delta^{10}(k_1+k_2+k_3) \varepsilon_{1,\mu_1\barred{\mu}_1}\varepsilon_2^{\alpha_2\barred{\alpha}_2}\varepsilon_3^{\alpha_3\barred{\alpha}_3}\gamma^{\mu_1}_{\alpha_2\alpha_3}\gamma^{\barred{\mu}_1}_{\barred{\alpha}_2\barred{\alpha}_3}
\end{equation}

\subsection{One Massive String State}

\begin{multline}
\mathcal{A}_3[\mathrm{NS_1^1 NS_1^1};\mathrm{NS_0 NS_0};\mathrm{NS_0 NS_0}] = \frac{8\pi ig_{\text{s}}}{\alpha'}(2\pi)^{10}\delta^{10}(k_1+k_2+k_3)\varepsilon_{1,\mu_1\nu_1\barred{\mu}_1\barred{\nu}_1}\varepsilon_{2,\mu_2\barred{\mu}_2}\varepsilon_{3,\mu_3\barred{\mu}_3}
\\ \times \left[\eta^{\mu_1\mu_2}\eta^{\nu_1\mu_3} + \frac{\alpha'}{8}\Big(\eta^{\mu_1\mu_2}k_{12}^{\mu_3} + \eta^{\mu_2\mu_3}k_{23}^{\mu_1} + \eta^{\mu_3\mu_1}k_{31}^{\mu_2}\Big)k_{23}^{\nu_1}\right]
\\ \times \left[\eta^{\barred{\mu}_1\barred{\mu}_2}\eta^{\barred{\nu}_1\barred{\mu}_3} + \frac{\alpha'}{8}\Big(\eta^{\barred{\mu}_1\barred{\mu}_2}k_{12}^{\barred{\mu}_3} + \eta^{\barred{\mu}_2\barred{\mu}_3}k_{23}^{\barred{\mu}_1} + \eta^{\barred{\mu}_3\barred{\mu}_1}k_{31}^{\barred{\mu}_2}\Big)k_{23}^{\barred{\nu}_1}\right]
\end{multline}

\begin{multline}
\mathcal{A}_3[\mathrm{NS_1^2 NS_1^2};\mathrm{NS_0 NS_0};\mathrm{NS_0 NS_0}]
\\ = -3!\pi ig_{\text{s}}(2\pi)^{10}\delta^{10}(k_1+k_2+k_3)\varepsilon_{1,\mu_1\nu_1\rho_1\barred{\mu}_1\barred{\nu}_1\barred{\rho}_1}\varepsilon_{2,\mu_2\barred{\mu}_2}\varepsilon_{3,\mu_3\barred{\mu}_3}\eta\up{8pt}{[\mu_1}{2pt}{\mu_2}\eta\up{8pt}{\nu_1}{2pt}{\mu_3}k\up{8pt}{\rho_1]}{-3pt}{23} \eta\up{8pt}{[\barred{\mu}_1}{0pt}{\barred{\mu}_2}\eta\up{8pt}{\barred{\nu}_1}{1pt}{\barred{\mu}_3}k\up{8pt}{\barred{\rho}_1]}{-3pt}{23}
\end{multline}

\begin{multline}
\mathcal{A}_3[\mathrm{NS_1^1 NS_1^1};\mathrm{R_0 R_0};\mathrm{R_0 R_0}] 
\\ = \frac{\pi i\alpha'g_{\text{s}}}{8}(2\pi)^{10}\delta^{10}(k_1+k_2+k_3)\varepsilon_{1,\mu_1\nu_1\barred{\mu}_1\barred{\nu}_1}\varepsilon_2^{\alpha_2\barred{\alpha}_2}\varepsilon_3^{\alpha_3\barred{\alpha}_3}k_{23}^{\mu_1}\gamma^{\nu_1}_{\alpha_2\alpha_3}k_{23}^{\barred{\mu}_1}\gamma^{\barred{\nu}_1}_{\barred{\alpha}_2\barred{\alpha}_3}
\end{multline}

\begin{multline}
\mathcal{A}_3[\mathrm{NS_1^2 NS_1^2};\mathrm{R_0 R_0};\mathrm{R_0 R_0}] 
\\ = -\frac{\pi ig_{\text{s}}}{3!4}(2\pi)^{10}\delta^{10}(k_1+k_2+k_3)\varepsilon_{1,\mu_1\nu_1\rho_1\barred{\mu}_1\barred{\nu}_1\barred{\rho}_1}\varepsilon_2^{\alpha_2\barred{\alpha}_2}\varepsilon_3^{\alpha_3\barred{\alpha}_3}(\gamma^{\mu_1\nu_1\rho_1})_{\alpha_2\alpha_3}(\gamma^{\barred{\mu}_1\barred{\nu}_1\barred{\rho}_1})_{\barred{\alpha}_2\barred{\alpha}_3}
\end{multline}

\begin{multline}
\mathcal{A}_3[\mathrm{NS_0 NS_0};\mathrm{R_0 R_0};\mathrm{R_1 R_1}] = -\frac{4\pi i\alpha'g_{\text{s}}}{9}(2\pi)^{10}\delta^{10}(k_1+k_2+k_3)\varepsilon_{1,\mu_1\barred{\mu}_1}\varepsilon_2^{\alpha_2\barred{\alpha}_2}\varepsilon_{3,\mu_3\barred{\mu}_3}^{\alpha_3\barred{\alpha}_3}
\\ \times \left[\frac{1}{2}k_{12}^{\mu_3}\gamma^{\mu_1}_{\alpha_2\alpha_3} - \eta^{\mu_1\mu_3}(k_1\cdot\gamma)_{\alpha_2\alpha_3}\right]\left[\frac{1}{2}k_{12}^{\barred{\mu}_3}\gamma^{\barred{\mu}_1}_{\barred{\alpha}_2\barred{\alpha}_3} - \eta^{\barred{\mu}_1\barred{\mu}_3}(k_1\cdot\gamma)_{\barred{\alpha}_2\barred{\alpha}_3}\right].
\end{multline}

\subsection{Two Massive String States}

\begin{align}
\notag \mathcal{A}_3[\mathrm{NS_1^1 NS_1^1};\mathrm{NS_1^1 NS_1^1};\mathrm{NS_0 NS_0}] & = \pi ig_{\text{s}}(2\pi)^{10}\delta^{10}(k_1+k_2+k_3)\varepsilon_{1,\mu_1\nu_1\barred{\mu}_1\barred{\nu}_1}\varepsilon_{2,\mu_2\nu_2\barred{\mu}_2\barred{\nu}_2}\varepsilon_{3,\mu_3\barred{\mu}_3}
\\ & \hspace{15pt} \times A^{\{\mu_1\nu_1\}\{\mu_2\nu_2\}\mu_3}A^{\{\barred{\mu}_1\barred{\nu}_1\}\{\barred{\mu}_2\barred{\nu}_2\}\barred{\mu}_3},
\end{align}
where
\begin{equation}
A^{\mu_1\nu_1\mu_2\nu_2\mu_3} = \eta^{\mu_1\mu_2}\Big(\eta^{\nu_1\nu_2}k_{12}^{\mu_3} + 2\eta^{\nu_1\mu_3}k_{31}^{\nu_2} + 2\eta^{\nu_2\mu_3}k_{23}^{\nu_1}\Big) + \frac{\alpha'}{8}\Big(\eta^{\mu_1\mu_2}k_{12}^{\mu_3} + 4\eta^{\mu_3[\mu_1}k_3^{\mu_2]}\Big)k_{23}^{\nu_1}k_{31}^{\nu_2}.
\end{equation}

\begin{multline}
\mathcal{A}_3[\mathrm{NS_1^1 NS_1^1};\mathrm{NS_1^2 NS_1^2};\mathrm{NS_0 NS_0}] = -\frac{3\pi i\alpha' g_{\text{s}}}{4} (2\pi)^{10}\delta^{10}(k_1+k_2+k_3)
\\ \times\varepsilon_{1,\mu_1\nu_1\barred{\mu}_1\barred{\nu}_1}\varepsilon_{2,\mu_2\nu_2\rho_2\barred{\mu}_2\barred{\nu}_2\barred{\rho}_2}\varepsilon_{3,\mu_3\barred{\mu}_3}B^{\{\mu_1\nu_1\}\mu_2\nu_2\rho_2\mu_3}B^{\{\barred{\mu}_1\barred{\nu}_1\}\barred{\mu}_2\barred{\nu}_2\barred{\rho}_2\barred{\mu}_3},
\end{multline}
where
\begin{equation}
B^{\mu_1\nu_1\mu_2\nu_2\rho_2\mu_3} = \eta^{\mu_1\mu_2}\eta^{\nu_2\mu_3}k_{23}^{\nu_1}k_{31}^{\rho_2}.
\end{equation}

\begin{multline}
\mathcal{A}_3[\mathrm{NS_1^2 NS_1^2};\mathrm{NS_1^2 NS_1^2};\mathrm{NS_0 NS_0}] = \pi ig_{\text{s}} (2\pi)^{10}\delta^{10}(k_1+k_2+k_3)
\\ \times \varepsilon_{1,\mu_1\nu_1\rho_1\barred{\mu}_1\barred{\nu}_1\barred{\rho}_1}\varepsilon_{2,\mu_2\nu_2\rho_2\barred{\mu}_2\barred{\nu}_2\barred{\rho}_2}\varepsilon_{3,\mu_3\barred{\mu}_3}C^{\mu_1\nu_1\rho_1\mu_2\nu_2\rho_2\mu_3}C^{\barred{\mu}_1\barred{\nu}_1\barred{\rho}_1\barred{\mu}_2\barred{\nu}_2\barred{\rho}_2\barred{\mu}_3},
\end{multline}
where
\begin{equation}
C^{\mu_1\nu_1\rho_1\mu_2\nu_2\rho_2\mu_3} = \eta\up{8pt}{[\mu_1}{2pt}{\mu_2}\eta\up{8pt}{\nu_1}{2pt}{\nu_2}\eta\up{8pt}{\rho_1]}{2pt}{\rho_2}k_{12}^{\mu_3} + 6\eta\up{0pt}{\mu_3[\mu_1}{0pt}{}\eta\up{8pt}{[\mu_2}{0pt}{\nu_1}\eta\up{8pt}{\nu_2}{0pt}{\rho_1]}k\up{8pt}{\rho_2]}{-3pt}{3} - 6\eta\up{0pt}{\mu_3[\mu_2}{0pt}{}\eta\up{8pt}{[\mu_1}{0pt}{\nu_2}\eta\up{8pt}{\nu_1}{0pt}{\rho_2]}k\up{8pt}{\rho_1]}{-3pt}{3}.
\end{equation}

\begin{align}
\notag \mathcal{A}_3[\mathrm{NS_1^1 NS_1^1};\mathrm{R_0 R_0};\mathrm{R_1 R_1}] & = -\frac{8\pi ig_{\text{s}}}{9}(2\pi)^{10}\delta^{10}(k_1+k_2+k_3)\varepsilon_{1,\mu_1\nu_1\barred{\mu}_1\barred{\nu}_1}\varepsilon_2^{\alpha_2\barred{\alpha}_2}\varepsilon_{3,\mu_3\barred{\mu}_3}^{\alpha_3\barred{\alpha}_3}
\\ & \hspace{15pt} \times D^{\{\mu_1\nu_1\}\mu_3}_{\alpha_2\alpha_3}D^{\{\barred{\mu}_1\barred{\nu}_1\}\barred{\mu}_3}_{\barred{\alpha}_2\barred{\alpha}_3},
\end{align}
where
\begin{equation}
D^{\mu_1\nu_1\mu_3}_{\alpha_2\alpha_3} = \eta^{\mu_1\mu_3}\gamma^{\nu_1}_{\alpha_2\alpha_3} + \frac{\alpha'}{8}k_{23}^{\mu_1}\Big(k_{12}^{\mu_3}\gamma^{\nu_1}_{\alpha_2\alpha_3} - 2\eta^{\nu_1\mu_3}(k_1\cdot\gamma)_{\alpha_2\alpha_3}\Big).
\end{equation}

\begin{align}
\notag \mathcal{A}_3[\mathrm{NS_1^2 NS_1^2};\mathrm{R_0 R_0};\mathrm{R_1 R_1}] & = \frac{\pi i\alpha'g_{\text{s}}}{3!9}(2\pi)^{10} \delta^{10}(k_1+k_2+k_3)\varepsilon_{1,\mu_1\nu_1\rho_1\barred{\mu}_1\barred{\nu}_1\barred{\rho}_1}\varepsilon_2^{\alpha_2\barred{\alpha}_2}\varepsilon_{3,\mu_3\barred{\mu}_3}^{\alpha_3\barred{\alpha}_3}
\\ & \hspace{15pt} \times E^{\mu_1\nu_1\rho_1\mu_3}_{\alpha_2\alpha_3}E^{\barred{\mu}_1\barred{\nu}_1\barred{\rho}_1\barred{\mu}_3}_{\barred{\alpha}_2\barred{\alpha}_3},
\end{align}
where
\begin{equation}
E^{\mu_1\nu_1\rho_1\mu_3}_{\alpha_2\alpha_3} = 12\eta^{\mu_3[\mu_1}k_2^{\nu_1}\gamma^{\rho_1]}_{\alpha_2\alpha_3} - 3k_1^{\mu}\eta^{\mu_3[\mu_1}(\gamma^{\nu_1\rho_1]}_{\phantom{\nu_1\rho_1]}\mu})_{\alpha_2\alpha_3} + \frac{1}{2}k_{12}^{\mu_3}(\gamma^{\mu_1\nu_1\rho_1})_{\alpha_2\alpha_3}.
\end{equation}

\begin{align}
\notag \mathcal{A}_3[\mathrm{NS_0 NS_0};\mathrm{R_1 R_1};\mathrm{R_1 R_1}] & = \frac{4\pi i\alpha'^2 g_{\text{s}}}{9^2}(2\pi)^{10}\delta^{10}(k_1+k_2+k_3)\varepsilon_{1,\mu_1\barred{\mu}_1}\varepsilon_{2,\mu_2\barred{\mu}_2}^{\alpha_2\barred{\alpha}_2}\varepsilon_{3,\mu_3\barred{\mu}_3}^{\alpha_3\barred{\alpha}_3}
\\* & \hspace{15pt} \times F^{\mu_1\mu_2\mu_3}_{\alpha_2\alpha_3}F^{\barred{\mu}_1\barred{\mu}_2\barred{\mu}_3}_{\barred{\alpha}_2\barred{\alpha}_3},
\end{align}
where
\begin{multline}
F^{\mu_1\mu_2\mu_3}_{\alpha_2\alpha_3} = \frac{1}{2}k_{31}^{\mu_2}k_{12}^{\mu_3}\gamma^{\mu_1}_{\alpha_2\alpha_3} - \eta^{\mu_1\mu_2}k_{12}^{\mu_3}(k_2\cdot\gamma)_{\alpha_2\alpha_3} + \eta^{\mu_1\mu_3}k_{31}^{\mu_2}(k_3\cdot\gamma)_{\alpha_2\alpha_3} 
\\ - \eta^{\mu_2\mu_3}\left(\frac{1}{2}k_{23}^{\mu_1}(k_{23}\cdot\gamma)_{\alpha_2\alpha_3} + k_2^{\mu}k_3^{\nu}(\gamma^{\mu_1}_{\phantom{\mu_1}\mu\nu})_{\alpha_2\alpha_3}\right).
\end{multline}
In the previous amplitude, one might expect a leading (in $\alpha'$) term proportional to $\eta^{\mu_2\mu_3}\gamma^{\mu_1}_{\alpha_2\alpha_3}$, but its coefficient cancels between the two terms in $\mathcal{J}^{\mu}_{\alpha}$ in the vertex operator \eqref{R1 operator}.

\subsection{All Massive String States}

\begin{multline}
\mathcal{A}_3[\mathrm{NS_1^1 NS_1^1};\mathrm{NS_1^1 NS_1^1};\mathrm{NS_1^1 NS_1^1}] = \frac{8\pi ig_{\text{s}}}{\alpha' }(2\pi)^{10}\delta^{10}(k_1+k_2+k_3)
\\ \times \varepsilon_{1,\mu_1\nu_1\barred{\mu}_1\barred{\nu}_1}\varepsilon_{2,\mu_2\nu_2\barred{\mu}_2\barred{\nu}_2}\varepsilon_{3,\mu_3\nu_3\barred{\mu}_3\barred{\nu}_3}G^{\{\mu_1\nu_1\}\{\mu_2\nu_2\}\{\mu_3\nu_3\}}G^{\{\barred{\mu}_1\barred{\nu}_1\}\{\barred{\mu}_2\barred{\nu}_2\}\{\barred{\mu}_3\barred{\nu}_3\}},
\end{multline}
where
\begin{multline}
G^{\mu_1\nu_1\mu_2\nu_2\mu_3\nu_3} = 3\eta^{\mu_1\mu_2}\eta^{\nu_1\mu_3}\eta^{\nu_2\nu_3} + \frac{\alpha'}{2}\bigg(\frac{1}{4}\eta^{\mu_1\mu_2}\eta^{\nu_1\nu_2}k_{12}^{\mu_3}k_{12}^{\nu_3} + \eta^{\mu_1\mu_3}\eta^{\nu_1\nu_3}k_3^{\mu_2}k_3^{\nu_2} 
\\ + \eta^{\mu_2\mu_3}\eta^{\nu_2\nu_3}k_3^{\mu_1}k_3^{\nu_1} - 3\eta^{\mu_1\mu_3}\eta^{\mu_2\nu_3}k_3^{\nu_1}k_3^{\nu_2} + 3\eta^{\mu_1\mu_2}\eta^{\mu_3[\nu_1}k_3^{\nu_2]}k_{12}^{\nu_3}\bigg)
\\ - \frac{\alpha'^2}{16}\Big(\eta^{\mu_1\mu_2}k_{12}^{\mu_3} + 4\eta^{\mu_3[\mu_1}k_3^{\mu_2]}\Big)k_3^{\nu_1}k_3^{\nu_2}k_{12}^{\nu_3}.
\end{multline}
The preceding amplitude, despite being totally symmetric in the three states, is written to distinguish the third particle for ease of its use in unitarity checks.

\begin{multline}
\mathcal{A}_3[\mathrm{NS_1^1 NS_1^1};\mathrm{NS_1^1 NS_1^1};\mathrm{NS_1^2 NS_1^2}] = -3!\pi i g_{\text{s}}(2\pi)^{10}\delta^{10}(k_1+k_2+k_3)
\\ \times \varepsilon_{1,\mu_1\nu_1\barred{\mu}_1\barred{\nu}_1}\varepsilon_{2,\mu_2\nu_2\barred{\mu}_2\barred{\nu}_2}\varepsilon_{3,\mu_3\nu_3\rho_3\barred{\mu}_3\barred{\nu}_3\barred{\rho}_3}H^{\{\mu_1\nu_1\}\{\mu_2\nu_2\}\mu_3\nu_3\rho_3}H^{\{\barred{\mu}_1\barred{\nu}_1\}\{\barred{\mu}_2\barred{\nu}_2\}\barred{\mu}_3\barred{\nu}_3\barred{\rho}_3},
\end{multline}
where
\begin{equation}
H^{\mu_1\nu_1\mu_2\nu_2\mu_3\nu_3\rho_3} = \eta\up{8pt}{[\mu_3}{2pt}{\mu_1}\eta\up{8pt}{\nu_3}{2pt}{\mu_2}k\up{8pt}{\rho_3]}{-3pt}{12}\left(\eta^{\nu_1\nu_2} + \frac{\alpha'}{8}k_{23}^{\nu_1}k_{31}^{\nu_2}\right).
\end{equation}

\begin{multline}
\mathcal{A}_3[\mathrm{NS_1^1 NS_1^1};\mathrm{NS_1^2 NS_1^2};\mathrm{NS_1^2 NS_1^2}] = \frac{8\pi i g_{\text{s}}}{\alpha'}(2\pi)^{10}\delta^{10}(k_1+k_2+k_3)
\\ \times \varepsilon_{1,\mu_1\nu_1\barred{\mu}_1\barred{\nu}_1}\varepsilon_{2,\mu_2\nu_2\rho_2\barred{\mu}_2\barred{\nu}_2\barred{\rho}_2}\varepsilon_{3,\mu_3\nu_3\rho_3\barred{\mu}_3\barred{\nu}_3\barred{\rho}_3}I^{\{\mu_1\nu_1\}\mu_2\nu_2\rho_2\mu_3\nu_3\rho_3}I^{\{\barred{\mu}_1\barred{\nu}_1\}\barred{\mu}_2\barred{\nu}_2\barred{\rho}_2\barred{\mu}_3\barred{\nu}_3\barred{\rho}_3},
\end{multline}
where
\begin{multline}
I^{\mu_1\nu_1\mu_2\nu_2\rho_2\mu_3\nu_3\rho_3} = 3\eta\up{8pt}{[\mu_2}{0pt}{\mu_1}\eta\up{0pt}{\nu_1[\mu_3}{0pt}{}\eta\up{8pt}{\nu_2}{0pt}{\nu_3}\eta\up{8pt}{\rho_2]}{0pt}{\rho_3]}
\\ + \frac{\alpha'}{8}k_{23}^{\mu_1}\left(\eta\up{8pt}{[\mu_2}{2pt}{\mu_3}\eta\up{8pt}{\nu_2}{2pt}{\nu_3}\eta\up{8pt}{\rho_2]}{2pt}{\rho_3}k_{23}^{\nu_1} + 3\eta\up{0pt}{\nu_1[\mu_3}{0pt}{}\eta\up{8pt}{[\mu_2}{0pt}{\nu_3}\eta\up{8pt}{\nu_2}{0pt}{\rho_3]}k\up{8pt}{\rho_2]}{-3pt}{31} + 3\eta\up{0pt}{\nu_1[\mu_2}{0pt}{}\eta\up{8pt}{[\mu_3}{0pt}{\nu_2}\eta\up{8pt}{\nu_3}{0pt}{\rho_2]}k\up{8pt}{\rho_3]}{-3pt}{12}\right).
\end{multline}

\begin{multline}
\mathcal{A}_3[\mathrm{NS_1^2 NS_1^2};\mathrm{NS_1^2 NS_1^2};\mathrm{NS_1^2 NS_1^2}] = -\frac{3!^3\pi ig_{\text{s}}}{4}(2\pi)^{10}\delta^{10}(k_1+k_2+k_3)
\\ \times \varepsilon_{1,\mu_1\nu_1\rho_1\barred{\mu}_1\barred{\nu}_1\barred{\rho}_1}\varepsilon_{2,\mu_2\nu_2\rho_2\barred{\mu}_2\barred{\nu}_2\barred{\rho}_2}\varepsilon_{3,\mu_3\nu_3\rho_3\barred{\mu}_3\barred{\nu}_3\barred{\rho}_3}J^{\mu_1\nu_1\rho_1\mu_2\nu_2\rho_2\mu_3\nu_3\rho_3}J^{\barred{\mu}_1\barred{\nu}_1\barred{\rho}_1\barred{\mu}_2\barred{\nu}_2\barred{\rho}_2\barred{\mu}_3\barred{\nu}_3\barred{\rho}_3},
\end{multline}
where
\begin{equation}
J^{\mu_1\nu_1\rho_1\mu_2\nu_2\rho_2\mu_3\nu_3\rho_3} = k\up{10pt}{[\mu_1}{-3pt}{23}\eta\up{10pt}{\nu_1}{2pt}{[\mu_2}\eta\up{10pt}{\rho_1]}{-6pt}{[\mu_3}\eta\up{2pt}{\nu_2}{-6pt}{\nu_3}\eta\up{2pt}{\rho_2]}{-6pt}{\rho_3]} - k\up{10pt}{[\mu_2}{-3pt}{31}\eta\up{10pt}{\nu_2}{2pt}{[\mu_1}\eta\up{10pt}{\rho_2]}{-6pt}{[\mu_3}\eta\up{2pt}{\nu_1}{-6pt}{\nu_3}\eta\up{2pt}{\rho_1]}{-6pt}{\rho_3]} - k\up{10pt}{[\mu_3}{-3pt}{12}\eta\up{10pt}{\nu_3}{2pt}{[\mu_2}\eta\up{10pt}{\rho_3]}{-6pt}{[\mu_1}\eta\up{2pt}{\nu_2}{-6pt}{\nu_1}\eta\up{2pt}{\rho_2]}{-6pt}{\rho_1]}.
\end{equation}

\begin{multline}
\mathcal{A}_3[\mathrm{NS_1^1 NS_1^1};\mathrm{R_1 R_1};\mathrm{R_1 R_1}] 
\\ = -\frac{8\pi i\alpha'^2 g_{\text{s}}}{9^2}(2\pi)^{10}\delta^{10}(k_1+k_2+k_3)\varepsilon_{1,\mu_1\nu_1\barred{\mu}_1\barred{\nu}_1}\varepsilon_{2,\mu_2\barred{\mu}_2}^{\alpha_2\barred{\alpha}_2}\varepsilon_{3,\mu_3\barred{\mu}_3}^{\alpha_3\barred{\alpha}_3}K^{\{\mu_1\nu_1\}\mu_2\mu_3}_{\alpha_2\alpha_3}K^{\{\barred{\mu}_1\barred{\nu}_1\}\barred{\mu}_2\barred{\mu}_3}_{\barred{\alpha}_2\barred{\alpha}_3},
\end{multline}
where
\begin{multline}
K^{\mu_1\nu_1\mu_2\mu_3}_{\alpha_2\alpha_3} = \left(\frac{1}{2}\eta^{\mu_2\mu_3}k_{23}^{\mu_1} + \eta^{\mu_1\mu_2}k_{12}^{\mu_3} + \eta^{\mu_1\mu_3}k_{31}^{\mu_2} + \frac{\alpha'}{8}k_{23}^{\mu_1}k_{31}^{\mu_2}k_{12}^{\mu_3}\right)\gamma^{\nu_1}_{\alpha_2\alpha_3} 
\\ - 2\left(\eta^{\mu_1\mu_3}\eta^{\nu_1\mu_2} + \frac{\alpha'}{16}\left(\eta^{\mu_1\mu_2}k_{12}^{\mu_3} + \eta^{\mu_1\mu_3}k_{31}^{\mu_2} + \eta^{\mu_2\mu_3}k_{23}^{\mu_1}\right)k_{23}^{\nu_1}\right)(k_{23}\cdot\gamma)_{\alpha_2\alpha_3}
\\ + \frac{\alpha'}{8}\left(\eta^{\mu_1\mu_2}k_{12}^{\mu_3} - \eta^{\mu_1\mu_3}k_{31}^{\mu_2}\right)k_{23}^{\nu_1}(k_1\cdot\gamma)_{\alpha_2\alpha_3} - \frac{\alpha'}{4}\eta^{\mu_2\mu_3}k_{23}^{\mu_1}k_2^{\mu}k_3^{\nu}(\gamma^{\nu_1}_{\phantom{\nu_1}\mu\nu})_{\alpha_2\alpha_3}.
\end{multline}

\begin{multline}
\mathcal{A}_3[\mathrm{NS_1^2 NS_1^2};\mathrm{R_1 R_1};\mathrm{R_1 R_1}] 
\\ = \frac{16\pi ig_{\text{s}}}{3!9^2}(2\pi)^{10}\delta^{10}(k_1+k_2+k_3)
\varepsilon_{1,\mu_1\nu_1\rho_1\barred{\mu}_1\barred{\nu}_1\barred{\rho}_1}\varepsilon_{2,\mu_2\barred{\mu}_2}^{\alpha_2\barred{\alpha}_2}\varepsilon_{3,\mu_3\barred{\mu}_3}^{\alpha_3\barred{\alpha}_3}L^{\mu_1\nu_1\rho_1\mu_2\mu_3}_{\alpha_2\alpha_3}L^{\barred{\mu}_1\barred{\nu}_1\barred{\rho}_1\barred{\mu}_2\barred{\mu}_3}_{\barred{\alpha}_2\barred{\alpha}_3},
\end{multline}
where
\begin{multline}
L^{\mu_1\nu_1\rho_1\mu_2\mu_3}_{\alpha_2\alpha_3} = 6\eta\up{8pt}{[\mu_1}{2pt}{\mu_2}\eta\up{8pt}{\nu_1}{2pt}{\mu_3}\gamma\up{8pt}{\rho_1]}{-3pt}{\alpha_2\alpha_3} - \frac{1}{2}\left(\eta^{\mu_2\mu_3} + \frac{\alpha'}{4}k_{31}^{\mu_2}k_{12}^{\mu_3}\right)(\gamma^{\mu_1\nu_1\rho_1})_{\alpha_2\alpha_3} 
\\ + \frac{\alpha'}{4}\bigg(6\eta^{\mu_2\mu_3}k_2^{[\mu_1}k_3^{\nu_1}\gamma^{\rho_1]}_{\alpha_2\alpha_3} + 6k_{31}^{\mu_2}\eta^{\mu_3[\mu_1}k_3^{\nu_1}\gamma^{\rho_1]}_{\alpha_2\alpha_3} + 6k_{12}^{\mu_3}\eta^{\mu_2[\mu_1}k_2^{\nu_1}\gamma^{\rho_1]}_{\alpha_2\alpha_3} - 12\eta\up{8pt}{[\mu_1}{2pt}{\mu_2}\eta\up{8pt}{\nu_1}{2pt}{\mu_3}k\up{8pt}{\rho_1]}{-3pt}{\hspace{-1pt}3}(k_2\cdot\gamma)_{\alpha_2\alpha_3} 
\\ - 12\eta\up{8pt}{[\mu_1}{2pt}{\mu_2}\eta\up{8pt}{\nu_1}{2pt}{\mu_3}k\up{8pt}{\rho_1]}{-3pt}{\hspace{-1pt}2}(k_3\cdot\gamma)_{\alpha_2\alpha_3} + \frac{3}{2}\eta^{\mu_2\mu_3}k_{23}^{\mu}k_{23}^{[\mu_1}(\gamma^{\nu_1\rho_1]}_{\phantom{\nu_1\rho_1]}\mu})_{\alpha_2\alpha_3} - 3k_{31}^{\mu_2}k_3^{\mu}\eta^{\mu_3[\mu_1}(\gamma^{\nu_1\rho_1]}_{\phantom{\nu_1\rho_1]}\mu})_{\alpha_2\alpha_3} 
\\ + 3k_{12}^{\mu_3}k_2^{\mu}\eta^{\mu_2[\mu_1}(\gamma^{\nu_1\rho_1]}_{\phantom{\nu_1\rho_1]}\mu})_{\alpha_2\alpha_3} + 12k_2^{\mu}k_3^{\nu}\eta\up{8pt}{[\mu_1}{2pt}{\mu_2}\eta\up{8pt}{\nu_1}{2pt}{\mu_3}(\gamma\up{8pt}{\rho_1]}{-3pt}{\phantom{\rho_1]}\mu\nu})_{\alpha_2\alpha_3} + k_2^{\mu}k_3^{\nu}\eta^{\mu_2\mu_3}(\gamma^{\mu_1\nu_1\rho_1}_{\phantom{\mu_1\nu_1\rho_1}\mu\nu})_{\alpha_2\alpha_3}\bigg).
\end{multline}

\section{4-Point Amplitudes}\label{4-point}

In all the amplitudes below we employ the usual Mandelstam variables $s \equiv -(k_1+k_2)^2$, $t \equiv -(k_1+k_3)^2$ and $u \equiv -(k_1+k_4)^2$, which satisfy $s+t+u = \sum_i m_i^2$.

\subsection{All Massless String States}

Completely standard.
\begin{multline}
\mathcal{A}_4[\mathrm{NS_0 NS_0};\mathrm{NS_0 NS_0};\mathrm{NS_0 NS_0};\mathrm{NS_0 NS_0}]
\\ = -\frac{\pi i\alpha'^3 g_{\text{s}}^2}{8}(2\pi)^{10}\delta^{10}(k_1+k_2+k_3+k_4)\varepsilon_{1,\mu_1\barred{\mu}_1}\varepsilon_{2,\mu_2\barred{\mu}_2}\varepsilon_{3,\mu_3\barred{\mu}_3}\varepsilon_{4,\mu_4\barred{\mu}_4}
\\ \times \frac{\Gamma(-\frac{\alpha's}{4})\Gamma(-\frac{\alpha't}{4})\Gamma(-\frac{\alpha'u}{4})}{\Gamma(\frac{\alpha's}{4}\hspace{-2pt}+\hspace{-2pt}1)\Gamma(\frac{\alpha't}{4}\hspace{-2pt}+\hspace{-2pt}1)\Gamma(\frac{\alpha'u}{4}\hspace{-2pt}+\hspace{-2pt}1)}M^{\mu_1\mu_2\mu_3\mu_4}M^{\barred{\mu}_1\barred{\mu}_2\barred{\mu}_3\barred{\mu}_4},
\end{multline}
where
\begin{multline}
M^{\mu_1\mu_2\mu_3\mu_4} = \frac{tu}{2}\eta^{\mu_1\mu_2}\eta^{\mu_3\mu_4} + \frac{su}{2}\eta^{\mu_1\mu_3}\eta^{\mu_2\mu_4} + \frac{st}{2}\eta^{\mu_1\mu_4}\eta^{\mu_2\mu_3} 
\\ - \eta^{\mu_1\mu_2}(uk_1^{\mu_3}k_2^{\mu_4} + tk_2^{\mu_3}k_1^{\mu_4}) - \eta^{\mu_1\mu_3}(uk_1^{\mu_2}k_3^{\mu_4} + sk_3^{\mu_2}k_1^{\mu_4}) - \eta^{\mu_1\mu_4}(tk_1^{\mu_2}k_4^{\mu_3} + sk_4^{\mu_2}k_1^{\mu_3}) 
\\ - \eta^{\mu_2\mu_3}(tk_2^{\mu_1}k_3^{\mu_4} + sk_3^{\mu_1}k_2^{\mu_4}) - \eta^{\mu_2\mu_4}(uk_2^{\mu_1}k_4^{\mu_3} + sk_4^{\mu_1}k_2^{\mu_3}) - \eta^{\mu_3\mu_4}(uk_3^{\mu_1}k_4^{\mu_2} + tk_4^{\mu_1}k_3^{\mu_2}).
\end{multline}

\begin{multline}
\mathcal{A}_4[\mathrm{NS_0 NS_0};\mathrm{NS_0 NS_0};\mathrm{R_0 R_0};\mathrm{R_0 R_0}]
\\ = -\frac{\pi i\alpha'^3 g_{\text{s}}^2}{2^7}(2\pi)^{10}\delta^{10}(k_1+k_2+k_3+k_4)\varepsilon_{1,\mu_1\barred{\mu}_1}\varepsilon_{2,\mu_2\barred{\mu}_2}\varepsilon_3^{\alpha_3\barred{\alpha}_3}\varepsilon_4^{\alpha_4\barred{\alpha}_4}
\\ \times \frac{\Gamma(-\frac{\alpha's}{4})\Gamma(-\frac{\alpha't}{4})\Gamma(-\frac{\alpha'u}{4})}{\Gamma(\frac{\alpha's}{4}\hspace{-2pt}+\hspace{-2pt}1)\Gamma(\frac{\alpha't}{4}\hspace{-2pt}+\hspace{-2pt}1)\Gamma(\frac{\alpha'u}{4}\hspace{-2pt}+\hspace{-2pt}1)}N^{\mu_1\mu_2}_{\alpha_3\alpha_4}N^{\barred{\mu}_1\barred{\mu}_2}_{\barred{\alpha}_3\barred{\alpha}_4},
\end{multline}
where
\begin{equation}
N^{\mu_1\mu_2}_{\alpha_3\alpha_4} = 2\left((t-u)k_{12}^{(\mu_1} + \frac{s}{2}k_{34}^{(\mu_1}\right)\gamma^{\mu_2)}_{\alpha_3\alpha_4} \hspace{-4pt}-\hspace{-2pt} \frac{t-u}{2}\eta^{\mu_1\mu_2}(k_{12}\hspace{-2pt}\cdot\hspace{-2pt}\gamma)_{\alpha_3\alpha_4} - \frac{s}{2}k_{12}^{\mu}(\gamma^{\mu_1\mu_2}_{\phantom{\mu_1\mu_2}\mu})_{\alpha_3\alpha_4}.
\end{equation}

\begin{multline}
\mathcal{A}_4[\mathrm{R_0 R_0};\mathrm{R_0 R_0};\mathrm{R_0 R_0};\mathrm{R_0 R_0}]
\\ = -\frac{\pi i\alpha'^3 g_{\text{s}}^2}{2^7}(2\pi)^{10}\delta^{10}(k_1+k_2+k_3+k_4)\varepsilon_1^{\alpha_1\barred{\alpha}_1}\varepsilon_2^{\alpha_2\barred{\alpha}_2}\varepsilon_3^{\alpha_3\barred{\alpha}_3}\varepsilon_4^{\alpha_4\barred{\alpha}_4}
\\ \times \frac{\Gamma(-\frac{\alpha's}{4})\Gamma(-\frac{\alpha't}{4})\Gamma(-\frac{\alpha'u}{4})}{\Gamma(\frac{\alpha's}{4}\hspace{-2pt}+\hspace{-2pt}1)\Gamma(\frac{\alpha't}{4}\hspace{-2pt}+\hspace{-2pt}1)\Gamma(\frac{\alpha'u}{4}\hspace{-2pt}+\hspace{-2pt}1)}O_{\alpha_1\alpha_2\alpha_3\alpha_4}O_{\barred{\alpha}_1\barred{\alpha}_2\barred{\alpha}_3\barred{\alpha}_4},
\end{multline}
where
\begin{equation}
O_{\alpha_1\alpha_2\alpha_3\alpha_4} = t\gamma_{\mu\alpha_1\alpha_2}\gamma^{\mu}_{\alpha_3\alpha_4} - s\gamma_{\mu\alpha_1\alpha_3}\gamma^{\mu}_{\alpha_2\alpha_4}.
\end{equation}
While not manifest, the above amplitude is indeed totally symmetric in all four RR states on account of the Fierz identity $\gamma_{\mu\alpha_1(\alpha_2}\gamma^{\mu}_{\alpha_3\alpha_4)} = 0$.

\subsection{One Massive String State}

\subsubsection{All NSNS States}

\begin{multline}
\mathcal{A}_4[\mathrm{NS_0 NS_0};\mathrm{NS_0 NS_0};\mathrm{NS_0 NS_0};\mathrm{NS_1^1 NS_1^1}]
\\ = -8\pi^2 ig_{\text{s}}^2(2\pi)^{10}\delta^{10}(k_1+k_2+k_3+k_4)\varepsilon_{1,\mu_1\barred{\mu}_1}\varepsilon_{2,\mu_2\barred{\mu}_2}\varepsilon_{3,\mu_3\barred{\mu}_3}\varepsilon_{4,\mu_4\nu_4\barred{\mu}_4\barred{\nu}_4}
\\ \times \frac{\Gamma(1\hspace{-2pt}-\hspace{-2pt}\frac{\alpha's}{4})\Gamma(1\hspace{-2pt}-\hspace{-2pt}\frac{\alpha't}{4})\Gamma(1\hspace{-2pt}-\hspace{-2pt}\frac{\alpha'u}{4})}{\Gamma(\frac{\alpha's}{4})\Gamma(\frac{\alpha't}{4})\Gamma(\frac{\alpha'u}{4})}P^{\mu_1\mu_2\mu_3\{\mu_4\nu_4\}}P^{\barred{\mu}_1\barred{\mu}_2\barred{\mu}_3\{\barred{\mu}_4\barred{\nu}_4\}},
\end{multline}
where
\begin{multline}
P^{\mu_1\mu_2\mu_3\mu_4\nu_4} = \eta^{\mu_1\mu_2}\eta^{\mu_3\mu_4}\left[\left(1+\frac{u}{s}\right)k_1^{\nu_4} - \left(1+\frac{t}{s}\right)k_2^{\nu_4}\right] - \eta^{\mu_1\mu_3}\eta^{\mu_2\mu_4}\left(\frac{s}{t}k_2^{\nu_4} - k_3^{\nu_4}\right)
\\ + \eta^{\mu_1\mu_4}\eta^{\mu_2\mu_3}\left(\frac{s}{u}k_1^{\nu_4} - k_3^{\nu_4}\right) - \eta^{\mu_1\nu_4}\eta^{\mu_3\mu_4}\left(\frac{m^2}{s}k_1^{\mu_2} - \frac{m^2}{u}k_3^{\mu_2}\right) + \frac{m^2}{u}\eta^{\mu_1\nu_4}(\eta^{\mu_2\mu_3}k_2^{\mu_4} - \eta^{\mu_2\mu_4}k_2^{\mu_3})
\\ + \eta^{\mu_2\nu_4}\eta^{\mu_3\mu_4}\left(\frac{m^2}{s}k_2^{\mu_1} - \frac{m^2}{t}k_3^{\mu_1}\right) - \frac{m^2}{t}\eta^{\mu_2\nu_4}(\eta^{\mu_1\mu_3}k_1^{\mu_4} - \eta^{\mu_1\mu_4}k_1^{\mu_3}) 
\\ - \frac{\alpha'}{2}\eta^{\mu_1\mu_2}\left[\frac{m^2}{s}(k_1^{\mu_3}k_2^{\mu_4}-k_2^{\mu_3}k_1^{\mu_4})k_3^{\nu_4} - \frac{m^2}{t}k_1^{\mu_3}k_2^{\mu_4}k_2^{\nu_4} + \frac{m^2}{u}k_2^{\mu_3}k_1^{\mu_4}k_1^{\nu_4}\right]
\\ - \frac{\alpha'}{2}\eta^{\mu_1\mu_3}\left[\frac{m^2}{s}k_1^{\mu_2}k_3^{\mu_4}k_3^{\nu_4} - \frac{m^2}{t}(k_4^{\mu_2}k_1^{\mu_4} - k_1^{\mu_2}k_2^{\mu_4})k_2^{\nu_4} - \frac{m^2}{u}k_3^{\mu_2}k_1^{\mu_4}k_1^{\nu_4}\right]
\\ + \frac{\alpha'}{2}\eta^{\mu_1\mu_4}\left[\frac{m^2}{s}k_1^{\mu_2}k_4^{\mu_3}k_3^{\nu_4} - \frac{m^2}{t}k_4^{\mu_2}k_1^{\mu_3}k_2^{\nu_4} - \frac{m^2}{u}(k_3^{\mu_2}k_1^{\mu_3} - k_1^{\mu_2}k_2^{\mu_3})k_1^{\nu_4}\right]
\\ + \frac{\alpha'}{2}\eta^{\mu_2\mu_3}\left[\frac{m^2}{s}k_2^{\mu_1}k_3^{\mu_4}k_3^{\nu_4} - \frac{m^2}{t}k_3^{\mu_1}k_2^{\mu_4}k_2^{\nu_4} - \frac{m^2}{u}(k_4^{\mu_1}k_2^{\mu_4} - k_2^{\mu_1}k_1^{\mu_4})k_1^{\nu_4}\right]
\\ - \frac{\alpha'}{2}\eta^{\mu_2\mu_4}\left[\frac{m^2}{s}k_2^{\mu_1}k_4^{\mu_3}k_3^{\nu_4} - \frac{m^2}{t}(k_3^{\mu_1}k_2^{\mu_3} - k_2^{\mu_1}k_1^{\mu_3})k_2^{\nu_4} - \frac{m^2}{u}k_4^{\mu_1}k_2^{\mu_3}k_1^{\nu_4}\right]
\\ - \frac{\alpha'}{2}\eta^{\mu_3\mu_4}\left[\frac{m^2}{s}(k_3^{\mu_1}k_4^{\mu_2} - k_4^{\mu_1}k_3^{\mu_2})k_3^{\nu_4} - \frac{m^2}{t}k_3^{\mu_1}k_4^{\mu_2}k_2^{\nu_4} + \frac{m^2}{u}k_4^{\mu_1}k_3^{\mu_2}k_1^{\nu_4}\right].
\end{multline}

\begin{multline}
\mathcal{A}_4[\mathrm{NS_0 NS_0};\mathrm{NS_0 NS_0};\mathrm{NS_0 NS_0};\mathrm{NS_1^2 NS_1^2}]
\\ = \frac{16\pi^2 ig_{\text{s}}^2}{3!\alpha'}(2\pi)^{10}\delta^{10}(k_1+k_2+k_3+k_4)\varepsilon_{1,\mu_1\barred{\mu}_1}\varepsilon_{2,\mu_2\barred{\mu}_2}\varepsilon_{3,\mu_3\barred{\mu}_3}\varepsilon_{4,\mu_4\nu_4\rho_4\barred{\mu}_4\barred{\nu}_4\barred{\rho}_4}
\\ \times \frac{\Gamma(1\hspace{-2pt}-\hspace{-2pt}\frac{\alpha's}{4})\Gamma(1\hspace{-2pt}-\hspace{-2pt}\frac{\alpha't}{4})\Gamma(1\hspace{-2pt}-\hspace{-2pt}\frac{\alpha'u}{4})}{\Gamma(\frac{\alpha's}{4})\Gamma(\frac{\alpha't}{4})\Gamma(\frac{\alpha'u}{4})}Q^{\mu_1\mu_2\mu_3\mu_4\nu_4\rho_4}Q^{\barred{\mu}_1\barred{\mu}_2\barred{\mu}_3\barred{\mu}_4\barred{\nu}_4\barred{\rho}_4},
\end{multline}
where
\begin{multline}
Q^{\mu_1\mu_2\mu_3\mu_4\nu_4\rho_4} = 6\eta\up{8pt}{[\mu_4}{1pt}{\mu_1}\eta\up{8pt}{\nu_4}{1pt}{\mu_2}\eta\up{8pt}{\rho_4]}{2pt}{\mu_3} - 3\alpha'\bigg(\frac{m^2}{s}\eta^{\mu_1\mu_2}\eta^{\mu_3[\mu_4}k_1^{\nu_4}k_2^{\rho_4]} + \frac{m^2}{t}\eta^{\mu_1\mu_3}\eta^{\mu_2[\mu_4}k_1^{\nu_4}k_2^{\rho_4]} 
\\ + \frac{m^2}{u}\eta^{\mu_2\mu_3}\eta^{\mu_1[\mu_4}k_1^{\nu_4}k_2^{\rho_4]} + \frac{m^2}{t}\eta\up{8pt}{[\mu_4}{1pt}{\mu_1}\eta\up{8pt}{\nu_4}{2pt}{\mu_2}k\up{8pt}{\rho_4]}{-2pt}{\hspace{-1pt}2}k_1^{\mu_3} + \frac{m^2}{u}\eta\up{8pt}{[\mu_4}{1pt}{\mu_1}\eta\up{8pt}{\nu_4}{2pt}{\mu_2}k\up{8pt}{\rho_4]}{-2pt}{\hspace{-1pt}1}k_2^{\rho_3} - \frac{m^2}{s}\eta\up{8pt}{[\mu_4}{1pt}{\mu_1}\eta\up{8pt}{\nu_4}{1pt}{\mu_3}k\up{8pt}{\rho_4]}{-2pt}{\hspace{-1pt}3}k_1^{\mu_2} 
\\ - \frac{m^2}{u}\eta\up{8pt}{[\mu_4}{2pt}{\mu_1}\eta\up{8pt}{\nu_4}{2pt}{\mu_3}k\up{8pt}{\rho_4]}{-2pt}{\hspace{-1pt}1}k_3^{\mu_2} + \frac{m^2}{s}\eta\up{8pt}{[\mu_4}{1pt}{\mu_2}\eta\up{8pt}{\nu_4}{1pt}{\mu_3}k\up{8pt}{\rho_4]}{-2pt}{\hspace{-1pt}3}k_2^{\mu_1} + \frac{m^2}{t}\eta\up{8pt}{[\mu_4}{1pt}{\mu_2}\eta\up{8pt}{\nu_4}{2pt}{\mu_3}k\up{8pt}{\rho_4]}{-2pt}{\hspace{-1pt}2}k_3^{\mu_1}\bigg).
\end{multline}

\subsubsection{Mixed NSNS and RR States}

\begin{multline}
\mathcal{A}_4[\mathrm{NS_0 NS_0};\mathrm{NS_1^1 NS_1^1};\mathrm{R_0 R_0};\mathrm{R_0 R_0}]
\\ = 2\pi^2 ig_{\text{s}}^2(2\pi)^{10}\delta^{10}(k_1+k_2+k_3+k_4)\varepsilon_{1,\mu_1\barred{\mu}_1}\varepsilon_{2,\mu_2\nu_2\barred{\mu}_2\barred{\nu}_2}\varepsilon_3^{\alpha_3\barred{\alpha}_3}\varepsilon_4^{\alpha_4\barred{\alpha}_4}
\\ \times \frac{\Gamma(1\hspace{-2pt}-\hspace{-2pt}\frac{\alpha's}{4})\Gamma(1\hspace{-2pt}-\hspace{-2pt}\frac{\alpha't}{4})\Gamma(1\hspace{-2pt}-\hspace{-2pt}\frac{\alpha'u}{4})}{\Gamma(\frac{\alpha's}{4})\Gamma(\frac{\alpha't}{4})\Gamma(\frac{\alpha'u}{4})}R^{\mu_1\{\mu_2\nu_2\}}_{\alpha_3\alpha_4}R^{\barred{\mu}_1\{\barred{\mu}_2\barred{\nu}_2\}}_{\barred{\alpha}_3\barred{\alpha}_4},
\end{multline}
where
\begin{multline}
R^{\mu_1\mu_2\nu_2}_{\alpha_3\alpha_4} = \frac{m^2}{s}\eta^{\mu_1\mu_2}\gamma^{\nu_2}_{\alpha_3\alpha_4} - \frac{\alpha'}{2}\left(\frac{m^2}{s}k_2^{\mu_1}k_1^{\mu_2} + \frac{m^2}{u}k_4^{\mu_1}k_3^{\mu_2} + \frac{m^2}{t}k_3^{\mu_1}k_4^{\mu_2}\right)\gamma^{\nu_2}_{\alpha_3\alpha_4}
\\ + \frac{\alpha'}{4}\left(-\frac{2m^2}{s}k_1^{\nu_2} + \frac{m^2}{t}k_4^{\nu_2} + \frac{m^2}{u}k_3^{\nu_2}\right)\left[\frac{1}{2}\eta^{\mu_1\mu_2}(k_{12}\cdot\gamma)_{\alpha_3\alpha_4} - k_1^{\mu_2}\gamma^{\mu_1}_{\alpha_3\alpha_4}\right]
\\ + \frac{\alpha'}{4}\left(\frac{m^2}{t}k_4^{\nu_2} - \frac{m^2}{u}k_3^{\nu_2}\right)\bigg[k_{34}^{[\mu_1}\gamma^{\mu_2]}_{\alpha_3\alpha_4} + \frac{1}{2}k_{12}^{\mu}(\gamma^{\mu_1\mu_2}_{\phantom{\mu_1\mu_2}\mu})_{\alpha_3\alpha_4}\bigg].
\end{multline}

\begin{multline}
\mathcal{A}_4[\mathrm{NS_0 NS_0};\mathrm{NS_1^2 NS_1^2};\mathrm{R_0 R_0};\mathrm{R_0 R_0}]
\\ = -\frac{\pi^2 i\alpha' g_{\text{s}}^2}{3!16}(2\pi)^{10}\delta^{10}(k_1+k_2+k_3+k_4)\varepsilon_{1,\mu_1\barred{\mu}_1}\varepsilon_{2,\mu_2\nu_2\rho_2\barred{\mu}_2\barred{\nu}_2\barred{\rho}_2}\varepsilon_3^{\alpha_3\barred{\alpha}_3}\varepsilon_4^{\alpha_4\barred{\alpha}_4}
\\ \times \frac{\Gamma(1\hspace{-2pt}-\hspace{-2pt}\frac{\alpha's}{4})\Gamma(1\hspace{-2pt}-\hspace{-2pt}\frac{\alpha't}{4})\Gamma(1\hspace{-2pt}-\hspace{-2pt}\frac{\alpha'u}{4})}{\Gamma(\frac{\alpha's}{4})\Gamma(\frac{\alpha't}{4})\Gamma(\frac{\alpha'u}{4})}S^{\mu_1\mu_2\nu_2\rho_2}_{\alpha_3\alpha_4}S^{\barred{\mu}_1\barred{\mu}_2\barred{\nu}_2\barred{\rho}_2}_{\barred{\alpha}_3\barred{\alpha}_4},
\end{multline}
where
\begin{multline}
S^{\mu_1\mu_2\nu_2\rho_2}_{\alpha_3\alpha_4} = 6\left(\frac{4m^2}{s}+\frac{m^2}{t} + \frac{m^2}{u}\right)\eta^{\mu_1[\mu_2}k_1^{\nu_2}\gamma^{\rho_2]}_{\alpha_3\alpha_4} - 2\left(\frac{m^2}{t}k_3^{\mu_1} - \frac{m^2}{u}k_4^{\mu_1}\right)(\gamma^{\mu_2\nu_2\rho_2})_{\alpha_3\alpha_4}
\\ + 3\left(\frac{m^2}{t}-\frac{m^2}{u}\right)\left[\eta^{\mu_1[\mu_2}k_{34}^{\nu_2}\gamma^{\rho_2]}_{\alpha_3\alpha_4} - k_1^{[\mu_2}(\gamma^{\nu_2\rho_2]\mu_1})_{\alpha_3\alpha_4} + \frac{1}{2}k_{12}^{\mu}\eta^{\mu_1[\mu_2}(\gamma^{\nu_2\rho_2]}_{\phantom{\nu_2\rho_2]}\mu})_{\alpha_3\alpha_4}\right]
\\  + \frac{1}{2}\left(\frac{m^2}{t}+\frac{m^2}{u}\right)\left[k_{34}^{\mu_1}(\gamma^{\mu_2\nu_2\rho_2})_{\alpha_3\alpha_4} - 3k_{34}^{[\mu_2}(\gamma^{\nu_2\rho_2]\mu_1})_{\alpha_3\alpha_4} + k_{12}^{\mu}(\gamma^{\mu_1\mu_2\nu_2\rho_2}_{\phantom{\mu_1\mu_2\nu_2\rho_2}\mu})_{\alpha_3\alpha_4}\right].
\end{multline}

\begin{multline}
\mathcal{A}_4[\mathrm{NS_0 NS_0};\mathrm{NS_0 NS_0};\mathrm{R_0 R_0};\mathrm{R_1 R_1}]
\\ = -\frac{4\pi^2 i\alpha'^2 g_{\text{s}}^2}{9}(2\pi)^{10}\delta^{10}(k_1+k_2+k_3+k_4)\varepsilon_{1,\mu_1\barred{\mu}_1}\varepsilon_{2,\mu_2\barred{\mu}_2}\varepsilon_3^{\alpha_3\barred{\alpha}_3}\varepsilon_{4,\mu_4\barred{\mu}_4}^{\alpha_4\barred{\alpha}_4}
\\ \times \frac{\Gamma(1\hspace{-2pt}-\hspace{-2pt}\frac{\alpha's}{4})\Gamma(1\hspace{-2pt}-\hspace{-2pt}\frac{\alpha't}{4})\Gamma(1\hspace{-2pt}-\hspace{-2pt}\frac{\alpha'u}{4})}{\Gamma(\frac{\alpha's}{4})\Gamma(\frac{\alpha't}{4})\Gamma(\frac{\alpha'u}{4})}T^{\mu_1\mu_2\mu_4}_{\alpha_3\alpha_4}T^{\barred{\mu}_1\barred{\mu}_2\barred{\mu}_4}_{\barred{\alpha}_3\barred{\alpha}_4},
\end{multline}
where
\begin{multline}
T^{\mu_1\mu_2\mu_4}_{\alpha_3\alpha_4} 
\\ = \frac{1}{4}\eta^{\mu_1\mu_2}\hspace{-4pt}\left[\hspace{-3pt}\left(\hspace{-2pt}\frac{2m^2}{s}k_{12}^{\mu_4} \hspace{-2pt}-\hspace{-2pt} \frac{m^2}{t}k_2^{\mu_4} \hspace{-2pt}+\hspace{-2pt} \frac{m^2}{u}k_1^{\mu_4}\hspace{-2pt}\right)\hspace{-4pt}(k_4\hspace{-2pt}\cdot\hspace{-2pt}\gamma)_{\alpha_3\alpha_4} \hspace{-3pt}- \hspace{-3pt}\left(\hspace{-2pt}\frac{2m^2}{s}k_3^{\mu_4} \hspace{-2pt}-\hspace{-2pt} \frac{m^2}{t}k_2^{\mu_4} \hspace{-2pt}-\hspace{-2pt} \frac{m^2}{u}k_1^{\mu_4}\hspace{-2pt}\right)\hspace{-4pt}(k_{12}\hspace{-2pt}\cdot\hspace{-2pt}\gamma)_{\alpha_3\alpha_4}\hspace{-2pt}\right]
\\ - \frac{m^2}{2t}\eta^{\mu_2\mu_4}\hspace{-3pt}\left[\hspace{-2pt}\frac{s\hspace{-3pt}+\hspace{-3pt}2t\hspace{-3pt}+\hspace{-3pt}m^2}{4}\hspace{-1pt}\gamma^{\mu_1}_{\alpha_3\alpha_4} \hspace{-4pt}-\hspace{-3pt} \frac{1}{2}k_4^{\mu_1}\hspace{-2pt}(\hspace{-1pt}k_{12}\hspace{-2pt}\cdot\hspace{-2pt}\gamma\hspace{-1pt})_{\alpha_3\alpha_4} \hspace{-3pt}+\hspace{-3pt} \left(\hspace{-2pt}\frac{1}{2}k_{34}^{\mu_1} \hspace{-3pt}-\hspace{-3pt}  \frac{s\hspace{-3pt}+\hspace{-3pt}2t}{s}k_2^{\mu_1}\hspace{-3pt}\right)\hspace{-4pt}(\hspace{-1pt}k_4\hspace{-2pt}\cdot\hspace{-2pt}\gamma\hspace{-1pt})_{\alpha_3\alpha_4} \hspace{-3pt}+\hspace{-3pt} \frac{1}{2}k_{12}^{\mu}k_4^{\nu}(\hspace{-1pt}\gamma^{\mu_1}_{\phantom{\mu_1}\mu\nu}\hspace{-1pt})_{\alpha_3\alpha_4}\hspace{-2pt}\right]
\\ + \frac{m^2}{2u}\eta^{\mu_1\mu_4}\hspace{-3pt}\left[\hspace{-2pt}\frac{s\hspace{-3pt}+\hspace{-3pt}2u\hspace{-3pt}+\hspace{-3pt}m^2}{4}\hspace{-1pt}\gamma^{\mu_2}_{\alpha_3\alpha_4} \hspace{-4pt}+\hspace{-3pt} \frac{1}{2}k_4^{\mu_2}\hspace{-2pt}(\hspace{-1pt}k_{12}\hspace{-2pt}\cdot\hspace{-2pt}\gamma\hspace{-1pt})_{\alpha_3\alpha_4} \hspace{-3pt}+\hspace{-3pt} \left(\hspace{-2pt}\frac{1}{2}k_{34}^{\mu_2} \hspace{-3pt}-\hspace{-3pt} \frac{s\hspace{-3pt}+\hspace{-3pt}2u}{s}k_1^{\mu_2}\hspace{-3pt}\right)\hspace{-4pt}(\hspace{-1pt}k_4\hspace{-2pt}\cdot\hspace{-2pt}\gamma\hspace{-1pt})_{\alpha_3\alpha_4} \hspace{-4pt}-\hspace{-3pt} \frac{1}{2}k_{12}^{\mu}k_4^{\nu}(\hspace{-1pt}\gamma^{\mu_2}_{\phantom{\mu_2}\mu\nu}\hspace{-1pt})_{\alpha_3\alpha_4}\hspace{-2pt}\right]
\\ + \left(\frac{m^2}{s}k_1^{\mu_2}k_3^{\mu_4} \hspace{-2pt} + \frac{m^2}{2t}k_4^{\mu_2}k_2^{\mu_4} \hspace{-2pt} + \frac{m^2}{2u}k_3^{\mu_2}k_1^{\mu_4}\right)\hspace{-3pt}\gamma^{\mu_1}_{\alpha_3\alpha_4} - \left(\frac{m^2}{s}k_2^{\mu_1}k_3^{\mu_4} \hspace{-2pt} + \frac{m^2}{2u}k_4^{\mu_1}k_1^{\mu_4} \hspace{-2pt} + \frac{m^2}{2t}k_3^{\mu_1}k_2^{\mu_4}\right)\hspace{-3pt}\gamma^{\mu_2}_{\alpha_3\alpha_4} 
\\ - \frac{1}{4}\hspace{-3pt}\left(\hspace{-2pt}\frac{m^2}{u}k_1^{\mu_4} \hspace{-2pt}-\hspace{-2pt} \frac{m^2}{t}k_2^{\mu_4}\hspace{-2pt}\right)\hspace{-4pt}k_{12}^{\mu}(\gamma^{\mu_1\mu_2}_{\phantom{\mu_1\mu_2}\mu})_{\alpha_3\alpha_4} - \frac{1}{2}\hspace{-2pt}\left(\hspace{-2pt}\frac{m^2}{u}k_1^{\mu_4} \hspace{-2pt}+\hspace{-2pt} \frac{m^2}{t}k_2^{\mu_4}\hspace{-2pt}\right)\hspace{-4pt}\left(\hspace{-2pt}k_3^{[\mu_1}\gamma^{\mu_2]}_{\alpha_3\alpha_4} \hspace{-2pt}+\hspace{-2pt} \frac{1}{2}k_4^{\mu}(\gamma^{\mu_1\mu_2}_{\phantom{\mu_1\mu_2}\mu})_{\alpha_3\alpha_4}\hspace{-2pt}\right).
\end{multline}

\subsubsection{All RR States}

\begin{multline}
\mathcal{A}_4[\mathrm{R_0 R_0};\mathrm{R_0 R_0};\mathrm{R_0 R_0};\mathrm{R_1 R_1}]
\\ = \frac{\pi^2 i\alpha'^2 g_{\text{s}}^2}{9}(2\pi)^{10}\delta^{10}(k_1+k_2+k_3+k_4)\varepsilon_1^{\alpha_1\barred{\alpha}_1}\varepsilon_2^{\alpha_2\barred{\alpha}_2}\varepsilon_3^{\alpha_3\barred{\alpha}_3}\varepsilon_{4,\mu_4\barred{\mu}_4}^{\alpha_4\barred{\alpha}_4}
\\ \times \frac{\Gamma(1\hspace{-2pt}-\hspace{-2pt}\frac{\alpha's}{4})\Gamma(1\hspace{-2pt}-\hspace{-2pt}\frac{\alpha't}{4})\Gamma(1\hspace{-2pt}-\hspace{-2pt}\frac{\alpha'u}{4})}{\Gamma(\frac{\alpha's}{4})\Gamma(\frac{\alpha't}{4})\Gamma(\frac{\alpha'u}{4})}U^{\mu_4}_{\alpha_1\alpha_2\alpha_3\alpha_4}U^{\barred{\mu}_4}_{\barred{\alpha}_1\barred{\alpha}_2\barred{\alpha}_3\barred{\alpha}_4},
\end{multline}
where
\begin{multline}
U^{\mu_4}_{\alpha_1\alpha_2\alpha_3\alpha_4} = \left(\frac{m^2}{u}k_1^{\mu_4} - \frac{m^2}{s}k_3^{\mu_4}\right)\gamma_{\mu\alpha_1\alpha_2}\gamma^{\mu}_{\alpha_3\alpha_4} + \left(\frac{m^2}{u}k_1^{\mu_4} - \frac{m^2}{t}k_2^{\mu_4}\right)\gamma_{\mu\alpha_1\alpha_3}\gamma^{\mu}_{\alpha_2\alpha_4}
\\ + \frac{m^2}{s}\gamma^{\mu_4}_{\alpha_1\alpha_2}(k_4\cdot\gamma)_{\alpha_3\alpha_4} + \frac{m^2}{t}\gamma^{\mu_4}_{\alpha_1\alpha_3}(k_4\cdot\gamma)_{\alpha_2\alpha_4} + \frac{m^2}{u}(k_4\cdot\gamma)_{\alpha_1\alpha_4}\gamma^{\mu_4}_{\alpha_2\alpha_3}.
\end{multline}

\subsection{Two Massive String States}

\subsubsection{With Two Massless NSNS States}

\begin{multline}
\mathcal{A}_4[\mathrm{NS_0 NS_0};\mathrm{NS_0 NS_0};\mathrm{NS_1^1 NS_1^1};\mathrm{NS_1^1 NS_1^1}]
\\ = \frac{16\pi^2 ig_{\text{s}}^2}{\alpha'}(2\pi)^{10}\delta^{10}(k_1+k_2+k_3+k_4)\varepsilon_{1,\mu_1\barred{\mu}_1}\varepsilon_{2,\mu_2\barred{\mu}_2}\varepsilon_{3,\mu_3\nu_3\barred{\mu}_3\barred{\nu}_3}\varepsilon_{4,\mu_4\nu_4\barred{\mu}_4\barred{\nu}_4}
\\ \times \frac{\Gamma(1\hspace{-2pt}-\hspace{-2pt}\frac{\alpha's}{4})\Gamma(1\hspace{-2pt}-\hspace{-2pt}\frac{\alpha't}{4})\Gamma(1\hspace{-2pt}-\hspace{-2pt}\frac{\alpha'u}{4})}{\Gamma(\frac{\alpha's}{4})\Gamma(\frac{\alpha't}{4})\Gamma(\frac{\alpha'u}{4})}V^{\mu_1\mu_2\{\mu_3\nu_3\}\{\mu_4\nu_4\}}V^{\barred{\mu}_1\barred{\mu}_2\{\barred{\mu}_3\barred{\nu}_3\}\{\barred{\mu}_4\barred{\nu}_4\}},
\end{multline}
where
\begin{multline}
V^{\mu_1\mu_2\mu_3\nu_3\mu_4\nu_4} 
\\ = \bigg(\hspace{-3pt}\frac{(\hspace{-1pt}t\hspace{-2pt}-\hspace{-2pt}m^2\hspace{-1pt})(\hspace{-1pt}u\hspace{-2pt}-\hspace{-2pt}m^2\hspace{-1pt})}{sm^2}\eta^{\mu_1\mu_2}\eta^{\mu_3\mu_4} + \frac{(\hspace{-1pt}t\hspace{-2pt}+\hspace{-2pt}m^2\hspace{-1pt})(\hspace{-1pt}u\hspace{-2pt}-\hspace{-2pt}m^2\hspace{-1pt})}{tm^2}\eta^{\mu_1\mu_3}\eta^{\mu_2\mu_4} + \frac{(\hspace{-1pt}t\hspace{-2pt}-\hspace{-2pt}m^2\hspace{-1pt})(\hspace{-1pt}u\hspace{-2pt}+\hspace{-2pt}m^2\hspace{-1pt})}{um^2}\eta^{\mu_1\mu_4}\eta^{\mu_2\mu_3}\hspace{-3pt}\bigg)\eta^{\nu_3\nu_4}
\\ - \frac{\alpha'}{2}\eta^{\mu_1\mu_2}\eta^{\nu_3\nu_4}\hspace{-4pt}\left[\hspace{-2pt}\frac{(t\hspace{-2pt}-\hspace{-2pt}m^2)(u\hspace{-2pt}-\hspace{-2pt}m^2)}{sm^2}k_4^{\mu_3}k_3^{\mu_4} \hspace{-2pt}+\hspace{-2pt} \frac{(s\hspace{-2pt}+\hspace{-2pt}2m^2)(t\hspace{-2pt}-\hspace{-2pt}m^2)}{sm^2}k_2^{\mu_3}k_1^{\mu_4} \hspace{-2pt}+\hspace{-2pt} \frac{(s\hspace{-2pt}+\hspace{-2pt}2m^2)(u\hspace{-2pt}-\hspace{-2pt}m^2)}{sm^2}k_1^{\mu_3}k_2^{\mu_4}\hspace{-2pt}\right]
\\ + \frac{\alpha'}{2}\eta^{\mu_1\mu_3}\eta^{\nu_3\nu_4}\left[\frac{2(t-m^2)}{s}k_3^{\mu_2}k_1^{\mu_4} + \frac{u-m^2}{s}(k_{13}^{\mu_2}k_2^{\mu_4} - k_4^{\mu_2}k_{13}^{\mu_4}) - \frac{u-m^2}{t}k_4^{\mu_2}k_2^{\mu_4}\right]
\\ + \frac{\alpha'}{2}\eta^{\mu_1\mu_4}\eta^{\nu_3\nu_4}\left[\frac{2(u-m^2)}{s}k_4^{\mu_2}k_1^{\mu_3} + \frac{t-m^2}{s}(k_{14}^{\mu_2}k_2^{\mu_3}-k_3^{\mu_2}k_{14}^{\mu_3}) - \frac{t-m^2}{u}k_3^{\mu_2}k_2^{\mu_3} \right]
\\ + \frac{\alpha'}{2}\eta^{\mu_2\mu_3}\eta^{\nu_3\nu_4}\left[\frac{2(u-m^2)}{s}k_3^{\mu_1}k_2^{\mu_4} + \frac{t-m^2}{s}(k_{23}^{\mu_1}k_1^{\mu_4} - k_4^{\mu_1}k_{23}^{\mu_4}) - \frac{t-m^2}{u}k_4^{\mu_1}k_1^{\mu_4}\right]
\\ + \frac{\alpha'}{2}\eta^{\mu_2\mu_4}\eta^{\nu_3\nu_4}\left[\frac{2(t-m^2)}{s}k_4^{\mu_1}k_2^{\mu_3} + \frac{u-m^2}{s}(k_{24}^{\mu_1}k_1^{\mu_3}-k_3^{\mu_1}k_{24}^{\mu_3}) - \frac{u-m^2}{t} k_3^{\mu_1}k_1^{\mu_3}\right]
\\ - \frac{\alpha'}{2}\eta^{\mu_3\mu_4}\eta^{\nu_3\nu_4}\left[\frac{t-m^2}{s}k_4^{\mu_1}k_3^{\mu_2} + \frac{u-m^2}{s}k_3^{\mu_1}k_4^{\mu_2}\right] 
\\ - \frac{\alpha'}{2}\eta^{\mu_1\mu_3}\eta^{\mu_2\mu_4}\bigg[\frac{(t\hspace{-2pt}+\hspace{-2pt}m^2)(u\hspace{-2pt}-\hspace{-2pt}m^2)}{tm^2}k_4^{\nu_3}k_3^{\nu_4} + \frac{(s\hspace{-2pt}-\hspace{-2pt}m^2)(u\hspace{-2pt}-\hspace{-2pt}m^2)}{tm^2}k_1^{\nu_3}k_2^{\nu_4} + \frac{(s\hspace{-2pt}-\hspace{-2pt}m^2)(t\hspace{-2pt}+\hspace{-2pt}m^2)}{tm^2}k_2^{\nu_3}k_1^{\nu_4}\bigg]
\\ + \frac{\alpha'}{2}\eta^{\mu_1\mu_3}\eta^{\mu_2\nu_3}\left[\frac{t-m^2}{u}k_1^{\mu_4}k_1^{\nu_4} - 2k_1^{\mu_4}k_2^{\nu_4} + \frac{u-m^2}{t}k_2^{\mu_4}k_2^{\nu_4}\right]
\\ - \frac{\alpha'}{2}\eta^{\mu_1\mu_4}\eta^{\mu_2\mu_3}\bigg[\frac{(t\hspace{-2pt}-\hspace{-2pt}m^2)(u\hspace{-2pt}+\hspace{-2pt}m^2)}{um^2}k_4^{\nu_3}k_3^{\nu_4} + \frac{(s\hspace{-2pt}-\hspace{-2pt}m^2)(u\hspace{-2pt}+\hspace{-2pt}m^2)}{um^2}k_1^{\nu_3}k_2^{\nu_4} + \frac{(t\hspace{-2pt}-\hspace{-2pt}m^2)(s\hspace{-2pt}-\hspace{-2pt}m^2)}{um^2}k_2^{\nu_3}k_1^{\nu_4}\bigg]
\\ + \frac{\alpha'}{2}\eta^{\mu_1\mu_4}\eta^{\mu_2\nu_4}\left[\frac{u-m^2}{t}k_1^{\mu_3}k_1^{\nu_3} - 2k_1^{\mu_3}k_2^{\nu_3} + \frac{t-m^2}{u}k_2^{\mu_3}k_2^{\nu_3}\right]
\\ + \frac{\alpha'^2}{4}\Big(\eta^{\mu_1\mu_2}k_1^{\mu_3}k_2^{\mu_4} + \eta^{\mu_1\mu_3}(k_4^{\mu_2}k_1^{\mu_4}-k_1^{\mu_2}k_2^{\mu_4}) - \eta^{\mu_1\mu_4}k_4^{\mu_2}k_1^{\mu_3} - \eta^{\mu_2\mu_3}k_3^{\mu_1}k_2^{\mu_4} 
\\ + \eta^{\mu_2\mu_4}(k_3^{\mu_1}k_2^{\mu_3} - k_2^{\mu_1}k_1^{\mu_3}) + \eta^{\mu_3\mu_4}k_3^{\mu_1}k_4^{\mu_2}\Big)\left[\frac{u-m^2}{t}k_1^{\nu_3}k_2^{\nu_4} + k_2^{\nu_3}k_1^{\nu_4} + \frac{u-m^2}{s}k_4^{\nu_3}k_3^{\nu_4}\right]
\\ +\frac{\alpha'^2}{4}\Big(\eta^{\mu_1\mu_2}k_2^{\mu_3}k_1^{\mu_4} - \eta^{\mu_1\mu_3}k_3^{\mu_2}k_1^{\mu_4} + \eta^{\mu_1\mu_4}(k_3^{\mu_2}k_1^{\mu_3}-k_1^{\mu_2}k_2^{\mu_3}) + \eta^{\mu_2\mu_3}(k_4^{\mu_1}k_2^{\mu_4} -k_2^{\mu_1}k_1^{\mu_4}) 
\\ - \eta^{\mu_2\mu_4}k_4^{\mu_1}k_2^{\mu_3} + \eta^{\mu_3\mu_4}k_4^{\mu_1}k_3^{\mu_2}\Big)\left[k_1^{\nu_3}k_2^{\nu_4} + \frac{t-m^2}{u}k_2^{\nu_3}k_1^{\nu_4} + \frac{t-m^2}{s}k_4^{\nu_3}k_3^{\nu_4}\right].
\end{multline}

\begin{multline}
\mathcal{A}_4[\mathrm{NS_0 NS_0};\mathrm{NS_0 NS_0};\mathrm{NS_1^1 NS_1^1};\mathrm{NS_1^2 NS_1^2}]
\\ = -3!8\pi^2 ig_{\text{s}}^2(2\pi)^{10}\delta^{10}(k_1+k_2+k_3+k_4)\varepsilon_{1,\mu_1\barred{\mu}_1}\varepsilon_{2,\mu_2\barred{\mu}_2}\varepsilon_{3,\mu_3\nu_3\barred{\mu}_3\barred{\nu}_3}\varepsilon_{4,\mu_4\nu_4\rho_4\barred{\mu}_4\barred{\nu}_4\barred{\rho}_4}
\\ \times \frac{\Gamma(1\hspace{-2pt}-\hspace{-2pt}\frac{\alpha's}{4})\Gamma(1\hspace{-2pt}-\hspace{-2pt}\frac{\alpha't}{4})\Gamma(1\hspace{-2pt}-\hspace{-2pt}\frac{\alpha'u}{4})}{\Gamma(\frac{\alpha's}{4})\Gamma(\frac{\alpha't}{4})\Gamma(\frac{\alpha'u}{4})}W^{\mu_1\mu_2\{\mu_3\nu_3\}\mu_4\nu_4\rho_4}W^{\barred{\mu}_1\barred{\mu}_2\{\barred{\mu}_3\barred{\nu}_3\}\barred{\mu}_4\barred{\nu}_4\barred{\rho}_4},
\end{multline}
where
\begin{multline}
W^{\mu_1\mu_2\mu_3\nu_3\mu_4\nu_4\rho_4} = -\frac{u-m^2}{t}\eta^{\mu_1\nu_3}\eta\up{8pt}{[\mu_4}{2pt}{\mu_2}\eta\up{8pt}{\nu_4}{2pt}{\mu_3}k\up{8pt}{\rho_4]}{-2pt}{\hspace{-1pt}2} - \frac{t-m^2}{u}\eta^{\mu_2\nu_3}\eta\up{8pt}{[\mu_4}{2pt}{\mu_1}\eta\up{8pt}{\nu_4}{2pt}{\mu_3}k\up{8pt}{\rho_4]}{-2pt}{\hspace{-1pt}1}
\\ + \frac{\alpha'}{4}\hspace{-2pt}\left(\hspace{-2pt}\frac{s}{2}\eta\up{8pt}{[\mu_4}{1pt}{\mu_1}\eta\up{8pt}{\nu_4}{1pt}{\mu_2}\eta\up{8pt}{\rho_4]}{2pt}{\mu_3} \hspace{-2pt}-\hspace{-2pt} \eta^{\mu_1\mu_2}\eta^{\mu_3[\mu_4}k_1^{\nu_4}k_2^{\rho_4]} \hspace{-2pt}-\hspace{-2pt} \eta\up{8pt}{[\mu_4}{2pt}{\mu_1}\eta\up{8pt}{\nu_4}{2pt}{\mu_3}k\up{8pt}{\rho_4]}{-3pt}{2}k_1^{\mu_2} \hspace{-2pt}+\hspace{-2pt} \eta\up{8pt}{[\mu_4}{1pt}{\mu_2}\eta\up{8pt}{\nu_4}{1pt}{\mu_3}k\up{8pt}{\rho_4]}{-3pt}{1}k_2^{\mu_1}\hspace{-2pt}\right)\hspace{-3pt}\left(\hspace{-2pt}k_{12}^{\nu_3} - \frac{t-u}{s}k_4^{\nu_3}\hspace{-2pt}\right)
\\ - \frac{\alpha'}{2}\left(\eta^{\mu_1\mu_3}\eta^{\mu_2[\mu_4}k_1^{\nu_4}k_2^{\rho_4]} + \eta\up{8pt}{[\mu_4}{1pt}{\mu_1}\eta\up{8pt}{\nu_4}{2pt}{\mu_2}k\up{8pt}{\rho_4]}{-3pt}{2}k_1^{\mu_3}\right)\left(k_2^{\nu_3} - \frac{u-m^2}{t}k_1^{\nu_3}\right)
\\ + \frac{\alpha'}{2}\left(\eta^{\mu_2\mu_3}\eta^{\mu_1[\mu_4}k_1^{\nu_4}k_2^{\rho_4]} + \eta\up{8pt}{[\mu_4}{1pt}{\mu_1}\eta\up{8pt}{\nu_4}{2pt}{\mu_2}k\up{8pt}{\rho_4]}{-3pt}{1}k_2^{\rho_3}\right)\left(k_1^{\nu_3} - \frac{t-m^2}{u}k_2^{\nu_3}\right)
\\ + \frac{\alpha'}{2}\eta\up{8pt}{[\mu_4}{2pt}{\mu_1}\eta\up{8pt}{\nu_4}{2pt}{\mu_3}k\up{8pt}{\rho_4]}{-3pt}{1}\left(k_4^{\mu_2}k_1^{\nu_3} + \frac{t-m^2}{s}k_1^{\mu_2}k_4^{\nu_3} + \frac{t-m^2}{u}k_3^{\mu_2}k_2^{\nu_3}\right)
\\ + \frac{\alpha'}{2}\eta\up{8pt}{[\mu_4}{2pt}{\mu_2}\eta\up{8pt}{\nu_4}{2pt}{\mu_3}k\up{8pt}{\rho_4]}{-3pt}{2}\left(k_4^{\mu_1}k_2^{\nu_3}+\frac{u-m^2}{s}k_2^{\mu_1}k_4^{\nu_3} + \frac{u-m^2}{t}k_3^{\mu_1}k_1^{\nu_3}\right).
\end{multline}

\begin{multline}
\mathcal{A}_4[\mathrm{NS_0 NS_0};\mathrm{NS_0 NS_0};\mathrm{NS_1^2 NS_1^2};\mathrm{NS_1^2 NS_1^2}]
\\ = \frac{16\pi^2 ig_{\text{s}}^2}{\alpha'}(2\pi)^{10}\delta^{10}(k_1+k_2+k_3+k_4)\varepsilon_{1,\mu_1\barred{\mu}_1}\varepsilon_{2,\mu_2\barred{\mu}_2}\varepsilon_{3,\mu_3\nu_3\rho_3\barred{\mu}_3\barred{\nu}_3\barred{\rho}_3}\varepsilon_{4,\mu_4\nu_4\rho_4\barred{\mu}_4\barred{\nu}_4\barred{\rho}_4}
\\ \times \frac{\Gamma(1\hspace{-2pt}-\hspace{-2pt}\frac{\alpha's}{4})\Gamma(1\hspace{-2pt}-\hspace{-2pt}\frac{\alpha't}{4})\Gamma(1\hspace{-2pt}-\hspace{-2pt}\frac{\alpha'u}{4})}{\Gamma(\frac{\alpha's}{4})\Gamma(\frac{\alpha't}{4})\Gamma(\frac{\alpha'u}{4})}X^{\mu_1\mu_2\mu_3\nu_3\rho_3\mu_4\nu_4\rho_4}X^{\barred{\mu}_1\barred{\mu}_2\barred{\mu}_3\barred{\nu}_3\barred{\rho}_3\barred{\mu}_4\barred{\nu}_4\barred{\rho}_4},
\end{multline}
where
\begin{multline}
X^{\mu_1\mu_2\mu_3\nu_3\rho_3\mu_4\nu_4\rho_4}
\\ = \left[\frac{(t\hspace{-2pt}-\hspace{-2pt}m^2)(u\hspace{-2pt}-\hspace{-2pt}m^2)}{sm^2}\eta^{\mu_1\mu_2} \hspace{-2pt}+\hspace{-2pt} \frac{\alpha'}{2}\hspace{-2pt}\left(\hspace{-2pt}\frac{u\hspace{-2pt}-\hspace{-2pt}m^2}{s}k_3^{\mu_1}k_4^{\mu_2} \hspace{-2pt}+\hspace{-2pt} \frac{t\hspace{-2pt}-\hspace{-2pt}m^2}{s}k_4^{\mu_1}k_{3}^{\mu_2}\hspace{-2pt}\right)\hspace{-2pt}\right]\hspace{-3pt}\eta\up{8pt}{[\mu_3}{2pt}{\mu_4}\eta\up{8pt}{\nu_3}{2pt}{\nu_4}\eta\up{8pt}{\rho_3]}{2pt}{\rho_4}
\\ +\frac{\alpha'}{2}\bigg[\frac{6(u-m^2)}{t}\eta^{\mu_1[\mu_3}k_1^{\nu_3}\eta^{\rho_3][\mu_4}k_2^{\nu_4}\eta^{\rho_4]\mu_2} + \frac{6(t-m^2)}{u}\eta^{\mu_1[\mu_4}k_1^{\nu_4}\eta^{\rho_4][\mu_3}k_2^{\nu_3}\eta^{\rho_3]\mu_2}
\\ + 6\left(\eta^{\mu_1[\mu_3}k_2^{\nu_3}\eta^{\rho_3][\mu_4}k_1^{\nu_4}\eta^{\rho_4]\mu_2} + \eta^{\mu_1[\mu_4}k_2^{\nu_4}\eta^{\rho_4][\mu_3}k_1^{\nu_3}\eta^{\rho_3]\mu_2}\right)
\\ - \frac{3(u-m^2)}{s}\bigg(\eta^{\mu_1\mu_2}k\up{2pt}{[\mu_3}{-5pt}{\hspace{-1pt}1}k\up{9pt}{[\mu_4}{-3pt}{2}\eta\up{9pt}{\nu_4}{1pt}{\nu_3}\eta\up{9pt}{\rho_4]}{1pt}{\rho_3]} - k_2^{\mu_1}\eta\up{9pt}{\mu_2}{0pt}{[\mu_4}\eta\up{9pt}{[\mu_3}{0pt}{\nu_4}\eta\up{9pt}{\nu_3}{0pt}{\rho_4]}k\up{9pt}{\rho_3]}{-3pt}{1} - k_1^{\mu_2}\eta\up{9pt}{\mu_1}{0pt}{[\mu_3}\eta\up{9pt}{[\mu_4}{0pt}{\nu_3}\eta\up{9pt}{\nu_4}{0pt}{\rho_3]}k\up{9pt}{\rho_4]}{-3pt}{2} 
\\ - 2\eta\up{8pt}{[\mu_4}{2pt}{\mu_1}\eta\up{8pt}{\nu_4}{2pt}{\mu_2}\eta\up{8pt}{\rho_4][\mu_3}{0pt}{}k\up{8pt}{\nu_3}{-3pt}{1}k\up{8pt}{\rho_3]}{-3pt}{2} - 2\eta\up{0pt}{[\mu_3}{8pt}{\mu_1}\eta\up{0pt}{\nu_3}{8pt}{\mu_2}\eta\up{0pt}{\rho_3]}{8pt}{[\mu_4}k\up{8pt}{\nu_4}{-3pt}{1}k\up{8pt}{\rho_4]}{-3pt}{2} - \frac{s}{2}\eta\up{2pt}{\mu_1}{-6pt}{[\mu_3}\eta\up{9pt}{[\mu_4}{1pt}{\mu_2}\eta\up{9pt}{\nu_4}{-6pt}{\nu_3}\eta\up{9pt}{\rho_4]}{-6pt}{\rho_3]}\bigg)
\\ - \frac{3(t-m^2)}{s}\bigg(\eta^{\mu_1\mu_2}k\up{2pt}{[\mu_4}{-5pt}{\hspace{-1pt}1}k\up{9pt}{[\mu_3}{-3pt}{2}\eta\up{9pt}{\nu_3}{1pt}{\nu_4}\eta\up{9pt}{\rho_3]}{1pt}{\rho_4]} - k_2^{\mu_1}\eta\up{9pt}{\mu_2}{0pt}{[\mu_3}\eta\up{9pt}{[\mu_4}{0pt}{\nu_3}\eta\up{9pt}{\nu_4}{0pt}{\rho_3]}k\up{9pt}{\rho_4]}{-3pt}{1} - k_1^{\mu_2}\eta\up{9pt}{\mu_1}{0pt}{[\mu_4}\eta\up{9pt}{[\mu_3}{0pt}{\nu_4}\eta\up{9pt}{\nu_3}{0pt}{\rho_4]}k\up{9pt}{\rho_3]}{-3pt}{2} 
\\ - 2\eta\up{8pt}{[\mu_3}{2pt}{\mu_1}\eta\up{8pt}{\nu_3}{2pt}{\mu_2}\eta\up{8pt}{\rho_3][\mu_4}{0pt}{}k\up{8pt}{\nu_4}{-3pt}{1}k\up{8pt}{\rho_4]}{-3pt}{2} - 2\eta\up{0pt}{[\mu_4}{8pt}{\mu_1}\eta\up{0pt}{\nu_4}{8pt}{\mu_2}\eta\up{0pt}{\rho_4]}{8pt}{[\mu_3}k\up{8pt}{\nu_3}{-3pt}{1}k\up{8pt}{\rho_3]}{-3pt}{2} - \frac{s}{2}\eta\up{2pt}{\mu_1}{-6pt}{[\mu_4}\eta\up{9pt}{[\mu_3}{1pt}{\mu_2}\eta\up{9pt}{\nu_3}{-6pt}{\nu_4}\eta\up{9pt}{\rho_3]}{-6pt}{\rho_4]}\bigg)
\\ + \left(\frac{t-m^2}{s}k_3^{\mu_2} - \frac{u-m^2}{s}k_4^{\mu_2}\right)\left(3\eta\up{0pt}{[\mu_3}{9pt}{\mu_1}\eta\up{9pt}{[\mu_4}{0pt}{\nu_3}\eta\up{9pt}{\nu_4}{0pt}{\rho_3]}k\up{9pt}{\rho_4]}{-3pt}{1} - 3\eta\up{0pt}{[\mu_4}{9pt}{\mu_1}\eta\up{9pt}{[\mu_3}{0pt}{\nu_4}\eta\up{9pt}{\nu_3}{0pt}{\rho_4]}k\up{9pt}{\rho_3]}{-3pt}{1}\right)
\\ + \left(\frac{u-m^2}{s}k_3^{\mu_1} - \frac{t-m^2}{s}k_4^{\mu_1}\right)\left(3\eta\up{0pt}{[\mu_3}{9pt}{\mu_2}\eta\up{9pt}{[\mu_4}{0pt}{\nu_3}\eta\up{9pt}{\nu_4}{0pt}{\rho_3]}k\up{9pt}{\rho_4]}{-3pt}{2} - 3\eta\up{0pt}{[\mu_4}{9pt}{\mu_2}\eta\up{9pt}{[\mu_3}{0pt}{\nu_4}\eta\up{9pt}{\nu_3}{0pt}{\rho_4]}k\up{9pt}{\rho_3]}{-3pt}{2}\right)\bigg].
\end{multline}

\noindent And then there is this beast:
\begin{multline}
\mathcal{A}_4[\mathrm{NS_0 NS_0};\mathrm{NS_0 NS_0};\mathrm{R_1 R_1};\mathrm{R_1 R_1}] 
\\ = - \frac{8^2\pi^2 i\alpha' g_{\text{s}}^2}{9^2}(2\pi)^{10}\delta^{10}(k_1+k_2+k_3+k_4)\varepsilon_{1,\mu_1\barred{\mu}_1}\varepsilon_{2,\mu_2\barred{\mu}_2}\varepsilon_{3,\mu_3\barred{\mu}_3}^{\alpha_3\barred{\alpha}_3}\varepsilon_{4,\mu_4\barred{\mu}_4}^{\alpha_4\barred{\alpha}_4}
\\ \times \frac{\Gamma(1\hspace{-2pt}-\hspace{-2pt}\frac{\alpha's}{4})\Gamma(1\hspace{-2pt}-\hspace{-2pt}\frac{\alpha't}{4})\Gamma(1\hspace{-2pt}-\hspace{-2pt}\frac{\alpha'u}{4})}{\Gamma(\frac{\alpha's}{4})\Gamma(\frac{\alpha't}{4})\Gamma(\frac{\alpha'u}{4})}Y^{\mu_1\mu_2\mu_3\mu_4}_{\alpha_3\alpha_4}Y^{\barred{\mu}_1\barred{\mu}_2\barred{\mu}_3\barred{\mu}_4}_{\barred{\alpha}_3\barred{\alpha}_4},
\end{multline}
where
\begin{multline}
Y^{\mu_1\mu_2\mu_3\mu_4}_{\alpha_3\alpha_4} = \frac{t-u}{4s}\eta^{\mu_1\mu_2}\eta^{\mu_3\mu_4}\left((k_{12}\cdot\gamma)_{\alpha_3\alpha_4} + \frac{\alpha'}{4}k_3^{\mu}k_{12}^{\nu}k_4^{\rho}(\gamma_{\mu}\gamma_{\nu}\gamma_{\rho})_{\alpha_3\alpha_4}\right)
\\ - \frac{u-m^2}{2t}\eta^{\mu_1\mu_3}\eta^{\mu_2\mu_4}\left((k_{34}\cdot\gamma)_{\alpha_3\alpha_4} + \frac{\alpha'}{4}k_3^{\mu}k_{12}^{\nu}k_4^{\rho}(\gamma_{\mu}\gamma_{\nu}\gamma_{\rho})_{\alpha_3\alpha_4}\right)
\\ - \frac{t-m^2}{2u}\eta^{\mu_1\mu_4}\eta^{\mu_2\mu_3}\left((k_{34}\cdot\gamma)_{\alpha_3\alpha_4} - \frac{\alpha'}{4}k_3^{\mu}k_{12}^{\nu}k_4^{\rho}(\gamma_{\mu}\gamma_{\nu}\gamma_{\rho})_{\alpha_3\alpha_4}\right)
\\ - \frac{\alpha'}{8}\eta^{\mu_1\mu_2}\left[\frac{t-u}{s}k_4^{\mu_3}k_3^{\mu_4} - \left(\frac{u-m^2}{t}-1\right)k_1^{\mu_3}k_2^{\mu_4} + \left(\frac{t-m^2}{u}-1\right)k_2^{\mu_3}k_1^{\mu_4}\right](k_{12}\cdot\gamma)_{\alpha_3\alpha_4}
\\ + \frac{\alpha'}{8}\eta^{\mu_1\mu_2}\left[\frac{t-u}{s}k_{12}^{\mu_3}k_3^{\mu_4} + \left(\frac{u-m^2}{t}-1\right)k_1^{\mu_3}k_2^{\mu_4} + \left(\frac{t-m^2}{u}-1\right)k_2^{\mu_3}k_1^{\mu_4}\right](k_3\cdot\gamma)_{\alpha_3\alpha_4}
\\ + \frac{\alpha'}{8}\eta^{\mu_1\mu_2}\left[\frac{t-u}{s}k_4^{\mu_3}k_{12}^{\mu_4} - \left(\frac{u-m^2}{t}-1\right)k_1^{\mu_3}k_2^{\mu_4} - \left(\frac{t-m^2}{u}-1\right)k_2^{\mu_3}k_1^{\mu_4}\right](k_4\cdot\gamma)_{\alpha_3\alpha_4}
\\ - \frac{1}{4}\eta^{\mu_3\mu_4}\left[\left(\frac{t-u}{m^2}k_{12}^{(\mu_1} + \frac{s}{m^2}k_{34}^{(\mu_1}\right)\gamma^{\mu_2)}_{\alpha_3\alpha_4} - \frac{1}{2}\left(\frac{s}{m^2}k_{12}^{\mu} + \frac{t-u}{m^2}k_{34}^{\mu}\right)(\gamma^{\mu_1\mu_2}_{\phantom{\mu_1\mu_2}\mu})_{\alpha_3\alpha_4}\right]
\\ - \frac{\alpha'}{8}\eta^{\mu_3\mu_4}k_3^{[\mu_1}k_4^{\mu_2]}(k_{12}\cdot\gamma)_{\alpha_3\alpha_4}
\\ + \frac{\alpha'}{4}\eta^{\mu_3\mu_4}\hspace{-3pt}\left[\frac{1}{2}k_{12}^{[\mu_1}\hspace{-3pt}\left(k_4^{\mu_2]}(k_3\hspace{-2pt}\cdot\hspace{-2pt}\gamma)_{\alpha_3\alpha_4}\hspace{-3pt}-\hspace{-3pt}k_3^{\mu_2]}(k_4\hspace{-2pt}\cdot\hspace{-2pt}\gamma)_{\alpha_3\alpha_4}\right) \hspace{-3pt}+\hspace{-3pt} \left(\hspace{-3pt}\frac{u\hspace{-3pt}-\hspace{-3pt}m^2}{s}k_3^{\mu_1}k_4^{\mu_2} \hspace{-3pt}+\hspace{-3pt} \frac{t\hspace{-3pt}-\hspace{-3pt}m^2}{s}k_4^{\mu_1}k_3^{\mu_2}\hspace{-3pt}\right)\hspace{-3pt}(k_{34}\hspace{-2pt}\cdot\hspace{-2pt}\gamma)_{\alpha_3\alpha_4}\hspace{-2pt}\right]
\\ + \frac{\alpha'}{8}\eta^{\mu_3\mu_4}\hspace{-4pt}\left[\hspace{-2pt}k_3^{\mu}k_4^{\nu}\hspace{-3pt}\left(\hspace{-2pt}\frac{2(t\hspace{-3pt}-\hspace{-3pt}u)}{s}\hspace{-1pt}k_{12}^{(\mu_1} \hspace{-4pt}+\hspace{-2pt} k_{34}^{(\mu_1}\hspace{-4pt}\right)\hspace{-3pt}(\hspace{-1pt}\gamma^{\mu_2)}_{\phantom{\mu_2)}\hspace{-3pt}\mu\nu}\hspace{-1pt})_{\hspace{-1pt}\alpha_3\alpha_4} \hspace{-4pt}+\hspace{-2pt} k_{12}^{\mu}\hspace{-1pt}(\hspace{-1pt}k_4^{\nu}k_3^{[\mu_1}\hspace{-5pt}+\hspace{-2pt}k_3^{\nu}k_4^{[\mu_1}\hspace{-1pt})\hspace{-1pt}(\hspace{-1pt}\gamma^{\mu_2]}_{\phantom{\mu_2]}\hspace{-2pt}\mu\nu}\hspace{-1pt})_{\hspace{-1pt}\alpha_3\alpha_4} \hspace{-4pt}+\hspace{-2pt} \frac{1}{2}k_3^{\mu}k_{12}^{\nu}k_4^{\rho}(\hspace{-1pt}\gamma^{\mu_1\mu_2}_{\phantom{\mu_1\mu_2}\hspace{-1pt}\mu\nu\rho}\hspace{-1pt})_{\hspace{-1pt}\alpha_3\alpha_4}\hspace{-2pt}\right]
\\ +\frac{\alpha'}{2}\eta^{\mu_1\mu_3}k_1^{\mu_2}\left(\frac{t-m^2}{s}k_1^{\mu_4} - \frac{u-m^2}{s}k_2^{\mu_4}\right)(k_3\cdot\gamma)_{\alpha_3\alpha_4} 
\\ + \frac{\alpha'}{4}\frac{u-m^2}{t}\eta^{\mu_1\mu_3}k_2^{\mu_4}\left(\frac{t+m^2}{2}\gamma^{\mu_2}_{\alpha_3\alpha_4} + k_3^{\mu_2}(k_1\cdot\gamma)_{\alpha_3\alpha_4} - k_1^{\mu_2}(k_3\cdot\gamma)_{\alpha_3\alpha_4} + k_1^{\mu}k_3^{\nu}(\gamma^{\mu_2}_{\phantom{\mu_2}\mu\nu})_{\alpha_3\alpha_4}\right)
\\ + \frac{\alpha'}{4}\eta^{\mu_1\mu_3}k_1^{\mu_4}\left(\frac{u-m^2}{2}\gamma^{\mu_2}_{\alpha_3\alpha_4} + k_3^{\mu_2}(k_2\cdot\gamma)_{\alpha_3\alpha_4} - 2k_4^{\mu_2}(k_3\cdot\gamma)_{\alpha_3\alpha_4} + k_2^{\mu}k_3^{\nu}(\gamma^{\mu_2}_{\phantom{\mu_2}\mu\nu})_{\alpha_3\alpha_4}\right)
\\ - \frac{\alpha'}{2}\eta^{\mu_1\mu_4}\left(\frac{u-m^2}{s}k_1^{\mu_3} - \frac{t-m^2}{s}k_2^{\mu_3}\right)k_1^{\mu_2}(k_4\cdot\gamma)_{\alpha_3\alpha_4}
\\ - \frac{\alpha'}{4}\frac{t-m^2}{u}\eta^{\mu_1\mu_4}k_2^{\mu_3}\left(\frac{u+m^2}{2}\gamma^{\mu_2}_{\alpha_3\alpha_4} + k_4^{\mu_2}(k_1\cdot\gamma)_{\alpha_3\alpha_4} - k_1^{\mu_2}(k_4\cdot\gamma)_{\alpha_3\alpha_4} - k_1^{\mu}k_4^{\nu}(\gamma^{\mu_2}_{\phantom{\mu_2}\mu\nu})_{\alpha_3\alpha_4}\right)
\\ - \frac{\alpha'}{4}\eta^{\mu_1\mu_4}k_1^{\mu_3}\left(\frac{t-m^2}{2}\gamma^{\mu_2}_{\alpha_3\alpha_4} + k_4^{\mu_2}(k_2\cdot\gamma)_{\alpha_3\alpha_4} - 2k_3^{\mu_2}(k_4\cdot\gamma)_{\alpha_3\alpha_4} - k_2^{\mu}k_4^{\nu}(\gamma^{\mu_2}_{\phantom{\mu_2}\mu\nu})_{\alpha_3\alpha_4}\right)
\\ - \frac{\alpha'}{2}\eta^{\mu_2\mu_3}k_2^{\mu_1}\left(\frac{t-m^2}{s}k_1^{\mu_4} - \frac{u-m^2}{s}k_2^{\mu_4}\right)(k_3\cdot\gamma)_{\alpha_3\alpha_4}
\\ + \frac{\alpha'}{4}\eta^{\mu_2\mu_3}k_2^{\mu_4}\left(\frac{t-m^2}{2}\gamma^{\mu_1}_{\alpha_3\alpha_4} + k_3^{\mu_1}(k_1\cdot\gamma)_{\alpha_3\alpha_4} - 2k_4^{\mu_1}(k_3\cdot\gamma)_{\alpha_3\alpha_4} + k_1^{\mu}k_3^{\nu}(\gamma^{\mu_1}_{\phantom{\mu_1}\mu\nu})_{\alpha_3\alpha_4}\right) 
\\ + \frac{\alpha'}{4}\frac{t-m^2}{u}\eta^{\mu_2\mu_3}k_1^{\mu_4}\left(\frac{u+m^2}{2}\gamma^{\mu_1}_{\alpha_3\alpha_4} + k_3^{\mu_1}(k_2\cdot\gamma)_{\alpha_3\alpha_4} - k_2^{\mu_1}(k_3\cdot\gamma)_{\alpha_3\alpha_4} + k_2^{\mu}k_3^{\nu}(\gamma^{\mu_1}_{\phantom{\mu_1}\mu\nu})_{\alpha_3\alpha_4}\right)
\\ - \frac{\alpha'}{2}\eta^{\mu_2\mu_4}k_2^{\mu_1}\left(\frac{t-m^2}{s}k_2^{\mu_3} - \frac{u-m^2}{s}k_1^{\mu_3}\right)(k_4\cdot\gamma)_{\alpha_3\alpha_4}
\\ - \frac{\alpha'}{4}\eta^{\mu_2\mu_4}k_2^{\mu_3}\left(\frac{u-m^2}{2}\gamma^{\mu_1}_{\alpha_3\alpha_4} + k_4^{\mu_1}(k_1\cdot\gamma)_{\alpha_3\alpha_4} - 2k_3^{\mu_1}(k_4\cdot\gamma)_{\alpha_3\alpha_4} - k_1^{\mu}k_4^{\nu}(\gamma^{\mu_1}_{\phantom{\mu_1}\mu\nu})_{\alpha_3\alpha_4}\right) 
\\ - \frac{\alpha'}{4}\frac{u-m^2}{t}\eta^{\mu_2\mu_4}k_1^{\mu_3}\left(\frac{t+m^2}{2}\gamma^{\mu_1}_{\alpha_3\alpha_4} + k_4^{\mu_1}(k_2\cdot\gamma)_{\alpha_3\alpha_4} - k_2^{\mu_1}(k_4\cdot\gamma)_{\alpha_3\alpha_4} - k_2^{\mu}k_4^{\nu}(\gamma^{\mu_1}_{\phantom{\mu_1}\mu\nu})_{\alpha_3\alpha_4}\right)
\\ + \frac{\alpha'}{4}\bigg[2k_1^{\mu_2}k_1^{[\mu_3}k_2^{\mu_4]} + k_{34}^{\mu_2}k_1^{\mu_3}k_1^{\mu_4} + \frac{t-u}{s}k_1^{\mu_2}k_4^{\mu_3}k_3^{\mu_4} + \frac{u-m^2}{t}k_4^{\mu_2}k_1^{\mu_3}k_2^{\mu_4} - \frac{t-m^2}{u}k_3^{\mu_2}k_2^{\mu_3}k_1^{\mu_4}\bigg]\gamma^{\mu_1}_{\alpha_3\alpha_4}
\\ + \frac{\alpha'}{4}\bigg[2k_2^{\mu_1}k_2^{[\mu_3}k_1^{\mu_4]} + k_{34}^{\mu_1}k_2^{\mu_3}k_2^{\mu_4} - \frac{t-u}{s}k_2^{\mu_1}k_4^{\mu_3}k_3^{\mu_4} + \frac{t-m^2}{u}k_4^{\mu_1}k_2^{\mu_3}k_1^{\mu_4} - \frac{u-m^2}{t}k_3^{\mu_1}k_1^{\mu_3}k_2^{\mu_4}  \bigg]\gamma^{\mu_2}_{\alpha_3\alpha_4}
\\ - \frac{\alpha'}{8}\left(k_4^{\mu_3}k_3^{\mu_4} + \frac{s-m^2}{t}k_1^{\mu_3}k_2^{\mu_4} + \frac{s-m^2}{u}k_2^{\mu_3}k_1^{\mu_4}\right)k_{12}^{\mu}(\gamma^{\mu_1\mu_2}_{\phantom{\mu_1\mu_2}\mu})_{\alpha_3\alpha_4}
\\ + \frac{\alpha'}{8}\left[k_{12}^{[\mu_3}k_{34}^{\mu_4]}(k_3^{\mu}+k_4^{\mu}) + \left(\frac{u-m^2}{t}k_1^{\mu_3}k_2^{\mu_4} - \frac{t-m^2}{u}k_2^{\mu_3}k_1^{\mu_4}\right)k_{34}^{\mu}\right](\gamma^{\mu_1\mu_2}_{\phantom{\mu_1\mu_2}\mu})_{\alpha_3\alpha_4}.
\end{multline}
Believe it or not, the symmetries under $1\rightleftharpoons 2$ and under $3\rightleftharpoons 4$ are both manifest.

\subsubsection{With One Massless NSNS State and One Massless RR State}

These amplitudes have no symmetry and so have no nice way to combine the terms.
\begin{multline}
\mathcal{A}_4[\mathrm{NS_0 NS_0};\mathrm{NS_1^1 NS_1^1};\mathrm{R_0 R_0};\mathrm{R_1 R_1}]
\\ = \frac{8\pi^2 i\alpha' g_{\text{s}}^2}{9}(2\pi)^{10}\delta^{10}(k_1+k_2+k_3+k_4)\varepsilon_{1,\mu_1\barred{\mu}_1}\varepsilon_{2,\mu_2\nu_2\barred{\mu}_2\barred{\nu}_2}\varepsilon_3^{\alpha_3\barred{\alpha}_3}\varepsilon_{4,\mu_4\barred{\mu}_4}^{\alpha_4\barred{\alpha}_4}
\\ \times \frac{\Gamma(1\hspace{-2pt}-\hspace{-2pt}\frac{\alpha's}{4})\Gamma(1\hspace{-2pt}-\hspace{-2pt}\frac{\alpha't}{4})\Gamma(1\hspace{-2pt}-\hspace{-2pt}\frac{\alpha'u}{4})}{\Gamma(\frac{\alpha's}{4})\Gamma(\frac{\alpha't}{4})\Gamma(\frac{\alpha'u}{4})}Z^{\mu_1\{\mu_2\nu_2\}\mu_4}_{\alpha_3\alpha_4}Z^{\barred{\mu}_1\{\barred{\mu}_2\barred{\nu}_2\}\barred{\mu}_4}_{\barred{\alpha}_3\barred{\alpha}_4},
\end{multline}
where
\begin{multline}
Z^{\mu_1\mu_2\nu_2\mu_4}_{\alpha_3\alpha_4} = \eta^{\mu_1\mu_2}\eta^{\nu_2\mu_4}\left[\left(1 + \frac{t-m^2}{s}\right)(k_4\cdot\gamma)_{\alpha_3\alpha_4} + \frac{1}{2}\left(\frac{u-m^2}{t}-1\right)(k_1\cdot\gamma)_{\alpha_3\alpha_4}\right] 
\\ + \bigg[\eta^{\mu_1\mu_2}\left(\frac{t-m^2}{s}k_2^{\mu_4}-k_3^{\mu_4}\right) + \eta^{\mu_1\mu_4}\left(\frac{t-m^2}{u}k_3^{\mu_2} - k_4^{\mu_2}\right) - \eta^{\mu_2\mu_4}\left(\frac{u-m^2}{t}k_3^{\mu_1} - k_4^{\mu_1}\right)\bigg]\gamma^{\nu_2}_{\alpha_3\alpha_4}
\\ - \frac{1}{2}\eta^{\nu_2\mu_4}\left[\left(1+\frac{2(t-m^2)}{s}\right)k_1^{\mu_2}\gamma^{\mu_1}_{\alpha_3\alpha_4} - \left(\frac{u-m^2}{t} + 1\right)k_1^{\mu}(\gamma^{\mu_1\mu_2}_{\phantom{\mu_1\mu_2}\mu})_{\alpha_3\alpha_4}\right]
\\ + \frac{\alpha'}{2}\eta^{\mu_1\mu_2}k_1^{\mu_4}\left[\left(\frac{t-m^2}{s} + \frac{t-m^2}{2u}\right)k_3^{\nu_2} + \left(\frac{1}{2}+\frac{t-m^2}{s}\right)k_4^{\nu_2}\right](k_4\cdot\gamma)_{\alpha_3\alpha_4}
\\ + \frac{\alpha'}{4}\eta^{\mu_1\mu_2} \left[\frac{t-u}{s}k_1^{\nu_2}k_3^{\mu_4} + \left(\frac{t-m^2}{u} - 1\right)k_3^{\nu_2}k_1^{\mu_4} - \left(\frac{u-m^2}{t} - 1\right)k_4^{\nu_2}k_2^{\mu_4}\right](k_1\cdot\gamma)_{\alpha_3\alpha_4}
\\ + \frac{\alpha'}{4}\eta^{\mu_1\mu_4}k_1^{\mu_2}\left[\frac{t-u}{s}k_1^{\nu_2} - \left(\frac{t-m^2}{u}-1\right)k_3^{\nu_2}\right](k_4\cdot\gamma)_{\alpha_3\alpha_4}
\\ + \frac{\alpha'}{4}\eta^{\mu_1\mu_4}\left(\frac{t-m^2}{u}k_3^{\nu_2} - k_4^{\nu_2}\right)\left(\frac{u-m^2}{2}\gamma^{\mu_2}_{\alpha_3\alpha_4} + k_4^{\mu_2}(k_1\cdot\gamma)_{\alpha_3\alpha_4} - k_1^{\mu}k_4^{\nu}(\gamma^{\mu_2}_{\phantom{\mu_2}\mu\nu})_{\alpha_3\alpha_4}\right)
\\ + \frac{\alpha'}{4}\eta^{\mu_2\mu_4}\left(k_3^{\nu_2} - \frac{u-m^2}{t}k_4^{\nu_2}\right)\left(\frac{u-m^2}{2}\gamma^{\mu_1}_{\alpha_3\alpha_4} + k_4^{\mu_1}(k_1\hspace{-2pt}\cdot\hspace{-2pt}\gamma)_{\alpha_3\alpha_4} - k_1^{\mu}k_4^{\nu}(\gamma^{\mu_1}_{\phantom{\mu_1}\mu\nu})_{\alpha_3\alpha_4}\right)
\\ - \frac{\alpha'}{2}\eta^{\mu_2\mu_4}\left[k_3^{\mu_1}k_3^{\nu_2} + k_4^{\mu_1}k_4^{\nu_2} + \frac{t-m^2}{s}k_2^{\mu_1}k_1^{\nu_2} - \left(\frac{u-m^2}{t} - 1\right)k_3^{\mu_1}k_4^{\nu_2}\right](k_4\cdot\gamma)_{\alpha_3\alpha_4}
\\ - \frac{\alpha'}{4}\left[\frac{t-u}{s}k_1^{\mu_2}k_3^{\mu_4} + \left(\frac{t-m^2}{u} - 1\right)k_3^{\mu_2}k_1^{\mu_4} - \left(\frac{u-m^2}{t} - 1\right)k_4^{\mu_2}k_2^{\mu_4}\right]k_1^{\nu_2}\gamma^{\mu_1}_{\alpha_3\alpha_4}
\\ - \frac{\alpha'}{4}k_1^{\mu_4}\left(\frac{t-m^2}{u}k_3^{\mu_2} - k_4^{\mu_2}\right)k_4^{\nu_2}\gamma^{\mu_1}_{\alpha_3\alpha_4} + \frac{\alpha'}{4}k_4^{\mu_1}\left(\frac{t-m^2}{u}k_3^{\mu_2} - k_4^{\mu_2}\right)k_1^{\mu_4}\gamma^{\nu_2}_{\alpha_3\alpha_4}
\\ + \frac{\alpha'}{2}\hspace{-2pt}\bigg[\hspace{-2pt}k_2^{\mu_1}\hspace{-2pt}k_1^{\mu_2}\hspace{-2pt}k_3^{\mu_4} - \frac{t\hspace{-3pt}-\hspace{-3pt}m^2}{u}k_4^{\mu_1}\hspace{-2pt}k_3^{\mu_2}\hspace{-2pt}k_1^{\mu_4} - \frac{t\hspace{-3pt}-\hspace{-3pt}m^2}{s}(k_2^{\mu_1}\hspace{-2pt}k_1^{\mu_2}\hspace{-2pt}k_2^{\mu_4} - k_4^{\mu_1}\hspace{-2pt}k_3^{\mu_2}\hspace{-2pt}k_2^{\mu_4} + k_3^{\mu_1}\hspace{-2pt}k_4^{\mu_2}\hspace{-2pt}k_1^{\mu_4}) + \frac{u\hspace{-3pt}-\hspace{-3pt}m^2}{t} k_3^{\mu_1}\hspace{-2pt}k_4^{\mu_2}\hspace{-2pt}k_2^{\mu_4}\hspace{-2pt}\bigg]\hspace{-2pt}\gamma^{\nu_2}_{\alpha_3\alpha_4}
\\ + \frac{\alpha'}{4}\left[k_3^{\nu_2}k_2^{\mu_4} + k_4^{\nu_2}k_1^{\mu_4} - \frac{t-m^2}{u}k_3^{\nu_2}k_1^{\mu_4} - \frac{u-m^2}{t}k_4^{\nu_2}k_2^{\mu_4}\right]k_1^{\mu}(\gamma^{\mu_1\mu_2}_{\phantom{\mu_1\mu_2}\mu})_{\alpha_3\alpha_4}
\\ - \frac{\alpha'}{4}k_1^{\mu_4}\left(\frac{t-m^2}{u}k_3^{\nu_2} - k_4^{\nu_2}\right)k_4^{\mu}(\gamma^{\mu_1\mu_2}_{\phantom{\mu_1\mu_2}\mu})_{\alpha_3\alpha_4}.
\end{multline}

\begin{multline}
\mathcal{A}_4[\mathrm{NS_0 NS_0};\mathrm{NS_1^2 NS_1^2};\mathrm{R_0 R_0};\mathrm{R_1 R_1}]
\\ = -\frac{2\pi^2 ig_{\text{s}}^2}{27}(2\pi)^{10}\delta^{10}(k_1+k_2+k_3+k_4)\varepsilon_{1,\mu_1\barred{\mu}_1}\varepsilon_{2,\mu_2\nu_2\rho_2\barred{\mu}_2\barred{\nu}_2\barred{\rho}_2}\varepsilon_3^{\alpha_3\barred{\alpha}_3}\varepsilon_{4,\mu_4\barred{\mu}_4}^{\alpha_4\barred{\alpha}_4}
\\ \times \frac{\Gamma(1\hspace{-2pt}-\hspace{-2pt}\frac{\alpha's}{4})\Gamma(1\hspace{-2pt}-\hspace{-2pt}\frac{\alpha't}{4})\Gamma(1\hspace{-2pt}-\hspace{-2pt}\frac{\alpha'u}{4})}{\Gamma(\frac{\alpha's}{4})\Gamma(\frac{\alpha't}{4})\Gamma(\frac{\alpha'u}{4})}\Delta^{\mu_1\mu_2\nu_2\rho_2\mu_4}_{\alpha_3\alpha_4}\Delta^{\barred{\mu}_1\barred{\mu}_2\barred{\nu}_2\barred{\rho}_2\barred{\mu}_4}_{\barred{\alpha}_3\barred{\alpha}_4},
\end{multline}
where\footnote{Alas, the English alphabet has ended.}
\begin{multline}
\Delta^{\mu_1\mu_2\nu_2\rho_2\mu_4}_{\alpha_3\alpha_4} = -\frac{s-m^2}{u}\eta^{\mu_1\mu_4}(\gamma^{\mu_2\nu_2\rho_2})_{\alpha_3\alpha_4}
\\ + \frac{\alpha'}{2}\left[k_3^{\mu_1}k_1^{\mu_4} + k_4^{\mu_1}k_2^{\mu_4} - \frac{t-m^2}{u}k_4^{\mu_1}k_2^{\mu_4} - \frac{u-m^2}{t}k_3^{\mu_1}k_2^{\mu_4}\right](\gamma^{\mu_2\nu_2\rho_2})_{\alpha_3\alpha_4}
\\ -\frac{3\alpha'}{2}\left[\left(\frac{4(t-m^2)}{s}+\frac{t-m^2}{u}+1\right)k_1^{\mu_4} + \left(\frac{4(t-m^2)}{s}+\frac{s-m^2}{t}+4\right)k_2^{\mu_4}\right]\eta^{\mu_1[\mu_2}k_1^{\nu_2}\gamma^{\rho_2]}_{\alpha_3\alpha_4}
\\ - \frac{3\alpha'}{4}\left[\left(\frac{t-m^2}{u}-1\right)k_1^{\mu_4} + \left(\frac{u-m^2}{t}-1\right)k_2^{\mu_4}\right]\left(k_1^{[\mu_2}(\gamma^{\nu_2\rho_2]\mu_1})_{\alpha_3\alpha_4} - k_1^{\mu}\eta^{\mu_1[\mu_2}(\gamma^{\nu_2\rho_2]}_{\phantom{\nu_2\rho_2]}\mu})_{\alpha_3\alpha_4}\right)
\\ + \frac{\alpha'}{4}\left(\frac{s-m^2}{u}k_1^{\mu_4} - \frac{s-m^2}{t}k_2^{\mu_4}\right)k_1^{\mu}(\gamma^{\mu_1\mu_2\nu_2\rho_2}_{\phantom{\mu_1\mu_2\nu_2\rho_2}\mu})_{\alpha_3\alpha_4}
\\ + \frac{\alpha'}{2}\left(\frac{u-m^2}{t}k_3^{\mu_1} - k_4^{\mu_1}\right)\left[6\eta^{\mu_4[\mu_2}k_4^{\nu_2}\gamma^{\rho_2]}_{\alpha_3\alpha_4} - 3k_4^{\mu}\eta^{\mu_4[\mu_2}(\gamma^{\nu_2\rho_2]}_{\phantom{\nu_2\rho_2]}\mu})_{\alpha_3\alpha_4}\right]
\\ + \alpha'\left(\frac{6(t-m^2)}{s} + \frac{3(s-m^2)}{2t}+14\right)\eta\up{8pt}{[\mu_2}{2pt}{\mu_1}\eta\up{8pt}{\nu_2}{2pt}{\mu_4}k\up{8pt}{\rho_2]}{-3pt}{1}(k_4\cdot\gamma)_{\alpha_3\alpha_4}
\\ - 3\alpha'\bigg[\eta^{\mu_1\mu_4}k_1^{[\mu_2}k_4^{\nu_2}\gamma^{\rho_2]}_{\alpha_3\alpha_4} - \frac{u-m^2}{2}\eta\up{2pt}{\mu_1}{8pt}{[\mu_2}\eta\up{8pt}{\nu_2}{2pt}{\mu_4}\gamma\up{8pt}{\rho_2]}{-3pt}{\alpha_3\alpha_4} + k_4^{\mu_1}\eta^{\mu_4[\mu_2}k_1^{\nu_2}\gamma^{\rho_2]}_{\alpha_3\alpha_4}
\\  - k_1^{\mu_4}\eta^{\mu_1[\mu_2}k_4^{\nu_2}\gamma^{\rho_2]}_{\alpha_3\alpha_4} + \eta^{\mu_4[\mu_2}k_4^{\nu_2}k_1^{\rho_2]}\gamma^{\mu_1}_{\alpha_3\alpha_4} + k_4^{\mu}\eta^{\mu_4[\mu_2}k_1^{\nu_2}(\gamma^{\rho_2]\mu_1}_{\phantom{\rho_2]\mu_1}\mu})_{\alpha_3\alpha_4}\bigg]
\\ - \frac{3\alpha'}{4}\left(\frac{t-m^2}{u}-1\right)\left[\eta^{\mu_1\mu_4}k_4^{\mu}k_1^{[\mu_2}(\gamma^{\nu_2\rho_2]}_{\phantom{\nu_2\rho_2]}\mu})_{\alpha_3\alpha_4} - k_1^{\mu_4}k_4^{\mu}\eta^{\mu_1[\mu_2}(\gamma^{\nu_2\rho_2]}_{\phantom{\nu_2\rho_2]}\mu})_{\alpha_3\alpha_4}\right]
\\ - \frac{\alpha'}{4}\hspace{-3pt}\left(\hspace{-3pt}\frac{s\hspace{-3pt}-\hspace{-3pt}m^2}{u}\hspace{-3pt}\right)\hspace{-4pt}\bigg[\eta^{\mu_1\mu_4}\hspace{-3pt}\left(\hspace{-3pt}6k_1^{[\mu_2}k_4^{\nu_2}\gamma^{\rho_2]}_{\alpha_3\alpha_4} \hspace{-3pt}+\hspace{-3pt} \frac{u\hspace{-3pt}-\hspace{-3pt}m^2}{2}(\gamma^{\mu_2\nu_2\rho_2})_{\alpha_3\alpha_4} \hspace{-3pt}+\hspace{-3pt} 3k_1^{\mu}k_4^{[\mu_2}(\gamma^{\nu_2\rho_2]}_{\phantom{\nu_2\rho_2]}\mu})_{\alpha_3\alpha_4} \hspace{-3pt}-\hspace{-3pt} k_1^{\mu}k_4^{\nu}(\gamma^{\mu_2\nu_2\rho_2}_{\phantom{\mu_2\nu_2\rho_2}\mu\nu})_{\alpha_3\alpha_4}\hspace{-3pt}\right)
\\ - k_1^{\mu_4}\hspace{-3pt}\left(6\eta^{\mu_1[\mu_2}k_4^{\nu_2}\gamma^{\rho_2]}_{\alpha_3\alpha_4} \hspace{-3pt}-\hspace{-3pt} k_4^{\mu_1}(\gamma^{\mu_2\nu_2\rho_2})_{\alpha_3\alpha_4} \hspace{-3pt}+\hspace{-3pt} 3k_4^{[\mu_2}(\gamma^{\nu_2\rho_2]\mu_1})_{\alpha_3\alpha_4} \hspace{-3pt}+\hspace{-3pt} k_4^{\mu}(\gamma^{\mu_1\mu_2\nu_2\rho_2}_{\phantom{\nu_1]\mu_2\nu_2\rho_2}\mu})_{\alpha_3\alpha_4}\right)\bigg]
\\ - \frac{3\alpha'}{2}\left(\frac{u-m^2}{t}-1\right)\left[\eta^{\mu_4[\mu_2}k_4^{\nu_2}\eta^{\rho_2]\mu_1}(k_1\cdot\gamma)_{\alpha_3\alpha_4} - k_1^{\mu}k_4^{\nu}\eta\up{8pt}{[\mu_2}{2pt}{\mu_1}\eta\up{8pt}{\nu_2}{2pt}{\mu_4}(\gamma\up{8pt}{\rho_2]}{0pt}{}\up{0pt}{}{-2pt}{\mu\nu})_{\alpha_3\alpha_4}\right]
\\ + \frac{3\alpha'(s\hspace{-3pt}-\hspace{-3pt}m^2)}{2t}\hspace{-2pt}\bigg[\hspace{-2pt}\eta^{\mu_4[\mu_2}\hspace{-2pt}k_1^{\nu_2}\hspace{-2pt}k_4^{\rho_2]}\hspace{-2pt}\gamma^{\mu_1}_{\alpha_3\alpha_4} \hspace{-2pt}- k_4^{\mu_1}\hspace{-2pt}\eta^{\mu_4[\mu_2}\hspace{-2pt}k_1^{\nu_2}\gamma^{\rho_2]}_{\alpha_3\alpha_4} + \frac{u\hspace{-3pt}-\hspace{-3pt}m^2}{2}\hspace{-3pt}\left(\hspace{-3pt}\eta\up{8pt}{[\mu_2}{2pt}{\mu_1}\eta\up{8pt}{\nu_2}{2pt}{\mu_4}\gamma\up{8pt}{\rho_2]}{-3pt}{\alpha_3\alpha_4} \hspace{-2pt}-\hspace{-2pt} \frac{1}{2}\eta^{\mu_4[\mu_2}(\gamma^{\nu_2\rho_2]\mu_1})_{\alpha_3\alpha_4}\hspace{-3pt}\right) 
\\ - (\hspace{-1pt}k_1^{\mu}\hspace{-2pt}\eta^{\mu_4[\mu_2}\hspace{-2pt}k_4^{\nu_2}\hspace{-1pt}+\hspace{-1pt}k_4^{\mu}\hspace{-2pt}\eta^{\mu_4[\mu_2}\hspace{-2pt}k_1^{\nu_2}\hspace{-1pt})(\hspace{-1pt}\gamma^{\rho_2]\mu_1}_{\phantom{\rho_2]\mu_1}\hspace{-1pt}\mu}\hspace{-1pt})_{\alpha_3\alpha_4} - \frac{1}{2}\hspace{-1pt}k_4^{\mu_1}\hspace{-2pt}k_1^{\mu}\hspace{-1pt}\eta^{\mu_4[\mu_2}\hspace{-1pt}(\hspace{-1pt}\gamma^{\nu_2\rho_2]}_{\phantom{\nu_2\rho_2]}\hspace{-1pt}\mu}\hspace{-1pt})_{\alpha_3\alpha_4} + \frac{1}{2}\hspace{-1pt}k_1^{\mu}\hspace{-2pt}k_4^{\nu}\hspace{-1pt}\eta^{\mu_4[\mu_2}\hspace{-1pt}(\hspace{-1pt}\gamma^{\nu_2\rho_2]\mu_1}_{\phantom{\nu_2\rho_2]\hspace{-1pt}\mu_1}\mu\nu}\hspace{-1pt})_{\alpha_3\alpha_4}\hspace{-2pt}\bigg].
\end{multline}

\subsubsection{With Two Massless RR States}

\begin{multline}
\mathcal{A}_{4}[\mathrm{NS_1^1 NS_1^1};\mathrm{NS_1^1 NS_1^1};\mathrm{R_0 R_0};\mathrm{R_0 R_0}]
\\ = \frac{\pi^2 i\alpha'g_{\text{s}}^2}{4} (2\pi)^{10}\delta^{10}(k_1+k_2+k_3+k_4)\varepsilon_{1,\mu_1\nu_1\barred{\mu}_1\barred{\nu}_1}\varepsilon_{2,\mu_2\nu_2\barred{\mu}_2\barred{\nu}_2}\varepsilon_3^{\alpha_3\barred{\alpha}_3}\varepsilon_4^{\alpha_4\barred{\alpha}_4}
\\ \times \frac{\Gamma(1\hspace{-2pt}-\hspace{-2pt}\frac{\alpha's}{4})\Gamma(1\hspace{-2pt}-\hspace{-2pt}\frac{\alpha't}{4})\Gamma(1\hspace{-2pt}-\hspace{-2pt}\frac{\alpha'u}{4})}{\Gamma(\frac{\alpha's}{4})\Gamma(\frac{\alpha't}{4})\Gamma(\frac{\alpha'u}{4})}\Theta^{\{\mu_1\nu_1\}\{\mu_2\nu_2\}}_{\alpha_3\alpha_4}\Theta^{\{\barred{\mu}_1\barred{\nu}_1\}\{\barred{\mu}_2\barred{\nu}_2\}}_{\barred{\alpha}_3\barred{\alpha}_4},
\end{multline}
where
\begin{multline}
\Theta^{\mu_1\nu_1\mu_2\nu_2}_{\alpha_3\alpha_4} = \eta^{\mu_1\mu_2}\left[\frac{4(t-u)}{s}\hspace{-2pt}\left(\hspace{-2pt}k_{12}^{(\nu_{1}}\hspace{-1pt}\gamma^{\nu_2)}_{\alpha_3\alpha_4} \hspace{-2pt}-\hspace{-2pt} \frac{1}{8}\eta^{\nu_1\nu_2}\hspace{-1pt}(\hspace{-1pt}k_{12}\hspace{-2pt}\cdot\hspace{-2pt}\gamma\hspace{-1pt})_{\alpha_3\alpha_4}\hspace{-2pt}\right) \hspace{-2pt}+\hspace{-2pt} 3\hspace{-2pt}\left(\hspace{-2pt}k_{34}^{(\nu_1}\hspace{-1pt}\gamma^{\nu_2)}_{\alpha_3\alpha_4} \hspace{-2pt}-\hspace{-2pt} \frac{1}{6}k_{12}^{\mu}(\hspace{-1pt}\gamma^{\nu_1\nu_2}_{\phantom{\nu_1\nu_2}\mu}\hspace{-1pt})_{\alpha_3\alpha_4}\hspace{-2pt}\right)\right] 
\\ -\alpha'\hspace{-3pt}\left[\hspace{-2pt}\frac{t\hspace{-3pt}-\hspace{-3pt}u}{s}\hspace{-1pt}(\hspace{-1pt}k_3^{\hspace{-0.5pt}\mu_{\hspace{-0.5pt}1}}\hspace{-2pt}k_3^{\hspace{-0.5pt}\mu_{\hspace{-0.5pt}2}}\hspace{-4pt}+\hspace{-2pt}k_4^{\hspace{-0.5pt}\mu_{\hspace{-0.5pt}1}}\hspace{-2pt}k_4^{\hspace{-0.5pt}\mu_{\hspace{-0.5pt}2}}\hspace{-1pt})\hspace{-2pt}+\hspace{-3pt}\left(\hspace{-3pt}\frac{2\hspace{-1pt}(\hspace{-1pt}m^2\hspace{-4pt}-\hspace{-3pt}u\hspace{-1pt})}{s}\hspace{-2pt}+\hspace{-2pt}\frac{m^2\hspace{-4pt}-\hspace{-3pt}u}{t}\hspace{-3pt}\right)\hspace{-4pt}k_3^{\hspace{-0.5pt}\mu_{\hspace{-0.5pt}1}}\hspace{-2pt}k_4^{\hspace{-0.5pt}\mu_{\hspace{-0.5pt}2}} \hspace{-3pt}-\hspace{-3pt} \left(\hspace{-3pt}\frac{2\hspace{-1pt}(\hspace{-1pt}m^2\hspace{-4pt}-\hspace{-3pt}t\hspace{-1pt})}{s}\hspace{-3pt}+\hspace{-2pt}\frac{m^2\hspace{-4pt}-\hspace{-3pt}t}{u}\hspace{-3pt}\right)\hspace{-3pt}k_4^{\hspace{-0.5pt}\mu_{\hspace{-0.5pt}1}}\hspace{-2pt}k_3^{\hspace{-0.5pt}\mu_{\hspace{-0.5pt}2}}\hspace{-2pt}\right]\hspace{-5pt}\left(\hspace{-3pt}k_{\hspace{-0.5pt}1\hspace{-0.5pt}2}^{\hspace{-2pt}(\hspace{-1pt}\nu_{\hspace{-0.5pt}1}}\hspace{-2pt}\gamma^{\hspace{-1pt}\nu_2\hspace{-1pt})}_{\hspace{-1pt}\alpha_3\alpha_4}\hspace{-3pt}-\hspace{-2pt}\frac{1}{4}\hspace{-1pt}\eta^{\hspace{-1pt}\nu_1\hspace{-1pt}\nu_2}\hspace{-1pt}(\hspace{-1pt}k_{12}\hspace{-2pt}\cdot\hspace{-2pt}\gamma\hspace{-1pt})_{\alpha_3\alpha_4}\hspace{-4pt}\right) 
\\ - \frac{\alpha'}{2}\hspace{-3pt}\left[\hspace{-2pt}k_3^{\mu_1}\hspace{-2pt}k_3^{\mu_2}\hspace{-3pt}+\hspace{-2pt}k_4^{\mu_1}\hspace{-2pt}k_4^{\mu_2} \hspace{-2pt}+\hspace{-2pt} \frac{m^2\hspace{-3pt}-\hspace{-2pt}u}{t}k_3^{\mu_1}\hspace{-2pt}k_4^{\mu_2} \hspace{-2pt}+\hspace{-2pt} \frac{m^2\hspace{-3pt}-\hspace{-2pt}t}{u}k_4^{\mu_1}\hspace{-2pt}k_3^{\mu_2}\hspace{-1pt}\right]\hspace{-4pt}\left(\hspace{-3pt}k_{34}^{\hspace{-1pt}(\hspace{-1pt}\nu_1}\hspace{-1pt}\gamma^{\nu_2\hspace{-1pt})}_{\alpha_3\alpha_4}\hspace{-2pt}-\hspace{-2pt}\frac{1}{2}k_{12}^{\mu}(\hspace{-1pt}\gamma^{\nu_1\nu_2}_{\phantom{\nu_1\nu_2}\mu})_{\alpha_3\alpha_4}\hspace{-3pt}\right).
\end{multline}

\begin{multline}\label{NS11NS12R0R0}
\mathcal{A}_{4}[\mathrm{NS_1^1 NS_1^1};\mathrm{NS_1^2 NS_1^2};\mathrm{R_0 R_0};\mathrm{R_0 R_0}]
\\ = -\frac{\pi^2 ig_{\text{s}}^2}{3!8} (2\pi)^{10}\delta^{10}(k_1+k_2+k_3+k_4)\varepsilon_{1,\mu_1\nu_1\barred{\mu}_1\barred{\nu}_1}\varepsilon_{2,\mu_2\nu_2\rho_2\barred{\mu}_2\barred{\nu}_2\barred{\rho}_2}\varepsilon_3^{\alpha_3\barred{\alpha}_3}\varepsilon_4^{\alpha_4\barred{\alpha}_4}
\\ \times \frac{\Gamma(1\hspace{-2pt}-\hspace{-2pt}\frac{\alpha's}{4})\Gamma(1\hspace{-2pt}-\hspace{-2pt}\frac{\alpha't}{4})\Gamma(1\hspace{-2pt}-\hspace{-2pt}\frac{\alpha'u}{4})}{\Gamma(\frac{\alpha's}{4})\Gamma(\frac{\alpha't}{4})\Gamma(\frac{\alpha'u}{4})}\Lambda^{\{\mu_1\nu_1\}\mu_2\nu_2\rho_2}_{\alpha_3\alpha_4}\Lambda^{\{\barred{\mu}_1\barred{\nu}_1\}\barred{\mu}_2\barred{\nu}_2\barred{\rho}_2}_{\barred{\alpha}_3\barred{\alpha}_4},
\end{multline}
where
\begin{multline}\label{Lambda}
\Lambda^{\mu_1\nu_1\mu_2\nu_2\rho_2}_{\alpha_3\alpha_4} = 6\eta^{\mu_1[\mu_2}(\gamma^{\nu_2\rho_2]\nu_1})_{\alpha_3\alpha_4} + 6\alpha'\left(\frac{t\hspace{-2pt}-\hspace{-2pt}u}{s}k_2^{\nu_1} - k_{34}^{\nu_1}\right)\eta^{\mu_1[\mu_2}k_1^{\nu_2}\gamma^{\rho_2]}_{\alpha_3\alpha_4}  
\\ \hspace{-25pt}-\frac{\alpha'}{8}\hspace{-4pt}\left(\hspace{-3pt}\frac{s\hspace{-3pt}-\hspace{-3pt}u}{t}\hspace{-1pt}k_3^{\nu_1} \hspace{-3pt}-\hspace{-3pt} \frac{s\hspace{-2pt}-\hspace{-2pt}t}{u}\hspace{-1pt}k_4^{\nu_1} \hspace{-3pt}-\hspace{-3pt} k_{34}^{\nu_1}\hspace{-3pt}\right)\hspace{-5pt}\left[\hspace{-2pt}1\hspace{-1pt}2\eta^{\mu_1\hspace{-1pt}[\hspace{-1pt}\mu_{\hspace{-0.5pt}2}}\hspace{-1pt}k_1^{\hspace{-1pt}\nu_{\hspace{-0.5pt}2}}\hspace{-1pt}\gamma^{\rho_2]}_{\alpha_3\alpha_4} \hspace{-3pt}-\hspace{-3pt} 3k_{34}^{\hspace{-1pt}[\hspace{-0.5pt}\mu_2}\hspace{-1pt}(\hspace{-1pt}\gamma^{\nu_2\hspace{-0.5pt}\rho_2\hspace{-1pt}]\hspace{-0.5pt}\mu_1}\hspace{-1pt})_{\hspace{-1pt}\alpha_3\alpha_4} \hspace{-3pt}-\hspace{-3pt} k_{34}^{\mu_1}\hspace{-1pt}(\hspace{-1pt}\gamma^{\mu_2\hspace{-0.5pt}\nu_2\hspace{-0.5pt}\rho_2}\hspace{-1pt})_{\hspace{-1pt}\alpha_3\alpha_4} \hspace{-3pt}+\hspace{-3pt} k_{12}^{\mu}\hspace{-1pt}(\hspace{-1pt}\gamma^{\mu_1\hspace{-0.5pt}\mu_2\hspace{-0.5pt}\nu_2\hspace{-0.5pt}\rho_2}_{\phantom{\mu_1\mu_2\nu_2\rho_2}\hspace{-2pt}\mu}\hspace{-1pt})_{\hspace{-1pt}\alpha_3\alpha_4}\hspace{-2pt}\right]
\\ \hspace{-25pt}+ \frac{\alpha'}{8}\hspace{-4pt}\left(\hspace{-4pt}3\hspace{-1pt}k_2^{\hspace{-1pt}\nu_1} \hspace{-3pt}-\hspace{-3pt} \frac{s\hspace{-3pt}-\hspace{-3pt}u}{t}\hspace{-1pt}k_3^{\hspace{-1pt}\nu_1}\hspace{-3pt}-\hspace{-3pt}\frac{s\hspace{-3pt}-\hspace{-3pt}t}{u}\hspace{-1pt}k_4^{\hspace{-1pt}\nu_1}\hspace{-5pt}\right)\hspace{-5pt}\left[\hspace{-2pt}6\eta^{\mu_1\hspace{-1pt}[\hspace{-0.5pt}\mu_{\hspace{-0.5pt}2}}\hspace{-1pt}k_{34}^{\hspace{-0.5pt}\nu_{\hspace{-0.5pt}2}}\hspace{-1pt}\gamma^{\rho_2]}_{\alpha_3\alpha_4} \hspace{-3pt}-\hspace{-3pt} 6k_{1}^{\hspace{-1pt}[\mu_2}\hspace{-1pt}(\hspace{-1pt}\gamma^{\nu_2\hspace{-0.5pt}\rho_2\hspace{-1pt}]\hspace{-0.5pt}\mu_1}\hspace{-1pt})_{\hspace{-1pt}\alpha_3\alpha_4} \hspace{-3pt}+\hspace{-3pt} 2k_{2}^{\mu_1}\hspace{-1pt}(\hspace{-1pt}\gamma^{\mu_2\hspace{-0.5pt}\nu_2\hspace{-0.5pt}\rho_2}\hspace{-1pt})_{\hspace{-1pt}\alpha_3\alpha_4} \hspace{-3pt}+\hspace{-3pt} 3k_{12}^{\mu}\hspace{-1pt}\eta^{\mu_1\hspace{-1pt}[\hspace{-0.5pt}\mu_2}\hspace{-1pt}(\hspace{-1pt}\gamma_{\phantom{\nu_2\rho_2]}\hspace{-3pt}\mu}^{\nu_2\hspace{-0.5pt}\rho_2\hspace{-1pt}]}\hspace{-1pt})_{\hspace{-1pt}\alpha_3\alpha_4}\hspace{-2pt}\right]\hspace{-3pt}.
\end{multline}

\begin{multline}
\mathcal{A}_{4}[\mathrm{NS_1^2 NS_1^2};\mathrm{NS_1^2 NS_1^2};\mathrm{R_0 R_0};\mathrm{R_0 R_0}]
\\ = \frac{\pi^2 i\alpha'g_{\text{s}}^2}{3!^2 2^7}(2\pi)^{10}\delta^{10}(k_1+k_2+k_3+k_4)\varepsilon_{1,\mu_1\nu_1\rho_1\barred{\mu}_1\barred{\nu}_1\barred{\rho}_1}\varepsilon_{2,\mu_2\nu_2\rho_2\barred{\mu}_2\barred{\nu}_2\barred{\rho}_2}\varepsilon_3^{\alpha_3\barred{\alpha}_3}\varepsilon_4^{\alpha_4\barred{\alpha}_4}
\\* \times \frac{\Gamma(1\hspace{-2pt}-\hspace{-2pt}\frac{\alpha's}{4})\Gamma(1\hspace{-2pt}-\hspace{-2pt}\frac{\alpha't}{4})\Gamma(1\hspace{-2pt}-\hspace{-2pt}\frac{\alpha'u}{4})}{\Gamma(\frac{\alpha's}{4})\Gamma(\frac{\alpha't}{4})\Gamma(\frac{\alpha'u}{4})}\Xi^{\mu_1\nu_1\rho_1\mu_2\nu_2\rho_2}_{\alpha_3\alpha_4}\Xi^{\barred{\mu}_1\barred{\nu}_1\barred{\rho}_1\barred{\mu}_2\barred{\nu}_2\barred{\rho}_2}_{\barred{\alpha}_3\barred{\alpha}_4},
\end{multline}
where
\begin{multline}
\Xi^{\mu_1\nu_1\rho_1\mu_2\nu_2\rho_2}_{\alpha_3\alpha_4}
\\ = 6\hspace{-2pt}\left(\hspace{-2pt}\frac{4\hspace{-1pt}(\hspace{-1pt}t\hspace{-3pt}-\hspace{-3pt}u\hspace{-1pt})}{s}\hspace{-2pt}+\hspace{-2pt}\frac{s\hspace{-3pt}-\hspace{-3pt}m^2}{t}\hspace{-2pt}-\hspace{-2pt}\frac{s\hspace{-3pt}-\hspace{-3pt}m^2}{u}\hspace{-3pt}\right)\hspace{-5pt}\Big[\hspace{-2pt}6k\up{9pt}{[\mu_1}{-4pt}{2}\hspace{-2pt}\eta\up{9pt}{\nu_1}{2pt}{[\mu_2}\eta\up{9pt}{\rho_1]}{2pt}{\nu_2}\gamma\up{2pt}{\rho_2\hspace{-1pt}]}{-6pt}{\hspace{-1pt}\alpha_3\alpha_4} \hspace{-3pt}-\hspace{-2pt} 6k\up{9pt}{[\mu_2}{-4pt}{1}\hspace{-2pt}\eta\up{9pt}{\nu_2}{2pt}{[\mu_1}\eta\up{9pt}{\rho_2]}{2pt}{\nu_1}\gamma\up{2pt}{\rho_1\hspace{-1pt}]}{-6pt}{\hspace{-1pt}\alpha_3\alpha_4} \hspace{-3pt}+\hspace{-2pt} \eta\up{8pt}{[\mu_1}{2pt}{\mu_2}\hspace{-1pt}\eta\up{8pt}{\nu_1}{2pt}{\nu_2}\eta\up{8pt}{\rho_1]}{2pt}{\rho_2}\hspace{-1pt}(\hspace{-1pt}k_{12}\hspace{-2pt}\cdot\hspace{-2pt}\gamma\hspace{-1pt})_{\alpha_3\alpha_4}\hspace{-2pt}\Big]
\\ -\hspace{-2pt}1\hspace{-0.5pt}8\hspace{-3pt}\left(\hspace{-4pt}\frac{s\hspace{-3pt}-\hspace{-3pt}m^{\hspace{-1pt}2}\hspace{-3pt}}{t}\hspace{-3pt}+\hspace{-3pt}\frac{s\hspace{-3pt}-\hspace{-3pt}m^{\hspace{-1pt}2}\hspace{-3pt}}{u}\hspace{-3pt}+\hspace{-3pt}4\hspace{-3pt}\right)\hspace{-6pt}\Big[\hspace{-2pt}k\hspace{-1pt}\up{9pt}{\hspace{-1pt}[\hspace{-1pt}\mu_{\hspace{-0.5pt}1}}{-4pt}{3\hspace{-0.5pt}4}\hspace{-2pt}\eta\up{9pt}{\nu_1}{2pt}{\hspace{-1pt}[\hspace{-1pt}\mu_{\hspace{-0.5pt}2}}\eta\up{9pt}{\hspace{-1pt}\rho_{\hspace{-0.5pt}1}\hspace{-1pt}]}{2pt}{\hspace{-0.5pt}\nu_{2}}\hspace{-1pt}\gamma\up{2pt}{\hspace{-0.5pt}\rho_2\hspace{-1pt}]}{-6pt}{\hspace{-2pt}\alpha_3\alpha_4} \hspace{-5pt}+\hspace{-2pt} k\hspace{-1pt}\up{9pt}{\hspace{-1pt}[\hspace{-0.5pt}\mu_{\hspace{-0.5pt}2}}{-4pt}{3\hspace{-0.5pt}4}\hspace{-2pt}\eta\up{9pt}{\nu_2}{2pt}{\hspace{-1pt}[\hspace{-1pt}\mu_{\hspace{-0.5pt}1}}\hspace{-1pt}\eta\up{9pt}{\hspace{-1pt}\rho_{2}\hspace{-1pt}]}{2pt}{\hspace{-0.5pt}\nu_{\hspace{-0.5pt}1}}\hspace{-2pt}\gamma\up{2pt}{\hspace{-0.5pt}\rho_{\hspace{-0.5pt}1}\hspace{-1pt}]}{-6pt}{\hspace{-2pt}\alpha_3\alpha_4} \hspace{-5pt}-\hspace{-2pt} 2\hspace{-1pt}k\hspace{-1pt}\up{9pt}{\hspace{-2pt}[\hspace{-1pt}\mu_{\hspace{-0.5pt}1}}{-4pt}{2}\hspace{-4pt}\eta\up{9pt}{\nu_1}{2pt}{\hspace{-1pt}[\hspace{-1pt}\mu_{\hspace{-0.5pt}2}}\hspace{-2pt}\big(\hspace{-2pt}\gamma\up{9pt}{\hspace{-1pt}\rho_{\hspace{-0.5pt}1}\hspace{-1pt}]}{2pt}{\phantom{\rho_1]}\hspace{-5pt}\nu_{\hspace{-0.5pt}2}\hspace{-1pt}\rho_{\hspace{-0.5pt}2}\hspace{-1pt}]}\hspace{-1pt}\big)_{\hspace{-2pt}\alpha_3\hspace{-1pt}\alpha_4} \hspace{-6pt}-\hspace{-2pt} 2\hspace{-0.5pt}k\hspace{-1pt}\up{9pt}{\hspace{-2pt}[\hspace{-1pt}\mu_{\hspace{-0.5pt}2}}{-4pt}{1}\hspace{-4pt}\eta\up{9pt}{\nu_2}{2pt}{\hspace{-1pt}[\hspace{-0.5pt}\mu_{\hspace{-0.5pt}1}}\hspace{-2pt}\big(\hspace{-2pt}\gamma\up{9pt}{\hspace{-1pt}\rho_{\hspace{-0.5pt}2}\hspace{-1pt}]}{2pt}{\phantom{\rho_2]}\hspace{-5pt}\nu_{\hspace{-0.5pt}1}\hspace{-1pt}\rho_{\hspace{-0.5pt}1}\hspace{-1pt}]}\hspace{-1pt}\big)_{\hspace{-2pt}\alpha_3\hspace{-1pt}\alpha_4} \hspace{-6pt}-\hspace{-1pt} k_{12}^{\mu}\hspace{-1pt}\eta\up{8pt}{\hspace{-1pt}[\hspace{-1pt}\mu_{\hspace{-0.5pt}1}}{2pt}{\mu_{\hspace{-0.5pt}2}}\hspace{-1pt}\eta\up{8pt}{\hspace{-1pt}\nu_1}{2pt}{\hspace{-0.5pt}\nu_{\hspace{-0.5pt}2}}\hspace{-1pt}\big(\hspace{-2pt}\gamma\up{8pt}{\hspace{-1pt}\rho_{\hspace{-0.5pt}1}\hspace{-1pt}]}{2pt}{\phantom{\rho_1]}\hspace{-4pt}\rho_2\hspace{-1pt}]\hspace{-0.5pt}\mu}\hspace{-1pt})_{\hspace{-1pt}\alpha_3\hspace{-1pt}\alpha_4} \hspace{-2pt}\Big] 
\\ \hspace{-30pt}-\hspace{-2pt}3\hspace{-3pt}\left(\hspace{-4pt}\frac{s\hspace{-3pt}-\hspace{-3pt}m^{\hspace{-1pt}2}\hspace{-3pt}}{t}\hspace{-3pt}-\hspace{-3pt}\frac{s\hspace{-3pt}-\hspace{-3pt}m^{\hspace{-2pt}2}\hspace{-2pt}}{u}\hspace{-1pt}\right)\hspace{-7pt}\Big[\hspace{-2pt}6\hspace{-1pt}k\hspace{-2pt}\up{9pt}{\hspace{-1pt}[\hspace{-1pt}\mu_{\hspace{-1pt}1}}{-4pt}{3\hspace{-0.5pt}4}\hspace{-2pt}\eta\up{9pt}{\nu_1}{2pt}{\hspace{-1pt}[\hspace{-1pt}\mu_{\hspace{-0.5pt}2}}\hspace{-2pt}\big(\hspace{-2pt}\gamma\up{9pt}{\hspace{-2pt}\rho_{\hspace{-0.5pt}1}\hspace{-1pt}]}{2pt}{\phantom{\rho_1]}\hspace{-6pt}\nu_{\hspace{-0.5pt}2}\hspace{-1pt}\rho_{\hspace{-0.5pt}2}\hspace{-1pt}]}\hspace{-1pt}\big)_{\hspace{-3pt}\alpha_3\hspace{-1pt}\alpha_4} \hspace{-5pt}-\hspace{-2pt} 6\hspace{-1pt}k\hspace{-2pt}\up{9pt}{\hspace{-1pt}[\hspace{-1pt}\mu_{\hspace{-0.5pt}2}}{-4pt}{3\hspace{-0.5pt}4}\hspace{-2pt}\eta\up{9pt}{\nu_2}{2pt}{\hspace{-1pt}[\hspace{-1pt}\mu_{\hspace{-1pt}1}}\hspace{-2pt}\big(\hspace{-2pt}\gamma\up{9pt}{\hspace{-1pt}\rho_{\hspace{-0.5pt}2}\hspace{-1pt}]}{2pt}{\phantom{\rho_2]}\hspace{-5pt}\nu_{\hspace{-0.5pt}1}\hspace{-1pt}\rho_{\hspace{-0.5pt}1}\hspace{-1pt}]}\hspace{-1pt}\big)_{\hspace{-3pt}\alpha_3\hspace{-1pt}\alpha_4} \hspace{-4pt}+\hspace{-2pt} 2\hspace{-1pt}k_{2}^{\hspace{-1pt}[\hspace{-1pt}\mu_{\hspace{-0.5pt}1}}\hspace{-2pt}(\hspace{-2pt}\gamma^{\nu_{\hspace{-0.5pt}1}\hspace{-1pt}\rho_{\hspace{-0.5pt}1}\hspace{-1pt}]\hspace{-1pt}\mu_{\hspace{-0.5pt}2}\hspace{-1pt}\nu_{\hspace{-0.5pt}2}\hspace{-1pt}\rho_{\hspace{-0.5pt}2}}\hspace{-2pt})_{\hspace{-2pt}\alpha_3\hspace{-1pt}\alpha_4} \hspace{-2pt}-\hspace{-2pt} 2\hspace{-1pt}k_{1}^{\hspace{-1pt}[\hspace{-1pt}\mu_{\hspace{-0.5pt}2}}\hspace{-3pt}(\hspace{-2pt}\gamma^{\hspace{-1pt}\nu_{\hspace{-0.5pt}2}\hspace{-1pt}\rho_{\hspace{-0.5pt}2}\hspace{-1pt}]\hspace{-1pt}\mu_{\hspace{-1pt}1}\hspace{-1pt}\nu_{\hspace{-1pt}1}\hspace{-1pt}\rho_{\hspace{-1pt}1}}\hspace{-3pt})_{\hspace{-2pt}\alpha_3\hspace{-1pt}\alpha_4} \hspace{-2pt}+\hspace{-2pt} 3\hspace{-1pt}k_{\hspace{-1pt}1\hspace{-1pt}2}^{\mu}\hspace{-1pt}\eta\up{9pt}{\hspace{-1pt}[\hspace{-1pt}\mu_{\hspace{-0.5pt}1}}{2pt}{\hspace{-1pt}[\hspace{-1pt}\mu_{\hspace{-0.5pt}2}}\hspace{-2pt}\big(\hspace{-2pt}\gamma_{\hspace{-1pt}\mu}\up{9pt}{\hspace{-3pt}\nu_{\hspace{-0.5pt}1}\hspace{-1pt}\rho_{\hspace{-0.5pt}1}\hspace{-1pt}]}{2pt}{\phantom{\nu_1\hspace{-1pt}\rho_1\hspace{-2pt}]\hspace{-5pt}}\nu_{\hspace{-0.5pt}2}\hspace{-1pt}\rho_{\hspace{-0.5pt}2}\hspace{-1pt}]}\hspace{-1pt}\big)_{\hspace{-3pt}\alpha_3\hspace{-1pt}\alpha_4}\hspace{-2pt}\Big] 
\\ +\hspace{-3pt}\left(\hspace{-3pt}\frac{s\hspace{-3pt}-\hspace{-3pt}m^2}{t}\hspace{-3pt}+\hspace{-3pt}\frac{s\hspace{-3pt}-\hspace{-3pt}m^2}{u}\hspace{-2pt}\right)\hspace{-4pt}\Big[\hspace{-1pt}3\hspace{-1pt}k_{34}^{\hspace{-1pt}[\mu_1}\hspace{-2pt}(\hspace{-1pt}\gamma^{\nu_1\hspace{-1pt}\rho_1\hspace{-1pt}]\hspace{-1pt}\mu_2\hspace{-1pt}\nu_2\hspace{-1pt}\rho_2}\hspace{-1pt})_{\hspace{-1pt}\alpha_3\alpha_4} + 3\hspace{-1pt}k_{34}^{\hspace{-1pt}[\mu_2}\hspace{-2pt}(\hspace{-1pt}\gamma^{\nu_2\hspace{-1pt}\rho_2\hspace{-1pt}]\hspace{-1pt}\mu_1\hspace{-1pt}\nu_1\hspace{-1pt}\rho_1}\hspace{-1pt})_{\hspace{-1pt}\alpha_3\alpha_4} - k_{12}^{\mu}\hspace{-1pt}(\hspace{-1pt}\gamma_{\mu}^{\phantom{\mu}\mu_1\hspace{-1pt}\nu_1\hspace{-1pt}\rho_1\hspace{-1pt}\mu_2\hspace{-1pt}\nu_2\hspace{-1pt}\rho_2}\hspace{-1pt})_{\hspace{-1pt}\alpha_3\alpha_4}\hspace{-1pt}\Big].
\end{multline}

\begin{multline}
\mathcal{A}_4[\mathrm{R_0 R_0}, \mathrm{R_0 R_0}, \mathrm{R_1 R_1}, \mathrm{R_1 R_1}]
\\ = \frac{\pi^2 i\alpha'^3 g_{\text{s}}^2}{18^2}(2\pi)^{10}\delta^{10}(k_1+k_2+k_3+k_4)\varepsilon_{1}^{\alpha_1\barred{\alpha}_1}\varepsilon_{2}^{\alpha_2\barred{\alpha}_2}\varepsilon_{3,\mu_3\barred{\mu}_3}^{\alpha_3\barred{\alpha}_3}\varepsilon_{4,\mu_4\barred{\mu}_4}^{\alpha_4\barred{\alpha}_4}
\\ \times \frac{\Gamma(1\hspace{-2pt}-\hspace{-2pt}\frac{\alpha's}{4})\Gamma(1\hspace{-2pt}-\hspace{-2pt}\frac{\alpha't}{4})\Gamma(1\hspace{-2pt}-\hspace{-2pt}\frac{\alpha'u}{4})}{\Gamma(\frac{\alpha's}{4})\Gamma(\frac{\alpha't}{4})\Gamma(\frac{\alpha'u}{4})}\Sigma^{\mu_3\mu_4}_{\alpha_1\alpha_2\alpha_3\alpha_4}\Sigma^{\barred{\mu}_3\barred{\mu}_4}_{\barred{\alpha}_1\barred{\alpha}_2\barred{\alpha}_3\barred{\alpha}_4},
\end{multline}
where
\begin{multline}
\Sigma^{\mu_3\mu_4}_{\alpha_1\alpha_2\alpha_3\alpha_4} = \frac{8}{\alpha'}\eta^{\mu_3\mu_4}\left(\frac{t-m^2}{s}\gamma_{\mu\alpha_1\alpha_2}\gamma^{\mu}_{\alpha_3\alpha_4} - \gamma_{\mu\alpha_1\alpha_3}\gamma^{\mu}_{\alpha_2\alpha_4}\right)
\\ - \left[\frac{s\hspace{-3pt}-\hspace{-3pt}m^2}{u}k_{12}^{\mu_3}k_{12}^{\mu_4} \hspace{-3pt}+\hspace{-3pt} \left(\hspace{-3pt}1\hspace{-3pt}+\hspace{-3pt}\frac{m^2\hspace{-3pt}-\hspace{-3pt}t}{u}\hspace{-3pt}\right)\hspace{-4pt}(k_4^{\mu_3}k_{12}^{\mu_4}\hspace{-3pt}-\hspace{-3pt}k_{12}^{\mu_3}k_3^{\mu_4}) \hspace{-3pt}-\hspace{-3pt} \left(\hspace{-3pt}\frac{4(m^2\hspace{-3pt}-\hspace{-3pt}t)}{s} \hspace{-3pt}+\hspace{-3pt} \frac{m^2\hspace{-3pt}-\hspace{-3pt}t}{u} \hspace{-3pt}-\hspace{-3pt} 1\hspace{-3pt}\right)\hspace{-5pt}k_4^{\mu_3}k_3^{\mu_4}\right]\hspace{-4pt}\gamma_{\mu\alpha_1\alpha_2}\hspace{-2pt}\gamma^{\mu}_{\alpha_3\alpha_4}
\\ - \left[\hspace{-3pt}\left(\hspace{-3pt}\frac{s\hspace{-3pt}-\hspace{-3pt}m^2}{t} \hspace{-3pt}+\hspace{-3pt} \frac{s\hspace{-3pt}-\hspace{-3pt}m^2}{u}\hspace{-3pt}\right)\hspace{-4pt}k_{12}^{\mu_3}k_{12}^{\mu_4} \hspace{-3pt}+\hspace{-3pt} \left(\hspace{-3pt}\frac{m^2\hspace{-3pt}-\hspace{-3pt}s}{t} \hspace{-3pt}+\hspace{-3pt} \frac{s\hspace{-3pt}-\hspace{-3pt}m^2}{u}\hspace{-3pt}\right)\hspace{-4pt}(k_4^{\mu_3}k_{12}^{\mu_4} \hspace{-3pt}-\hspace{-3pt} k_{12}^{\mu_3}k_3^{\mu_4}) \hspace{-3pt}-\hspace{-3pt} \left(\hspace{-3pt}2 \hspace{-3pt}+\hspace{-3pt} \frac{m^2\hspace{-3pt}-\hspace{-3pt}t}{u} \hspace{-3pt}+\hspace{-3pt} \frac{m^2 \hspace{-3pt}-\hspace{-3pt} u}{t}\hspace{-3pt}\right)\hspace{-5pt}k_4^{\mu_3}k_3^{\mu_4}\right]\hspace{-4pt}\gamma_{\mu\alpha_1\alpha_3}\hspace{-2pt}\gamma^{\mu}_{\alpha_2\alpha_4}
\\ - 2k_{12}^{\mu_3}\left[\gamma^{\mu_4}_{\alpha_1\alpha_2}(k_4\cdot\gamma)_{\alpha_3\alpha_4} - \left(\frac{m^2-s}{t}\right)\gamma^{\mu_4}_{\alpha_1\alpha_3}(k_4\cdot\gamma)_{\alpha_2\alpha_4} + \left(\frac{s-m^2}{u}\right)(k_4\cdot\gamma)_{\alpha_1\alpha_4}\gamma^{\mu_4}_{\alpha_2\alpha_3}\right]
\\ - 2k_4^{\mu_3} \hspace{-3pt}\left[\hspace{-3pt}\left(\hspace{-3pt}\frac{u\hspace{-3pt}-\hspace{-3pt}t}{s}\hspace{-3pt}\right)\hspace{-4pt}\gamma^{\mu_4}_{\alpha_1\alpha_2}(k_4\hspace{-3pt}\cdot\hspace{-3pt}\gamma)_{\alpha_3\alpha_4} \hspace{-3pt}-\hspace{-3pt} \left(\hspace{-3pt}1 \hspace{-3pt}+\hspace{-3pt} \frac{m^2\hspace{-3pt}-\hspace{-3pt}u}{t}\hspace{-3pt}\right)\hspace{-4pt}\gamma^{\mu_4}_{\alpha_1\alpha_3}(k_4\hspace{-3pt}\cdot\hspace{-3pt}\gamma)_{\alpha_2\alpha_4} \hspace{-3pt}+\hspace{-3pt} \left(\hspace{-3pt}1 \hspace{-3pt}+\hspace{-3pt} \frac{m^2\hspace{-3pt}-\hspace{-3pt}t}{u}\hspace{-3pt}\right)\hspace{-3pt}(k_4\hspace{-3pt}\cdot\hspace{-3pt}\gamma)_{\alpha_1\alpha_4}\gamma^{\mu_4}_{\alpha_2\alpha_3}\right]
\\ + 2k_{12}^{\mu_4}\left[\gamma^{\mu_3}_{\alpha_1\alpha_2}(k_3\cdot\gamma)_{\alpha_3\alpha_4} + \left(\frac{s-m^2}{u}\right)\gamma^{\mu_3}_{\alpha_1\alpha_4}(k_3\cdot\gamma)_{\alpha_2\alpha_3} - \left(\frac{m^2-s}{t}\right)(k_3\cdot\gamma)_{\alpha_1\alpha_3}\gamma^{\mu_3}_{\alpha_2\alpha_4}\right]
\\ - 2k_3^{\mu_4}\left[\hspace{-3pt}\left(\hspace{-3pt}\frac{u\hspace{-3pt}-\hspace{-3pt}t}{s}\hspace{-3pt}\right)\hspace{-4pt}\gamma^{\mu_3}_{\alpha_1\alpha_2}(k_3\hspace{-3pt}\cdot\hspace{-3pt}\gamma)_{\alpha_3\alpha_4} \hspace{-3pt}+\hspace{-3pt} \left(\hspace{-3pt}1 \hspace{-3pt}+\hspace{-3pt} \frac{m^2\hspace{-3pt}-\hspace{-3pt}t}{u}\hspace{-3pt}\right)\hspace{-4pt}\gamma^{\mu_3}_{\alpha_1\alpha_4}(k_3\hspace{-3pt}\cdot\hspace{-3pt}\gamma)_{\alpha_2\alpha_3} \hspace{-3pt}-\hspace{-3pt} \left(\hspace{-3pt}1 \hspace{-3pt}+\hspace{-3pt} \frac{m^2\hspace{-3pt}-\hspace{-3pt}u}{t}\hspace{-3pt}\right)\hspace{-3pt}(k_3\hspace{-3pt}\cdot\hspace{-3pt}\gamma)_{\alpha_1\alpha_3}\gamma^{\mu_3}_{\alpha_2\alpha_4}\right]
\\ + 4\left(\frac{m^2-u}{t}\right)\eta^{\mu_3\mu_4}k_3^{\mu}k_4^{\nu}\gamma_{\mu\alpha_1\alpha_3}\gamma_{\nu\alpha_2\alpha_4} - 4\left(\frac{m^2-t}{u}\right)\eta^{\mu_3\mu_4}k_4^{\mu}k_3^{\nu}\gamma_{\mu\alpha_1\alpha_4}\gamma_{\nu\alpha_2\alpha_3}
\\ - \eta^{\mu_3\mu_4}\left(k_3^{\mu}k_4^{\nu}\gamma_{\mu\alpha_1\alpha_3}\gamma_{\nu\alpha_2\alpha_4} - k_3^{\nu}k_4^{\mu}\gamma_{\mu\alpha_1\alpha_4}\gamma_{\nu\alpha_2\alpha_3}\right)
\\ - \frac{1}{2}\left(\frac{m^2-t}{s}-2\right)\left[\eta^{\mu_3\mu_4}k_3^{\rho}k_4^{\lambda}(\gamma_{\nu}\gamma_{\rho}\gamma_{\lambda})_{\alpha_2\alpha_4}\gamma^{\nu}_{\alpha_1\alpha_3} - k_3^{\rho}k_4^{\lambda}(\gamma^{\mu_3}\gamma_{\rho}\gamma_{\lambda})_{\alpha_2\alpha_4}\gamma^{\mu_4}_{\alpha_1\alpha_3} \right. 
\\ \left. + k_3^{\mu_4}k_4^{\lambda}(\gamma^{\mu_3}\gamma_{\nu}\gamma_{\lambda})_{\alpha_2\alpha_4}\gamma^{\nu}_{\alpha_1\alpha_3} - k_3^{\rho}k_4^{\lambda}(\gamma^{\mu_3}\gamma_{\rho}\gamma^{\mu_4})_{\alpha_1\alpha_2}\gamma_{\lambda\alpha_3\alpha_4}\right]
\\ - \frac{1}{2}\left(\frac{m^2-t}{s}+1\right)\left[\eta^{\mu_3\mu_4}k_3^{\rho}k_4^{\lambda}(\gamma_{\nu}\gamma_{\lambda}\gamma_{\rho})_{\alpha_2\alpha_3}\gamma^{\nu}_{\alpha_1\alpha_4} - k_3^{\rho}k_4^{\lambda}(\gamma^{\mu_4}\gamma_{\lambda}\gamma_{\rho})_{\alpha_2\alpha_3}\gamma^{\mu_3}_{\alpha_1\alpha_4} \right. 
\\ \left. + k_3^{\rho}k_4^{\mu_3}(\gamma^{\mu_4}\gamma_{\nu}\gamma_{\rho})_{\alpha_2\alpha_3}\gamma^{\nu}_{\alpha_1\alpha_4} - k_3^{\rho}k_4^{\lambda}(\gamma^{\mu_4}\gamma_{\lambda}\gamma^{\mu_3})_{\alpha_1\alpha_2}\gamma_{\rho\alpha_3\alpha_4}\right]
\\ - \frac{3}{4}\left(\frac{u-t}{s}\right)\eta^{\mu_3\mu_4}k_3^{\rho}k_4^{\lambda}(\gamma_{\rho}\gamma_{\nu}\gamma_{\lambda})_{\alpha_3\alpha_4}\gamma^{\nu}_{\alpha_1\alpha_2} + 3\eta^{\mu_3\mu_4}k_3^{\rho}k_4^{\lambda}\gamma_{\rho\alpha_3[\alpha_1}\gamma_{|\lambda|\alpha_2]\alpha_4} 
\\ - \frac{1}{2}k_3^{\rho}k_4^{\mu_3}(\gamma_{\rho}\gamma^{\mu_4}\gamma_{\nu})_{\alpha_3[\alpha_1}\gamma^{\nu}_{\alpha_2]\alpha_4} + k_3^{\rho}k_4^{\lambda}(\gamma_{\rho}\gamma^{\mu_4}\gamma_{\lambda})_{\alpha_3[\alpha_1}\gamma^{\mu_3}_{\alpha_2]\alpha_4} - \frac{1}{2}\eta^{\mu_3\mu_4}k_3^{\rho}k_4^{\lambda}(\gamma_{\rho}\gamma_{\nu}\gamma_{\lambda})_{\alpha_3[\alpha_1}\gamma^{\nu}_{\alpha_2]\alpha_4}.
\end{multline}
The preceding amplitude can be written in many equivalent ways due to the Fierz identities; we have not simplified it as much as possible.

\section{Example Unitarity Check}\label{unitarity}

Unitarity of the S-matrix at tree-level relates the poles of amplitudes to factorized subamplitudes exchanging an on-shell particle. For four-point amplitudes, this factorization in the $s$-channel, for instance, takes the form
\begin{equation}
\up{0pt}{\displaystyle \text{Pole}}{-8pt}{s\rightarrow m_*^2} \mathcal{A}_4(\{\varepsilon_i\},\hspace{-1pt}\{k_i\}) = \int \frac{d^d k}{(2\pi)^d}\mathcal{A}_3(\varepsilon_1,k_1;\varepsilon_2,k_2;k)^{\{\mu\}}\frac{i\Pi_{\{\mu\},\{\nu\}}(k)}{s-m_*^2}\mathcal{A}_3(-k;\varepsilon_3,k_3;\varepsilon_4,k_4)^{\{\nu\}},
\end{equation}
where $\mathcal{A}(\{\varepsilon_j\},\{k_j\};k)^{\{\mu\}}$ denotes an amplitude with the relevant polarization tensor stripped off leaving however many free indices are used to describe the representation of the exchanged particle, and $\Pi_{\{\mu\},\{\nu\}}(k)$ is the appropriate propagator numerator. Here, we explicitly check all the massless and first-massive pole factorizations of the amplitude $\mathcal{A}_4[\mathrm{NS_1^1 NS_1^1};\mathrm{NS_1^2 NS_1^2};\mathrm{R_0 R_0};\mathrm{R_0 R_0}]$ given in \eqref{NS11NS12R0R0} and \eqref{Lambda}. For this section, let us keep the spacetime dimension $d$ explicit and write the vertex operators as
\begin{align}
\mathcal{V}^{(-1,-1)}_{\mathrm{NS_0 NS_0}}(\varepsilon,k) & = A_1 g_{\text{s}}\varepsilon_{\mu\barred{\mu}}\hspace{2pt}\normal{e^{-\phi}}\psi^{\mu}\hspace{2pt}\normal{e^{-\widetilde{\phi}}}\widetilde{\psi}^{\barred{\mu}}\hspace{2pt}\normal{e^{ik\cdot X}}
\\ \mathcal{V}^{(-\frac{1}{2},-\frac{1}{2})}_{\mathrm{R_0 R_0}}(\varepsilon,k) & = \frac{iA_2 \sqrt{\alpha'}}{2}g_{\text{s}}\varepsilon^{\alpha\barred{\alpha}}\hspace{2pt}\mathcal{S}_{\alpha}\hspace{2pt}\widetilde{\mathcal{S}}_{\barred{\alpha}}\hspace{2pt}\normal{e^{ik\cdot X}}
\\ \mathcal{V}^{(-1,-1)}_{\mathrm{NS_1^1 NS_1^1}}(\varepsilon,k) & = -\frac{2B_1 g_{\text{s}}}{\alpha'}\varepsilon_{\mu\nu\barred{\mu}\barred{\nu}}\hspace{2pt}\normal{e^{-\phi}}\psi^{\mu}\hspace{2pt}\normal{e^{-\widetilde{\phi}}}\widetilde{\psi}^{\barred{\mu}}\hspace{2pt}\normal{\partial X^{\nu}\barred{\partial}X^{\barred{\nu}}e^{ik\cdot X}}
\\ \mathcal{V}^{(-1,-1)}_{\mathrm{NS_1^2 NS_1^2}}(\varepsilon,k) & = -\frac{B_2 g_{\text{s}}}{3!}\varepsilon_{\mu\nu\rho\barred{\mu}\barred{\nu}\barred{\rho}}\hspace{2pt}\normal{e^{-\phi}}\normal{\psi^{\mu}\psi^{\nu}\psi^{\rho}}\hspace{2pt}\normal{e^{-\widetilde{\phi}}}\normal{\widetilde{\psi}^{\barred{\mu}}\widetilde{\psi}^{\barred{\nu}}\widetilde{\psi}^{\barred{\rho}}}\hspace{2pt}\normal{e^{ik\cdot X}}
\\ \mathcal{V}^{(-1,-1)}_{\mathrm{NS_1^1 NS_1^2}}(\varepsilon,k) & = \frac{B_3 g_{\text{s}}}{\sqrt{3\alpha'}}\varepsilon_{\mu\nu\barred{\mu}\barred{\nu}\barred{\rho}}\hspace{2pt}\normal{e^{-\phi}}\psi^{\mu}\normal{e^{-\widetilde{\phi}}}\normal{\widetilde{\psi}^{\barred{\mu}}\widetilde{\psi}^{\barred{\nu}}\widetilde{\psi}^{\barred{\rho}}}\normal{\partial X^{\nu}e^{ik\cdot X}}
\\ \mathcal{V}^{(-1,-1)}_{\mathrm{NS_1^2 NS_1^1}}(\varepsilon,k) & = \frac{B_4 g_{\text{s}}}{\sqrt{3\alpha'}}\varepsilon_{\mu\nu\rho\barred{\mu}\barred{\nu}}\hspace{2pt}\normal{e^{-\phi}}\normal{\psi^{\mu}\psi^{\nu}\psi^{\rho}}\normal{e^{-\widetilde{\phi}}}\widetilde{\psi}^{\barred{\mu}}\normal{\barred{\partial}X^{\barred{\nu}}e^{ik\cdot X}}
\\ \mathcal{V}^{(-\frac{1}{2},-\frac{1}{2})}_{\mathrm{R_1 R_1}}(\varepsilon,k) & = \frac{i(d-2)Cg_{\text{s}}}{(d-1)\sqrt{\alpha'}}\varepsilon_{\mu\barred{\mu}}^{\alpha\barred{\alpha}}\hspace{2pt}\normal{\mathcal{J}^{\mu}_{\alpha}\widetilde{\mathcal{J}}^{\barred{\mu}}_{\barred{\alpha}}e^{ik\cdot X}},
\end{align}
where $A_i$, $B_i$ and $C$ are positive constants which are to be determined from unitarity. Moreover, we write the operator appearing in the massive RR vertex operator as
\begin{equation}\label{modified J}
\mathcal{J}^{\mu}_{\alpha} = \partial X^{\mu}\mathcal{S}_{\alpha} + \frac{c\alpha'}{2(d-2)}k^{\rho}(\gamma_{\rho}\gamma_{\nu})_{\alpha}^{\phantom{\alpha}\beta}\normal{j^{\mu\nu}\mathcal{S}_{\beta}},
\end{equation}
allowing for any constant $c$. Both $d$ and $c$ will also be determined from unitarity. The only propagators which have $d$ dependence are those for the $\mathrm{NS_1^1 NS_1^1}$ and $\mathrm{R_1 R_1}$ states, being
\begin{align}
\Pi^{\mu_1\nu_1\barred{\mu}_1\barred{\nu}_1}_{\mu_2\nu_2\barred{\mu}_2\barred{\nu}_2}(k) & = \left(\hspace{-2pt}\hat{\eta}^{(\mu_1}_{\mu_2}(k)\hat{\eta}^{\nu_1)}_{\nu_2}\hspace{-1pt}(\hspace{-1pt}k\hspace{-1pt}) \hspace{-1pt}-\hspace{-1pt} \frac{1}{d\hspace{-2pt}-\hspace{-2pt}1}\hat{\eta}^{\mu_1\nu_1}\hspace{-1pt}(\hspace{-1pt}k\hspace{-1pt})\hat{\eta}_{\mu_2\nu_2}\hspace{-1pt}(\hspace{-1pt}k\hspace{-1pt})\hspace{-2pt}\right)\hspace{-3pt}\left(\hspace{-2pt}\hat{\eta}^{(\barred{\mu}_1}_{\barred{\mu}_2}\hspace{-1pt}(\hspace{-1pt}k\hspace{-1pt})\hat{\eta}^{\barred{\nu}_1)}_{\barred{\nu}_2}\hspace{-1pt}(\hspace{-1pt}k\hspace{-1pt}) \hspace{-1pt}-\hspace{-1pt} \frac{1}{d\hspace{-2pt}-\hspace{-2pt}1}\hat{\eta}^{\barred{\mu}_1\barred{\nu}_1}\hspace{-1pt}(\hspace{-1pt}k\hspace{-1pt})\hat{\eta}_{\barred{\mu}_2\barred{\nu}_2}\hspace{-1pt}(\hspace{-1pt}k\hspace{-1pt})\hspace{-2pt}\right)
\\ \Pi^{\mu\barred{\mu},\nu\barred{\nu}}_{\alpha\barred{\alpha},\beta\barred{\beta}} & = \left[\hat{\eta}^{\mu\nu}(k)(k\cdot\gamma)_{\alpha\beta} - \frac{1}{d-2}k^{\rho}(\gamma_{\rho}^{\phantom{\rho}\mu\nu})_{\alpha\beta}\right]\left[\hat{\eta}^{\barred{\mu}\barred{\nu}}(k)(k\cdot\gamma)_{\barred{\alpha}\barred{\beta}} - \frac{1}{d-2}k^{\barred{\rho}}(\gamma_{\barred{\rho}}^{\phantom{\rho}\barred{\mu}\barred{\nu}})_{\barred{\alpha}\barred{\beta}}\right],
\end{align}
respectively.

\subsection{Factorization onto the Massless Poles}

For the amplitude $\mathcal{A}_4[\mathrm{NS_1^1 NS_1^1};\mathrm{NS_1^2 NS_1^2};\mathrm{R_0 R_0};\mathrm{R_0 R_0}]$, the massless $s$-channel pole is due to the exchange of an $\mathrm{NS_0 NS_0}$ state, while the $t$- and $u$-channel massless poles are due to the exchange of an $\mathrm{R_0 R_0}$ state. The $u$-channel result here follows immediately from the $t$-channel result. The required $s$- and $t$-channel factorizations read 
\begin{align}
\notag \hspace{25pt}&\hspace{-25pt} \int \hspace{-2pt}\frac{d^d k}{(2\pi)^d}\mathcal{A}\up{11pt}{\mathrm{NS_1^1 NS_1^1};\mathrm{NS_1^2 NS_1^2};\mathrm{NS_0 NS_0}}{-3pt}{3}\hspace{-65pt}(\hspace{-1pt}\varepsilon_1,\hspace{-1pt}k_1;\hspace{-1pt}\varepsilon_2,\hspace{-1pt}k_2;\hspace{-1pt}k\hspace{-1pt})^{\mu\barred{\mu}}\frac{i\Pi_{\mu\barred{\mu},\nu\barred{\nu}}(k)}{s}\mathcal{A}\up{11pt}{\mathrm{NS_0 NS_0};\mathrm{R_0 R_0};\mathrm{R_0 R_0}}{-3pt}{3}\hspace{-70pt}(\hspace{-1pt}-k;\hspace{-1pt}\varepsilon_3,\hspace{-1pt}k_3;\hspace{-1pt}\varepsilon_4,\hspace{-1pt}k_4)^{\nu\barred{\nu}}
\\ \notag & = \frac{3!2A_1^2 A_2^2 B_1 B_2\pi^2 i\alpha'g_{\text{s}}^2}{s}(2\pi)^d\delta^d(k_1\hspace{-2pt}+\hspace{-2pt}k_2\hspace{-2pt}+\hspace{-2pt}k_3\hspace{-2pt}+\hspace{-2pt}k_4)\varepsilon_{1,\mu_1\nu_1\barred{\mu}_1\barred{\nu}_1}\varepsilon_{2,\mu_2\nu_2\rho_2\barred{\mu}_2\barred{\nu}_2\barred{\rho}_2}\varepsilon_3^{\alpha_3\barred{\alpha}_3}\varepsilon_4^{\alpha_4\barred{\alpha}_4}
\\ & \hspace{15pt} \times \hspace{-3pt} \Big(k_2^{\nu_1}\eta^{\mu_1[\mu_2}k_1^{\nu_2}\gamma_{\alpha_3\alpha_4}^{\rho_2]}\Big)\left(k_2^{\barred{\nu}_1}\eta^{\barred{\mu}_1[\barred{\mu}_2}k_1^{\barred{\nu}_2}\gamma^{\barred{\rho}_2]}_{\barred{\alpha}_3\barred{\alpha}_4}\right)
\\ \notag \hspace{25pt}&\hspace{-25pt} \int \hspace{-2pt}\frac{d^d k}{(2\pi)^d}\mathcal{A}\up{11pt}{\mathrm{NS_1^1 NS_1^1};\mathrm{R_0 R_0};\mathrm{R_0 R_0}}{-3pt}{3}\hspace{-65pt}(\hspace{-1pt}\varepsilon_1,\hspace{-1pt}k_1;\hspace{-1pt}\varepsilon_3,\hspace{-1pt}k_3;\hspace{-1pt}k\hspace{-1pt})_{\alpha\barred{\alpha}}\frac{i\Pi^{\alpha\barred{\alpha};\beta\barred{\beta}}(k)}{t}\mathcal{A}\up{11pt}{\mathrm{NS_1^2 NS_1^2};\mathrm{R_0 R_0};\mathrm{R_0 R_0}}{-3pt}{3}\hspace{-72pt}(\varepsilon_2,\hspace{-1pt}k_2;\varepsilon_4,\hspace{-1pt}k_4;-k)_{\beta\barred{\beta}}
\\ \notag & = \frac{A_2^4 B_1 B_2\pi^2 i\alpha'g_{\text{s}}^2}{3!8t}(2\pi)^d \delta^d(k_1\hspace{-2pt}+\hspace{-2pt}k_2\hspace{-2pt}+\hspace{-2pt}k_3\hspace{-2pt}+\hspace{-2pt}k_4)\varepsilon_{1,\mu_1\nu_1\barred{\mu}_1\barred{\nu}_1}\varepsilon_{2,\mu_2\nu_2\rho_2\barred{\mu}_2\barred{\nu}_2\barred{\rho}_2}\varepsilon_3^{\alpha_3\barred{\alpha}_3}\varepsilon_4^{\alpha_4\barred{\alpha}_4}
\\ & \hspace{15pt} \times \left[-k_3^{\mu_1}(k_2^{\mu}\hspace{-2pt}+\hspace{-2pt}k_4^{\mu})\hspace{-2pt}\left(\gamma^{\nu_1}\gamma_{\mu}\gamma^{\mu_2\nu_2\rho_2}\right)_{\alpha_3\alpha_4}\right] \left[-k_3^{\barred{\mu}_1}(k_2^{\barred{\mu}}\hspace{-2pt}+\hspace{-2pt}k_4^{\barred{\mu}})\hspace{-2pt}\left(\gamma^{\barred{\nu}_1}\gamma_{\barred{\mu}}\gamma^{\barred{\mu}_2\barred{\nu}_2\barred{\rho}_2}\right)_{\barred{\alpha}_3\barred{\alpha}_4}\right]
\\ \notag & = \frac{A_2^4 B_1 B_2\pi^2 i\alpha'g_{\text{s}}^2}{3!32t}(2\pi)^d \delta^d(k_1\hspace{-2pt}+\hspace{-2pt}k_2\hspace{-2pt}+\hspace{-2pt}k_3\hspace{-2pt}+\hspace{-2pt}k_4)\varepsilon_{1,\mu_1\nu_1\barred{\mu}_1\barred{\nu}_1}\varepsilon_{2,\mu_2\nu_2\rho_2\barred{\mu}_2\barred{\nu}_2\barred{\rho}_2}\varepsilon_3^{\alpha_3\barred{\alpha}_3}\varepsilon_4^{\alpha_4\barred{\alpha}_4}
\\ \notag & \hspace{15pt} \times k_3^{\mu_1}\hspace{-3pt}\left[6\eta^{\nu_1[\mu_2}\hspace{-2pt}\left(2k_1^{\nu_2} \hspace{-3pt}+\hspace{-2pt} k_{34}^{\nu_2}\right)\hspace{-2pt}\gamma^{\rho_2]}_{\alpha_3\alpha_4} + \left(2k_2^{\nu_1} \hspace{-3pt}-\hspace{-2pt} k_{34}^{\nu_1}\right)\hspace{-2pt}(\gamma^{\mu_2\nu_2\rho_2})_{\alpha_3\alpha_4} \right.
\\ \notag & \hspace{45pt} \left. - 3\hspace{-2pt}\left(\hspace{-2pt}2k_1^{[\mu_2} \hspace{-4pt}+\hspace{-2pt} k_{34}^{[\mu_2}\hspace{-2pt}\right)\hspace{-2pt}(\gamma^{\nu_2\rho_2]\nu_1})_{\alpha_3\alpha_4} \hspace{-3pt}+\hspace{-2pt} 3k_{12}^{\mu}\eta^{\nu_1[\mu_2}(\gamma^{\nu_2\rho_2]}_{\phantom{\nu_2\rho_2]}\mu})_{\alpha\beta} \hspace{-3pt}+\hspace{-2pt} k_{12}^{\mu}(\gamma_{\mu}^{\phantom{\mu}\nu_1\mu_2\nu_2\rho_2})_{\alpha_3\alpha_4}\right]
\\ \notag & \hspace{15pt} \times k_3^{\barred{\mu}_1}\hspace{-3pt}\left[6\eta^{\barred{\nu}_1[\barred{\mu}_2}\hspace{-2pt}\left(2k_1^{\barred{\nu}_2} \hspace{-3pt}+\hspace{-2pt} k_{34}^{\barred{\nu}_2}\right)\hspace{-2pt}\gamma^{\barred{\rho}_2]}_{\barred{\alpha}_3\barred{\alpha}_4} + \left(2k_2^{\barred{\nu}_1} \hspace{-3pt}-\hspace{-2pt} k_{34}^{\barred{\nu}_1}\right)\hspace{-2pt}(\gamma^{\barred{\mu}_2\barred{\nu}_2\barred{\rho}_2})_{\barred{\alpha}_3\barred{\alpha}_4} \right.
\\ & \hspace{45pt} \left. - 3\hspace{-2pt}\left(\hspace{-2pt}2k_1^{[\barred{\mu}_2} \hspace{-4pt}+\hspace{-2pt} k_{34}^{[\barred{\mu}_2}\hspace{-2pt}\right)\hspace{-2pt}(\gamma^{\barred{\nu}_2\barred{\rho}_2]\barred{\nu}_1})_{\barred{\alpha}_3\barred{\alpha}_4} \hspace{-3pt}+\hspace{-2pt} 3k_{12}^{\barred{\mu}}\eta^{\barred{\nu}_1[\barred{\mu}_2}(\gamma^{\barred{\nu}_2\barred{\rho}_2]}_{\phantom{\barred{\nu}_2\barred{\rho}_2]}\barred{\mu}})_{\barred{\alpha}_3\barred{\alpha}_4} \hspace{-3pt}+\hspace{-2pt} k_{12}^{\barred{\mu}}(\gamma_{\barred{\mu}}^{\phantom{\barred{\mu}}\barred{\nu}_1\barred{\mu}_2\barred{\nu}_2\barred{\rho}_2})_{\barred{\alpha}_3\barred{\alpha}_4}\right]\hspace{-3pt}.
\end{align}
On the other hand, extracting the actual $s$- and $t$-channel massless poles of the four-point amplitude \eqref{NS11NS12R0R0} results in
\begin{align}
\notag \hspace{50pt}&\hspace{-50pt}\up{0pt}{\displaystyle\text{Pole}}{-7pt}{\hspace{2pt}s\rightarrow 0} \ \mathcal{A}\up{8pt}{\mathrm{NS_1^1 NS_1^1};\mathrm{NS_1^2 NS_1^2};\mathrm{R_0 R_0};\mathrm{R_0 R_0}}{-3pt}{4}(\{\varepsilon_i\},\{k_i\})
\\* \notag & = \frac{3!2A_2^2 B_1 B_2\pi^2 i\alpha'g_{\text{s}}^2}{s} (2\pi)^d\delta^{d}\hspace{-1pt}(\hspace{-1pt}k_{\hspace{-0.5pt}1}\hspace{-2pt}+\hspace{-2pt}k_{\hspace{-0.5pt}2}\hspace{-2pt}+\hspace{-2pt}k_{\hspace{-0.5pt}3}\hspace{-2pt}+\hspace{-2pt}k_{\hspace{-0.5pt}4}\hspace{-1pt})\varepsilon_{1,\mu_{1}\nu_{1}\barred{\mu}_{1}\barred{\nu}_{1}}\varepsilon_{2,\mu_{2}\nu_{2}\rho_{2}\barred{\mu}_{2}\barred{\nu}_{2}\barred{\rho}_{2}}\varepsilon_3^{\alpha_3\barred{\alpha}_3}\varepsilon_4^{\alpha_4\barred{\alpha}_4}
\\* & \hspace{15pt} \times \left(k_2^{\nu_1}\eta^{\mu_1[\mu_2}k_1^{\nu_2}\gamma^{\rho_2]}_{\alpha_3\alpha_4}\right)\left(k_2^{\barred{\nu}_1} \eta^{\barred{\mu}_1[\barred{\mu}_2}k_1^{\barred{\nu}_2}\gamma^{\barred{\rho}_2]}_{\barred{\alpha}_3\barred{\alpha}_4}\right)
\\ \notag \hspace{50pt}&\hspace{-50pt}\up{0pt}{\displaystyle\text{Pole}}{-7pt}{\hspace{2pt}t\rightarrow 0} \ \mathcal{A}^{\mathrm{NS_1^1 NS_1^1};\mathrm{NS_1^2 NS_1^2};\mathrm{R_0 R_0};\mathrm{R_0 R_0}}_{4}(\{\varepsilon_i\},\{k_i\}) 
\\ \notag & = \frac{A_2^2 B_1 B_2\pi^2 i\alpha'g_{\text{s}}^2}{3!32t}(2\pi)^d\delta^{d}\hspace{-1pt}(\hspace{-1pt}k_{\hspace{-0.5pt}1}\hspace{-2pt}+\hspace{-2pt}k_{\hspace{-0.5pt}2}\hspace{-2pt}+\hspace{-2pt}k_{\hspace{-0.5pt}3}\hspace{-2pt}+\hspace{-2pt}k_{\hspace{-0.5pt}4}\hspace{-1pt})\varepsilon_{1,\mu_{1}\nu_{1}\barred{\mu}_{1}\barred{\nu}_{1}}\varepsilon_{2,\mu_{2}\nu_{2}\rho_{2}\barred{\mu}_{2}\barred{\nu}_{2}\barred{\rho}_{2}}\varepsilon_3^{\alpha_3\barred{\alpha}_3}\varepsilon_4^{\alpha_4\barred{\alpha}_4}
\\ \notag & \hspace{-50pt}\times k_3^{\nu_1}\hspace{-2pt} \Big[6\eta^{\mu_1\hspace{-1pt}[\hspace{-0.5pt}\mu_{2}}\hspace{-3pt}\left(\hspace{-1pt}2\hspace{-1pt}k_1^{\nu_{2}}\hspace{-3pt}+\hspace{-2pt}k_{34}^{\nu_{2}}\hspace{-1pt}\right)\hspace{-2pt}\gamma^{\rho_2]}_{\alpha_3\alpha_4} \hspace{-2pt}-\hspace{-2pt} 3\hspace{-2pt}\left(\hspace{-2pt}2k_{1}^{\hspace{-0.5pt}[\hspace{-0.5pt}\mu_2}\hspace{-4pt}+\hspace{-3pt}k_{34}^{\hspace{-0.5pt}[\mu_2}\hspace{-2pt}\right)\hspace{-3pt}(\hspace{-2pt}\gamma^{\nu_2\rho_2\hspace{-1pt}]\hspace{-0.5pt}\mu_1}\hspace{-1pt})_{\hspace{-1pt}\alpha_3\alpha_4} \hspace{-3pt}+\hspace{-2pt} \left(\hspace{-1pt}2k_{2}^{\mu_1}\hspace{-3pt}-\hspace{-2pt}k_{34}^{\mu_1}\hspace{-1pt}\right)\hspace{-3pt}(\hspace{-2pt}\gamma^{\mu_2\nu_2\rho_2}\hspace{-1pt})_{\hspace{-1pt}\alpha_3\alpha_4} 
\\ \notag & \hspace{-35pt} + \hspace{-2pt} 3k_{12}^{\mu}\eta^{\mu_1\hspace{-1pt}[\hspace{-0.5pt}\mu_2}\hspace{-1pt}(\hspace{-1pt}\gamma_{\phantom{\nu_2\rho_2]}\hspace{-2pt}\mu}^{\nu_2\rho_2\hspace{-1pt}]}\hspace{-1pt})_{\hspace{-1pt}\alpha_3\alpha_4}\hspace{-3pt}+\hspace{-3pt} k_{12}^{\mu}\hspace{-1pt}(\hspace{-1pt}\gamma_{\mu}^{\phantom{\mu}\hspace{-1pt}\mu_1\mu_2\nu_2\rho_2}\hspace{-1pt})_{\hspace{-1pt}\alpha_3\alpha_4}\hspace{-2pt}\Big]\hspace{-2pt}k_3^{\barred{\nu}_1}\hspace{-3pt} \Big[\hspace{-1pt}6\eta^{\barred{\mu}_1\hspace{-1pt}[\hspace{-0.5pt}\barred{\mu}_{2}}\hspace{-4pt}\left(\hspace{-3pt}2\hspace{-1pt}k_1^{\barred{\nu}_{2}}\hspace{-4pt}+\hspace{-3pt}k_{34}^{\barred{\nu}_{2}}\hspace{-3pt}\right)\hspace{-3pt}\gamma^{\barred{\rho}_2]}_{\barred{\alpha}_3\barred{\alpha}_4} \hspace{-4pt}-\hspace{-2pt} 3\hspace{-3pt}\left(\hspace{-3pt}2k_{1}^{\hspace{-0.5pt}[\hspace{-0.5pt}\barred{\mu}_2}\hspace{-4pt}+\hspace{-3pt}k_{34}^{\hspace{-0.5pt}[\barred{\mu}_2}\hspace{-3pt}\right)\hspace{-4pt}(\hspace{-2pt}\gamma^{\barred{\nu}_2\barred{\rho}_2\hspace{-1pt}]\hspace{-0.5pt}\barred{\mu}_1}\hspace{-2pt})_{\hspace{-1pt}\barred{\alpha}_3\barred{\alpha}_4} 
\\ & \hspace{10pt} +\hspace{-3pt} \left(\hspace{-2pt}2k_{2}^{\barred{\mu}_1}\hspace{-3pt}-\hspace{-2pt}k_{34}^{\barred{\mu}_1}\hspace{-3pt}\right)\hspace{-3pt}(\hspace{-2pt}\gamma^{\barred{\mu}_2\barred{\nu}_2\barred{\rho}_2}\hspace{-1pt})_{\hspace{-1pt}\barred{\alpha}_3\barred{\alpha}_4} \hspace{-3pt} + \hspace{-2pt} 3k_{12}^{\barred{\mu}}\eta^{\barred{\mu}_1\hspace{-1pt}[\hspace{-0.5pt}\barred{\mu}_2}\hspace{-1pt}(\hspace{-1pt}\gamma_{\phantom{\nu_2\rho_2]}\hspace{-2pt}\barred{\mu}}^{\barred{\nu}_2\barred{\rho}_2\hspace{-1pt}]}\hspace{-1pt})_{\hspace{-1pt}\barred{\alpha}_3\barred{\alpha}_4}\hspace{-3pt}+\hspace{-3pt} k_{12}^{\barred{\mu}}\hspace{-1pt}(\hspace{-1pt}\gamma_{\barred{\mu}}^{\phantom{\mu}\hspace{-1pt}\barred{\mu}_1\barred{\mu}_2\barred{\nu}_2\barred{\rho}_2}\hspace{-1pt})_{\hspace{-1pt}\barred{\alpha}_3\barred{\alpha}_4}\hspace{-2pt}\Big].
\end{align}
The massless poles in all three channels thus exactly match the unitarity computations provided that $A_1 = 1$ and $A_2 = 1$.

\subsection{Factorization onto the Massive Poles}

The factorizations due to the on-shell exchange of massive string states constitute highly nontrivial checks on the veracity of the amplitudes involved.

\subsubsection{$s$-channel}

The first-massive poles for the amplitude $\mathcal{A}_4[\mathrm{NS_1^1 NS_1^1};\mathrm{NS_1^2 NS_1^2};\mathrm{R_0 R_0};\mathrm{R_0 R_0}]$ in the $s$-channel come from all four massive NSNS states. Let us first compute the requisite factorizations due to the intermediate states $\mathrm{NS_1^1 NS_1^1}$ and $\mathrm{NS_1^2 NS_1^2}$. These contributions read
\begin{multline}
\int\frac{d^d k}{(2\pi)^d}\mathcal{A}_{3}\up{12pt}{\hspace{-5pt}\mathrm{NS_1^1 NS_1^1};\mathrm{NS_1^2 NS_1^2};\mathrm{NS_1^1 NS_1^1}}{0pt}{}\hspace{-75pt}(\varepsilon_1,k_1;\varepsilon_2,k_2;k)_{\mu\nu\barred{\mu}\barred{\nu}} \ \frac{i\Pi^{\mu\nu\barred{\mu}\barred{\nu}}_{\rho\lambda\barred{\rho}\barred{\lambda}}(k)}{s-m^2} \ \mathcal{A}_{3}\up{12pt}{\hspace{-5pt}\mathrm{NS_1^1 NS_1^1};\mathrm{R_0 R_0};\mathrm{R_0 R_0}}{0pt}{}\hspace{-72pt}(-k;\varepsilon_3,k_3;\varepsilon_4,k_4)^{\rho\lambda\barred{\rho}\barred{\lambda}}
\\ = \left(-3!B_1^2 B_2\pi i g_{\text{s}}\right)\hspace{-3pt}\frac{i}{s\hspace{-2pt}-\hspace{-2pt}m^2}\hspace{-3pt}\left(\hspace{-3pt}\frac{A_2^2 B_1\pi i\alpha' g_{\text{s}}}{8}\hspace{-3pt}\right)\hspace{-3pt}(2\pi)^d\delta^d(k_1\hspace{-1pt}+\hspace{-1pt}k_2\hspace{-1pt}+\hspace{-1pt}k_3\hspace{-1pt}+\hspace{-1pt}k_4)\varepsilon_{1,\mu_1\nu_1\barred{\mu}_1\barred{\nu}_1}\varepsilon_{2,\mu_2\nu_2\rho_2\barred{\mu}_2\barred{\nu}_2\barred{\rho}_2}\varepsilon_3^{\alpha_3\barred{\alpha}_3}\varepsilon_4^{\alpha_4\barred{\alpha}_4}
\\ \times \left[2\eta\up{6pt}{[\mu_2}{-2pt}{\mu_1}\eta\up{6pt}{\nu_2}{-2pt}{\mu}k_{1}^{\rho_2]}\hspace{-3pt}\left(\hspace{-3pt}\eta^{\nu_1\nu}\hspace{-2pt}+\hspace{-2pt}\frac{\alpha'}{4}k_2^{\nu_1}k_{12}^{\nu}\hspace{-3pt}\right)\right]\hat{\eta}_{\rho\{\mu}(k_3\hspace{-2pt}+\hspace{-2pt}k_4)\hat{\eta}_{\nu\}\lambda}(k_3\hspace{-2pt}+\hspace{-2pt}k_4)\left(-k_{34}^{\rho}\gamma^{\lambda}_{\alpha_3\alpha_4}\right)
\\ \times \left[2\eta\up{6pt}{[\barred{\mu}_2}{-2pt}{\barred{\mu}_1}\eta\up{6pt}{\barred{\nu}_2}{-2pt}{\barred{\mu}}k_{1}^{\barred{\rho}_2]}\hspace{-3pt}\left(\hspace{-3pt}\eta^{\barred{\nu}_1\barred{\nu}}\hspace{-2pt}+\hspace{-2pt}\frac{\alpha'}{4}k_2^{\barred{\nu}_1}k_{12}^{\barred{\nu}}\hspace{-3pt}\right)\right]\hat{\eta}_{\barred{\rho}\{\barred{\mu}}(k_3\hspace{-2pt}+\hspace{-2pt}k_4)\hat{\eta}_{\barred{\nu}\}\barred{\lambda}}(k_3\hspace{-2pt}+\hspace{-2pt}k_4)\left(-k_{34}^{\barred{\rho}}\gamma^{\barred{\lambda}}_{\barred{\alpha}_3\barred{\alpha}_4}\right),
\end{multline}
while the second reads
\begin{multline}
\int\frac{d^d k}{(2\pi)^d}\mathcal{A}_{3}\up{12pt}{\hspace{-5pt}\mathrm{NS_1^1 NS_1^1};\mathrm{NS_1^2 NS_1^2};\mathrm{NS_1^2 NS_1^2}}{0pt}{}\hspace{-73pt}(\varepsilon_1,k_1;\varepsilon_2,k_2;k)_{\mu\nu\rho\barred{\mu}\barred{\nu}\barred{\rho}} \ \frac{i\Pi^{\mu\nu\rho\barred{\mu}\barred{\nu}\barred{\rho}}_{\lambda\sigma\kappa\barred{\lambda}\barred{\sigma}\barred{\kappa}}(k)}{s-m^2} \ \mathcal{A}_{3}\up{12pt}{\hspace{-5pt}\mathrm{NS_1^2 NS_1^2};\mathrm{R_0 R_0};\mathrm{R_0 R_0}}{0pt}{}\hspace{-72pt}(-k;\varepsilon_3,k_3;\varepsilon_4,k_4)^{\lambda\sigma\kappa\barred{\lambda}\barred{\sigma}\barred{\kappa}}
\\ = \left(\hspace{-3pt}\frac{8B_1 B_2^2\pi i g_{\text{s}}}{\alpha'}\hspace{-3pt}\right)\hspace{-3pt}\frac{i}{s\hspace{-2pt}-\hspace{-2pt}m^2}\hspace{-3pt}\left(\hspace{-3pt}-\frac{A_2^2 B_2\pi i g_{\text{s}}}{3!4}\hspace{-3pt}\right)\hspace{-3pt}(2\pi)^d\delta^d(k_1\hspace{-1pt}+\hspace{-1pt}k_2\hspace{-1pt}+\hspace{-1pt}k_3\hspace{-1pt}+\hspace{-1pt}k_4)\varepsilon_{1,\mu_1\nu_1\barred{\mu}_1\barred{\nu}_1}\varepsilon_{2,\mu_2\nu_2\rho_2\barred{\mu}_2\barred{\nu}_2\barred{\rho}_2}\varepsilon_3^{\alpha_3\barred{\alpha}_3}\varepsilon_4^{\alpha_4\barred{\alpha}_4}
\\ \hspace{-40pt} \times \hspace{-2pt}(\hspace{-1pt}-\hspace{-1pt}1\hspace{-1pt})\hspace{-4pt}\left[\hspace{-2pt}3\eta\up{9pt}{[\mu_2}{2pt}{\mu_1}\hspace{-2pt}\eta\up{2pt}{[\mu}{-4pt}{\nu_1}\eta\up{10pt}{\nu_2}{1pt}{\nu}\eta\up{10pt}{\rho_2]}{1pt}{\rho]} \hspace{-3pt}+\hspace{-3pt} \frac{\alpha'}{2}\hspace{-1pt}k_2^{\mu_1}\hspace{-4pt}\left(\hspace{-4pt}k_2^{\nu_1}\eta\up{8pt}{[\mu_2}{2pt}{\mu}\eta\up{8pt}{\nu_2}{2pt}{\nu}\eta\up{8pt}{\rho_2]}{2pt}{\rho} \hspace{-4pt}-\hspace{-3pt} 3k\up{9pt}{\hspace{-2pt}[\mu_2}{-3pt}{1}\hspace{-2pt}\eta\up{1pt}{[\mu}{-5pt}{\nu_1}\eta\up{9pt}{\nu_2}{1pt}{\nu}\eta\up{10pt}{\rho_2]}{1pt}{\rho]} \hspace{-3pt}+\hspace{-3pt} \frac{3}{2}k\up{9pt}{\hspace{-2pt}[\mu}{-3pt}{12}\hspace{-2pt}\eta\up{1pt}{[\mu_2}{-5pt}{\nu_1}\eta\up{9pt}{\nu}{1pt}{\nu_2}\eta\up{10pt}{\rho]}{1pt}{\rho_2]}\hspace{-2pt}\right)\hspace{-2pt}\right]\hspace{-4pt}\hat{\eta}^{[\mu}_{\lambda}\hspace{-2pt}(\hspace{-1pt}k_3\hspace{-3pt}+\hspace{-3pt}k_4\hspace{-1pt})\hspace{-1pt}\hat{\eta}^{\nu}_{\sigma}\hspace{-2pt}(\hspace{-1pt}k_3\hspace{-3pt}+\hspace{-3pt}k_4\hspace{-1pt})\hspace{-1pt}\hat{\eta}^{\rho]}_{\kappa}\hspace{-2pt}(\hspace{-1pt}k_3\hspace{-3pt}+\hspace{-3pt}k_4\hspace{-1pt})\hspace{-1pt}(\hspace{-1pt}\gamma^{\lambda\sigma\kappa}\hspace{-1pt})_{\hspace{-1pt}\alpha_3\alpha_4}
\\ \hspace{-40pt}\times \hspace{-2pt}(\hspace{-1pt}-\hspace{-1pt}1\hspace{-1pt})\hspace{-4pt}\left[\hspace{-2pt}3\eta\up{9pt}{[\barred{\mu}_2}{2pt}{\barred{\mu}_1}\hspace{-2pt}\eta\up{2pt}{[\barred{\mu}}{-4pt}{\barred{\nu}_1}\eta\up{10pt}{\barred{\nu}_2}{1pt}{\barred{\nu}}\eta\up{10pt}{\barred{\rho}_2]}{1pt}{\barred{\rho}]} \hspace{-3pt}+\hspace{-3pt} \frac{\alpha'}{2}\hspace{-1pt}k_2^{\barred{\mu}_1}\hspace{-4pt}\left(\hspace{-4pt}k_2^{\barred{\nu}_1}\eta\up{8pt}{[\barred{\mu}_2}{2pt}{\barred{\mu}}\eta\up{8pt}{\barred{\nu}_2}{2pt}{\barred{\nu}}\eta\up{8pt}{\barred{\rho}_2]}{2pt}{\barred{\rho}} \hspace{-4pt}-\hspace{-3pt} 3k\up{9pt}{\hspace{-2pt}[\barred{\mu}_2}{-3pt}{1}\hspace{-2pt}\eta\up{1pt}{[\barred{\mu}}{-5pt}{\barred{\nu}_1}\eta\up{9pt}{\barred{\nu}_2}{1pt}{\barred{\nu}}\eta\up{10pt}{\barred{\rho}_2]}{1pt}{\barred{\rho}]} \hspace{-3pt}+\hspace{-3pt} \frac{3}{2}k\up{9pt}{\hspace{-2pt}[\barred{\mu}}{-3pt}{12}\hspace{-2pt}\eta\up{1pt}{[\barred{\mu}_2}{-5pt}{\barred{\nu}_1}\eta\up{9pt}{\barred{\nu}}{1pt}{\barred{\nu}_2}\eta\up{10pt}{\barred{\rho}]}{1pt}{\barred{\rho}_2]}\hspace{-2pt}\right)\hspace{-2pt}\right]\hspace{-4pt}\hat{\eta}^{[\barred{\mu}}_{\barred{\lambda}}\hspace{-2pt}(\hspace{-1pt}k_3\hspace{-3pt}+\hspace{-3pt}k_4\hspace{-1pt})\hspace{-1pt}\hat{\eta}^{\barred{\nu}}_{\barred{\sigma}}\hspace{-2pt}(\hspace{-1pt}k_3\hspace{-3pt}+\hspace{-3pt}k_4\hspace{-1pt})\hspace{-1pt}\hat{\eta}^{\barred{\rho}]}_{\barred{\kappa}}\hspace{-2pt}(\hspace{-1pt}k_3\hspace{-3pt}+\hspace{-3pt}k_4\hspace{-1pt})\hspace{-1pt}(\hspace{-1pt}\gamma^{\barred{\lambda}\barred{\sigma}\barred{\kappa}}\hspace{-1pt})_{\hspace{-1pt}\barred{\alpha}_3\barred{\alpha}_4}\hspace{-1pt}.
\end{multline}
On the support of the delta function and the physicality conditions, we compute
\begin{align}
\hat{\eta}^{\{\mu}_{\rho}(k_3\hspace{-2pt}+\hspace{-2pt}k_4)\hat{\eta}^{\nu\}}_{\lambda}(k_3\hspace{-2pt}+\hspace{-2pt}k_4)k_{34}^{\rho}\gamma^{\lambda}_{\alpha_3\alpha_4} & = k_{34}^{(\mu}\gamma^{\nu)}_{\alpha_3\alpha_4}
\\ \hat{\eta}^{[\mu}_{\lambda}\hspace{-2pt}(\hspace{-1pt}k_3\hspace{-3pt}+\hspace{-3pt}k_4\hspace{-1pt})\hspace{-1pt}\hat{\eta}^{\nu}_{\sigma}\hspace{-1pt}(\hspace{-1pt}k_3\hspace{-3pt}+\hspace{-3pt}k_4\hspace{-1pt})\hat{\eta}^{\rho]}_{\kappa}\hspace{-1pt}(\hspace{-1pt}k_3\hspace{-3pt}+\hspace{-3pt}k_4\hspace{-1pt})(\gamma^{\lambda\sigma\kappa})_{\alpha_3\alpha_4} & = (\gamma^{\mu\nu\rho})_{\alpha\beta} + \frac{12}{s}k_3^{[\mu}k_4^{\nu}\gamma^{\rho]}_{\alpha_3\alpha_4}.
\end{align}
Performing the remaining contractions (for which we use $k_1^{[\mu_2}k_3^{\nu_2}k_4^{\rho_2]} = 0$, again valid on momentum conservation and the physicality conditions), the above pole factorizations are 
\begin{align}
\notag \hspace{25pt}&\hspace{-25pt}\int\frac{d^d k}{(2\pi)^d}\mathcal{A}_{3}\up{12pt}{\hspace{-5pt}\mathrm{NS_1^1 NS_1^1};\mathrm{NS_1^2 NS_1^2};\mathrm{NS_1^1 NS_1^1}}{0pt}{}\hspace{-75pt}(\varepsilon_1,k_1;\varepsilon_2,k_2;k)_{\mu\nu\barred{\mu}\barred{\nu}} \ \frac{i\Pi^{\mu\nu\barred{\mu}\barred{\nu}}_{\rho\lambda\barred{\rho}\barred{\lambda}}(k)}{s-m^2} \ \mathcal{A}_{3}\up{12pt}{\hspace{-5pt}\mathrm{NS_1^1 NS_1^1};\mathrm{R_0 R_0};\mathrm{R_0 R_0}}{0pt}{}\hspace{-72pt}(-k;\varepsilon_3,k_3;\varepsilon_4,k_4)^{\rho\lambda\barred{\rho}\barred{\lambda}}
\\ \notag & = \frac{3!A_2^2 B_1^3 B_2 \pi^2 i\alpha'g_{\text{s}}^2}{8(s\hspace{-2pt}-\hspace{-2pt}m^2)}(2\pi)^d\delta^d(k_1\hspace{-1pt}+\hspace{-1pt}k_2\hspace{-1pt}+\hspace{-1pt}k_3\hspace{-1pt}+\hspace{-1pt}k_4)\varepsilon_{1,\mu_1\nu_1\barred{\mu}_1\barred{\nu}_1}\varepsilon_{2,\mu_2\nu_2\rho_2\barred{\mu}_2\barred{\nu}_2\barred{\rho}_2}\varepsilon_3^{\alpha_3\barred{\alpha}_3}\varepsilon_4^{\alpha_4\barred{\alpha}_4}
\\ \notag & \hspace{15pt} \times \hspace{-2pt}\left[\hspace{-1pt}\eta^{\mu_1\hspace{-1pt}[\hspace{-1pt}\mu_2}\hspace{-2pt}k_1^{\nu_2}\hspace{-2pt}k_{34}^{\rho_2\hspace{-1pt}]}\hspace{-4pt}\left(\hspace{-4pt}\gamma^{\nu_1}_{\alpha_3\alpha_4}\hspace{-3pt}+\hspace{-2pt}\frac{\alpha'}{4}\hspace{-1pt}k_2^{\nu_1}\hspace{-2pt}(\hspace{-1pt}k_{12}\hspace{-3pt}\cdot\hspace{-3pt}\gamma\hspace{-1pt})_{\hspace{-1pt}\alpha_3\alpha_4}\hspace{-4pt}\right) \hspace{-2pt}-\hspace{-2pt} \left(\hspace{-3pt}\frac{t\hspace{-2pt}-\hspace{-2pt}u}{m^2}k_2^{\nu_1}\hspace{-3pt}-\hspace{-3pt}k_{34}^{\nu_1}\hspace{-3pt}\right)\hspace{-4pt}\eta^{\mu_1\hspace{-1pt}[\hspace{-1pt}\mu_2}\hspace{-2pt}k_1^{\nu_2}\hspace{-1pt}\gamma^{\rho_2\hspace{-1pt}]}_{\hspace{-1pt}\alpha_3\alpha_4}\hspace{-1pt}\right]
\\ & \hspace{15pt} \times \hspace{-2pt}\left[\hspace{-1pt}\eta^{\barred{\mu}_1\hspace{-1pt}[\hspace{-1pt}\barred{\mu}_2}\hspace{-2pt}k_1^{\barred{\nu}_2}\hspace{-2pt}k_{34}^{\barred{\rho}_2\hspace{-1pt}]}\hspace{-4pt}\left(\hspace{-4pt}\gamma^{\barred{\nu}_1}_{\barred{\alpha}_3\barred{\alpha}_4}\hspace{-3pt}+\hspace{-2pt}\frac{\alpha'}{4}\hspace{-1pt}k_2^{\barred{\nu}_1}\hspace{-2pt}(\hspace{-1pt}k_{12}\hspace{-3pt}\cdot\hspace{-3pt}\gamma\hspace{-1pt})_{\hspace{-1pt}\barred{\alpha}_3\barred{\alpha}_4}\hspace{-4pt}\right) \hspace{-2pt}-\hspace{-2pt} \left(\hspace{-3pt}\frac{t\hspace{-2pt}-\hspace{-2pt}u}{m^2}k_2^{\barred{\nu}_1}\hspace{-3pt}-\hspace{-3pt}k_{34}^{\barred{\nu}_1}\hspace{-3pt}\right)\hspace{-4pt}\eta^{\barred{\mu}_1\hspace{-1pt}[\hspace{-1pt}\barred{\mu}_2}\hspace{-2pt}k_1^{\barred{\nu}_2}\hspace{-1pt}\gamma^{\barred{\rho}_2\hspace{-1pt}]}_{\hspace{-1pt}\barred{\alpha}_3\barred{\alpha}_4}\hspace{-2pt}\right]
\\ \notag \hspace{25pt}&\hspace{-25pt}\int\frac{d^d k}{(2\pi)^d}\mathcal{A}_{3}\up{12pt}{\hspace{-5pt}\mathrm{NS_1^1 NS_1^1};\mathrm{NS_1^2 NS_1^2};\mathrm{NS_1^2 NS_1^2}}{0pt}{}\hspace{-73pt}(\varepsilon_1,k_1;\varepsilon_2,k_2;k)_{\mu\nu\rho\barred{\mu}\barred{\nu}\barred{\rho}} \ \frac{i\Pi^{\mu\nu\rho\barred{\mu}\barred{\nu}\barred{\rho}}_{\lambda\sigma\kappa\barred{\lambda}\barred{\sigma}\barred{\kappa}}(k)}{s-m^2} \ \mathcal{A}_{3}\up{12pt}{\hspace{-5pt}\mathrm{NS_1^2 NS_1^2};\mathrm{R_0 R_0};\mathrm{R_0 R_0}}{0pt}{}\hspace{-72pt}(-k;\varepsilon_3,k_3;\varepsilon_4,k_4)^{\lambda\sigma\kappa\barred{\lambda}\barred{\sigma}\barred{\kappa}}
\\ \notag & = \frac{2A_2^2 B_1 B_2^3\pi^2 ig_{\text{s}}^2}{3!\alpha'(s\hspace{-2pt}-\hspace{-2pt}m^2)}(2\pi)^d\delta^d(k_1\hspace{-1pt}+\hspace{-1pt}k_2\hspace{-1pt}+\hspace{-1pt}k_3\hspace{-1pt}+\hspace{-1pt}k_4)\varepsilon_{1,\mu_1\nu_1\barred{\mu}_1\barred{\nu}_1}\varepsilon_{2,\mu_2\nu_2\rho_2\barred{\mu}_2\barred{\nu}_2\barred{\rho}_2}\varepsilon_3^{\alpha_3\barred{\alpha}_3}\varepsilon_4^{\alpha_4\barred{\alpha}_4}
\\ \notag & \hspace{-35pt}\times\hspace{-3pt}(\hspace{-1pt}-\hspace{-1pt}1\hspace{-1pt})\hspace{-4pt} \left[\hspace{-2pt}\frac{6}{m^2}\hspace{-1pt}\eta^{\mu_1\hspace{-1pt}[\mu_2}\hspace{-5pt}\left(\hspace{-3pt}k_{34}^{\nu_2}\hspace{-1pt}\gamma^{\rho_2\hspace{-1pt}]}_{\hspace{-1pt}\alpha_3\alpha_4}\hspace{-1pt}k_2^{\hspace{-1pt}\nu_1} \hspace{-4pt}-\hspace{-3pt} k_1^{\hspace{-1pt}\nu_2}\hspace{-1pt}\gamma^{\rho_2\hspace{-1pt}]}_{\hspace{-1pt}\alpha_3\alpha_4}\hspace{-1pt}k_{34}^{\nu_1} \hspace{-3pt}+\hspace{-3pt} k_1^{\nu_2}\hspace{-2pt}k_{34}^{\rho_2\hspace{-1pt}]}\hspace{-1pt}\gamma^{\nu_1}_{\hspace{-1pt}\alpha_3\alpha_4}\hspace{-3pt}\right) \hspace{-3pt}+\hspace{-3pt} \frac{3\alpha'}{2m^2}\hspace{-1pt}k_2^{\mu_1}\hspace{-2pt}\eta^{\nu_1\hspace{-1pt}[\mu_2}\hspace{-5pt}\left(\hspace{-3pt}(\hspace{-1pt}t\hspace{-3pt}-\hspace{-3pt}u\hspace{-1pt})\hspace{-2pt}k_1^{\nu_2}\hspace{-1pt}\gamma^{\rho_2\hspace{-1pt}]}_{\hspace{-1pt}\alpha_3\alpha_4} \hspace{-3pt}+\hspace{-3pt} k_1^{\nu_2}\hspace{-2pt}k_{34}^{\rho_2\hspace{-1pt}]}\hspace{-2pt}(\hspace{-2pt}k_{12}\hspace{-3pt}\cdot\hspace{-3pt}\gamma\hspace{-2pt})_{\hspace{-1pt}\alpha_3\alpha_4}\hspace{-4pt}\right) \right. 
\\ \notag & \hspace{-10pt} \left. + 3\eta^{\mu_1\hspace{-1pt}[\hspace{-1pt}\mu_2}\hspace{-1pt}(\hspace{-1pt}\gamma^{\nu_2\rho_2\hspace{-1pt}]\nu_1}\hspace{-2pt})_{\hspace{-1pt}\alpha_3\alpha_4} \hspace{-2pt}+\hspace{-2pt} \frac{\alpha'}{2}\hspace{-1pt}k_2^{\mu_1}\hspace{-5pt}\left(\hspace{-4pt}k_2^{\nu_1}\hspace{-2pt}(\hspace{-1pt}\gamma^{\mu_2\nu_2\rho_2}\hspace{-1pt})_{\hspace{-1pt}\alpha_3\alpha_4} \hspace{-3pt}-\hspace{-3pt} 3k_1^{\hspace{-1pt}[\mu_2}\hspace{-2pt}(\hspace{-1pt}\gamma^{\nu_2\rho_2\hspace{-1pt}]\nu_1}\hspace{-2pt})_{\hspace{-1pt}\alpha_3\alpha_4} \hspace{-3pt}+\hspace{-3pt} \frac{3}{2}\hspace{-1pt}k_{12}^{\mu}\hspace{-1pt}\eta^{\nu_1\hspace{-1pt}[\mu_2}\hspace{-2pt}(\hspace{-1pt}\gamma^{\nu_2\rho_2\hspace{-1pt}]}_{\phantom{\nu_2\rho_2]}\hspace{-2pt}\mu}\hspace{-1pt})_{\hspace{-1pt}\alpha_3\alpha_4}\hspace{-4pt}\right)\hspace{-2pt}\right]
\\ \notag & \hspace{-35pt}\times\hspace{-3pt}(\hspace{-1pt}-\hspace{-1pt}1\hspace{-1pt})\hspace{-4pt} \left[\hspace{-2pt}\frac{6}{m^2}\hspace{-1pt}\eta^{\barred{\mu}_1\hspace{-1pt}[\barred{\mu}_2}\hspace{-5pt}\left(\hspace{-3pt}k_{34}^{\barred{\nu}_2}\hspace{-1pt}\gamma^{\barred{\rho}_2\hspace{-1pt}]}_{\hspace{-1pt}\barred{\alpha}_3\barred{\alpha}_4}\hspace{-1pt}k_2^{\hspace{-1pt}\barred{\nu}_1} \hspace{-4pt}-\hspace{-3pt} k_1^{\hspace{-1pt}\barred{\nu}_2}\hspace{-1pt}\gamma^{\barred{\rho}_2\hspace{-1pt}]}_{\hspace{-1pt}\barred{\alpha}_3\barred{\alpha}_4}\hspace{-1pt}k_{34}^{\barred{\nu}_1} \hspace{-3pt}+\hspace{-3pt} k_1^{\barred{\nu}_2}\hspace{-2pt}k_{34}^{\barred{\rho}_2\hspace{-1pt}]}\hspace{-1pt}\gamma^{\barred{\nu}_1}_{\hspace{-1pt}\barred{\alpha}_3\barred{\alpha}_4}\hspace{-3pt}\right) \hspace{-3pt}+\hspace{-3pt} \frac{3\alpha'}{2m^2}\hspace{-1pt}k_2^{\barred{\mu}_1}\hspace{-2pt}\eta^{\barred{\nu}_1\hspace{-1pt}[\barred{\mu}_2}\hspace{-5pt}\left(\hspace{-3pt}(\hspace{-1pt}t\hspace{-3pt}-\hspace{-3pt}u\hspace{-1pt})\hspace{-2pt}k_1^{\barred{\nu}_2}\hspace{-1pt}\gamma^{\barred{\rho}_2\hspace{-1pt}]}_{\hspace{-1pt}\barred{\alpha}_3\barred{\alpha}_4} \hspace{-3pt}+\hspace{-3pt} k_1^{\barred{\nu}_2}\hspace{-2pt}k_{34}^{\barred{\rho}_2\hspace{-1pt}]}\hspace{-2pt}(\hspace{-2pt}k_{12}\hspace{-3pt}\cdot\hspace{-3pt}\gamma\hspace{-2pt})_{\hspace{-1pt}\barred{\alpha}_3\barred{\alpha}_4}\hspace{-4pt}\right) \right. 
\\ & \hspace{-10pt} \left. + 3\eta^{\barred{\mu}_1\hspace{-1pt}[\hspace{-1pt}\barred{\mu}_2}\hspace{-1pt}(\hspace{-1pt}\gamma^{\barred{\nu}_2\barred{\rho}_2\hspace{-1pt}]\barred{\nu}_1}\hspace{-2pt})_{\hspace{-1pt}\barred{\alpha}_3\barred{\alpha}_4} \hspace{-2pt}+\hspace{-2pt} \frac{\alpha'}{2}\hspace{-1pt}k_2^{\barred{\mu}_1}\hspace{-5pt}\left(\hspace{-4pt}k_2^{\barred{\nu}_1}\hspace{-2pt}(\hspace{-1pt}\gamma^{\barred{\mu}_2\barred{\nu}_2\barred{\rho}_2}\hspace{-1pt})_{\hspace{-1pt}\barred{\alpha}_3\barred{\alpha}_4} \hspace{-3pt}-\hspace{-3pt} 3k_1^{\hspace{-1pt}[\barred{\mu}_2}\hspace{-2pt}(\hspace{-1pt}\gamma^{\barred{\nu}_2\barred{\rho}_2\hspace{-1pt}]\barred{\nu}_1}\hspace{-2pt})_{\hspace{-1pt}\barred{\alpha}_3\barred{\alpha}_4} \hspace{-3pt}+\hspace{-3pt} \frac{3}{2}\hspace{-1pt}k_{12}^{\barred{\mu}}\hspace{-1pt}\eta^{\barred{\nu}_1\hspace{-1pt}[\barred{\mu}_2}\hspace{-2pt}(\hspace{-1pt}\gamma^{\barred{\nu}_2\barred{\rho}_2\hspace{-1pt}]}_{\phantom{\barred{\nu}_2\barred{\rho}_2]}\hspace{-2pt}\barred{\mu}}\hspace{-1pt})_{\hspace{-1pt}\barred{\alpha}_3\barred{\alpha}_4}\hspace{-4pt}\right)\hspace{-2pt}\right].
\end{align}
The contributions from the other two massive NSNS vertex operators are then obtained via holomorphic factorization (which is why phases from the two sides were kept separate above).
Then, in order for the sum of all four contributions to itself holomorphically factorize, it is first necessary that $B_1 B_2 = B_3 B_4$. Then, the total $s$-channel pole factorization is
\begin{multline}
\int\frac{d^d k}{(2\pi)^d}\mathcal{A}_{3}\up{12pt}{\hspace{-5pt}\mathrm{NS_1^1 NS_1^1};\mathrm{NS_1^2 NS_1^2};\mathrm{NS_1^1 NS_1^1}}{0pt}{}\hspace{-75pt}(\varepsilon_1,k_1;\varepsilon_2,k_2;k)_{\mu\nu\barred{\mu}\barred{\nu}} \ \frac{i\Pi^{\mu\nu\barred{\mu}\barred{\nu}}_{\rho\lambda\barred{\rho}\barred{\lambda}}(k)}{s-m^2} \ \mathcal{A}_{3}\up{12pt}{\hspace{-5pt}\mathrm{NS_1^1 NS_1^1};\mathrm{R_0 R_0};\mathrm{R_0 R_0}}{0pt}{}\hspace{-72pt}(-k;\varepsilon_3,k_3;\varepsilon_4,k_4)^{\rho\lambda\barred{\rho}\barred{\lambda}}
\\* + \int\frac{d^d k}{(2\pi)^d}\mathcal{A}_{3}\up{12pt}{\hspace{-5pt}\mathrm{NS_1^1 NS_1^1};\mathrm{NS_1^2 NS_1^2};\mathrm{NS_1^2 NS_1^2}}{0pt}{}\hspace{-73pt}(\varepsilon_1,k_1;\varepsilon_2,k_2;k)_{\mu\nu\rho\barred{\mu}\barred{\nu}\barred{\rho}} \ \frac{i\Pi^{\mu\nu\rho\barred{\mu}\barred{\nu}\barred{\rho}}_{\lambda\sigma\kappa\barred{\lambda}\barred{\sigma}\barred{\kappa}}(k)}{s-m^2} \ \mathcal{A}_{3}\up{12pt}{\hspace{-5pt}\mathrm{NS_1^2 NS_1^2};\mathrm{R_0 R_0};\mathrm{R_0 R_0}}{0pt}{}\hspace{-72pt}(-k;\varepsilon_3,k_3;\varepsilon_4,k_4)^{\lambda\sigma\kappa\barred{\lambda}\barred{\sigma}\barred{\kappa}}
\\ + \int\frac{d^d k}{(2\pi)^d}\mathcal{A}_{3}\up{12pt}{\hspace{-5pt}\mathrm{NS_1^1 NS_1^1};\mathrm{NS_1^2 NS_1^2};\mathrm{NS_1^1 NS_1^2}}{0pt}{}\hspace{-73pt}(\varepsilon_1,k_1;\varepsilon_2,k_2;k)_{\mu\nu\barred{\mu}\barred{\nu}\barred{\rho}} \ \frac{i\Pi^{\mu\nu\barred{\mu}\barred{\nu}\barred{\rho}}_{\lambda\sigma\barred{\lambda}\barred{\sigma}\barred{\kappa}}(k)}{s-m^2} \ \mathcal{A}_{3}\up{12pt}{\hspace{-5pt}\mathrm{NS_1^1 NS_1^2};\mathrm{R_0 R_0};\mathrm{R_0 R_0}}{0pt}{}\hspace{-72pt}(-k;\varepsilon_3,k_3;\varepsilon_4,k_4)^{\lambda\sigma\kappa\barred{\lambda}\barred{\sigma}\barred{\kappa}}
\\ + \int\frac{d^d k}{(2\pi)^d}\mathcal{A}_{3}\up{12pt}{\hspace{-5pt}\mathrm{NS_1^1 NS_1^1};\mathrm{NS_1^2 NS_1^2};\mathrm{NS_1^2 NS_1^1}}{0pt}{}\hspace{-75pt}(\varepsilon_1,k_1;\varepsilon_2,k_2;k)_{\mu\nu\rho\barred{\mu}\barred{\nu}} \ \frac{i\Pi^{\mu\nu\rho\barred{\mu}\barred{\nu}}_{\lambda\sigma\kappa\barred{\lambda}\barred{\sigma}}(k)}{s-m^2} \ \mathcal{A}_{3}\up{12pt}{\hspace{-5pt}\mathrm{NS_1^2 NS_1^1};\mathrm{R_0 R_0};\mathrm{R_0 R_0}}{0pt}{}\hspace{-72pt}(-k;\varepsilon_3,k_3;\varepsilon_4,k_4)^{\lambda\sigma\kappa\barred{\lambda}\barred{\sigma}}
\\ = \frac{3!A_2^2 B_1^3 B_2\pi^2 i\alpha'g_{\text{s}}^2}{8(s\hspace{-2pt}-\hspace{-2pt}m^2)}(2\pi)^d\delta^d(k_1\hspace{-2pt}+\hspace{-2pt}k_2\hspace{-2pt}+\hspace{-2pt}k_3\hspace{-2pt}+\hspace{-2pt}k_4)\varepsilon_{1,\mu_1\nu_1\barred{\mu}_1\barred{\nu}_1}\varepsilon_{2,\mu_2\nu_2\rho_2\barred{\mu}_2\barred{\nu}_2\barred{\rho}_2}\varepsilon_3^{\alpha_3\barred{\alpha}_3}\varepsilon_4^{\alpha_4\barred{\alpha}_4}
\\ \times \hspace{-4pt}\left\{\hspace{-2pt}\eta^{\mu_1\hspace{-1pt}[\hspace{-1pt}\mu_2}\hspace{-2pt}k_1^{\nu_2}\hspace{-2pt}k_{34}^{\rho_2\hspace{-1pt}]}\hspace{-4pt}\left(\hspace{-4pt}\gamma^{\nu_1}_{\alpha_3\alpha_4}\hspace{-3pt}+\hspace{-2pt}\frac{\alpha'}{4}\hspace{-1pt}k_2^{\nu_1}\hspace{-2pt}(\hspace{-1pt}k_{12}\hspace{-3pt}\cdot\hspace{-3pt}\gamma\hspace{-1pt})_{\hspace{-1pt}\alpha_3\alpha_4}\hspace{-4pt}\right) \hspace{-2pt}-\hspace{-2pt} \left(\hspace{-3pt}\frac{t\hspace{-2pt}-\hspace{-2pt}u}{m^2}k_2^{\nu_1}\hspace{-3pt}-\hspace{-3pt}k_{34}^{\nu_1}\hspace{-3pt}\right)\hspace{-4pt}\eta^{\mu_1\hspace{-1pt}[\hspace{-1pt}\mu_2}\hspace{-2pt}k_1^{\nu_2}\hspace{-1pt}\gamma^{\rho_2\hspace{-1pt}]}_{\hspace{-1pt}\alpha_3\alpha_4} \right. 
\\ \left. \hspace{-25pt} - \frac{4B_3^2}{3!\alpha'B_1^2}\hspace{-4pt}\left[\hspace{-2pt}\frac{6}{m^2}\hspace{-1pt}\eta^{\mu_1\hspace{-1pt}[\mu_2}\hspace{-5pt}\left(\hspace{-3pt}k_{34}^{\nu_2}\hspace{-1pt}\gamma^{\rho_2\hspace{-1pt}]}_{\hspace{-1pt}\alpha_3\alpha_4}\hspace{-1pt}k_2^{\hspace{-1pt}\nu_1} \hspace{-4pt}-\hspace{-3pt} k_1^{\hspace{-1pt}\nu_2}\hspace{-1pt}\gamma^{\rho_2\hspace{-1pt}]}_{\hspace{-1pt}\alpha_3\alpha_4}\hspace{-1pt}k_{34}^{\nu_1} \hspace{-3pt}+\hspace{-3pt} k_1^{\nu_2}\hspace{-2pt}k_{34}^{\rho_2\hspace{-1pt}]}\hspace{-1pt}\gamma^{\nu_1}_{\hspace{-1pt}\alpha_3\alpha_4}\hspace{-3pt}\right) \hspace{-3pt}+\hspace{-3pt} \frac{3\alpha'}{2m^2}\hspace{-1pt}k_2^{\mu_1}\hspace{-2pt}\eta^{\nu_1\hspace{-1pt}[\mu_2}\hspace{-5pt}\left(\hspace{-3pt}(\hspace{-1pt}t\hspace{-3pt}-\hspace{-3pt}u\hspace{-1pt})\hspace{-2pt}k_1^{\nu_2}\hspace{-1pt}\gamma^{\rho_2\hspace{-1pt}]}_{\hspace{-1pt}\alpha_3\alpha_4} \hspace{-3pt}+\hspace{-3pt} k_1^{\nu_2}\hspace{-2pt}k_{34}^{\rho_2\hspace{-1pt}]}\hspace{-2pt}(\hspace{-2pt}k_{12}\hspace{-3pt}\cdot\hspace{-3pt}\gamma\hspace{-2pt})_{\hspace{-1pt}\alpha_3\alpha_4}\hspace{-4pt}\right) \right. \right.
\\ \left. \left. + 3\eta^{\mu_1\hspace{-1pt}[\hspace{-1pt}\mu_2}\hspace{-1pt}(\hspace{-1pt}\gamma^{\nu_2\rho_2\hspace{-1pt}]\nu_1}\hspace{-2pt})_{\hspace{-1pt}\alpha_3\alpha_4} \hspace{-2pt}+\hspace{-2pt} \frac{\alpha'}{2}\hspace{-1pt}k_2^{\mu_1}\hspace{-5pt}\left(\hspace{-4pt}k_2^{\nu_1}\hspace{-2pt}(\hspace{-1pt}\gamma^{\mu_2\nu_2\rho_2}\hspace{-1pt})_{\hspace{-1pt}\alpha_3\alpha_4} \hspace{-3pt}-\hspace{-3pt} 3k_1^{\hspace{-1pt}[\mu_2}\hspace{-2pt}(\hspace{-1pt}\gamma^{\nu_2\rho_2\hspace{-1pt}]\nu_1}\hspace{-2pt})_{\hspace{-1pt}\alpha_3\alpha_4} \hspace{-3pt}+\hspace{-3pt} \frac{3}{2}\hspace{-1pt}k_{12}^{\mu}\hspace{-1pt}\eta^{\nu_1\hspace{-1pt}[\mu_2}\hspace{-2pt}(\hspace{-1pt}\gamma^{\nu_2\rho_2\hspace{-1pt}]}_{\phantom{\nu_2\rho_2]}\hspace{-2pt}\mu}\hspace{-1pt})_{\hspace{-1pt}\alpha_3\alpha_4}\hspace{-4pt}\right)\hspace{-2pt}\right]\hspace{-2pt}\right\}
\\ \times \hspace{-4pt}\left\{\hspace{-2pt}\eta^{\barred{\mu}_1\hspace{-1pt}[\hspace{-1pt}\barred{\mu}_2}\hspace{-2pt}k_1^{\barred{\nu}_2}\hspace{-2pt}k_{34}^{\barred{\rho}_2\hspace{-1pt}]}\hspace{-4pt}\left(\hspace{-4pt}\gamma^{\barred{\nu}_1}_{\barred{\alpha}_3\barred{\alpha}_4}\hspace{-3pt}+\hspace{-2pt}\frac{\alpha'}{4}\hspace{-1pt}k_2^{\barred{\nu}_1}\hspace{-2pt}(\hspace{-1pt}k_{12}\hspace{-3pt}\cdot\hspace{-3pt}\gamma\hspace{-1pt})_{\hspace{-1pt}\barred{\alpha}_3\barred{\alpha}_4}\hspace{-4pt}\right) \hspace{-2pt}-\hspace{-2pt} \left(\hspace{-3pt}\frac{\alpha'\hspace{-1pt}(\hspace{-1pt}t\hspace{-3pt}-\hspace{-3pt}u\hspace{-1pt})}{4}k_2^{\barred{\nu}_1}\hspace{-3pt}-\hspace{-3pt}k_{34}^{\barred{\nu}_1}\hspace{-3pt}\right)\hspace{-4pt}\eta^{\barred{\mu}_1\hspace{-1pt}[\hspace{-1pt}\barred{\mu}_2}\hspace{-2pt}k_1^{\barred{\nu}_2}\hspace{-1pt}\gamma^{\barred{\rho}_2\hspace{-1pt}]}_{\hspace{-1pt}\barred{\alpha}_3\barred{\alpha}_4} \right. 
\\ \left. \hspace{-25pt} - \frac{4B_4^2}{3!\alpha'B_1^2}\hspace{-4pt}\left[\hspace{-2pt}\frac{6}{m^2}\hspace{-1pt}\eta^{\barred{\mu}_1\hspace{-1pt}[\barred{\mu}_2}\hspace{-5pt}\left(\hspace{-3pt}k_{34}^{\barred{\nu}_2}\hspace{-1pt}\gamma^{\barred{\rho}_2\hspace{-1pt}]}_{\hspace{-1pt}\barred{\alpha}_3\barred{\alpha}_4}\hspace{-1pt}k_2^{\hspace{-1pt}\barred{\nu}_1} \hspace{-4pt}-\hspace{-3pt} k_1^{\hspace{-1pt}\barred{\nu}_2}\hspace{-1pt}\gamma^{\barred{\rho}_2\hspace{-1pt}]}_{\hspace{-1pt}\barred{\alpha}_3\barred{\alpha}_4}\hspace{-1pt}k_{34}^{\barred{\nu}_1} \hspace{-3pt}+\hspace{-3pt} k_1^{\barred{\nu}_2}\hspace{-2pt}k_{34}^{\barred{\rho}_2\hspace{-1pt}]}\hspace{-1pt}\gamma^{\barred{\nu}_1}_{\hspace{-1pt}\barred{\alpha}_3\barred{\alpha}_4}\hspace{-3pt}\right) \hspace{-3pt}+\hspace{-3pt} \frac{3\alpha'}{2m^2}\hspace{-1pt}k_2^{\barred{\mu}_1}\hspace{-2pt}\eta^{\barred{\nu}_1\hspace{-1pt}[\barred{\mu}_2}\hspace{-5pt}\left(\hspace{-3pt}(\hspace{-1pt}t\hspace{-3pt}-\hspace{-3pt}u\hspace{-1pt})\hspace{-2pt}k_1^{\barred{\nu}_2}\hspace{-1pt}\gamma^{\barred{\rho}_2\hspace{-1pt}]}_{\hspace{-1pt}\barred{\alpha}_3\barred{\alpha}_4} \hspace{-3pt}+\hspace{-3pt} k_1^{\barred{\nu}_2}\hspace{-2pt}k_{34}^{\barred{\rho}_2\hspace{-1pt}]}\hspace{-2pt}(\hspace{-2pt}k_{12}\hspace{-3pt}\cdot\hspace{-3pt}\gamma\hspace{-2pt})_{\hspace{-1pt}\barred{\alpha}_3\barred{\alpha}_4}\hspace{-4pt}\right) \right.\right.
\\ \left.\left. + 3\eta^{\barred{\mu}_1\hspace{-1pt}[\hspace{-1pt}\barred{\mu}_2}\hspace{-1pt}(\hspace{-1pt}\gamma^{\barred{\nu}_2\barred{\rho}_2\hspace{-1pt}]\barred{\nu}_1}\hspace{-2pt})_{\hspace{-1pt}\barred{\alpha}_3\barred{\alpha}_4} \hspace{-2pt}+\hspace{-2pt} \frac{\alpha'}{2}\hspace{-1pt}k_2^{\barred{\mu}_1}\hspace{-5pt}\left(\hspace{-4pt}k_2^{\barred{\nu}_1}\hspace{-2pt}(\hspace{-1pt}\gamma^{\barred{\mu}_2\barred{\nu}_2\barred{\rho}_2}\hspace{-1pt})_{\hspace{-1pt}\barred{\alpha}_3\barred{\alpha}_4} \hspace{-3pt}-\hspace{-3pt} 3k_1^{\hspace{-1pt}[\barred{\mu}_2}\hspace{-2pt}(\hspace{-1pt}\gamma^{\barred{\nu}_2\barred{\rho}_2\hspace{-1pt}]\barred{\nu}_1}\hspace{-2pt})_{\hspace{-1pt}\barred{\alpha}_3\barred{\alpha}_4} \hspace{-3pt}+\hspace{-3pt} \frac{3}{2}\hspace{-1pt}k_{12}^{\barred{\mu}}\hspace{-1pt}\eta^{\barred{\nu}_1\hspace{-1pt}[\barred{\mu}_2}\hspace{-2pt}(\hspace{-1pt}\gamma^{\barred{\nu}_2\barred{\rho}_2\hspace{-1pt}]}_{\phantom{\barred{\nu}_2\barred{\rho}_2]}\hspace{-2pt}\barred{\mu}}\hspace{-1pt})_{\hspace{-1pt}\barred{\alpha}_3\barred{\alpha}_4}\hspace{-4pt}\right)\hspace{-2pt}\right]\hspace{-2pt}\right\}.
\end{multline}
In order for this expression to give the correct tensor structure in the actual four-point amplitude pole written below, the relative coefficient between the pole contributions from $\mathrm{NS_1^1}$ and from $\mathrm{NS_1^2}$ in braces must be $-\frac{4}{3!\alpha'}$, in which case the above total pole factorization becomes
\begin{multline}
\int\frac{d^d k}{(2\pi)^d}\mathcal{A}_{3}\up{12pt}{\hspace{-5pt}\mathrm{NS_1^1 NS_1^1};\mathrm{NS_1^2 NS_1^2};\mathrm{NS_1^1 NS_1^1}}{0pt}{}\hspace{-80pt}(\varepsilon_1,k_1;\varepsilon_2,k_2;k)_{\mu\nu\barred{\mu}\barred{\nu}}\frac{i\Pi^{\mu\nu\barred{\mu}\barred{\nu}}_{\rho\lambda\barred{\rho}\barred{\lambda}}(k)}{s-m^2}\mathcal{A}_{3}\up{12pt}{\hspace{-5pt}\mathrm{NS_1^1 NS_1^1};\mathrm{R_0 R_0};\mathrm{R_0 R_0}}{0pt}{}\hspace{-70pt}(-k;\varepsilon_3,k_3;\varepsilon_4,k_4)^{\rho\lambda\barred{\rho}\barred{\lambda}}
\\ + \int\frac{d^d k}{(2\pi)^d}\mathcal{A}_{3}\up{12pt}{\hspace{-5pt}\mathrm{NS_1^1 NS_1^1};\mathrm{NS_1^2 NS_1^2};\mathrm{NS_1^2 NS_1^2}}{0pt}{}\hspace{-80pt}(\varepsilon_1,k_1;\varepsilon_2,k_2;k)_{\mu\nu\rho\barred{\mu}\barred{\nu}\barred{\rho}}\frac{i\Pi^{\mu\nu\rho\barred{\mu}\barred{\nu}\barred{\rho}}_{\lambda\sigma\kappa\barred{\lambda}\barred{\sigma}\barred{\kappa}}(k)}{s-m^2}\mathcal{A}_{3}\up{12pt}{\hspace{-5pt}\mathrm{NS_1^2 NS_1^2};\mathrm{R_0 R_0};\mathrm{R_0 R_0}}{0pt}{}\hspace{-70pt}(-k;\varepsilon_3,k_3;\varepsilon_4,k_4)^{\lambda\sigma\kappa\barred{\lambda}\barred{\sigma}\barred{\kappa}}
\\ + \int\frac{d^d k}{(2\pi)^d}\mathcal{A}_{3}\up{12pt}{\hspace{-5pt}\mathrm{NS_1^1 NS_1^1};\mathrm{NS_1^2 NS_1^2};\mathrm{NS_1^1 NS_1^2}}{0pt}{}\hspace{-80pt}(\varepsilon_1,k_1;\varepsilon_2,k_2;k)_{\mu\nu\barred{\mu}\barred{\nu}\barred{\rho}}\frac{i\Pi^{\mu\nu\barred{\mu}\barred{\nu}\barred{\rho}}_{\lambda\sigma\barred{\lambda}\barred{\sigma}\barred{\kappa}}(k)}{s-m^2}\mathcal{A}_{3}\up{12pt}{\hspace{-5pt}\mathrm{NS_1^1 NS_1^2};\mathrm{R_0 R_0};\mathrm{R_0 R_0}}{0pt}{}\hspace{-70pt}(-k;\varepsilon_3,k_3;\varepsilon_4,k_4)^{\lambda\sigma\kappa\barred{\lambda}\barred{\sigma}\barred{\kappa}}
\\ + \int\frac{d^d k}{(2\pi)^d}\mathcal{A}_{3}\up{12pt}{\hspace{-5pt}\mathrm{NS_1^1 NS_1^1};\mathrm{NS_1^2 NS_1^2};\mathrm{NS_1^2 NS_1^1}}{0pt}{}\hspace{-80pt}(\varepsilon_1,k_1;\varepsilon_2,k_2;k)_{\mu\nu\rho\barred{\mu}\barred{\nu}}\frac{i\Pi^{\mu\nu\rho\barred{\mu}\barred{\nu}}_{\lambda\sigma\kappa\barred{\lambda}\barred{\sigma}}(k)}{s-m^2}\mathcal{A}_{3}\up{12pt}{\hspace{-5pt}\mathrm{NS_1^2 NS_1^1};\mathrm{R_0 R_0};\mathrm{R_0 R_0}}{0pt}{}\hspace{-70pt}(-k;\varepsilon_3,k_3;\varepsilon_4,k_4)^{\lambda\sigma\kappa\barred{\lambda}\barred{\sigma}}
\\ = \frac{3!A_2^2 B_1^3 B_2\pi^2 i\alpha'g_{\text{s}}^2}{8(s\hspace{-2pt}-\hspace{-2pt}m^2)}(2\pi)^d\delta^d(k_1\hspace{-2pt}+\hspace{-2pt}k_2\hspace{-2pt}+\hspace{-2pt}k_3\hspace{-2pt}+\hspace{-2pt}k_4)\varepsilon_{1,\mu_1\nu_1\barred{\mu}_1\barred{\nu}_1}\varepsilon_{2,\mu_2\nu_2\rho_2\barred{\mu}_2\barred{\nu}_2\barred{\rho}_2}\varepsilon_3^{\alpha_3\barred{\alpha}_3}\varepsilon_4^{\alpha_4\barred{\alpha}_4}
\\ \times \hspace{-2pt}\left[\hspace{-1pt}2\hspace{-3pt}\left(\hspace{-3pt}\frac{t\hspace{-2pt}-\hspace{-2pt}u}{m^2}\hspace{-1pt}k_2^{\nu_1}\hspace{-4pt}-\hspace{-3pt}k_{34}^{\nu_1}\hspace{-3pt}\right)\hspace{-4pt}\eta^{\mu_1\hspace{-1pt}[\hspace{-1pt}\mu_2}\hspace{-2pt}k_1^{\nu_2}\hspace{-1pt}\gamma^{\rho_2\hspace{-1pt}]}_{\hspace{-1pt}\alpha_3\alpha_4} \hspace{-3pt}+\hspace{-3pt} k_2^{\hspace{-1pt}\nu_1}\hspace{-1pt}\eta^{\mu_1\hspace{-1pt}[\mu_2}\hspace{-1pt}k_{34}^{\nu_2}\hspace{-1pt}\gamma^{\rho_2\hspace{-1pt}]}_{\hspace{-1pt}\alpha_3\alpha_4} \hspace{-3pt}+\hspace{-3pt} \frac{m^{\hspace{-1pt}2}\hspace{-2pt}}{2}\hspace{-1pt}\eta^{\mu_1\hspace{-1pt}[\hspace{-1pt}\mu_2}\hspace{-2pt}(\hspace{-1pt}\gamma^{\nu_2\rho_2\hspace{-1pt}]\hspace{-1pt}\nu_1}\hspace{-2pt})_{\hspace{-1pt}\alpha_3\alpha_4} \right. 
\\ \left. + k_2^{\mu_1}\hspace{-5pt}\left(\hspace{-3pt}\frac{1}{3}\hspace{-1pt}k_2^{\nu_1}\hspace{-2pt}(\hspace{-1pt}\gamma^{\mu_2\nu_2\rho_2}\hspace{-1pt})_{\hspace{-1pt}\alpha_3\alpha_4} \hspace{-3pt}-\hspace{-3pt} k_1^{\hspace{-1pt}[\mu_2}\hspace{-2pt}(\hspace{-1pt}\gamma^{\nu_2\rho_2\hspace{-1pt}]\nu_1}\hspace{-2pt})_{\hspace{-1pt}\alpha_3\alpha_4} \hspace{-3pt}+\hspace{-3pt} \frac{1}{2}\hspace{-1pt}k_{12}^{\mu}\hspace{-1pt}\eta^{\nu_1\hspace{-1pt}[\mu_2}\hspace{-2pt}(\hspace{-1pt}\gamma^{\nu_2\rho_2\hspace{-1pt}]}_{\phantom{\nu_2\rho_2]}\hspace{-2pt}\mu}\hspace{-1pt})_{\hspace{-1pt}\alpha_3\alpha_4}\hspace{-4pt}\right)\hspace{-2pt}\right]
\\ \times \hspace{-2pt}\left[\hspace{-1pt}2\hspace{-3pt}\left(\hspace{-3pt}\frac{t\hspace{-2pt}-\hspace{-2pt}u}{m^2}\hspace{-1pt}k_2^{\barred{\nu}_1}\hspace{-4pt}-\hspace{-3pt}k_{34}^{\barred{\nu}_1}\hspace{-3pt}\right)\hspace{-4pt}\eta^{\barred{\mu}_1\hspace{-1pt}[\hspace{-1pt}\barred{\mu}_2}\hspace{-2pt}k_1^{\barred{\nu}_2}\hspace{-1pt}\gamma^{\barred{\rho}_2\hspace{-1pt}]}_{\hspace{-1pt}\barred{\alpha}_3\barred{\alpha}_4} \hspace{-3pt}+\hspace{-3pt} k_2^{\hspace{-1pt}\barred{\nu}_1}\hspace{-1pt}\eta^{\barred{\mu}_1\hspace{-1pt}[\barred{\mu}_2}\hspace{-1pt}k_{34}^{\barred{\nu}_2}\hspace{-1pt}\gamma^{\barred{\rho}_2\hspace{-1pt}]}_{\hspace{-1pt}\barred{\alpha}_3\barred{\alpha}_4} \hspace{-3pt}+\hspace{-3pt} \frac{m^{\hspace{-1pt}2}\hspace{-2pt}}{2}\hspace{-1pt}\eta^{\barred{\mu}_1\hspace{-1pt}[\hspace{-1pt}\barred{\mu}_2}\hspace{-2pt}(\hspace{-1pt}\gamma^{\barred{\nu}_2\barred{\rho}_2\hspace{-1pt}]\hspace{-1pt}\barred{\nu}_1}\hspace{-2pt})_{\hspace{-1pt}\barred{\alpha}_3\barred{\alpha}_4} \right. 
\\ \left. + k_2^{\barred{\mu}_1}\hspace{-5pt}\left(\hspace{-3pt}\frac{1}{3}\hspace{-1pt}k_2^{\barred{\nu}_1}\hspace{-2pt}(\hspace{-1pt}\gamma^{\barred{\mu}_2\barred{\nu}_2\barred{\rho}_2}\hspace{-1pt})_{\hspace{-1pt}\barred{\alpha}_3\barred{\alpha}_4} \hspace{-3pt}-\hspace{-3pt} k_1^{\hspace{-1pt}[\barred{\mu}_2}\hspace{-2pt}(\hspace{-1pt}\gamma^{\barred{\nu}_2\barred{\rho}_2\hspace{-1pt}]\barred{\nu}_1}\hspace{-2pt})_{\hspace{-1pt}\barred{\alpha}_3\barred{\alpha}_4} \hspace{-3pt}+\hspace{-3pt} \frac{1}{2}\hspace{-1pt}k_{12}^{\barred{\mu}}\hspace{-1pt}\eta^{\barred{\nu}_1\hspace{-1pt}[\barred{\mu}_2}\hspace{-2pt}(\hspace{-1pt}\gamma^{\barred{\nu}_2\barred{\rho}_2\hspace{-1pt}]}_{\phantom{\barred{\nu}_2\barred{\rho}_2]}\hspace{-2pt}\barred{\mu}}\hspace{-1pt})_{\hspace{-1pt}\barred{\alpha}_3\barred{\alpha}_4}\hspace{-4pt}\right)\hspace{-2pt}\right].
\end{multline}

On the other hand, explicitly extracting the first-massive $s$-channel pole from the four-point amplitude \eqref{NS11NS12R0R0} gives 
\begin{multline}
\up{0pt}{\displaystyle\text{Pole}}{-8pt}{\hspace{1pt}s\rightarrow m^2} \ \mathcal{A}_4^{\mathrm{NS_1^1 NS_1^1};\mathrm{NS_1^2 NS_1^2};\mathrm{R_0 R_0};\mathrm{R_0 R_0}}(\{\varepsilon_i\},\{k_i\}) 
\\ = \frac{3!A_2^2 B_1 B_2\pi^2 ig_{\text{s}}^2}{2\alpha'(s\hspace{-2pt}-\hspace{-2pt}m^2)} (2\pi)^d\delta^{d}(k_{1}\hspace{-1pt}+\hspace{-1pt}k_{2}\hspace{-1pt}+\hspace{-1pt}k_{3}\hspace{-1pt}+\hspace{-1pt}k_{4})\varepsilon_{1,\mu_{1}\nu_{1}\rho_{1}\barred{\mu}_{1}\barred{\nu}_{1}\barred{\rho}_{1}}\varepsilon_{2,\mu_{2}\nu_{2}\barred{\mu}_{2}\barred{\nu}_{2}}\varepsilon_3^{\alpha_3\barred{\alpha}_3}\varepsilon_4^{\alpha_4\barred{\alpha}_4}
\\ \times \left[\eta^{\mu_1\hspace{-1pt}[\mu_2}\hspace{-1pt}(\hspace{-1pt}\gamma^{\nu_2\rho_2]\nu_1}\hspace{-1pt})_{\alpha_3\alpha_4} \hspace{-2pt}+\hspace{-2pt} \alpha'\hspace{-3pt}\left(\hspace{-2pt}\frac{t\hspace{-3pt}-\hspace{-3pt}u}{m^2}\hspace{-1pt}k_2^{\nu_1} \hspace{-3pt}-\hspace{-3pt} k_{34}^{\nu_1}\hspace{-3pt}\right)\hspace{-3pt}\eta^{\mu_1[\mu_2}k_1^{\nu_2}\gamma^{\rho_2]}_{\alpha_3\alpha_4} \right. 
\\ \left. + \frac{\alpha'}{2}\hspace{-1pt}k_2^{\nu_1}\hspace{-5pt}\left(\hspace{-4pt}\eta^{\mu_1\hspace{-1pt}[\hspace{-1pt}\mu_{2}}\hspace{-2pt}k_{34}^{\nu_{2}}\hspace{-2pt}\gamma^{\rho_2\hspace{-1pt}]}_{\alpha_3\alpha_4} \hspace{-3pt}-\hspace{-3pt} k_{1}^{\hspace{-1pt}[\mu_2}\hspace{-2pt}(\hspace{-2pt}\gamma^{\nu_2\rho_2\hspace{-1pt}]\hspace{-1pt}\mu_1}\hspace{-2pt})_{\hspace{-1pt}\alpha_3\alpha_4} \hspace{-3pt}+\hspace{-3pt} \frac{1}{3}k_{2}^{\mu_1}\hspace{-2pt}(\hspace{-2pt}\gamma^{\mu_2\nu_2\rho_2}\hspace{-2pt})_{\hspace{-1pt}\alpha_3\alpha_4} \hspace{-3pt}+\hspace{-3pt} \frac{1}{2}\hspace{-1pt}k_{12}^{\mu}\hspace{-1pt}\eta^{\mu_1\hspace{-1pt}[\hspace{-1pt}\mu_2}\hspace{-2pt}(\hspace{-2pt}\gamma_{\phantom{\nu_2\rho_2]}\hspace{-2pt}\mu}^{\nu_2\rho_2\hspace{-1pt}]}\hspace{-1pt})_{\hspace{-1pt}\alpha_3\alpha_4}\hspace{-4pt}\right)\hspace{-2pt}\right]
\\ \times \hspace{-4pt}\left[\eta^{\barred{\mu}_1\hspace{-1pt}[\barred{\mu}_2}\hspace{-1pt}(\hspace{-1pt}\gamma^{\barred{\nu}_2\barred{\rho}_2]\barred{\nu}_1}\hspace{-1pt})_{\barred{\alpha}_3\barred{\alpha}_4} \hspace{-2pt}+\hspace{-2pt} \alpha'\hspace{-3pt}\left(\hspace{-2pt}\frac{t\hspace{-3pt}-\hspace{-3pt}u}{m^2}\hspace{-1pt}k_2^{\barred{\nu}_1} \hspace{-3pt}-\hspace{-3pt} k_{34}^{\barred{\nu}_1}\hspace{-3pt}\right)\hspace{-3pt}\eta^{\barred{\mu}_1[\barred{\mu}_2}k_1^{\barred{\nu}_2}\gamma^{\barred{\rho}_2]}_{\barred{\alpha}_3\barred{\alpha}_4} \right. 
\\ \left. + \frac{\alpha'}{2}\hspace{-1pt}k_2^{\barred{\nu}_1}\hspace{-5pt}\left(\hspace{-4pt}\eta^{\barred{\mu}_1\hspace{-1pt}[\hspace{-1pt}\barred{\mu}_2}\hspace{-2pt}k_{34}^{\hspace{-1pt}\barred{\nu}_{2}}\hspace{-2pt}\gamma^{\barred{\rho}_2\hspace{-1pt}]}_{\barred{\alpha}_3\barred{\alpha}_4} \hspace{-3pt}-\hspace{-3pt} k_{1}^{\hspace{-1pt}[\barred{\mu}_2}\hspace{-2pt}(\hspace{-2pt}\gamma^{\barred{\nu}_2\barred{\rho}_2\hspace{-1pt}]\hspace{-1pt}\barred{\mu}_1}\hspace{-2pt})_{\hspace{-1pt}\barred{\alpha}_3\barred{\alpha}_4} \hspace{-3pt}+\hspace{-3pt} \frac{1}{3}k_{2}^{\barred{\mu}_1}\hspace{-2pt}(\hspace{-2pt}\gamma^{\barred{\mu}_2\barred{\nu}_2\barred{\rho}_2}\hspace{-2pt})_{\hspace{-1pt}\barred{\alpha}_3\barred{\alpha}_4} \hspace{-3pt}+\hspace{-3pt} \frac{1}{2}\hspace{-1pt}k_{12}^{\barred{\mu}}\hspace{-1pt}\eta^{\barred{\mu}_1\hspace{-1pt}[\hspace{-1pt}\barred{\mu}_2}\hspace{-2pt}(\hspace{-2pt}\gamma_{\phantom{\barred{\nu}_2\barred{\rho}_2]}\hspace{-2pt}\mu}^{\barred{\nu}_2\barred{\rho}_2\hspace{-1pt}]}\hspace{-1pt})_{\hspace{-1pt}\barred{\alpha}_3\barred{\alpha}_4}\hspace{-4pt}\right)\hspace{-2pt}\right].
\end{multline}
Therefore, this pole exactly matches the preceding unitarity computation provided that $B_1 = B_2 = B_3 = B_4 = 1$.

\subsubsection{$t$-channel}

Finally, we compute the required $t$-channel first-massive pole factorization of the four-point amplitude $\mathcal{A}_4[\mathrm{NS_1^1 NS_1^1};\mathrm{NS_1^2 NS_1^2};\mathrm{R_0 R_0};\mathrm{R_0 R_0}]$, which is due to the on-shell exchange of the massive RR state. Keeping the explicit factor of $c$ in \eqref{modified J}, this pole factorization reads
\begin{multline}
\int\frac{d^d k}{(2\pi)^d}\mathcal{A}_{3}\up{12pt}{\hspace{-5pt}\mathrm{NS_1^1 NS_1^1};\mathrm{R_0 R_0};\mathrm{R_1 R_1}}{0pt}{}\hspace{-70pt}(\varepsilon_1,k_1;\varepsilon_3,k_3;k)^{\mu\barred{\mu}}_{\alpha\barred{\alpha}} \ \frac{i\Pi^{\alpha\barred{\alpha},\beta\barred{\beta}}_{\mu\barred{\mu},\nu\barred{\nu}}(k)}{t-m^2} \ \mathcal{A}_{3}\up{12pt}{\hspace{-5pt}\mathrm{NS_1^2 NS_1^2};\mathrm{R_0 R_0};\mathrm{R_1 R_1}}{0pt}{}\hspace{-70pt}(\varepsilon_2,k_2;\varepsilon_4,k_4;-k)^{\nu\barred{\nu}}_{\beta\barred{\beta}}
\\ = \hspace{-1pt}\left(-\frac{A_2 B_1 C(d-2)\pi i g_{\text{s}}}{d-1}\right)\frac{i}{t\hspace{-2pt}-\hspace{-2pt}m^2}\left(\frac{A_2 B_2 C(d-2)\pi i\alpha' g_{\text{s}}^2}{3!8(d-1)}\right)
\\ (2\pi)^d\delta^d(k_1\hspace{-1pt}+\hspace{-1pt}k_2\hspace{-1pt}+\hspace{-1pt}k_3\hspace{-1pt}+\hspace{-1pt}k_4)\varepsilon_{1,\mu_1\nu_1\barred{\mu}_1\barred{\nu}_1}\varepsilon_{2,\mu_2\nu_2\rho_2\barred{\mu}_2\barred{\nu}_2\barred{\rho}_2}\varepsilon_3^{\alpha_3\barred{\alpha}_3}\varepsilon_4^{\alpha_4\barred{\alpha}_4}
\\ \times \hspace{-4pt}\left[\hspace{-2pt}\eta^{\nu_1\mu}\hspace{-1pt}\gamma^{\mu_1}_{\alpha_3\alpha}\hspace{-3pt}+\hspace{-3pt}\frac{\alpha'}{4}\hspace{-1pt}k_3^{\nu_1}\hspace{-3pt}\Big(\hspace{-2pt}k_{13}^{\mu}\hspace{-1pt}\gamma^{\mu_1}_{\alpha_3\alpha} \hspace{-3pt}-\hspace{-3pt} 2c\eta^{\mu_1\mu}\hspace{-1pt}(\hspace{-1pt}k_1\hspace{-3pt}\cdot\hspace{-3pt}\gamma\hspace{-1pt})_{\hspace{-1pt}\alpha_3\alpha}\hspace{-3pt}\Big)\hspace{-3pt}\right]\hspace{-4pt}\left[\hspace{-2pt}\eta^{\barred{\nu}_1\barred{\mu}}\hspace{-1pt}\gamma^{\barred{\mu}_1}_{\barred{\alpha}_3\barred{\alpha}} \hspace{-3pt}+\hspace{-3pt} \frac{\alpha'}{4}\hspace{-1pt}k_3^{\barred{\nu}_1}\hspace{-3pt}\Big(\hspace{-2pt}k_{13}^{\barred{\mu}}\hspace{-1pt}\gamma^{\barred{\mu}_1}_{\barred{\alpha}_3\barred{\alpha}} \hspace{-3pt}-\hspace{-3pt} 2c\eta^{\barred{\mu}_1\barred{\mu}}\hspace{-1pt}(\hspace{-1pt}k_1\hspace{-3pt}\cdot\hspace{-3pt}\gamma\hspace{-1pt})_{\hspace{-1pt}\barred{\alpha}_3\barred{\alpha}}\hspace{-3pt}\Big)\hspace{-2pt}\right]
\\ \times \Pi_{\mu\barred{\mu},\nu\barred{\nu}}^{\alpha\barred{\alpha},\beta\barred{\beta}}(k_2\hspace{-2pt}+\hspace{-2pt}k_4)\left[\hspace{-1pt}\frac{1}{2}\hspace{-1pt}k_{24}^{\nu}\hspace{-1pt}(\hspace{-1pt}\gamma^{\mu_2\nu_2\rho_2}\hspace{-1pt})_{\hspace{-1pt}\alpha_4\beta}\hspace{-3pt}+\hspace{-3pt}c\hspace{-2pt}\left(\hspace{-3pt}12\eta^{\nu[\mu_2}\hspace{-2pt}k_4^{\nu_2}\hspace{-2pt}\gamma^{\rho_2]}_{\alpha_4\beta} \hspace{-3pt}-\hspace{-3pt} 3k_2^{\rho}\hspace{-2pt}\eta^{\nu[\mu_2}\hspace{-2pt}(\hspace{-1pt}\gamma^{\nu_2\rho_2]}_{\phantom{\nu_2\rho_2]}\hspace{-1pt}\rho}\hspace{-1pt})_{\hspace{-1pt}\alpha_4\beta}\hspace{-4pt}\right)\hspace{-2pt}\right]
\\ \times\left[\hspace{-1pt}\frac{1}{2}\hspace{-1pt}k_{24}^{\barred{\nu}}\hspace{-1pt}(\hspace{-1pt}\gamma^{\barred{\mu}_2\barred{\nu}_2\barred{\rho}_2}\hspace{-1pt})_{\hspace{-1pt}\barred{\alpha}_4\barred{\beta}}\hspace{-3pt}+\hspace{-3pt}c\hspace{-2pt}\left(\hspace{-3pt}12\eta^{\barred{\nu}[\barred{\mu}_2}\hspace{-2pt}k_4^{\barred{\nu}_2}\hspace{-2pt}\gamma^{\barred{\rho}_2]}_{\barred{\alpha}_4\barred{\beta}} \hspace{-3pt}-\hspace{-3pt} 3k_2^{\barred{\rho}}\hspace{-2pt}\eta^{\barred{\nu}[\barred{\mu}_2}\hspace{-2pt}(\hspace{-1pt}\gamma^{\barred{\nu}_2\barred{\rho}_2]}_{\phantom{\barred{\nu}_2\barred{\rho}_2]}\hspace{-1pt}\barred{\rho}}\hspace{-1pt})_{\hspace{-1pt}\barred{\alpha}_4\barred{\beta}}\hspace{-4pt}\right)\hspace{-2pt}\right],
\end{multline}
where as usual the massive RR propagator is
\begin{multline}
\Pi_{\mu\barred{\mu},\nu\barred{\nu}}^{\alpha\barred{\alpha},\beta\barred{\beta}}(k) = \left[\left(\eta_{\mu\nu} - \frac{k_{\mu}k_{\nu}}{k^2}\right)(k\cdot\gamma)^{\alpha\beta} - \frac{1}{d\hspace{-2pt}-\hspace{-2pt}2}k^{\rho}(\gamma_{\rho\mu\nu})^{\alpha\beta}\right]
\\ \times \left[\left(\eta_{\barred{\mu}\barred{\nu}} - \frac{k_{\barred{\mu}}k_{\barred{\nu}}}{k^2}\right)(k\cdot\gamma)^{\barred{\alpha}\barred{\beta}} - \frac{1}{d\hspace{-2pt}-\hspace{-2pt}2}k^{\barred{\rho}}(\gamma_{\barred{\rho}\barred{\mu}\barred{\nu}})^{\barred{\alpha}\barred{\beta}}\right].
\end{multline}
After a lengthy calculation (as usual on the support of the delta function and physicality conditions), the total contraction with the holomorphic propagator appearing here is
\begin{multline}
\left[\hspace{-2pt}\eta^{\mu_1\mu}\hspace{-1pt}\gamma^{\nu_1}_{\alpha_3\alpha}\hspace{-3pt}+\hspace{-3pt}\frac{\alpha'}{4}\hspace{-1pt}k_3^{\mu_1}\hspace{-3pt}\Big(\hspace{-2pt}k_{13}^{\mu}\hspace{-1pt}\gamma^{\nu_1}_{\alpha_3\alpha} \hspace{-3pt}-\hspace{-3pt} 2c\eta^{\nu_1\mu}\hspace{-1pt}(\hspace{-1pt}k_1\hspace{-3pt}\cdot\hspace{-3pt}\gamma\hspace{-1pt})_{\hspace{-1pt}\alpha_3\alpha}\hspace{-3pt}\Big)\hspace{-3pt}\right]
\\ \times \left[\hspace{-3pt}\left(\hspace{-3pt}\eta_{\mu\nu} \hspace{-3pt}-\hspace{-3pt} \frac{(\hspace{-1pt}k_2\hspace{-3pt}+\hspace{-3pt}k_4\hspace{-1pt})_{\hspace{-1pt}\mu}\hspace{-1pt}(\hspace{-1pt}k_2\hspace{-3pt}+\hspace{-3pt}k_4\hspace{-1pt})_{\hspace{-1pt}\nu}}{(k_2\hspace{-2pt}+\hspace{-2pt}k_4)^2}\right)\hspace{-3pt}\big((k_2\hspace{-2pt}+\hspace{-2pt}k_4)\hspace{-2pt}\cdot\hspace{-2pt}\gamma\big)^{\alpha\beta} \hspace{-3pt}-\hspace{-3pt} \frac{1}{d\hspace{-3pt}-\hspace{-3pt}2}(k_2\hspace{-1pt}+\hspace{-1pt}k_4)^{\lambda}(\gamma_{\lambda\mu\nu})^{\alpha\beta}\right]
\\ \times \left[\frac{1}{2}k_{24}^{\nu}(\gamma^{\mu_2\nu_2\rho_2})_{\alpha_4\beta} + c\left(12\eta^{\nu[\mu_2}k_4^{\nu_2}\gamma^{\rho_2]}_{\alpha_4\beta} \hspace{-1pt}-\hspace{-1pt} 3k_2^{\rho}\eta^{\nu[\mu_2}(\gamma^{\nu_2\rho_2]}_{\phantom{\nu_2\rho_2]}\rho})_{\alpha_4\beta}\right)\right]
\\ = \frac{d\hspace{-2pt}-\hspace{-2pt}1}{d\hspace{-2pt}-\hspace{-2pt}2}\left\{\vphantom{\frac{\alpha'}{4}}-3ct\eta^{\mu_1[\mu_2}(\gamma^{\nu_2\rho_2]\nu_1})_{\alpha_3\alpha_4}  \right. 
\\ + 6\left[\left(\frac{s}{m^2} + \frac{(1+3c+4c^2)t}{2m^2} - \frac{1}{2} + \frac{(1\hspace{-2pt}+\hspace{-2pt}2c)}{d\hspace{-2pt}-\hspace{-2pt}1}\hspace{-2pt}\left(\hspace{-2pt}\frac{(1\hspace{-2pt}-\hspace{-2pt}5c)t}{2m^2} \hspace{-2pt}+\hspace{-2pt} \frac{1}{2}\hspace{-2pt}\right)\hspace{-2pt}\right)k_3^{\mu_1} + k_4^{\mu_1}\right]\eta^{\nu_1[\mu_2}k_1^{\nu_2}\gamma^{\rho_2]}_{\alpha_3\alpha_4}
\\ + 3\left[\left(\frac{s}{m^2} + \frac{(1-2c+4c^2)t}{2m^2} - \frac{1}{2} + \frac{(1\hspace{-2pt}+\hspace{-2pt}2c)}{d\hspace{-2pt}-\hspace{-2pt}1}\hspace{-2pt}\left(\hspace{-2pt}\frac{(3\hspace{-2pt}-\hspace{-2pt}4c)t}{2m^2} \hspace{-2pt}+\hspace{-2pt} \frac{1}{2}\hspace{-2pt}\right)\hspace{-2pt}\right)k_3^{\mu_1} + k_4^{\mu_1}\right]\eta^{\nu_1[\mu_2}k_{34}^{\nu_2}\gamma^{\rho_2]}_{\alpha_3\alpha_4}
\\ + \left[\left(\frac{s}{m^2} + \frac{(1+3c)t}{2m^2} - \frac{(1+c)}{2} + \frac{(1\hspace{-2pt}-\hspace{-2pt}c)}{d\hspace{-2pt}-\hspace{-2pt}1}\hspace{-2pt}\left(\hspace{-2pt}\frac{(1+6c)t}{2m^2} \hspace{-2pt}-\hspace{-2pt} \frac{1}{2}\hspace{-2pt}\right)\hspace{-2pt}\right)k_3^{\mu_1} + k_4^{\mu_1}\right]k_2^{\nu_1}(\gamma^{\mu_2\nu_2\rho_2})_{\alpha_3\alpha_4}
\\ - \frac{1}{2}\left[\left(\frac{s}{m^2} + \frac{(1-2c)t}{2m^2} - \frac{(1+2c)}{2} + \frac{3}{d\hspace{-2pt}-\hspace{-2pt}1}\hspace{-2pt}\left(\hspace{-2pt}\frac{(1\hspace{-2pt}+\hspace{-2pt}6c)t}{2m^2} \hspace{-2pt}-\hspace{-2pt} \frac{1}{2}\hspace{-2pt}\right)\hspace{-2pt}\right)k_3^{\mu_1} + k_4^{\mu_1}\right]k_{34}^{\nu_1}(\gamma^{\mu_2\nu_2\rho_2})_{\alpha_3\alpha_4}
\\ - 3\left[\left(\frac{s}{m^2} + \frac{(1+3c)t}{2m^2} - \frac{1}{2} + \frac{(1\hspace{-2pt}+\hspace{-2pt}2c)}{d\hspace{-2pt}-\hspace{-2pt}1}\hspace{-2pt}\left(\hspace{-2pt}\frac{(1\hspace{-2pt}-\hspace{-2pt}5c)t}{2m^2} \hspace{-2pt}+\hspace{-2pt} \frac{1}{2}\hspace{-2pt}\right)\hspace{-2pt}\right)k_3^{\mu_1} + k_4^{\mu_1}\right]k_1^{[\mu_2}(\gamma^{\nu_2\rho_2]\nu_1})_{\alpha_3\alpha_4}
\\ - \frac{3}{2}\left[\left(\frac{s}{m^2} + \frac{(1-2c)t}{2m^2} - \frac{1}{2} + \frac{(1\hspace{-2pt}+\hspace{-2pt}2c)}{d\hspace{-2pt}-\hspace{-2pt}1}\hspace{-2pt}\left(\hspace{-2pt}\frac{(3\hspace{-2pt}-\hspace{-2pt}4c)t}{2m^2} \hspace{-2pt}+\hspace{-2pt} \frac{1}{2}\hspace{-2pt}\right)\hspace{-2pt}\right)k_3^{\mu_1} + k_4^{\mu_1}\right]k_{34}^{[\mu_2}(\gamma^{\nu_2\rho_2]\nu_1})_{\alpha_3\alpha_4}
\\ + \frac{3}{2}\left[\left(\frac{s}{m^2} + \frac{(1+4c^2)t}{2m^2} - \frac{1}{2} - \frac{(1\hspace{-2pt}+\hspace{-2pt}2c)}{d\hspace{-2pt}-\hspace{-2pt}1}\hspace{-2pt}\left(\hspace{-2pt}\frac{(1\hspace{-2pt}+\hspace{-2pt}6c)t}{2m^2} \hspace{-2pt}-\hspace{-2pt} \frac{1}{2}\hspace{-2pt}\right)\hspace{-2pt}\right)k_3^{\mu_1} + k_4^{\mu_1}\right]k_{12}^{\mu}\eta^{\nu_1[\mu_2}(\gamma^{\nu_2\rho_2]}_{\phantom{\nu_2\rho_2]}\mu})_{\alpha_3\alpha_4}
\\ \left. + \frac{1}{2}\left[\left(\frac{s}{m^2} + \frac{t}{2m^2} - \frac{1}{2} - \frac{(1\hspace{-2pt}+\hspace{-2pt}2c)}{d\hspace{-2pt}-\hspace{-2pt}1}\hspace{-2pt}\left(\hspace{-2pt}\frac{(1\hspace{-2pt}+\hspace{-2pt}6c)t}{2m^2} \hspace{-2pt}-\hspace{-2pt} \frac{1}{2}\hspace{-2pt}\right)\hspace{-2pt}\right)k_3^{\mu_1} + k_4^{\mu_1}\right]k_{12}^{\mu}(\gamma_{\mu}^{\phantom{\mu}\nu_1\mu_2\nu_2\rho_2})_{\alpha_3\alpha_4}  \right\}
\\ - 3\alpha'\left(\frac{c^2}{2} - \frac{2(1+c)}{d-2}\right)k_3^{\mu_1}k_3^{\nu_1}\left(k_3^{[\mu_2}k_4^{\nu_2}\gamma^{\rho_2]}_{\alpha_3\alpha_4} + \frac{1}{2}k_{12}^{\mu}k_4^{[\mu_2}(\gamma^{\nu_2\rho_2]}_{\phantom{\nu_2\rho_2]}\mu})_{\alpha_3\alpha_4}\right).
\end{multline}
That the final term above with four momenta vanish when $d=10$ requires\footnote{Or $c = \frac{1}{2}$, but it is only the other solution which matches the rest of the amplitude pole.} $c = 1$. Then, evaluating the residue of the above $t = m^2$ pole at $c = 1$ gives the necessary unitarity factorization as
\begin{multline}
\int\frac{d^d k}{(2\pi)^d}\mathcal{A}_{3}\up{12pt}{\hspace{-5pt}\mathrm{NS_1^1 NS_1^1};\mathrm{R_0 R_0};\mathrm{R_1 R_1}}{0pt}{}\hspace{-70pt}(\varepsilon_1,k_1;\varepsilon_3,k_3;k)^{\mu\barred{\mu}}_{\alpha\barred{\alpha}} \ \frac{i\Pi_{\mu\barred{\mu},\nu\barred{\nu}}^{\alpha\barred{\alpha},\beta\barred{\beta}}(k)}{t-m^2} \ \mathcal{A}_{3}\up{12pt}{\hspace{-5pt}\mathrm{NS_1^2 NS_1^2};\mathrm{R_0 R_0};\mathrm{R_1 R_1}}{0pt}{}\hspace{-70pt}(\varepsilon_2,k_2;\varepsilon_4,k_4;-k)^{\nu\barred{\nu}}_{\beta\barred{\beta}}
\\ = \frac{A_2^2 B_1 B_2 C^2\pi^2 ig_{\text{s}}^2}{3!2\alpha'(t\hspace{-2pt}-\hspace{-2pt}m^2)}(2\pi)^d\delta^d(k_1\hspace{-1pt}+\hspace{-1pt}k_2\hspace{-1pt}+\hspace{-1pt}k_3\hspace{-1pt}+\hspace{-1pt}k_4)\varepsilon_{1,\mu_1\nu_1\barred{\mu}_1\barred{\nu}_1}\varepsilon_{2,\mu_2\nu_2\rho_2\barred{\mu}_2\barred{\nu}_2\barred{\rho}_2}\varepsilon_3^{\alpha_3\barred{\alpha}_3}\varepsilon_4^{\alpha_4\barred{\alpha}_4}
\\ \times \left\{\vphantom{\frac{\alpha'}{4}}6\eta^{\mu_1[\mu_2}(\gamma^{\nu_2\rho_2]\nu_1})_{\alpha_3\alpha_4} - 12\alpha'k_3^{\mu_1}\eta^{\nu_1[\mu_2}k_1^{\nu_2}\gamma^{\rho_2]}_{\alpha_3\alpha_4}  \right. 
\\ - \frac{\alpha'}{4}\hspace{-4pt}\left[\hspace{-3pt}\left(\hspace{-2pt}\frac{s}{m^{\hspace{-1pt}2}} \hspace{-3pt}-\hspace{-3pt} 1\hspace{-3pt}\right)\hspace{-4pt}k_3^{\mu_1} \hspace{-4pt}+\hspace{-3pt} k_4^{\mu_1}\hspace{-2pt}\right]\hspace{-5pt}\left[\hspace{-1pt}12\eta^{\nu_1\hspace{-1pt}[\mu_2}\hspace{-1pt}k_1^{\nu_2}\hspace{-1pt}\gamma^{\rho_2]}_{\alpha_3\alpha_4} \hspace{-3pt}-\hspace{-3pt} k_{34}^{\nu_1}\hspace{-1pt}(\hspace{-1pt}\gamma^{\mu_2\nu_2\rho_2}\hspace{-1pt})_{\alpha_3\alpha_4} \hspace{-3pt}-\hspace{-3pt} 3k_{34}^{\hspace{-1pt}[\mu_2}\hspace{-1pt}(\hspace{-1pt}\gamma^{\nu_2\rho_2]\nu_1}\hspace{-1pt})_{\alpha_3\alpha_4} \hspace{-3pt}+\hspace{-3pt} k_{12}^{\mu}\hspace{-1pt}(\hspace{-1pt}\gamma_{\mu}^{\phantom{\mu}\hspace{-1pt}\nu_1\mu_2\nu_2\rho_2}\hspace{-1pt})_{\alpha_3\alpha_4} \hspace{-1pt}\right]
\\ - \frac{\alpha'}{4}\hspace{-4pt}\left[\hspace{-3pt}\left(\hspace{-2pt}\frac{s}{m^{\hspace{-1pt}2}} \hspace{-3pt}+\hspace{-3pt} 1\hspace{-3pt}\right)\hspace{-4pt}k_3^{\mu_1} \hspace{-5pt}+\hspace{-3pt} k_4^{\mu_1}\hspace{-2pt}\right]\hspace{-5pt}\left[\hspace{-1pt}6\eta^{\nu_1\hspace{-1pt}[\mu_2}\hspace{-2pt}k_{34}^{\nu_2}\hspace{-1pt}\gamma^{\rho_2]}_{\alpha_3\alpha_4} \hspace{-3pt}+\hspace{-3pt} 2k_2^{\nu_1}\hspace{-2pt}(\hspace{-2pt}\gamma^{\mu_2\nu_2\rho_2}\hspace{-2pt})_{\hspace{-1pt}\alpha_3\alpha_4} \hspace{-3pt}-\hspace{-3pt} 6k_1^{\hspace{-1pt}[\mu_2}\hspace{-2pt}(\hspace{-2pt}\gamma^{\nu_2\rho_2]\nu_1}\hspace{-2pt})_{\hspace{-1pt}\alpha_3\alpha_4} \hspace{-3pt}+\hspace{-3pt} 3k_{12}^{\mu}\hspace{-1pt}\eta^{\nu_1\hspace{-1pt}[\mu_2}\hspace{-1pt}(\hspace{-2pt}\gamma^{\nu_2\rho_2]}_{\phantom{\nu_2\rho_2]}\hspace{-1pt}\mu}\hspace{-1pt})_{\hspace{-1pt}\alpha_3\alpha_4}\hspace{-2pt}\right]
\\ - \frac{\alpha'}{4}\hspace{-3pt}\left(\hspace{-2pt}\frac{d\hspace{-3pt}-\hspace{-3pt}10}{d\hspace{-3pt}-\hspace{-3pt}1}\hspace{-2pt}\right)\hspace{-3pt}k_3^{\mu_1}\hspace{-4pt}\left[\vphantom{\frac{3\alpha'}{2}}\hspace{-1pt}6\eta^{\nu_1[\mu_2}k_1^{\nu_2}\gamma^{\rho_2]}_{\alpha_3\alpha_4} \hspace{-2pt}+\hspace{-2pt} k_{34}^{\nu_1}(\gamma^{\mu_2\nu_2\rho_2})_{\alpha_3\alpha_4} \hspace{-2pt}-\hspace{-2pt} 3k_1^{[\mu_2}(\gamma^{\nu_2\rho_2]\nu_1})_{\alpha_3\alpha_4} \hspace{-2pt}+\hspace{-2pt} 3k_{12}^{\mu}\eta^{\nu_1[\mu_2}(\gamma^{\nu_2\rho_2]}_{\phantom{\nu_2\rho_2]}\mu})_{\alpha_3\alpha_4} \right.
\\ \left.\left. +  k_{12}^{\mu}(\gamma_{\mu}^{\phantom{\mu}\nu_1\mu_2\nu_2\rho_2})_{\alpha_3\alpha_4} \hspace{-2pt}-\hspace{-2pt} 3\alpha'k_3^{\nu_1}k_3^{[\mu_2}k_4^{\nu_2}\gamma^{\rho_2]}_{\alpha_3\alpha_4} \hspace{-2pt}-\hspace{-2pt} \frac{3\alpha'}{2}k_3^{\nu_1}k_{12}^{\mu}k_4^{[\mu_2}(\gamma^{\nu_2\rho_2]}_{\phantom{\nu_2\rho_2]}\mu})_{\alpha_3\alpha_4}\right]\right\}
\\ \times (\text{antiholomorphic}).
\end{multline}

On the other hand, extracting the actual $t$-channel first-massive pole of the four-point amplitude \eqref{NS11NS12R0R0} results in
\begin{multline}
\up{0pt}{\displaystyle\text{Pole}}{-8pt}{\hspace{1pt}t\rightarrow m^2} \ \mathcal{A}_{4}[\mathrm{NS_1^1 NS_1^1};\mathrm{NS_1^2 NS_1^2};\mathrm{R_0 R_0};\mathrm{R_0 R_0}]
\\ = \frac{A_2^2 B_1 B_2\pi^2 ig_{\text{s}}^2}{3!2\alpha'(t\hspace{-2pt}-\hspace{-2pt}m^2)} (2\pi)^{d}\delta^{d}(k_1\hspace{-1pt}+\hspace{-1pt}k_2\hspace{-1pt}+\hspace{-1pt}k_3\hspace{-1pt}+\hspace{-1pt}k_4)\varepsilon_{1,\mu_{1}\nu_{1}\barred{\mu}_{1}\barred{\nu}_{1}}\varepsilon_{2,\mu_{2}\nu_{2}\rho_{2}\barred{\mu}_{2}\barred{\nu}_{2}\barred{\rho}_{2}}\varepsilon_3^{\alpha_3\barred{\alpha}_3}\varepsilon_4^{\alpha_4\barred{\alpha}_4}
\\ \times \left\{6\eta^{\mu_1[\mu_2}(\gamma^{\nu_2\rho_2]\nu_1})_{\alpha_3\alpha_4} - 12\alpha'k_3^{\nu_1}\eta^{\mu_1[\mu_2}k_1^{\nu_2}\gamma^{\rho_2]}_{\alpha_3\alpha_4}   \right. 
\\ \left. -\frac{\alpha'}{4}\hspace{-4pt}\left[\hspace{-2pt}\left(\hspace{-2pt}\frac{s}{m^2}\hspace{-3pt}-\hspace{-3pt}1\hspace{-3pt}\right)\hspace{-3pt}k_3^{\nu_1} \hspace{-3pt}+\hspace{-3pt} k_4^{\nu_1}\hspace{-2pt}\right]\hspace{-6pt}\left[\hspace{-2pt}1\hspace{-1pt}2\eta^{\mu_1\hspace{-1pt}[\hspace{-1pt}\mu_{\hspace{-0.5pt}2}}\hspace{-2pt}k_1^{\hspace{-1pt}\nu_{\hspace{-0.5pt}2}}\hspace{-2pt}\gamma^{\rho_2\hspace{-1pt}]}_{\alpha_3\alpha_4} \hspace{-3pt}-\hspace{-3pt} 3k_{34}^{\hspace{-1pt}[\hspace{-0.5pt}\mu_2}\hspace{-2pt}(\hspace{-2pt}\gamma^{\nu_2\hspace{-1pt}\rho_2\hspace{-2pt}]\hspace{-1pt}\mu_1}\hspace{-2pt})_{\hspace{-2pt}\alpha_3\alpha_4} \hspace{-3pt}-\hspace{-3pt} k_{34}^{\mu_1}\hspace{-2pt}(\hspace{-2pt}\gamma^{\mu_2\hspace{-1pt}\nu_2\hspace{-1pt}\rho_2}\hspace{-2pt})_{\hspace{-1pt}\alpha_3\alpha_4} \hspace{-3pt}+\hspace{-3pt} k_{12}^{\mu}\hspace{-2pt}(\hspace{-2pt}\gamma_{\hspace{-1pt}\mu}^{\phantom{\mu}\hspace{-2pt}\mu_1\hspace{-1pt}\mu_2\hspace{-1pt}\nu_2\hspace{-1pt}\rho_2}\hspace{-2pt})_{\hspace{-1pt}\alpha_3\alpha_4}\hspace{-2pt}\right]   \right.
\\ \left. - \frac{\alpha'}{4}\hspace{-4pt}\left[\hspace{-2pt} \left(\hspace{-2pt}\frac{s}{m^2}\hspace{-3pt}+\hspace{-3pt}1\hspace{-3pt}\right)\hspace{-3pt}k_3^{\nu_1}\hspace{-3pt}+\hspace{-3pt}k_4^{\nu_1}\hspace{-2pt}\right]\hspace{-6pt}\left[\hspace{-2pt}6\eta^{\mu_1\hspace{-1pt}[\hspace{-1pt}\mu_{\hspace{-0.5pt}2}}\hspace{-2pt}k_{\hspace{-1pt}34}^{\hspace{-1pt}\nu_{\hspace{-0.5pt}2}}\hspace{-2pt}\gamma^{\rho_2\hspace{-1pt}]}_{\alpha_3\alpha_4} \hspace{-3pt}-\hspace{-3pt} 6k_{1}^{\hspace{-1pt}[\hspace{-0.5pt}\mu_2}\hspace{-2pt}(\hspace{-2pt}\gamma^{\nu_2\hspace{-1pt}\rho_2\hspace{-2pt}]\hspace{-1pt}\mu_1}\hspace{-2pt})_{\hspace{-2pt}\alpha_3\alpha_4} \hspace{-3pt}+\hspace{-3pt} 2k_{2}^{\mu_1}\hspace{-2pt}(\hspace{-2pt}\gamma^{\mu_2\hspace{-1pt}\nu_2\hspace{-1pt}\rho_2}\hspace{-2pt})_{\hspace{-1pt}\alpha_3\alpha_4} \hspace{-4pt}+\hspace{-3pt} 3\hspace{-1pt}k_{\hspace{-1pt}1\hspace{-1pt}2}^{\mu}\hspace{-1pt}\eta^{\mu_1\hspace{-1pt}[\hspace{-1pt}\mu_2}\hspace{-2pt}(\hspace{-2pt}\gamma_{\phantom{\nu_2\rho_2]}\hspace{-6pt}\mu}^{\hspace{-1pt}\nu_2\hspace{-1pt}\rho_2\hspace{-1pt}]}\hspace{-1pt})_{\hspace{-1pt}\alpha_3\alpha_4}\hspace{-2pt}\right]\hspace{-3pt}\right\}
\\ \times (\text{antiholomorphic}).
\end{multline}
These two expressions agree exactly when $d = 10$ and $C = 1$. Thus, the four-point amplitude \eqref{NS11NS12R0R0} indeed satisfies tree-level unitarity, which moreover confirms the normalizations provided in Section \ref{vertex operators}.

\section*{Acknowledgments}

I would like to thank Nathan Agmon and Andrea Dei for collaboration on early stages of this project. This work is supported in part by DOE grant DE-SC0007870.

\appendix
\section{Identities}\label{identities}

The following gamma matrix identities hold ---

\noindent (Anti)commutators:
\begin{align}
\{\gamma^{\mu},\gamma^{\nu_1\nu_2\cdots\nu_{2n+1}}\}^{\alpha}_{\phantom{\alpha}\beta} & = 2(2n+1)\eta^{\mu[\nu_1}(\gamma^{\nu_2\cdots\nu_{2n+1}]})^{\alpha}_{\phantom{\alpha}\beta}
\\ [\gamma^{\mu},\gamma^{\nu_1\nu_2\cdots\nu_{2n}}]_{\alpha\beta} & = 4n\eta^{\mu[\nu_1}(\gamma^{\nu_2\cdots\nu_{2n}]})_{\alpha\beta}
\end{align}
Parity:
\begin{equation}
\left(\gamma^{\mu_1\cdots\mu_n}\right)^{\mathrm{T}} = (-1)^{\frac{n(n-1)}{2}}\gamma^{\mu_1\cdots\mu_n}
\end{equation}
Multiplication:
\begin{equation}
\gamma^{\mu_1\cdots\mu_m}\gamma_{\nu_1\cdots\nu_n} = \sum_{i=0}^m \frac{(-1)^{i(m-\frac{i+1}{2})}m! n!}{i!(m-i)!(n-i)!}\eta^{[\mu_1}_{[\nu_1}\cdots \eta^{\mu_i}_{\nu_i}\gamma^{\mu_{i+1}\cdots\mu_m]}_{\phantom{\mu_{i+1}\cdots\mu_m]}\nu_{i+1}\cdots\nu_n]}
\end{equation}
Fierz identities:
\begin{align}
\gamma_{\mu\alpha(\beta}\gamma^{\mu}_{\gamma\delta)} & = 0
\\ (\gamma_{\mu\nu})^{\alpha}_{\phantom{\alpha}\beta}(\gamma^{\mu\nu})^{\gamma}_{\phantom{\gamma}\delta} & = 4\gamma_{\mu}^{\alpha\gamma}\gamma^{\mu}_{\beta\delta} - 2\delta^{\alpha}_{\phantom{\alpha}\beta}\delta^{\gamma}_{\phantom{\gamma}\delta} - 8\delta^{\alpha}_{\phantom{\alpha}\delta}\delta^{\gamma}_{\phantom{\gamma}\beta}
\\ (\gamma^{\mu\nu}_{\phantom{\mu\nu}\rho})_{(\alpha|(\beta}\gamma^{\rho}_{\gamma)|\delta)} & = -\gamma^{[\mu}_{\alpha\delta}\gamma^{\nu]}_{\beta\gamma},
\end{align}
which are valid only in ten dimensions.

All tree-level four-point worldsheet moduli integrals are computed via
\begin{multline}
\int_{\mathds{C}}d^2 z \hspace{-2pt}\left(\hspace{-2pt}\sum_{i,j}A_{ij}z^{n_i}(1\hspace{-2pt}-\hspace{-2pt}z)^{m_j}\hspace{-2pt}\right)\hspace{-3pt}\left(\hspace{-2pt}\sum_{k,\ell}\hspace{-2pt}B_{k\ell}\barred{z}^{p_k}(1\hspace{-2pt}-\hspace{-2pt}\barred{z})^{q_{\ell}}\right)\hspace{-2pt}|z|^{-2a}|1\hspace{-2pt}-\hspace{-2pt}z|^{-2b} 
\\ = 2\pi\frac{\Gamma(1\hspace{-2pt}-\hspace{-2pt}a)\Gamma(1\hspace{-2pt}-\hspace{-2pt}b)\Gamma(a\hspace{-2pt}+\hspace{-2pt}b\hspace{-2pt}-\hspace{-2pt}1)}{\Gamma(a)\Gamma(b)\Gamma(2\hspace{-2pt}-\hspace{-2pt}a\hspace{-2pt}-\hspace{-2pt}b)}\hspace{-2pt}\left(\hspace{-2pt}\sum_{i,j}\hspace{-2pt}A_{ij}\frac{(1\hspace{-2pt}-\hspace{-2pt}a)_{n_i} (1\hspace{-2pt}-\hspace{-2pt}b)_{m_j}}{(2\hspace{-2pt}-\hspace{-2pt}a\hspace{-2pt}-\hspace{-2pt}b)_{n_i+m_j}}\hspace{-2pt}\right)\hspace{-4pt}\left(\hspace{-2pt}\sum_{k,\ell}\hspace{-2pt}B_{k\ell}\frac{(1\hspace{-2pt}-\hspace{-2pt}a)_{p_k}(1\hspace{-2pt}-\hspace{-2pt}b)_{q_{\ell}}}{(2\hspace{-2pt}-\hspace{-2pt}a\hspace{-2pt}-\hspace{-2pt}b)_{p_k+q_{\ell}}}\hspace{-2pt}\right),
\end{multline}
valid for arbitrary constants $A_{ij}$ and $B_{k\ell}$ and sets of integer powers $\{n_i,m_j,p_k,q_{\ell}\}$. Here, $(a)_n \equiv \frac{\Gamma(a+n)}{\Gamma(a)}$ is the ascending Pochhammer symbol.

\section{OPEs}\label{OPEs}

\begin{align}
\partial X^{\mu}(z)\partial X^{\nu}(0) & \sim -\frac{\alpha'\eta^{\mu\nu}}{2z^2}
\\ \partial X^{\mu}(z)\normal{e^{ik\cdot X(0)}} & \sim -\frac{i\alpha' k^{\mu}}{2z}\normal{e^{ik\cdot X(0)}}
\\ \normal{e^{ik_i\cdot X(z_i,\barred{z}_i)}}\normal{e^{ik_j\cdot X(z_j,\barred{z}_j)}} & = |z|^{\alpha' k_i\cdot k_j}\normal{e^{i[k_i\cdot X(z_i,\barred{z}_i)+k_j\cdot X(z_j,\barred{z}_j)]}}
\end{align}

\begin{align}
\normal{e^{\frac{n}{2}\phi(z)}}\normal{e^{\frac{m}{2}\phi(0)}} & = \frac{1}{z^{\frac{nm}{4}}}\normal{e^{\frac{n}{2}\phi(z)+\frac{m}{2}\phi(0)}}
\end{align}

\begin{align}
\psi^{\mu}(z)\psi^{\nu}(0) & \sim  \frac{\eta^{\mu\nu}}{z}
\\ \normal{e^{-\phi(z)}}\psi^{\mu}(z)\normal{e^{-\phi(0)}}\psi^{\nu}(0) & \sim -\frac{\eta^{\mu\nu}}{z^2}\normal{e^{-2\phi(0)}} - \frac{1}{z}\normal{\left(\psi^{\mu}\psi^{\nu} - \eta^{\mu\nu}\partial\phi\right)e^{-2\phi}}(0)
\\ \notag \normal{e^{-\phi(z)}}\normal{\psi^{\mu_1}\psi^{\nu_1}\psi^{\rho_1}}(z)\normal{e^{-\phi(0)}}\normal{\psi^{\mu_2}\psi^{\nu_2}\psi^{\rho_2}}(0) \hspace{-75pt} & \hspace{75pt} \sim \frac{6}{z^4}\eta\up{2pt}{\mu_2}{9pt}{[\mu_1}\eta\up{2pt}{\nu_2}{9pt}{\nu_1}\eta\up{2pt}{\rho_2}{9pt}{\rho_1]}\normal{e^{-2\phi(0)}}
\\ & + \frac{6}{z^3}\eta\up{2pt}{[\mu_2}{9pt}{[\mu_1}\eta\up{2pt}{\nu_2}{9pt}{\nu_1}\normal{\big(3\psi\up{9pt}{\rho_1]}{2pt}{}\psi\up{9pt}{}{2pt}{\rho_2]} - \eta\up{2pt}{\rho_2]}{9pt}{\rho_1]}\partial\phi\big)e^{-2\phi}}(0) + \mathcal{O}\left(\frac{1}{z^2}\right)
\end{align}

\begin{align}
j^{\mu\nu}(z)\normal{e^{-\phi}}\psi^{\rho}(0) & \sim -\frac{2i}{z}\eta^{\rho[\mu}\normal{e^{-\phi}}\psi^{\nu]}(0)
\\ j^{\mu_1\nu_1}(z)j^{\mu_2\nu_2}(0) & \sim \frac{2}{z^2}\eta\up{8pt}{[\mu_1}{2pt}{\mu_2}\eta\up{8pt}{\nu_1]}{2pt}{\nu_2} - \frac{4i}{z}\eta\up{8pt}{[\mu_1}{0pt}{[\mu_2}j\up{8pt}{\nu_1]}{0pt}{\phantom{\nu_1]}\nu_2]}(0)
\\ \notag j^{\mu_1\nu_1}(z)\normal{e^{-\phi}}\normal{\psi^{\mu_2}\psi^{\nu_2}\psi^{\rho_2}}(0) & \sim -\frac{6i}{z^2}\normal{e^{-\phi}}\eta\up{8pt}{[\mu_2}{2pt}{\mu_1}\eta\up{8pt}{\nu_2}{2pt}{\nu_1}\psi\up{8pt}{\rho_2]}{0pt}{}(0)
\\ & \hspace{-20pt} - \frac{2}{z}\normal{e^{-\phi}}\eta\up{8pt}{[\mu_2}{0pt}{[\mu_1}\normal{\psi\up{0pt}{\nu_1]}{0pt}{}j\up{8pt}{\nu_2\rho_2]}{0pt}{}}(0) + \frac{4}{z}\normal{e^{-\phi}}\eta\up{8pt}{[\mu_2}{0pt}{[\mu_1}\normal{\psi\up{8pt}{\nu_2}{0pt}{}j\up{0pt}{\nu_1]}{8pt}{\phantom{\nu_1]}\rho_2]}}(0)
\end{align}

\begin{align}
\normal{e^{-\phi(z)}}\psi^{\mu}(z)\mathcal{S}_{\alpha}(0) & \sim -\frac{\gamma^{\mu}_{\alpha\beta}}{\sqrt{2}z}\mathcal{S}^{\beta}(0)
\\ \normal{e^{-\phi(z)}}\psi^{\mu}(z)\mathcal{S}^{\alpha}(0) & \sim \frac{\gamma^{\mu\alpha\beta}}{\sqrt{2}z^2}\normal{e^{-\frac{5}{2}\phi(0)}}S_{\beta}(0) + \mathcal{O}\left(\frac{1}{z}\right)
\\ \normal{e^{\phi(z)}}\psi^{\mu}(z)\mathcal{S}_{\alpha}(0) & \sim -\frac{\gamma^{\mu}_{\alpha\beta}}{\sqrt{2}}\normal{e^{\frac{1}{2}\phi(0)}}S^{\beta}(0)
\\ \normal{e^{\phi(z)}}\psi^{\mu}(z)\mathcal{S}^{\alpha}(0) & \sim \frac{\gamma^{\mu\alpha\beta}}{\sqrt{2}}z \mathcal{S}_{\beta}(0)
\end{align}

\begin{align}
\mathcal{S}_{\alpha}(z)\mathcal{S}_{\beta}(0) & \sim \frac{\gamma^{\mu}_{\alpha\beta}}{\sqrt{2}z}\normal{e^{-\phi}}\psi_{\mu}(0)
\\ \mathcal{S}^{\alpha}(z)\mathcal{S}_{\beta}(0) & \sim \frac{\delta^{\alpha}_{\phantom{\alpha}\beta}}{z^2}\normal{e^{-2\phi(0)}} + \mathcal{O}\left(\frac{1}{z}\right)
\end{align}

\begin{align}
j^{\mu\nu}(z)\mathcal{S}_{\alpha}(0) & \sim -\frac{i}{2z}(\gamma^{\mu\nu})_{\alpha}^{\phantom{\alpha}\beta}\mathcal{S}_{\beta}(0)
\\ j^{\mu\nu}(z)\mathcal{S}^{\alpha}(0) & \sim -\frac{i}{2z}(\gamma^{\mu\nu})^{\alpha}_{\phantom{\alpha}\beta}\mathcal{S}^{\beta}(0)
\end{align}

\begin{align}
\normal{e^{-\phi}}\psi^{\mu}(z)\normal{j^{\nu\rho}\mathcal{S}_{\alpha}}(0) & \sim -\frac{\sqrt{2}i}{z^2}\eta^{\mu[\nu}\gamma^{\rho]}_{\alpha\beta}\mathcal{S}^{\beta}(0) + \frac{2i}{z}\eta^{\mu[\nu}\normal{\normal{e^{-\phi}}\psi^{\rho]}\mathcal{S}_{\alpha}}(0) - \frac{\gamma^{\mu}_{\alpha\beta}}{\sqrt{2}z}\normal{j^{\nu\rho}\mathcal{S}^{\beta}}(0)
\\ \normal{e^{\phi}}\psi^{\mu}(z)\normal{j^{\nu\rho}\mathcal{S}_{\alpha}}(0) & \sim -\frac{i\sqrt{2}}{z}\eta^{\mu[\nu}\gamma^{\rho]}_{\alpha\beta}\normal{e^{\frac{1}{2}\phi}}S^{\beta}(0)
\end{align}

\begin{equation}
j^{\mu\nu}(z)\normal{j^{\rho\lambda}\mathcal{S}_{\alpha}}(0) \sim \frac{2}{z^2}\eta\up{10pt}{[\mu}{2pt}{[\rho}(\gamma\up{2pt}{\lambda]}{0pt}{}\gamma\up{10pt}{\nu]}{0pt}{})_{\alpha}^{\phantom{\alpha}\beta}\mathcal{S}_{\beta}(0) - \frac{i}{z}\left(4\eta\up{10pt}{[\mu}{2pt}{[\rho}\normal{j\up{10pt}{\nu]}{0pt}{}\up{2pt}{\lambda]}{0pt}{}\mathcal{S}_{\alpha}}(0) + \frac{1}{2}(\gamma^{\mu\nu})_{\alpha}^{\phantom{\alpha}\beta}\normal{j^{\rho\lambda}\mathcal{S}_{\beta}}(0)\right)
\end{equation}

\begin{multline}
j^{\mu_1\nu_1}(z)k_2^{\rho}(\gamma_{\rho}\gamma_{\nu_2})_{\alpha_2}^{\phantom{\alpha_2}\beta_2}\normal{j^{\mu_2\nu_2}\mathcal{S}_{\beta_2}}(0) \sim \frac{d-2}{z^2}k_2^{\rho}\eta^{\mu_2[\mu_1}(\gamma_{\rho}\gamma^{\nu_1]})_{\alpha_2}^{\phantom{\alpha_2}\beta_2}\mathcal{S}_{\beta_2}(0)
\\ - \frac{i}{z}\left(2k_2^{\rho}(\gamma_{\rho}\gamma_{\nu_2})_{\alpha_2}^{\phantom{\alpha_2}\beta_2}\eta\up{8pt}{[\mu_1}{2pt}{\mu_2}\normal{j\up{8pt}{\nu_1]}{0pt}{}\up{2pt}{\nu_2}{0pt}{}\mathcal{S}_{\beta_2}}(0) + \frac{1}{2}k_2^{\rho}(\gamma_{\rho}\gamma^{\mu_1\nu_1}\gamma_{\nu_2})_{\alpha_2}^{\phantom{\alpha_2}\beta_2}\normal{j^{\mu_2\nu_2}\mathcal{S}_{\beta_2}}(0)\right)
\end{multline}
In the preceding expression, it is implicit that $k_2$ is the momentum corresponding to the massive RR vertex operator containing $\normal{j^{\mu_2\nu_2}\mathcal{S}_{\beta_2}}$ and that the physicality conditions are used.

\begin{equation}
\mathcal{S}_{\alpha}(z)\normal{j^{\mu\nu}\mathcal{S}_{\beta}}(0) \sim \frac{i(\gamma^{\mu\nu}\gamma_{\rho})_{\alpha\beta}}{2\sqrt{2}z^2}\normal{e^{-\phi}}\psi^{\rho}(0) + \frac{1}{z}\left[\frac{i}{2}(\gamma^{\mu\nu})_{\alpha}^{\phantom{\alpha}\gamma}\normal{\mathcal{S}_{\gamma}\mathcal{S}_{\beta}}(0) + \frac{1}{\sqrt{2}}\gamma^{\rho}_{\alpha\beta}\normal{e^{-\phi}}\normal{\psi_{\rho}j^{\mu\nu}}(0)\right]
\end{equation}

\section{Correlators}\label{correlators}

Here we collect all the worldsheet sphere correlators relevant to the results presented in this paper. We use the canonical normalizations of the individual CFTs, as the string normalization is taken into account via \eqref{string normalization}. Furthermore, we define the unique four-point conformal cross-ratio as
\begin{equation}
z \equiv \frac{z_{12}z_{34}}{z_{13}z_{24}},
\end{equation}
from which it immediately follows that
\begin{equation}
1-z = \frac{z_{14}z_{23}}{z_{13}z_{24}}.
\end{equation}
The bosonic correlators are trivial to compute; they all stem from the fundamental four-point correlator
\begin{multline}
\left\langle\normal{e^{ik_1\cdot X(z_1,\barred{z}_1)}}\normal{e^{ik_2\cdot X(z_2,\barred{z}_2)}}\normal{e^{ik_3\cdot X(z_3,\barred{z}_3)}}\normal{e^{ik_4\cdot X(z_4,\barred{z}_4)}}\right\rangle
\\ = \left|\frac{z_{12}^{n_1+n_2}z_{14}^{n_1+n_4}z_{23}^{n_2+n_3}z_{34}^{n_3+n_4}}{z_{13}^{n_2+n_4}z_{24}^{n_1+n_3}}\right|^2\frac{(2\pi)^d i\delta^d(k_1+k_2+k_3+k_4)}{|z|^{\frac{\alpha's}{2}}|1-z|^{\frac{\alpha'u}{2}}},
\end{multline}
where $n_i$ is the integer mass-level of the $i^{\text{th}}$ state appearing in the relation $m_i^2 = \frac{4n_i}{\alpha'}$. 

Now we record the holomorphic sphere correlators in the fermion plus picture ghost CFT in the normalization $\langle\normal{e^{-2\phi}}\rangle = 1$. Only the correlators involving spin fields are nontrivial to compute. The five-point correlators included below appear in computations for which a PCO is also inserted.

\subsection{Fermion 2-Point Correlators}

\begin{align}
\left\langle \normal{e^{-\phi}}\psi^{\mu_1}(z_1)\normal{e^{-\phi}}\psi^{\mu_2}(z_2)\right\rangle & = -\frac{1}{z_{12}^2}\eta^{\mu_1\mu_2}
\\ \left\langle \normal{e^{-\phi}}\normal{\psi^{\mu_1}\psi^{\nu_1}\psi^{\rho_1}}(z_1)\normal{e^{-\phi}}\normal{\psi^{\mu_2}\psi^{\nu_2}\psi^{\rho_2}}(z_2)\right\rangle & = \frac{3!}{z_{12}^4}\eta\up{8pt}{[\mu_1}{2pt}{\mu_2}\eta\up{8pt}{\nu_1}{2pt}{\nu_2}\eta\up{8pt}{\rho_1]}{2pt}{\rho_2}
\\ \left\langle \mathcal{S}^{\alpha}(z_1)\mathcal{S}_{\beta}(z_2)\right\rangle & = \frac{1}{z_{12}^2}\delta^{\alpha}_{\phantom{\alpha}\beta}
\\ \left\langle \normal{e^{\frac{1}{2}\phi}}S^{\alpha}(z_1)\normal{e^{-\frac{5}{2}\phi}}S_{\beta}(z_2)\right\rangle & = -\delta^{\alpha}_{\phantom{\alpha}\beta}
\end{align}

\subsection{Fermion 3-Point Correlators}

\begin{equation}
\left\langle j^{\mu_1\nu_1}(z_1)\normal{e^{-\phi}}\psi^{\mu_2}(z_2)\normal{e^{-\phi}}\psi^{\mu_3}(z_3)\right\rangle = \frac{2i}{z_{12}z_{13}z_{23}}\eta^{\mu_2[\mu_1}\eta^{\nu_1]\mu_3}
\end{equation}

\begin{equation}
\left\langle j^{\mu_1\nu_1}(z_1)\normal{e^{-\phi}}\psi^{\mu_2}(z_2)\normal{e^{-\phi}}\normal{\psi^{\mu_3}\psi^{\nu_3}\psi^{\rho_3}}(z_3)\right\rangle = \frac{6i}{z_{13}^2 z_{23}^2}\eta\up{8pt}{[\mu_3}{2pt}{\mu_1}\eta\up{8pt}{\nu_3}{2pt}{\nu_1}\eta\up{8pt}{\rho_3]}{2pt}{\mu_2}
\end{equation}

\begin{equation}
\left\langle j^{\mu_1\nu_1}(z_1)\normal{e^{-\phi}}\normal{\psi^{\mu_2}\psi^{\nu_2}\psi^{\rho_2}}(z_2)\normal{e^{-\phi}}\normal{\psi^{\mu_3}\psi^{\nu_3}\psi^{\rho_3}}(z_3)\right\rangle = -\frac{3!^2 i}{z_{12}z_{13}z_{23}^3}\eta\up{0pt}{[\mu_2}{9pt}{[\mu_1}\eta\up{9pt}{\nu_1][\mu_3}{0pt}{}\eta\up{9pt}{\nu_3}{0pt}{\nu_2}\eta\up{9pt}{\rho_3]}{0pt}{\rho_2]}
\end{equation}

\begin{align}
\left\langle \normal{e^{-\phi}}\psi^{\mu_1}(z_1)\mathcal{S}_{\alpha}(z_2)\mathcal{S}_{\beta}(z_3)\right\rangle & = -\frac{1}{\sqrt{2}z_{12}z_{13}z_{23}}\gamma^{\mu_1}_{\alpha\beta}
\\ \left\langle \normal{e^{-\phi}}\psi^{\mu_1}(z_1)\normal{e^{\frac{1}{2}\phi}}S^{\alpha}(z_2)\mathcal{S}^{\beta}(z_3)\right\rangle & = -\frac{1}{\sqrt{2}z_{13}^2}\gamma^{\mu_1\alpha\beta}
\end{align}

\begin{align}
\left\langle \normal{e^{\phi}}\psi^{\mu_1}(z_1)\normal{e^{-\frac{5}{2}\phi}}S_{\alpha}(z_2)\mathcal{S}_{\beta}(z_3)\right\rangle & = -\frac{z_{12}^2}{\sqrt{2}z_{23}^2}\gamma^{\mu_1}_{\alpha\beta}
\\ \left\langle \normal{e^{\phi}}\psi^{\mu_1}(z_1)\mathcal{S}^{\alpha}(z_2)\mathcal{S}^{\beta}(z_3)\right\rangle & = -\frac{z_{12}z_{13}}{\sqrt{2}z_{23}^2}\gamma^{\mu_1\alpha\beta}
\end{align}

\begin{align}
\left\langle \normal{e^{-2\phi}}(z_1)\normal{e^{\frac{1}{2}\phi}}S^{\alpha}(z_2)\mathcal{S}_{\beta}(z_3)\right\rangle & = \frac{z_{12}}{z_{13}z_{23}}\delta^{\alpha}_{\phantom{\alpha}\beta}
\end{align}

\begin{align}
\left\langle j^{\mu_1\nu_1}(z_1)\mathcal{S}_{\alpha}(z_2)\mathcal{S}^{\beta}(z_3)\right\rangle & = \frac{i}{2z_{12}z_{13}z_{23}}\left(\gamma^{\mu_1\nu_1}\right)_{\alpha}^{\phantom{\alpha}\beta}
\\ \left\langle\normal{e^{-2\phi}}j^{\mu_1\nu_1}(z_1)\normal{e^{\frac{1}{2}\phi}}S^{\alpha}(z_2)\mathcal{S}_{\beta}(z_3)\right\rangle & = -\frac{i}{2z_{13}^2}\left(\gamma^{\mu_1\nu_1}\right)^{\alpha}_{\phantom{\alpha}\beta}
\end{align}

\begin{align}
\left\langle \normal{e^{-\phi}}\normal{\psi^{\mu_1}\psi^{\nu_1}\psi^{\rho_1}}(z_1)\mathcal{S}_{\alpha}(z_2)\mathcal{S}_{\beta}(z_3)\right\rangle & = \frac{1}{2\sqrt{2}z_{12}^2 z_{13}^2}\left(\gamma^{\mu_1\nu_1\rho_1}\right)_{\alpha\beta}
\\ \left\langle \normal{e^{-\phi}}\normal{\psi^{\mu_1}\psi^{\nu_1}\psi^{\rho_1}}(z_1)\normal{e^{\frac{1}{2}\phi(z_2)}}S^{\alpha}(z_2)\mathcal{S}^{\beta}(z_3)\right\rangle & = \frac{z_{23}}{2\sqrt{2}z_{12}z_{13}^3}\left(\gamma^{\mu_1\nu_1\rho_1}\right)^{\alpha\beta}
\end{align}

\begin{multline}
\left\langle \normal{e^{-\phi}}\psi^{\mu_1}(z_1)\mathcal{S}_{\alpha_2}(z_2)\normal{j^{\mu_3\nu_3}\mathcal{S}_{\alpha_3}}(z_3) \right\rangle 
\\ = -\frac{i}{2\sqrt{2}z_{12}z_{13}z_{23}^2}\left[2\eta^{\mu_1[\mu_3}\gamma^{\nu_3]}_{\alpha_2\alpha_3} + (\gamma^{\mu_1\mu_3\nu_3})_{\alpha_2\alpha_3}\right] + \frac{2i}{\sqrt{2}z_{13}^2 z_{23}^2}\eta^{\mu_1[\mu_3}\gamma^{\nu_3]}_{\alpha_2\alpha_3}
\end{multline}

\begin{equation}
\left\langle \normal{e^{-\phi}}\psi^{\mu_1}(z_1)\mathcal{S}_{\alpha_2}(z_2)k_3^{\rho}(\gamma_{\rho}\gamma_{\nu_3})_{\alpha_3}^{\phantom{\alpha_3}\beta_3}\normal{j^{\mu_3\nu_3}\mathcal{S}_{\beta_3}}(z_3) \right\rangle = \frac{i(d-2)}{\sqrt{2}z_{13}^2 z_{23}^2}\eta^{\mu_1\mu_3}(k_3\cdot\gamma)_{\alpha_2\alpha_3}
\end{equation}
In the preceding expression, and in future ones like it, it is understood that the operator $k_i^{\rho}(\gamma_{\rho}\gamma_{\nu_i})_{\alpha_i}^{\phantom{\alpha_i}\beta_i}\normal{j^{\mu_i\nu_i}\mathcal{S}_{\beta_i}}$ is contracted against a massive RR polarization vector-spinor satisfying the physicality conditions.

\begin{multline}
\left\langle \normal{e^{-\phi}}\psi^{\mu_1}(z_1)\normal{j^{\mu_2\nu_2}\mathcal{S}_{\alpha_2}}(z_2)k_3^{\rho}(\gamma_{\rho}\gamma_{\nu_3})_{\alpha_3}^{\phantom{\alpha_3}\beta_3}\normal{j^{\mu_3\nu_3}\mathcal{S}_{\beta_3}}(z_3)\right\rangle 
\\ = \frac{d-2}{\sqrt{2}z_{12}z_{13}z_{23}^3}\bigg[\frac{z_{12}}{z_{23}}\eta^{\mu_1\mu_3}\left(k_3^{[\mu_2}\gamma^{\nu_2]}_{\alpha_2\alpha_3} - \frac{1}{2}k_3^{\mu}(\gamma^{\mu_2\nu_2}_{\phantom{\mu_2\nu_2}\mu})_{\alpha_2\alpha_3}\right)
\\ - \eta^{\mu_3[\mu_2}k_3^{\nu_2]}\gamma^{\mu_1}_{\alpha_2\alpha_3} + k_3^{\mu_1}\eta^{\mu_3[\mu_2}\gamma^{\nu_2]}_{\alpha_2\alpha_3} - \eta^{\mu_1[\mu_2}\eta^{\nu_2]\mu_3}(k_3\cdot\gamma)_{\alpha_2\alpha_3} + k_3^{\mu}\eta^{\mu_3[\mu_2}(\gamma^{\nu_2]\mu_1}_{\phantom{\nu_2]\mu_1}\mu})_{\alpha_2\alpha_3}\bigg]
\end{multline}

\begin{multline}
\left\langle \normal{e^{-\phi}}\psi^{\mu_1}(z_1)k_2^{\rho}(\gamma_{\rho}\gamma_{\nu_2})_{\alpha_2}^{\phantom{\alpha_2}\beta_2}\normal{j^{\mu_2\nu_2}\mathcal{S}_{\beta_2}}(z_2)k_3^{\lambda}(\gamma_{\lambda}\gamma_{\nu_3})_{\alpha_3}^{\phantom{\alpha_3}\beta_3}\normal{j^{\mu_3\nu_3}\mathcal{S}_{\beta_3}}(z_3)\right\rangle 
\\ = -\frac{(d-2)^2}{2\sqrt{2}z_{12}z_{13}z_{23}^3}\eta^{\mu_2\mu_3}\bigg[k_2\cdot k_3\gamma^{\mu_1}_{\alpha_2\alpha_3} - k_3^{\mu_1}(k_2\cdot\gamma)_{\alpha_2\alpha_3} - k_2^{\mu_1}(k_3\cdot\gamma)_{\alpha_2\alpha_3} + k_2^{\mu}k_3^{\nu}(\gamma^{\mu_1}_{\phantom{\mu_1}\mu\nu})_{\alpha_2\alpha_3}\bigg]
\end{multline}

\begin{multline}
\left\langle \normal{e^{-\phi}}\normal{\psi^{\mu_1}\psi^{\nu_1}\psi^{\rho_1}}(z_1)\mathcal{S}_{\alpha_2}(z_2)\normal{j^{\mu_3\nu_3}\mathcal{S}_{\alpha_3}}(z_3) \right\rangle
\\ = -\frac{i}{4\sqrt{2}z_{12}^2 z_{13}^2 z_{23}}\left[6\eta\up{8pt}{[\mu_1}{2pt}{\mu_3}\eta\up{8pt}{\nu_1}{2pt}{\nu_3}\gamma\up{8pt}{\rho_1]}{-2pt}{\alpha_2\alpha_3} - 6\eta\up{10pt}{[\mu_1}{2pt}{[\mu_3}(\gamma\up{10pt}{\nu_1\rho_1]}{2pt}{\phantom{\nu_1\rho_1]}\nu_3]})_{\alpha_2\alpha_3} - (\gamma^{\mu_1\nu_1\rho_1\mu_3\nu_3})_{\alpha_2\alpha_3}\right] 
\\ + \frac{3i}{\sqrt{2}z_{12}z_{13}^3 z_{23}}\left[2\eta\up{8pt}{[\mu_1}{2pt}{\mu_3}\eta\up{8pt}{\nu_1}{2pt}{\nu_3}\gamma\up{8pt}{\rho_1]}{-2pt}{\alpha_2\alpha_3} - \eta\up{10pt}{[\mu_1}{2pt}{[\mu_3}(\gamma\up{10pt}{\nu_1\rho_1]}{2pt}{\phantom{\nu_1\rho_1]}\nu_3]})_{\alpha_2\alpha_3}\right]
\end{multline}

\begin{multline}
\left\langle \normal{e^{-\phi}}\normal{\psi^{\mu_1}\psi^{\nu_1}\psi^{\rho_1}}(z_1)\mathcal{S}_{\alpha_2}(z_2)k_3^{\rho}(\gamma_{\rho}\gamma_{\nu_3})_{\alpha_3}^{\phantom{\alpha_3}\beta_3}\normal{j^{\mu_3\nu_3}\mathcal{S}_{\beta_3}}(z_3) \right\rangle 
\\ = \frac{3i(d-2)}{2\sqrt{2}z_{12}z_{13}^3 z_{23}}\left[2\eta^{\mu_3[\mu_1}k_3^{\nu_1}\gamma^{\rho_1]}_{\alpha_2\alpha_3} - k_3^{\mu}\eta^{\mu_3[\mu_1}(\gamma^{\nu_1\rho_1]}_{\phantom{\nu_1\rho_1]}\mu})_{\alpha_2\alpha_3}\right]
\end{multline}

\begin{multline}
\left\langle \normal{e^{-\phi}}\normal{\psi^{\mu_1}\psi^{\nu_1}\psi^{\rho_1}}(z_1)k_2^{\rho}(\gamma_{\rho}\gamma_{\nu_2})_{\alpha_2}^{\phantom{\alpha_2}\beta_2}\normal{j^{\mu_2\nu_2}\mathcal{S}_{\beta_2}}(z_2)\normal{j^{\mu_3\nu_3}\mathcal{S}_{\alpha_3}}(z_3) \right\rangle 
\\ = -\frac{d-2}{\sqrt{2}z_{12}^2 z_{13}^2 z_{23}^2}\Bigg\{14\eta\up{8pt}{[\mu_1}{2pt}{\mu_2}\eta\up{8pt}{\nu_1}{2pt}{\mu_3}\eta\up{8pt}{\rho_1]}{2pt}{\nu_3}(k_2\cdot\gamma)_{\alpha_2\alpha_3} + 3\eta^{\mu_2[\mu_3}\eta^{\nu_3][\mu_1}k_2^{\nu_1}\gamma^{\rho_1]}_{\alpha_2\alpha_3} - 6\eta\up{10pt}{[\mu_1}{2pt}{\mu_2}\eta\up{10pt}{\nu_1}{2pt}{[\mu_3}k\up{3pt}{\nu_3]}{-4pt}{\hspace{-1pt}2}\gamma\up{10pt}{\rho_1]}{-3pt}{\hspace{-1pt}\alpha_2\alpha_3}
\\ + 6\eta\up{10pt}{[\mu_1}{2pt}{\mu_2}\eta\up{10pt}{\nu_1}{2pt}{[\mu_3}k\up{8pt}{\rho_1]}{-3pt}{2}\gamma\up{3pt}{\nu_3]}{-5pt}{\hspace{-1pt}\alpha_2\alpha_3} - \frac{1}{2}\eta^{\mu_2[\mu_3}k_2^{\nu_3]}(\gamma^{\mu_1\nu_1\rho_1})_{\alpha_2\alpha_3} + \frac{3}{2}\eta\up{2pt}{\mu_2[\mu_3}{0pt}{}k\up{8pt}{[\mu_1}{-3pt}{2}(\gamma\up{2pt}{\nu_3]}{0pt}{}\up{8pt}{\nu_1\rho_1]}{0pt}{})_{\alpha_2\alpha_3}
\\ + 6k_2^{\mu}\eta\up{10pt}{[\mu_1}{2pt}{\mu_2}\eta\up{10pt}{\nu_1}{2pt}{[\mu_3}(\gamma\up{10pt}{\rho_1]}{0pt}{}\up{2pt}{\nu_3]}{-2pt}{\phantom{\nu_0]}\mu})_{\alpha_2\alpha_3} + \frac{3}{2}k_2^{\mu}\eta^{\mu_2[\mu_3}\eta^{\nu_3][\mu_1}(\gamma^{\nu_1\rho_1]}_{\phantom{\nu_1\rho_1]}\mu})_{\alpha_2\alpha_3} - \frac{1}{2}k_2^{\mu}\eta^{\mu_2[\mu_3}(\gamma^{\nu_3]\mu_1\nu_1\rho_1}_{\phantom{\nu_0]\mu_1\nu_1\rho_1}\mu})_{\alpha_2\alpha_3}\Bigg\}
\\ \hspace{-25pt} - \frac{3(d-2)}{4\sqrt{2}z_{12}^3 z_{13}z_{23}^2}\hspace{-3pt}\left[\hspace{-2pt}4\eta^{\mu_2[\mu_1}\hspace{-2pt}k_2^{\nu_1}\hspace{-1pt}\eta^{\rho_1][\mu_3}\gamma^{\nu_3]}_{\alpha_2\alpha_3} \hspace{-4pt}-\hspace{-2pt} 4\eta\up{8pt}{[\mu_1}{2pt}{\mu_2}\hspace{-1pt}k\up{3pt}{[\mu_3}{-4pt}{\hspace{-1pt}2}\eta\up{8pt}{\nu_1}{2pt}{\nu_3]}\gamma\up{8pt}{\rho_1]}{-3pt}{\hspace{-1pt}\alpha_2\alpha_3} \hspace{-4pt}-\hspace{-2pt} 2\eta\up{8pt}{[\mu_1}{2pt}{\mu_2}\eta\up{8pt}{\nu_1}{2pt}{\mu_3}\eta\up{8pt}{\rho_1]}{2pt}{\nu_3}(k_2\hspace{-2pt}\cdot\hspace{-2pt}\gamma)_{\alpha_2\alpha_3} \hspace{-3pt}+\hspace{-2pt} 2\eta^{\mu_2[\mu_1}\hspace{-2pt}k_2^{\nu_1}\hspace{-1pt}(\gamma^{\rho_1]\mu_3\nu_3})_{\alpha_2\alpha_3} \right. 
\\ \left. + 2\eta\up{9pt}{[\mu_1}{2pt}{\mu_2}k_2^{[\mu_3}(\gamma\up{4pt}{\nu_3]}{0pt}{}\up{9pt}{\nu_1\rho_1]}{0pt}{})_{\alpha_2\alpha_3} - 4k_2^{\mu}\eta\up{8pt}{[\mu_1}{2pt}{\mu_2}\eta\up{8pt}{\nu_1}{2pt}{[\mu_3}(\gamma\up{8pt}{\rho_1]}{0pt}{}\up{2pt}{\nu_3]}{-2pt}{\phantom{\nu_0]}\mu})_{\alpha_2\alpha_3} + k_2^{\mu}\eta\up{8pt}{[\mu_1}{2pt}{\mu_2}(\gamma\up{8pt}{\nu_1\rho_1]}{0pt}{}\up{2pt}{\mu_3\nu_3}{-2pt}{\phantom{\mu_3\nu_3}\mu})_{\alpha_2\alpha_3}\right]
\end{multline}

\begin{multline}
\left\langle \normal{e^{-\phi}}\normal{\psi^{\mu_1}\psi^{\nu_1}\psi^{\rho_1}}(z_1)k_2^{\rho}(\gamma_{\rho}\gamma_{\nu_2})_{\alpha_2}^{\phantom{\alpha_2}\beta_2}\normal{j^{\mu_2\nu_2}\mathcal{S}_{\beta_2}}(z_2)k_3^{\lambda}(\gamma_{\lambda}\gamma_{\nu_3})_{\alpha_3}^{\phantom{\alpha_3}\beta_3}\normal{j^{\mu_3\nu_3}\mathcal{S}_{\beta_3}}(z_3) \right\rangle 
\\ = \frac{(d-2)^2}{4\sqrt{2}z_{12}^2 z_{13}^2 z_{23}^2}\Bigg[6\eta^{\mu_2\mu_3}k_2^{[\mu_1}k_3^{\nu_1}\gamma^{\rho_1]}_{\alpha_2\alpha_3} + 12k_2\cdot k_3 \eta\up{8pt}{[\mu_1}{2pt}{\mu_2}\eta\up{8pt}{\nu_1}{2pt}{\mu_3}\gamma\up{8pt}{\rho_1]}{-3pt}{\alpha_2\alpha_3} - 12\eta\up{8pt}{[\mu_1}{2pt}{\mu_2}\eta\up{8pt}{\nu_1}{2pt}{\mu_3}k\up{8pt}{\rho_1]}{-3pt}{\hspace{-1pt}3}(k_2\cdot\gamma)_{\alpha_2\alpha_3} 
\\ - 12\eta\up{8pt}{[\mu_1}{2pt}{\mu_2}\eta\up{8pt}{\nu_1}{2pt}{\mu_3}k\up{8pt}{\rho_1]}{-3pt}{\hspace{-1pt}2}(k_3\cdot\gamma)_{\alpha_2\alpha_3} + k_2\cdot k_3 \eta^{\mu_2\mu_3}(\gamma^{\mu_1\nu_1\rho_1})_{\alpha_2\alpha_3} - 3k_2^{\mu}\eta^{\mu_2\mu_3}k_3^{[\mu_1}(\gamma^{\nu_1\rho_1]}_{\phantom{\nu_1\rho_1]}\mu})_{\alpha_2\alpha_3} 
\\ - 3k_3^{\mu}\eta^{\mu_2\mu_3}k_2^{[\mu_1}(\gamma^{\nu_1\rho_1]}_{\phantom{\nu_1\rho_1]}\mu})_{\alpha_2\alpha_3} + 12k_2^{\mu}k_3^{\nu}\eta\up{8pt}{[\mu_1}{2pt}{\mu_2}\eta\up{8pt}{\nu_1}{2pt}{\mu_3}(\gamma\up{8pt}{\rho_1]}{-3pt}{\phantom{\rho_1]}\mu\nu})_{\alpha_2\alpha_3} + k_2^{\mu}k_3^{\nu}\eta^{\mu_2\mu_3}(\gamma^{\mu_1\nu_1\rho_1}_{\phantom{\mu_1\nu_1\rho_1}\mu\nu})_{\alpha_2\alpha_3}\Bigg]
\end{multline}

\subsection{Fermion 4-Point Correlators}

\begin{multline}
\left\langle j^{\mu_1\nu_1}(z_1)j^{\mu_2\nu_2}(z_2)\normal{e^{-\phi}}\psi^{\mu_3}(z_3)\normal{e^{-\phi}}\psi^{\mu_4}(z_4)\right\rangle
\\ = -\frac{2}{z_{12}^2 z_{34}^2}\left(\eta^{\mu_2[\mu_1}\eta^{\nu_1]\nu_2}\eta^{\mu_3\mu_4} + 2z\eta^{\mu_3[\mu_1}\eta^{\nu_1][\mu_2}\eta^{\nu_2]\mu_4} - \frac{2z}{1-z}\eta^{\mu_4[\mu_1}\eta^{\nu_1][\mu_2}\eta^{\nu_2]\mu_3}\right)
\end{multline}

\begin{multline}
\left\langle j^{\mu_1\nu_1}(z_1)j^{\mu_2\nu_2}(z_2)\normal{e^{-\phi}}\psi^{\mu_3}(z_3)\normal{e^{-\phi}}\normal{\psi^{\mu_4}\psi^{\nu_4}\psi^{\rho_4}}(z_4)\right\rangle
\\ = -\frac{12}{z_{12}z_{14}z_{24}z_{34}^2}\left(2\eta\up{10pt}{[\mu_4}{1pt}{[\mu_1}\eta\up{1pt}{\nu_1][\mu_2}{0pt}{}\eta\up{9pt}{\nu_4}{1pt}{\nu_2]}\eta\up{9pt}{\rho_4]}{2pt}{\mu_3} - z\eta\up{9pt}{\mu_3}{1pt}{[\mu_1}\eta\up{9pt}{[\mu_4}{1pt}{\nu_1]}\eta\up{8pt}{\nu_4}{2pt}{\mu_2}\eta\up{8pt}{\rho_4]}{2pt}{\nu_2} + \frac{z}{1-z}\eta\up{8pt}{[\mu_4}{2pt}{\mu_1}\eta\up{8pt}{\nu_4}{2pt}{\nu_1}\eta\up{9pt}{\rho_4]}{1pt}{[\mu_2}\eta\up{9pt}{\mu_3}{1pt}{\nu_2]}\right)
\end{multline}

\begin{multline}
\left\langle j^{\mu_1\nu_1}(z_1)j^{\mu_2\nu_2}(z_2)\normal{e^{-\phi}}\normal{\psi^{\mu_3}\psi^{\nu_3}\psi^{\rho_3}}(z_3)\normal{e^{-\phi}}\normal{\psi^{\mu_4}\psi^{\nu_4}\psi^{\rho_4}}(z_4)\right\rangle
\\ = \frac{12}{z_{12}^2 z_{34}^4}\bigg[\eta\up{8pt}{[\mu_1}{2pt}{\mu_2}\eta\up{8pt}{\nu_1]}{2pt}{\nu_2}\eta\up{8pt}{[\mu_3}{2pt}{\mu_4}\eta\up{8pt}{\nu_3}{2pt}{\nu_4}\eta\up{8pt}{\rho_3]}{2pt}{\rho_4}
+ 3z^2\eta\up{8pt}{[\mu_3}{2pt}{\mu_1}\eta\up{8pt}{\nu_3}{2pt}{\nu_1}\eta\up{8pt}{\rho_3]}{0pt}{[\mu_4}\eta\up{8pt}{\mu_2}{0pt}{\nu_4}\eta\up{8pt}{\nu_2}{0pt}{\rho_4]} + \frac{3z^2}{(1-z)^2}\eta\up{8pt}{[\mu_4}{2pt}{\mu_1}\eta\up{8pt}{\nu_4}{2pt}{\nu_1}\eta\up{8pt}{\rho_4]}{0pt}{[\mu_3}\eta\up{8pt}{\mu_2}{0pt}{\nu_3}\eta\up{8pt}{\nu_2}{0pt}{\rho_3]}
\\ - 6z\left(2\eta\up{-6pt}{[\mu_3}{2pt}{[\mu_1}\eta\up{10pt}{[\mu_4}{2pt}{\nu_1]}\eta\up{10pt}{\nu_4}{2pt}{[\mu_2}\eta\up{2pt}{\nu_2]}{-5pt}{\nu_3}\eta\up{9pt}{\rho_4]}{-5pt}{\rho_3]} - \eta\up{9pt}{[\mu_1}{-5pt}{[\mu_3}\eta\up{9pt}{\nu_1]}{1pt}{[\mu_2}\eta\up{9pt}{[\mu_4}{1pt}{\nu_2]}\eta\up{9pt}{\nu_4}{-5pt}{\nu_3}\eta\up{9pt}{\rho_4]}{-5pt}{\rho_3]}\right)
\\ + \frac{6z}{1-z}\left(2\eta\up{-6pt}{[\mu_4}{2pt}{[\mu_1}\eta\up{10pt}{[\mu_3}{2pt}{\nu_1]}\eta\up{10pt}{\nu_3}{2pt}{[\mu_2}\eta\up{2pt}{\nu_2]}{-5pt}{\nu_4}\eta\up{9pt}{\rho_3]}{-5pt}{\rho_4]} - \eta\up{9pt}{[\mu_1}{-5pt}{[\mu_4}\eta\up{9pt}{\nu_1]}{1pt}{[\mu_2}\eta\up{9pt}{[\mu_3}{1pt}{\nu_2]}\eta\up{9pt}{\nu_3}{-5pt}{\nu_4}\eta\up{9pt}{\rho_3]}{-5pt}{\rho_4]}\right)\bigg]
\end{multline}

\begin{equation}
\left\langle\normal{e^{-\phi}}\psi^{\mu_1}(z_1)\normal{e^{-\phi}}\psi^{\mu_2}(z_2)\normal{e^{\frac{1}{2}\phi}}S^{\alpha}(z_3)\mathcal{S}_{\beta}(z_4)\right\rangle = -\frac{1}{2z_{12} z_{14}z_{24}}\left[\left(\frac{2-z}{z}\right)\eta^{\mu_1\mu_2}\delta^{\alpha}_{\phantom{\alpha}\beta} - \left(\gamma^{\mu_1\mu_2}\right)^{\alpha}_{\phantom{\alpha}\beta}\right]
\end{equation}

\begin{equation}
\left\langle\normal{e^{\phi}}\psi^{\mu_1}(z_1)\normal{e^{-\phi}}\psi^{\mu_2}(z_2)\mathcal{S}^{\alpha}(z_3)\mathcal{S}_{\beta}(z_4)\right\rangle = -\frac{z_{12}z_{13}}{2z_{23}^2 z_{24}z_{34}}\left[\left(\frac{2-z}{z}\right)\eta^{\mu_1\mu_2}\delta^{\alpha}_{\phantom{\alpha}\beta} - \left(\gamma^{\mu_1\mu_2}\right)^{\alpha}_{\phantom{\alpha}\beta}\right]
\end{equation}

\begin{equation}
\left\langle\normal{e^{\phi}}\psi^{\mu_1}(z_1)\normal{e^{-2\phi}}(z_2)\mathcal{S}_{\alpha}(z_3)\mathcal{S}_{\beta}(z_4)\right\rangle = -\frac{z_{12}^2}{\sqrt{2}z_{23}z_{24}z_{34}}\gamma^{\mu_1}_{\alpha\beta}
\end{equation}

\begin{equation}
\left\langle j^{\mu_1\nu_1}(z_1)\normal{e^{-\phi}}\psi^{\mu_2}(z_2)\mathcal{S}_{\alpha}(z_3)\mathcal{S}_{\beta}(z_4)\right\rangle 
\\ = \frac{i}{2\sqrt{2}z_{13}z_{14}z_{23}z_{24}}\hspace{-2pt}\left[2\hspace{-2pt}\left(\hspace{-2pt}\frac{2\hspace{-2pt}-\hspace{-2pt}z}{z}\hspace{-2pt}\right)\hspace{-2pt}\eta^{\mu_2[\mu_1}\gamma^{\nu_1]}_{\alpha\beta} + \left(\gamma^{\mu_1\nu_1\mu_2}\right)_{\alpha\beta}\right]
\end{equation}

\begin{equation}
\left\langle j^{\mu_1\nu_1}(z_1)\normal{e^{-\phi}}\psi^{\mu_2}(z_2)\normal{e^{\frac{1}{2}\phi}}S^{\alpha}(z_3)\mathcal{S}^{\beta}(z_4)\right\rangle  = \frac{iz_{34}}{2\sqrt{2}z_{13}z_{14}z_{24}^2}\hspace{-2pt}\left[2\hspace{-2pt}\left(\hspace{-2pt}\frac{2\hspace{-2pt}-\hspace{-2pt}z}{z}\hspace{-2pt}\right)\hspace{-2pt}\eta^{\mu_2[\mu_1}\gamma^{\nu_1]\alpha\beta} + \left(\gamma^{\mu_1\nu_1\mu_2}\right)^{\alpha\beta}\right]
\end{equation}

\begin{equation}
\left\langle\normal{e^{\phi}}\psi^{\mu_1}(z_1)\normal{e^{-2\phi}}j^{\mu_2\nu_2}(z_2)\mathcal{S}_{\alpha}(z_3)\mathcal{S}_{\beta}(z_4)\right\rangle = -\frac{iz_{12}^2}{2\sqrt{2}z_{23}^2 z_{24}^2}\hspace{-2pt}\left[2\hspace{-2pt}\left(\hspace{-2pt}\frac{2\hspace{-2pt}-\hspace{-2pt}z}{z}\hspace{-2pt}\right)\hspace{-2pt}\eta^{\mu_1[\mu_2}\gamma^{\nu_2]}_{\alpha\beta} - \left(\gamma^{\mu_1\mu_2\nu_2}\right)_{\alpha\beta}\right]
\end{equation}

\begin{align}
\notag\hspace{100pt}&\hspace{-100pt}\left\langle\normal{e^{-\phi}}\normal{\psi^{\mu_1}\psi^{\nu_1}\psi^{\rho_1}}(z_1)\normal{e^{-\phi}}\psi^{\mu_2}(z_2)\normal{e^{\frac{1}{2}\phi}}S^{\alpha}(z_3)\mathcal{S}_{\beta}(z_4)\right\rangle
\\ & = \frac{z_{34}}{4z_{12}z_{13}z_{14}^2 z_{24}}\left[3\hspace{-2pt}\left(\hspace{-2pt}\frac{2\hspace{-2pt}-\hspace{-2pt}z}{z}\hspace{-2pt}\right)\hspace{-2pt}\eta^{\mu_2[\mu_1}(\gamma^{\nu_1\rho_1]})^{\alpha}_{\phantom{\alpha}\beta} - \left(\gamma^{\mu_1\nu_1\rho_1\mu_2}\right)^{\alpha}_{\phantom{\alpha}\beta}\right]
\\ \notag\hspace{100pt}&\hspace{-100pt}\left\langle\normal{e^{\phi}}\psi^{\mu_1}(z_1)\normal{e^{-\phi}}\normal{\psi^{\mu_2}\psi^{\nu_2}\psi^{\rho_2}}(z_2)\mathcal{S}^{\alpha}(z_3)\mathcal{S}_{\beta}(z_4)\right\rangle
\\ & = -\frac{z_{12}z_{13}}{4z_{23}^3 z_{24}^2}\left[3\hspace{-2pt}\left(\hspace{-2pt}\frac{2\hspace{-2pt}-\hspace{-2pt}z}{z}\hspace{-2pt}\right)\hspace{-2pt}\eta^{\mu_1[\mu_2}(\gamma^{\nu_2\rho_2]})_{\beta}^{\phantom{\beta}\alpha} + \left(\gamma^{\mu_1\mu_2\nu_2\rho_2}\right)_{\beta}^{\phantom{\beta}\alpha}\right]
\end{align}

\begin{multline}
\left\langle\ j^{\mu_1\nu_1}(z_1)\normal{e^{-2\phi}}j^{\mu_2\nu_2}(z_2)\normal{e^{\frac{1}{2}\phi}}S^{\alpha}(z_3)\mathcal{S}_{\beta}(z_4)\right\rangle
\\ = \frac{z_{34}}{4z_{13}z_{14}z_{24}^2}\hspace{-2pt}\left[\hspace{-2pt}2\hspace{-2pt}\left(\hspace{-2pt}\frac{2\hspace{-2pt}-\hspace{-2pt}z}{z}\hspace{-2pt}\right)^{\hspace{-2pt}2}\hspace{-3pt}\eta^{\mu_1[\mu_2}\eta^{\nu_2]\nu_1}\delta^{\alpha}_{\phantom{\alpha}\beta} \hspace{-2pt}+\hspace{-2pt} 4\hspace{-2pt}\left(\hspace{-2pt}\frac{2\hspace{-2pt}-\hspace{-2pt}z}{z}\hspace{-2pt}\right)\hspace{-3pt}\eta^{\mu_1][\mu_2}\hspace{-1pt}(\gamma^{\nu_2][\nu_1})^{\alpha}_{\phantom{\alpha}\beta} \hspace{-2pt}-\hspace{-2pt} \left(\gamma^{\mu_1\nu_1\mu_2\nu_2}\right)^{\alpha}_{\phantom{\alpha}\beta}\hspace{-2pt}\right]
\end{multline}

\begin{multline}
\left\langle j^{\mu_1\nu_1}(z_1)\normal{e^{-\phi}}\normal{\psi^{\mu_2}\psi^{\nu_2}\psi^{\rho_2}}(z_2)\mathcal{S}_{\alpha}(z_3)\mathcal{S}_{\beta}(z_4)\right\rangle
\\ = \frac{iz_{34}}{4\sqrt{2}z_{13}z_{14}z_{23}^2 z_{24}^2}\hspace{-3pt}\left[\hspace{-2pt}6\hspace{-3pt}\left(\hspace{-3pt}\frac{2\hspace{-3pt}-\hspace{-3pt}z}{z}\hspace{-3pt}\right)^{\hspace{-3pt}2}\hspace{-3pt}\eta\up{8pt}{[\mu_2}{2pt}{\mu_1}\eta\up{8pt}{\nu_2}{2pt}{\nu_1}\gamma\up{8pt}{\rho_2]}{-3pt}{\alpha\beta} \hspace{-2pt}-\hspace{-2pt} 6\hspace{-3pt}\left(\hspace{-3pt}\frac{2\hspace{-3pt}-\hspace{-3pt}z}{z}\hspace{-3pt}\right)\hspace{-4pt}\eta^{\mu_1][\mu_2}\hspace{-1pt}(\gamma^{\nu_2\rho_2][\nu_1})_{\alpha\beta} \hspace{-2pt}-\hspace{-2pt} \left(\gamma^{\mu_1\nu_1\mu_2\nu_2\rho_2}\right)_{\alpha\beta}\hspace{-2pt}\right]
\end{multline}

\begin{multline}
\left\langle j^{\mu_1\nu_1}(z_1)\normal{e^{-\phi}}\normal{\psi^{\mu_2}\psi^{\nu_2}\psi^{\rho_2}}(z_2)\normal{e^{\frac{1}{2}\phi}}S^{\alpha}(z_3)\mathcal{S}^{\beta}(z_4)\right\rangle
\\ = \frac{iz_{34}^2}{4\sqrt{2}z_{13}z_{14}z_{23} z_{24}^3}\hspace{-3pt}\left[\hspace{-2pt}6\hspace{-3pt}\left(\hspace{-3pt}\frac{2\hspace{-3pt}-\hspace{-3pt}z}{z}\hspace{-3pt}\right)^{\hspace{-3pt}2}\hspace{-3pt}\eta\up{8pt}{[\mu_2}{2pt}{\mu_1}\eta\up{8pt}{\nu_2}{2pt}{\nu_1}\gamma\up{8pt}{\rho_2]}{2pt}{\phantom{\rho_2]}\alpha\beta} \hspace{-2pt}-\hspace{-2pt} 6\hspace{-3pt}\left(\hspace{-3pt}\frac{2\hspace{-3pt}-\hspace{-3pt}z}{z}\hspace{-3pt}\right)\hspace{-4pt}\eta^{\mu_1][\mu_2}\hspace{-1pt}(\gamma^{\nu_2\rho_2][\nu_1})^{\alpha\beta} \hspace{-2pt}-\hspace{-2pt} \left(\gamma^{\mu_1\nu_1\mu_2\nu_2\rho_2}\right)^{\alpha\beta}\hspace{-2pt}\right]
\end{multline}

\begin{multline}
\left\langle\normal{e^{-\phi}}\normal{\psi^{\mu_1}\psi^{\nu_1}\psi^{\rho_1}}(z_1)\normal{e^{-\phi}}\normal{\psi^{\mu_2}\psi^{\nu_2}\psi^{\rho_2}}(z_2)\normal{e^{\frac{1}{2}\phi}}S^{\alpha}(z_3)\mathcal{S}_{\beta}(z_4)\right\rangle
\\ = \frac{z_{34}^2}{8z_{12}z_{13}z_{14}^2 z_{23}z_{24}^2}\left[6\hspace{-2pt}\left(\hspace{-2pt}\frac{2\hspace{-2pt}-\hspace{-2pt}z}{z}\hspace{-2pt}\right)^{\hspace{-2pt}3}\hspace{-2pt}\eta\up{8pt}{[\mu_1}{2pt}{\mu_2}\eta\up{8pt}{\nu_1}{2pt}{\nu_2}\eta\up{8pt}{\rho_1]}{2pt}{\rho_2}\delta^{\alpha}_{\phantom{\alpha}\beta} - 18\hspace{-2pt}\left(\hspace{-2pt}\frac{2\hspace{-2pt}-\hspace{-2pt}z}{z}\hspace{-2pt}\right)^{\hspace{-2pt}2}\hspace{-2pt}\eta\up{10pt}{[\mu_1}{2pt}{[\mu_2}\eta\up{10pt}{\nu_1}{2pt}{\nu_2}\big(\gamma\up{10pt}{\rho_1]}{2pt}{}\up{10pt}{}{2pt}{\rho_2]}\big)^{\alpha}_{\phantom{\alpha}\beta} \right. 
\\ \left. - 9\hspace{-2pt}\left(\hspace{-2pt}\frac{2\hspace{-2pt}-\hspace{-2pt}z}{z}\hspace{-2pt}\right)\hspace{-2pt}\eta^{\mu_1][\mu_2}(\gamma^{\nu_2\rho_2][\nu_1\rho_1})^{\alpha}_{\phantom{\alpha}\beta} + (\gamma^{\mu_1\nu_1\rho_1\mu_2\nu_2\rho_2})^{\alpha}_{\phantom{\alpha}\beta}\right]
\end{multline}

\begin{multline}
\left\langle j^{\mu_1\nu_1}(z_1)\normal{e^{-\phi}}\psi^{\mu_2}(z_2)\mathcal{S}_{\alpha_3}(z_3)k_4^{\lambda}(\gamma_{\lambda}\gamma_{\nu_4})_{\alpha_4}^{\phantom{\alpha_4}\beta_4}\normal{j^{\mu_4\nu_4}\mathcal{S}_{\beta_4}}(z_4)\right\rangle 
\\ = \frac{d-2}{\sqrt{2}z_{12}z_{14}z_{24}z_{34}^2}\bigg[2\eta^{\mu_2[\mu_1}\eta^{\nu_1]\mu_4}(k_4\cdot\gamma)_{\alpha_3\alpha_4} - z\left(\eta^{\mu_2\mu_4}k_4^{[\mu_1}\gamma^{\nu_1]}_{\alpha_3\alpha_4} - \frac{1}{2}\eta^{\mu_2\mu_4}k_4^{\mu}(\gamma^{\mu_1\nu_1}_{\phantom{\mu_1\nu_1}\mu})_{\alpha_3\alpha_4}\right)
\\ - \frac{z}{1-z}\left(\eta^{\mu_4[\mu_1}k_4^{\nu_1]}\gamma^{\mu_2}_{\alpha_3\alpha_4} - k_4^{\mu_2}\eta^{\mu_4[\mu_1}\gamma^{\nu_1]}_{\alpha_3\alpha_4} + \eta^{\mu_4[\mu_1}\eta^{\nu_1]\mu_2}(k_4\cdot\gamma)_{\alpha_3\alpha_4} - k_4^{\mu}\eta^{\mu_4[\mu_1}(\gamma^{\nu_1]\mu_2}_{\phantom{\nu_1]\mu_2}\mu})_{\alpha_3\alpha_4}\right)\bigg]
\end{multline}

\begin{multline}
\left\langle j^{\mu_1\nu_1}(z_1)\normal{e^{-\phi}}\psi^{\mu_2}(z_2)k_3^{\rho}(\gamma_{\rho}\gamma_{\nu_3})_{\alpha_3}^{\phantom{\alpha_3}\beta_3}\normal{j^{\mu_3\nu_3}\mathcal{S}_{\beta_3}}(z_3)k_4^{\lambda}(\gamma_{\lambda}\gamma_{\nu_4})_{\alpha_4}^{\phantom{\alpha_4}\beta_4}\normal{j^{\mu_4\nu_4}\mathcal{S}_{\beta_4}}(z_4)\right\rangle
\\ = -\frac{i(d-2)^2}{\sqrt{2}z_{12}z_{14}z_{23}z_{34}^3}\bigg[k_3^{\mu}k_4^{\nu}\hspace{-2pt}\left(\hspace{-2pt}\eta^{\mu_3\mu_4}\eta^{\mu_2[\mu_1} \hspace{-2pt}+\hspace{-2pt} \frac{z}{1\hspace{-2pt}-\hspace{-2pt}z}\eta^{\mu_2\mu_3}\eta^{\mu_4[\mu_1} \hspace{-2pt}-\hspace{-2pt} z(1\hspace{-2pt}-\hspace{-2pt}z)\eta^{\mu_2\mu_4}\eta^{\mu_3[\mu_1}\hspace{-2pt}\right)\hspace{-3pt}(\gamma_{\mu}\gamma^{\nu_1]}\gamma_{\nu})_{\alpha_3\alpha_4} 
\\ + z\bigg(\eta^{\mu_3[\mu_1}\eta^{\nu_1]\mu_4}k_3^{\mu}k_4^{\nu}(\gamma_{\mu}\gamma^{\mu_2}\gamma_{\nu})_{\alpha_3\alpha_4} + \frac{1}{4}\eta^{\mu_3\mu_4}k_3^{\mu}k_4^{\nu}(\gamma_{\mu}\gamma^{\mu_1\nu_1}\gamma^{\mu_2}\gamma_{\nu})_{\alpha_3\alpha_4}\bigg)\bigg]
\end{multline}

\begin{multline}
\left\langle j^{\mu_1\nu_1}(z_1)\normal{e^{-\phi}}\normal{\psi^{\mu_2}\psi^{\nu_2}\psi^{\rho_2}}(z_2)\mathcal{S}_{\alpha_3}(z_3)k_4^{\lambda}(\gamma_{\lambda}\gamma_{\nu_4})_{\alpha_4}^{\phantom{\alpha_4}\beta_4}\normal{j^{\mu_4\nu_4}\mathcal{S}_{\beta_4}}(z_4)\right\rangle 
\\ = \frac{d-2}{\sqrt{2}z_{12}^2 z_{24}^2 z_{34}^2}\bigg\{\left(6-\frac{8z}{1-z} + \frac{3z^2}{2(1-z)}\right)\eta\up{8pt}{[\mu_2}{2pt}{\mu_1}\eta\up{8pt}{\nu_2}{2pt}{\nu_1}\eta\up{8pt}{\rho_2]}{2pt}{\mu_4}(k_4\cdot\gamma)_{\alpha_3\alpha_4}
\\ + \frac{z}{1\hspace{-3pt}-\hspace{-3pt}z}\hspace{-2pt}\bigg[6k_4^{\mu}\eta\up{10pt}{[\mu_2}{2pt}{\mu_4}\eta\up{10pt}{\nu_2}{2pt}{[\mu_1}(\gamma\up{10pt}{\rho_2]}{0pt}{}\up{2pt}{\nu_1]}{-2pt}{\phantom{\nu_1]}\mu})_{\alpha_3\alpha_4} \hspace{-2pt}+ 6\eta^{\mu_4[\mu_2}k_4^{\nu_2}\eta^{\rho_2][\mu_1}\gamma^{\nu_1]}_{\alpha_3\alpha_4} \hspace{-2pt}+ 6k\up{2pt}{[\mu_1}{-4pt}{\hspace{-1pt}4}\eta\up{2pt}{\nu_1]}{10pt}{[\mu_2}\eta\up{10pt}{\nu_2}{2pt}{\mu_4}\gamma\up{10pt}{\rho_2]}{-3pt}{\alpha_3\alpha_4} \hspace{-2pt}- 6\eta^{\mu_4[\mu_1}\eta^{\nu_1][\mu_2}k_4^{\nu_2}\gamma^{\rho_2]}_{\alpha_3\alpha_4}\hspace{-2pt}\bigg]
\\ + \frac{3z(2-z)}{2(1-z)^2}k_4^{\mu}\eta^{\mu_4[\mu_1}\eta^{\nu_1][\mu_2}(\gamma^{\nu_2\rho_2]}_{\phantom{\nu_2\rho_2]}\mu})_{\alpha_3\alpha_4} - \frac{z^2}{2(1-z)^2}\bigg[6\eta^{\mu_4[\mu_1}\eta^{\nu_1][\mu_2}k_4^{\nu_2}\gamma^{\rho_2]}_{\alpha_3\alpha_4} 
\\ - \eta^{\mu_4[\mu_1}k_4^{\nu_1]}(\gamma^{\mu_2\nu_2\rho_2})_{\alpha_3\alpha_4} + 3\eta\up{2pt}{\mu_4[\mu_1}{0pt}{}k\up{8pt}{[\mu_2}{-3pt}{4}(\gamma\up{2pt}{\nu_1]}{0pt}{}\up{8pt}{\nu_2\rho_2]}{0pt}{})_{\alpha_3\alpha_4} + k_4^{\mu}\eta^{\mu_4[\mu_1}(\gamma^{\nu_1]\mu_2\nu_2\rho_2}_{\phantom{\nu_1]\mu_2\nu_2\rho_2}\mu})_{\alpha_3\alpha_4}\bigg]
\\ - \frac{3z^2}{2(1-z)}\left[2\eta^{\mu_4[\mu_2}k_4^{\nu_2}\eta^{\rho_2][\mu_1}\gamma^{\nu_1]}_{\alpha_3\alpha_4} + 2k\up{8pt}{[\mu_1}{-3pt}{4}\eta\up{8pt}{\nu_1][\mu_2}{0pt}{}\eta\up{8pt}{\nu_2}{2pt}{\mu_4}\gamma\up{8pt}{\rho_2]}{-3pt}{\alpha_3\alpha_4} - \eta^{\mu_4[\mu_2}k_4^{\nu_2}(\gamma^{\rho_2]\mu_1\nu_1})_{\alpha_3\alpha_4} \right. 
\\ \left. - \eta\up{8pt}{[\mu_2}{2pt}{\mu_4}k\up{4pt}{[\mu_1}{-3pt}{4}(\gamma\up{3pt}{\nu_1]}{0pt}{}\up{8pt}{\nu_2\rho_2]}{0pt}{})_{\alpha_3\alpha_4} + 2k_4^{\mu}\eta\up{8pt}{[\mu_2}{2pt}{\mu_4}\eta\up{8pt}{\nu_2}{1pt}{[\mu_1}(\gamma\up{8pt}{\rho_2]}{0pt}{}\up{2pt}{\nu_1]}{-2pt}{\phantom{\nu_1]}\mu})_{\alpha_3\alpha_4} + \frac{1}{2}k_4^{\mu}\eta^{\mu_4[\mu_2}(\gamma^{\nu_2\rho_2]\mu_1\nu_1}_{\phantom{\nu_2\rho_2]\mu_1\nu_1}\mu})_{\alpha_3\alpha_4}\right]\bigg\}
\end{multline}

\begin{equation}
\left\langle \mathcal{S}_{\alpha_1}(z_1)\mathcal{S}_{\alpha_2}(z_2)\mathcal{S}_{\alpha_3}(z_3)\mathcal{S}_{\alpha_4}(z_4)\right\rangle = -\frac{1}{2z_{12}z_{13}z_{24}z_{34}}\hspace{-2pt}\left(\hspace{-2pt}\frac{1}{1\hspace{-2pt}-\hspace{-2pt}z}\gamma_{\mu\alpha_1\alpha_2}\gamma^{\mu}_{\alpha_3\alpha_4} \hspace{-2pt}+\hspace{-2pt} \frac{z}{1\hspace{-2pt}-\hspace{-2pt}z}\gamma_{\mu\alpha_1\alpha_3}\gamma^{\mu}_{\alpha_2\alpha_4}\hspace{-2pt}\right)
\end{equation}

\begin{multline}
\left\langle \mathcal{S}_{\alpha_1}(z_1)\mathcal{S}_{\alpha_2}(z_2)\mathcal{S}_{\alpha_3}(z_3)\normal{j^{\mu_4\nu_4}\mathcal{S}_{\alpha_4}}(z_4)\right\rangle
\\ = -\frac{i}{4z_{12}z_{13}z_{14}z_{24}z_{34}}\left(\frac{1}{1-z}(\gamma^{\mu_4\nu_4}\gamma_{\mu})_{\alpha_1\alpha_2}\gamma^{\mu}_{\alpha_3\alpha_4} + \frac{z}{1-z}(\gamma^{\mu_4\nu_4}\gamma_{\mu})_{\alpha_1\alpha_3}\gamma^{\mu}_{\alpha_2\alpha_4}\right)
\\ - \frac{i}{4z_{12}z_{13}z_{24}^2 z_{34}}\left(\frac{1}{1-z}(\gamma^{\mu_4\nu_4}\gamma_{\mu})_{\alpha_2\alpha_1}\gamma^{\mu}_{\alpha_3\alpha_4} + \frac{z}{1-z}\gamma_{\mu\alpha_1\alpha_3}(\gamma^{\mu_4\nu_4}\gamma^{\mu})_{\alpha_2\alpha_4}\right)
\\ - \frac{i}{4z_{12}z_{13}z_{24}z_{34}^2}\left(\frac{1}{1-z}\gamma_{\mu\alpha_1\alpha_2}(\gamma^{\mu_4\nu_4}\gamma^{\mu})_{\alpha_3\alpha_4} + \frac{z}{1-z}(\gamma^{\mu_4\nu_4}\gamma_{\mu})_{\alpha_3\alpha_1}\gamma^{\mu}_{\alpha_2\alpha_4}\right)
\end{multline}

\begin{multline}
\left\langle \mathcal{S}_{\alpha_1}(z_1)\mathcal{S}_{\alpha_2}(z_2)\mathcal{S}_{\alpha_3}(z_3)k_4^{\lambda}(\gamma_{\lambda}\gamma_{\nu_4})_{\alpha_4}^{\phantom{\alpha_4}\beta_4}\normal{j^{\mu_4\nu_4}\mathcal{S}_{\beta_4}}(z_4)\right\rangle
\\ = \frac{i(d-2)}{2z_{12}z_{14}z_{24}z_{34}^2}\left(\gamma^{\mu_4}_{\alpha_1\alpha_2}(k_4\cdot\gamma)_{\alpha_3\alpha_4} - z\gamma^{\mu_4}_{\alpha_1\alpha_3}(k_4\cdot\gamma)_{\alpha_2\alpha_4} + \frac{z}{1-z}(k_4\cdot\gamma)_{\alpha_1\alpha_4}\gamma^{\mu_4}_{\alpha_2\alpha_3}\right)
\end{multline}

\begin{multline}
\left\langle \mathcal{S}_{\alpha_1}(z_1)\mathcal{S}_{\alpha_2}(z_2)k_3^{\rho}(\gamma_{\rho}\gamma_{\nu_3})_{\alpha_3}^{\phantom{\alpha_3}\beta_3}\normal{j^{\mu_3\nu_3}\mathcal{S}_{\beta_3}}(z_3)k_4^{\lambda}(\gamma_{\lambda}\gamma_{\nu_4})_{\alpha_4}^{\phantom{\alpha_4}\beta_4}\normal{j^{\mu_4\nu_4}\mathcal{S}_{\beta_4}}(z_4)\right\rangle
\\ = \frac{d-2}{2z_{13}^2 z_{24}^2 z_{34}^2}\Bigg\{-2^{\frac{d-4}{2}}\eta^{\mu_3\mu_4}k_3^{\mu}k_4^{\nu}\gamma_{\mu\alpha_1\alpha_3}\gamma_{\nu\alpha_2\alpha_4} + \frac{2^{\frac{d-4}{2}}}{(1-z)^2}\eta^{\mu_3\mu_4}k_4^{\mu}k_3^{\nu}\gamma_{\mu\alpha_1\alpha_4}\gamma_{\nu\alpha_2\alpha_3}
\\ + \frac{2}{1-z}\eta^{\mu_3\mu_4}\left(k_3^{\mu}k_4^{\nu}\gamma_{\mu\alpha_1\alpha_3}\gamma_{\nu\alpha_2\alpha_4} - k_3^{\nu}k_4^{\mu}\gamma_{\mu\alpha_1\alpha_4}\gamma_{\nu\alpha_2\alpha_3}\right)
\\ + \left(\frac{1}{z}-\frac{1}{1-z}\right)\left[\eta^{\mu_3\mu_4}k_3^{\rho}k_4^{\lambda}(\gamma_{\nu}\gamma_{\rho}\gamma_{\lambda})_{\alpha_2\alpha_4}\gamma^{\nu}_{\alpha_1\alpha_3} - k_3^{\rho}k_4^{\lambda}(\gamma^{\mu_3}\gamma_{\rho}\gamma_{\lambda})_{\alpha_2\alpha_4}\gamma^{\mu_4}_{\alpha_1\alpha_3} \right. 
\\ \left. + k_3^{\mu_4}k_4^{\lambda}(\gamma^{\mu_3}\gamma_{\nu}\gamma_{\lambda})_{\alpha_2\alpha_4}\gamma^{\nu}_{\alpha_1\alpha_3} - k_3^{\rho}k_4^{\lambda}(\gamma^{\mu_3}\gamma_{\rho}\gamma^{\mu_4})_{\alpha_1\alpha_2}\gamma_{\lambda\alpha_3\alpha_4}\right]
\\ + \left(\frac{1}{z(1-z)} + \frac{1}{1-z}\right)\left[\eta^{\mu_3\mu_4}k_3^{\rho}k_4^{\lambda}(\gamma_{\nu}\gamma_{\lambda}\gamma_{\rho})_{\alpha_2\alpha_3}\gamma^{\nu}_{\alpha_1\alpha_4} - k_3^{\rho}k_4^{\lambda}(\gamma^{\mu_4}\gamma_{\lambda}\gamma_{\rho})_{\alpha_2\alpha_3}\gamma^{\mu_3}_{\alpha_1\alpha_4} \right. 
\\ \left. + k_3^{\rho}k_4^{\mu_3}(\gamma^{\mu_4}\gamma_{\nu}\gamma_{\rho})_{\alpha_2\alpha_3}\gamma^{\nu}_{\alpha_1\alpha_4} - k_3^{\rho}k_4^{\lambda}(\gamma^{\mu_4}\gamma_{\lambda}\gamma^{\mu_3})_{\alpha_1\alpha_2}\gamma_{\rho\alpha_3\alpha_4}\right]
\\ + (2^{\frac{d-6}{2}}-1)\left(\frac{1}{z(1-z)} - \frac{1}{2(1-z)}\right)\eta^{\mu_3\mu_4}k_3^{\rho}k_4^{\lambda}(\gamma_{\rho}\gamma_{\nu}\gamma_{\lambda})_{\alpha_3\alpha_4}\gamma^{\nu}_{\alpha_1\alpha_2}
\\ - \frac{1}{1-z}\bigg[\frac{d+2}{2}\eta^{\mu_3\mu_4}k_3^{\rho}k_4^{\lambda}\gamma_{\rho\alpha_3[\alpha_1}\gamma_{|\lambda|\alpha_2]\alpha_4} - k_3^{\rho}k_4^{\mu_3}(\gamma_{\rho}\gamma^{\mu_4}\gamma_{\nu})_{\alpha_3[\alpha_1}\gamma^{\nu}_{\alpha_2]\alpha_4}
\\ + 2k_3^{\rho}k_4^{\lambda}(\gamma_{\rho}\gamma^{\mu_4}\gamma_{\lambda})_{\alpha_3[\alpha_1}\gamma^{\mu_3}_{\alpha_2]\alpha_4} - \eta^{\mu_3\mu_4}k_3^{\rho}k_4^{\lambda}(\gamma_{\rho}\gamma_{\nu}\gamma_{\lambda})_{\alpha_3[\alpha_1}\gamma^{\nu}_{\alpha_2]\alpha_4}\bigg]\Bigg\}
\end{multline}

\subsection{Fermion 5-Point Correlators}

For five-point functions, we take the two conformal cross-ratios to be the same $z$ as before as well as
\begin{equation}
\zeta \equiv \frac{z_{01}z_{23}}{z_{02}z_{13}}.
\end{equation}

\begin{multline}
\left\langle \normal{e^{\phi}}\psi^{\mu_0}(z_0)\normal{e^{-\phi}}\psi^{\mu_1}(z_1)\normal{e^{-\phi}}\psi^{\mu_2}(z_2)\mathcal{S}_{\alpha}(z_3)\mathcal{S}_{\beta}(z_4)\right\rangle
\\ = \frac{z_{01}z_{02}}{2\sqrt{2}z_{12}z_{13}z_{14}z_{23}z_{24}}\hspace{-2pt}\left[\hspace{-2pt}\left(\hspace{-2pt}\frac{2(1\hspace{-2pt}-\hspace{-2pt}\zeta)(1\hspace{-2pt}-\hspace{-2pt}z)}{\zeta z}\hspace{-2pt}-\hspace{-2pt}1\hspace{-2pt}\right)\hspace{-2pt}\eta^{\mu_0\mu_{1}}\gamma^{\mu_2}_{\alpha\beta} \hspace{-2pt}-\hspace{-2pt} \left(\hspace{-2pt}\frac{2(1\hspace{-2pt}-\hspace{-2pt}\zeta)}{z}\hspace{-2pt}-\hspace{-2pt}1\hspace{-2pt}\right)\hspace{-2pt}\eta^{\mu_0\mu_2}\gamma^{\mu_1}_{\alpha\beta} \right. 
\\ \left. + \left(\frac{2-z}{z}\right)\hspace{-2pt}\eta^{\mu_{1}\mu_2}\gamma^{\mu_0}_{\alpha\beta} - \gamma^{\mu_0\mu_{1}\mu_{2}}_{\alpha\beta}\right]
\end{multline}

\begin{multline}
\left\langle \normal{e^{\phi}}\psi^{\mu_0}(z_0)\normal{e^{-\phi}}\psi^{\mu_1}(z_1)\normal{e^{-\phi}}\normal{\psi^{\mu_2}\psi^{\nu_2}\psi^{\rho_2}}(z_2)\mathcal{S}_{\alpha}(z_3)\mathcal{S}_{\beta}(z_4)\right\rangle 
\\ = -\frac{z_{01}z_{02}z_{34}}{4\sqrt{2}z_{12}z_{13}z_{14} z_{23}^2 z_{24}^2}\hspace{-2pt}\left[6\hspace{-2pt}\left(\hspace{-2pt}\frac{2(1\hspace{-2pt}-\hspace{-2pt}\zeta)}{z}\hspace{-2pt}-\hspace{-2pt}1\hspace{-2pt}\right)\hspace{-3pt}\left(\hspace{-2pt}\frac{2\hspace{-2pt}-\hspace{-2pt}z}{z}\hspace{-2pt}\right)\eta\up{8pt}{[\mu_2}{2pt}{\mu_0}\eta\up{8pt}{\nu_2}{2pt}{\mu_1}\gamma\up{8pt}{\rho_2]}{-3pt}{\alpha\beta} - 3\hspace{-2pt}\left(\hspace{-2pt}\frac{2(1\hspace{-2pt}-\hspace{-2pt}\zeta)}{z}\hspace{-2pt}-\hspace{-2pt}1\hspace{-2pt}\right)\hspace{-2pt}\eta^{\mu_0[\mu_2}\gamma^{\nu_2\rho_2]\mu_1}_{\alpha\beta}  \right.
\\ \left. + 3\hspace{-2pt}\left(\hspace{-2pt}\frac{2\hspace{-2pt}-\hspace{-2pt}z}{z}\hspace{-2pt}\right)\hspace{-2pt}\eta^{\mu_{1}[\mu_2}\gamma^{\nu_2\rho_2]\mu_0}_{\alpha\beta} + \left(\hspace{-2pt}\frac{2(1\hspace{-2pt}-\hspace{-2pt}\zeta)(1\hspace{-2pt}-\hspace{-2pt}z)}{\zeta z}\hspace{-2pt}-\hspace{-2pt}1\hspace{-2pt}\right)\hspace{-2pt}\eta^{\mu_0\mu_{1}}\gamma^{\mu_2\nu_2\rho_2}_{\alpha\beta} - \gamma^{\mu_0\mu_1\mu_2\nu_2\rho_2}_{\alpha\beta}\hspace{-2pt}\right]
\end{multline}

\begin{multline}
\left\langle \normal{e^{\phi}}\psi^{\mu_0}(z_0)\normal{e^{-\phi}}\normal{\psi^{\mu_1}\psi^{\nu_1}\psi^{\rho_1}}(z_1)\normal{e^{-\phi}}\normal{\psi^{\mu_2}\psi^{\nu_2}\psi^{\rho_2}}(z_2)\mathcal{S}_{\alpha}(z_3)\mathcal{S}_{\beta}(z_4)\right\rangle
\\ = \frac{z_{01}z_{02}z_{34}^2}{8\sqrt{2}z_{12}z_{13}^2 z_{14}^2 z_{23}^2 z_{24}^2}\hspace{-2pt}\Bigg\{\hspace{-1pt}6\hspace{-3pt}\left(\hspace{-2.5pt}\frac{2\hspace{-3pt}-\hspace{-3pt}z}{z}\hspace{-3pt}\right)^{\hspace{-3pt}3}\hspace{-4pt}\left[\hspace{-2pt}3\hspace{-3pt}\left(\hspace{-3pt}1 \hspace{-3pt}-\hspace{-3pt} \frac{2\hspace{-1pt}(\hspace{-1pt}1\hspace{-3pt}-\hspace{-3pt}z\hspace{-1pt})}{\zeta\hspace{-1pt}(\hspace{-1pt}2\hspace{-3pt}-\hspace{-3pt}z\hspace{-1pt})}\hspace{-3pt}\right)\hspace{-4pt}\eta\up{9pt}{[\hspace{-0.5pt}\mu_{\hspace{-0.5pt}1}}{2pt}{\mu_0}\hspace{-2pt}\eta\up{9pt}{\nu_1}{2pt}{[\hspace{-0.5pt}\mu_{\hspace{-0.5pt}2}}\eta\up{9pt}{\rho_{\hspace{-0.5pt}1}\hspace{-0.5pt}]}{2pt}{\nu_{2}}\hspace{-1pt}\gamma\up{2pt}{\rho_{\hspace{-0.5pt}2}\hspace{-0.5pt}]}{-7pt}{\hspace{-2pt}\alpha\beta} \hspace{-3pt}+\hspace{-3pt} 3\hspace{-3pt}\left(\hspace{-3pt}1 \hspace{-3pt}-\hspace{-3pt} \frac{2\zeta}{2\hspace{-3pt}-\hspace{-3pt}z}\hspace{-3pt}\right)\hspace{-4pt}\eta\up{9pt}{[\hspace{-0.5pt}\mu_{\hspace{-0.5pt}2}}{2pt}{\mu_0}\hspace{-2pt}\eta\up{9pt}{\nu_2}{2pt}{[\hspace{-0.5pt}\mu_{\hspace{-0.5pt}1}}\hspace{-1pt}\eta\up{9pt}{\rho_{\hspace{-0.5pt}2}\hspace{-0.5pt}]}{2pt}{\nu_{1}}\hspace{-1pt}\gamma\up{2pt}{\rho_{\hspace{-0.5pt}1}\hspace{-0.5pt}]}{-7pt}{\hspace{-2pt}\alpha\beta} \hspace{-3pt}-\hspace{-3pt} \eta\up{9pt}{[\hspace{-0.5pt}\mu_{\hspace{-0.5pt}1}}{2pt}{\mu_{\hspace{-0.5pt}2}}\hspace{-1pt}\eta\up{9pt}{\nu_1}{2pt}{\nu_{\hspace{-0.5pt}2}}\hspace{-1pt}\eta\up{9pt}{\rho_{\hspace{-0.5pt}1}\hspace{-0.5pt}]}{2pt}{\rho_{2}}\hspace{-1pt}\gamma^{\mu_0}_{\alpha\beta}\hspace{-2pt}\right] 
\\ - 18\hspace{-2pt}\left(\hspace{-2pt}\frac{2\hspace{-2pt}-\hspace{-2pt}z}{z}\hspace{-2pt}\right)^{\hspace{-2pt}2}\hspace{-2pt}\left[\hspace{-1pt}\left(\hspace{-2pt}1 \hspace{-2pt}-\hspace{-2pt} \frac{2(\hspace{-0.5pt}1\hspace{-2pt}-\hspace{-2pt}z\hspace{-0.5pt})}{\zeta(\hspace{-0.5pt}2\hspace{-2pt}-\hspace{-2pt}z\hspace{-0.5pt})}\hspace{-2pt}\right)\hspace{-3pt}\eta\up{9pt}{[\hspace{-0.5pt}\mu_{\hspace{-0.5pt}1}}{2pt}{\mu_0}\hspace{-2pt}\eta\up{9pt}{\nu_1}{2pt}{[\hspace{-0.5pt}\mu_{\hspace{-0.5pt}2}}\big(\gamma\up{9pt}{\rho_{1}\hspace{-0.5pt}]}{0pt}{}\up{9pt}{}{2pt}{\hspace{-1pt}\nu_{\hspace{-0.5pt}2}\rho_{\hspace{-0.5pt}2}\hspace{-0.5pt}]}\big)_{\alpha\beta} \hspace{-2pt}-\hspace{-2pt} \left(\hspace{-2pt}1 \hspace{-2pt}-\hspace{-2pt} \frac{2\zeta}{2\hspace{-2pt}-\hspace{-2pt}z}\hspace{-2pt}\right)\hspace{-2pt}\eta\up{9pt}{[\hspace{-0.5pt}\mu_{\hspace{-0.5pt}2}}{2pt}{\mu_0}\hspace{-2pt}\eta\up{9pt}{\nu_2}{2pt}{[\hspace{-0.5pt}\mu_{\hspace{-0.5pt}1}}\hspace{-1pt}\big(\gamma\up{9pt}{\rho_{2}\hspace{-0.5pt}]}{0pt}{}\up{9pt}{}{2pt}{\hspace{-1pt}\nu_{\hspace{-0.5pt}1}\rho_{\hspace{-0.5pt}1}\hspace{-0.5pt}]}\big)_{\alpha\beta} \hspace{-2pt}-\hspace{-2pt} \eta\up{10pt}{[\hspace{-0.5pt}\mu_{\hspace{-0.5pt}1}}{2pt}{[\hspace{-0.5pt}\mu_{\hspace{-0.5pt}2}}\eta\up{10pt}{\nu_{\hspace{-0.5pt}1}}{2pt}{\nu_{\hspace{-0.5pt}2}}\big(\gamma\up{10pt}{\rho_{\hspace{-0.5pt}1}\hspace{-0.5pt}]}{0pt}{}\up{10pt}{}{2pt}{\hspace{-1pt}\rho_{\hspace{-0.5pt}2}\hspace{-0.5pt}]\hspace{-0.5pt}\mu_0}\big)_{\alpha\beta}\right]
\\ - 3\hspace{-3pt}\left(\hspace{-3pt}\frac{2\hspace{-3pt}-\hspace{-3pt}z}{z}\hspace{-3pt}\right)\hspace{-5pt}\left[\hspace{-2pt}\left(\hspace{-3pt}1 \hspace{-3pt}-\hspace{-3pt} \frac{2\hspace{-0.5pt}(\hspace{-1pt}1\hspace{-3pt}-\hspace{-3pt}z\hspace{-0.5pt})}{\zeta\hspace{-0.5pt}(\hspace{-1pt}2\hspace{-3pt}-\hspace{-3pt}z\hspace{-0.5pt})}\hspace{-2.5pt}\right)\hspace{-3pt}\eta^{\mu_0[\hspace{-0.5pt}\mu_{1}}\hspace{-1pt}(\gamma^{\nu_{1}\hspace{-1pt}\rho_{1}\hspace{-0.5pt}]\mu_{2}\hspace{-0.5pt}\nu_{2}\hspace{-0.5pt}\rho_{2}})_{\alpha\beta} \hspace{-3pt}+\hspace{-3pt} \left(\hspace{-3pt}1 \hspace{-3pt}-\hspace{-3pt} \frac{2\zeta}{2\hspace{-3pt}-\hspace{-3pt}z}\hspace{-3pt}\right)\hspace{-3pt}\eta^{\mu_0[\mu_{2}}\hspace{-1pt}(\gamma^{\nu_{2}\hspace{-0.5pt}\rho_{2}\hspace{-0.5pt}]\mu_{1}\hspace{-1pt}\nu_{1}\hspace{-1pt}\rho_{1}})_{\alpha\beta} \hspace{-3pt}-\hspace{-2pt} 3\eta^{\mu_{1}][\mu_{2}}\hspace{-1pt}(\gamma^{\nu_{2}\rho_{2}]\mu_0[\nu_{1}\rho_{1}})_{\alpha\beta}\hspace{-2pt}\right]
\\ - (\gamma^{\mu_0\mu_{1}\nu_{1}\rho_{1}\mu_2\nu_2\rho_2})_{\alpha\beta}\Bigg\}
\end{multline}

\begin{multline}
\left\langle\normal{e^{-\phi}}\psi^{\mu_0}(z_0)j^{\mu_1\nu_1}(z_1)\normal{e^{-\phi}}\normal{\psi^{\mu_2}\psi^{\nu_2}\psi^{\rho_2}}(z_2)\normal{e^{\frac{1}{2}\phi}}S^{\alpha}(z_3)\mathcal{S}_{\beta}(z_4)\right\rangle
\\ = \frac{iz_{34}^2}{8z_{02}z_{04}z_{13} z_{14}z_{23} z_{24}^2}\hspace{-2pt}\Bigg[6\hspace{-2pt}\left(\hspace{-2pt}\frac{2\hspace{-2pt}-\hspace{-2pt}z}{z}\hspace{-2pt}\right)^{\hspace{-2pt}3}\hspace{-2pt}\left(\hspace{-2pt}1\hspace{-2pt}-\hspace{-2pt}\frac{2\zeta}{2\hspace{-2pt}-\hspace{-2pt}z}\hspace{-2pt}\right)\hspace{-2pt}\eta\up{9pt}{[\mu_0}{2pt}{\mu_2}\eta\up{9pt}{\mu_1}{2pt}{\nu_2}\eta\up{9pt}{\nu_1]}{2pt}{\rho_2}\delta^{\alpha}_{\phantom{\alpha}\beta} 
\\ - 6\hspace{-3pt}\left(\hspace{-3pt}\frac{2\hspace{-3pt}-\hspace{-3pt}z}{z}\hspace{-3pt}\right)^{\hspace{-3pt}2}\hspace{-3pt}\left(\hspace{-3pt}1\hspace{-3pt}-\hspace{-3pt}\frac{2(1\hspace{-3pt}-\hspace{-3pt}z)}{\zeta(2\hspace{-3pt}-\hspace{-3pt}z)}\hspace{-3pt}\right)\hspace{-3pt}\eta^{\mu_0[\mu_1}\eta^{\nu_1][\mu_2}\hspace{-1pt}(\gamma^{\nu_2\rho_2]})^{\alpha}_{\phantom{\alpha}\beta} - 12\hspace{-2pt}\left(\hspace{-3pt}\frac{2\hspace{-3pt}-\hspace{-3pt}z}{z}\hspace{-3pt}\right)^{\hspace{-3pt}2}\hspace{-3pt}\left(\hspace{-3pt}1\hspace{-3pt}-\hspace{-3pt}\frac{2\zeta}{2\hspace{-3pt}-\hspace{-3pt}z}\hspace{-3pt}\right)\hspace{-3pt}\eta\up{9pt}{\mu_0}{1pt}{[\mu_2}\eta\up{9pt}{[\mu_1}{1pt}{\nu_2}\big(\gamma\up{9pt}{\nu_1]}{1pt}{}\up{9pt}{}{1pt}{\rho_2]}\big)^{\alpha}_{\phantom{\alpha}\beta}
\\ + 6\hspace{-2pt}\left(\hspace{-3pt}\frac{2\hspace{-3pt}-\hspace{-3pt}z}{z}\hspace{-3pt}\right)^{\hspace{-3pt}2}\hspace{-2pt}\eta\up{2pt}{\mu_1}{9pt}{[\mu_2}\eta\up{2pt}{\nu_1}{9pt}{\nu_2}\big(\gamma\up{9pt}{\rho_2]}{2pt}{}\up{9pt}{}{2pt}{\mu_0}\big)^{\alpha}_{\phantom{\alpha}\beta} + 2\hspace{-2pt}\left(\hspace{-3pt}\frac{2\hspace{-3pt}-\hspace{-3pt}z}{z}\hspace{-3pt}\right)\hspace{-3pt}\left(\hspace{-3pt}1 \hspace{-3pt}-\hspace{-3pt} \frac{2(1\hspace{-3pt}-\hspace{-3pt}z)}{\zeta(2\hspace{-3pt}-\hspace{-3pt}z)}\hspace{-3pt}\right)\hspace{-2pt}\eta^{\mu_0[\mu_1}\hspace{-1pt}\big(\gamma^{\nu_1]\mu_2\nu_2\rho_2}\big)^{\alpha}_{\phantom{\alpha}\beta}
\\ - 3\hspace{-3pt}\left(\hspace{-3pt}\frac{2\hspace{-3pt}-\hspace{-3pt}z}{z}\hspace{-3pt}\right)\hspace{-3pt}\left(\hspace{-3pt}1\hspace{-3pt}-\hspace{-3pt}\frac{2\zeta}{2\hspace{-3pt}-\hspace{-3pt}z}\hspace{-3pt}\right)\hspace{-3pt}\eta^{\mu_0[\mu_2}\hspace{-1pt}(\gamma^{\nu_2\rho_2]\mu_1\nu_1})^{\alpha}_{\phantom{\alpha}\beta} + 6\hspace{-3pt}\left(\hspace{-3pt}\frac{2\hspace{-3pt}-\hspace{-3pt}z}{z}\hspace{-3pt}\right)\hspace{-3pt}\eta^{\mu_1][\mu_2}\hspace{-1pt}(\gamma^{\nu_2\rho_2]\mu_0[\nu_1})^{\alpha}_{\phantom{\alpha}\beta} + (\gamma^{\mu_0\mu_1\nu_1\mu_2\nu_2\rho_2})^{\alpha}_{\phantom{\alpha}\beta} \Bigg]
\end{multline}

\end{document}